\documentclass[11pt]{llncs}
\usepackage{geometry}
\usepackage{Xiao-llncs}
\newcommand{\msf}{\mathsf}
\newcommand{\mcal}{\mathcal}

\newcommand{\Adv}{\ensuremath{\mathcal{A}}\xspace}

\renewcommand{\bar}{\overline}
\newcommand{\Bad}{\ensuremath{\mathsf{Bad}}\xspace}

\newcommand{\bits}{\ensuremath{\{0,1\}}\xspace}

\newcommand{\CCY}{\mathsf{CCY}}

\newcommand{\cind}{\ensuremath{\stackrel{\text{c}}{\approx}}\xspace}

\newcommand{\com}{\ensuremath{\mathsf{com}}\xspace}
\newcommand{\Com}{\ensuremath{\mathsf{Com}}\xspace}

\newcommand{\decom}{\ensuremath{\mathsf{decom}}\xspace}

\renewcommand{\epsilon}{\varepsilon}

\newcommand{\ExtCom}{\ensuremath{\mathsf{ExtCom}}\xspace}
\newcommand{\extcomindex}{\ensuremath{J}\xspace}

\renewcommand{\hat}{\widehat}

\newcommand{\idind}{\ensuremath{\stackrel{\text{i.d.}}{=\mathrel{\mkern-3mu}=}}\xspace}

\newcommand{\Lang}{\ensuremath{\mathcal{L}}\xspace}

\newcommand{\Naturals}{\ensuremath{\mathbb{N}}\xspace}

\newcommand{\negl}{\ensuremath{\mathsf{negl}}\xspace}

\newcommand{\NP}{\ensuremath{\mathbf{NP}}\xspace}

\newcommand{\OUT}{\ensuremath{\mathsf{OUT}}\xspace}

\renewcommand{\paragraph}{\para}

\renewcommand{\P}{\ensuremath{\mathcal{P}}\xspace}

\newcommand{\pick}{\ensuremath{\xleftarrow{\$}}\xspace}
\newcommand{\picks}{\pick}

\newcommand{\poly}{\ensuremath{\mathsf{poly}}\xspace}

\newcommand{\red}[1]{{\color{red} #1}}
\newcommand{\Relation}{\ensuremath{\mathcal{R}}\xspace}

\newcommand{\scheduleset}{\ensuremath{\mathcal{S}}\xspace}
\newcommand{\SimExt}{\ensuremath{\mathcal{SE}}\xspace}

\newcommand{\Set}[1]{\ensuremath{\{#1\}}\xspace}

\newcommand{\sind}{\ensuremath{\stackrel{\text{s}}{\approx}}\xspace}
\newcommand{\ST}{\ensuremath{\mathsf{ST}}\xspace}

\renewcommand{\tilde}{\widetilde}

\newcommand{\UpdateLine}{\red{\noindent\makebox[\linewidth]{\rule{\paperwidth-8em}{2pt}Update upto here}}}

\newcommand{\Val}{\msf{Val}}

\newcommand{\Verify}{\ensuremath{\mathsf{Verify}}\xspace}

\newcommand{\WIAoK}{\ensuremath{\mathsf{WIAoK}}\xspace}
\newcommand{\WIPoK}{\ensuremath{\mathsf{WIPoK}}\xspace}

\newcommand{\sample}{\gets}

\newcommand{\ra}{\rightarrow}

\newcommand{\secpar}{\lambda}

\newcommand{\st}{\mathsf{st}}
\newcommand{\out}{\mathsf{out}}
\newcommand{\bit}{\{0,1\}}

\newcommand{\hil}{\mathcal{H}}

\newcommand{\defeq}{:=}

\newcommand{\A}{\mathcal{A}}
\newcommand{\B}{\mathcal{B}}
\newcommand{\C}{\mathcal{C}}

\newcommand{\ext}{\mathsf{Ext}}

\newcommand{\calX}{\mathcal{X}}
\newcommand{\calY}{\mathcal{Y}}

\newcommand{\ot}{\otimes}
\newcommand{\fail}{\mathsf{Fail}}

\newcommand{\wiaok}{\mathsf{wiaok}}

\newcommand{\TD}{\mathsf{TD}}
\newcommand{\reginp}{\mathsf{Inp}}

\newcommand*{\regX}{\mathbf{X}}
\newcommand*{\regY}{\mathbf{Y}}
\newcommand*{\regZ}{\mathbf{Z}}
\newcommand{\regW}{\mathbf{W}}

\newcommand*{\regM}{\mathbf{M}}
\newcommand*{\regout}{\mathbf{Out}}
\newcommand*{\reganc}{\mathbf{Anc}}

\newcommand*{\regA}{\mathbf{A}}

\newcommand*{\regB}{\mathbf{B}}
\newcommand{\regother}{\mathbf{Other}}

\newcommand*{\regD}{\mathbf{D}}
\newcommand{\siml}{\mathsf{Sim}}

\newcommand{\nonabort}{\mathsf{na}}
\newcommand{\comb}{\mathsf{comb}}

\newcommand{\compind}{\cind}
\newcommand{\statind}{\sind}

\newcommand{\val}{\mathsf{val}}

\newcommand{\Exp}{\mathsf{Exp}}

\newcommand{\sfQ}{\mathsf{Q}}
\newcommand{\sfR}{\mathsf{R}}
\newcommand{\amp}{\mathsf{amp}}
\newcommand{\Amp}{\mathsf{Amp}}
\newenvironment{boxfig}[2]{\begin{figure}[#1]\fbox{\begin{minipage}{0.97\linewidth}
                        \vspace{0.2em}
                        \makebox[0.025\linewidth]{}
                        \begin{minipage}{0.95\linewidth}
            {{
                        #2 }}
                        \end{minipage}
                        \vspace{0.2em}
                        \end{minipage}}}{\end{figure}}

\makeatletter
\def\@fnsymbol#1{\ensuremath{\ifcase#1\or *\or \dagger\or \ddagger\or
   \mathsection\or \mathparagraph\or \|\or **\or \dagger\dagger
   \or \ddagger\ddagger \else\@ctrerr\fi}}
\makeatother

\begin{document}

% \listoffixmes
% \addcontentsline{toc}{section}{List of Corrections}
% \newpage

\title{A New Approach to Post-Quantum Non-Malleability}
% \title{}
% \titlerunning{}
% \toctitle{}

\author{
Xiao Liang\inst{1}
\and 
Omkant Pandey\inst{2}\thanks{Supported in part by DARPA SIEVE Award HR00112020026, NSF CAREER Award 2144303, NSF grants 1907908, 2028920, 2106263, and 2128187. Any opinions, findings and conclusions or recommendations expressed in this material are those of the author(s) and do not necessarily reflect the views of the United States Government, DARPA, or NSF.}
\and
Takashi Yamakawa\inst{3}
}

\institute{
Rice University, Houston, USA\\ \email{xiao.crypto@gmail.com}
 \and
Stony Brook University, Stony Brook, USA\\ \email{omkant@cs.stonybrook.edu}
 \and
NTT Social Informatics Laboratories, Tokyo, Japan\\ \email{takashi.yamakawa.ga@hco.ntt.co.jp}
}

% -- The wrap-up around "\maketitle" is used to remove paper title from ToC
\let\oldaddcontentsline\addcontentsline
\def\addcontentsline#1#2#3{}
\maketitle
\def\addcontentsline#1#2#3{\oldaddcontentsline{#1}{#2}{#3}}

%!TEX root = ../main.tex
\begin{abstract}
We provide the first {\em constant-round} construction of post-quantum non-malleable commitments under the minimal assumption that {\em post-quantum one-way functions} exist. We achieve the standard notion of non-malleability with respect to commitments. Prior constructions required $\Omega(\log^*\secpar)$ rounds under the same assumption.
\\[-0.5em]

\hspace{1.2em}We achieve our results through a new technique for constant-round non-malleable commitments which is easier to use in the post-quantum setting. The technique also yields an almost elementary proof of security for constant-round non-malleable commitments in the classical setting, which may be of independent interest.
\\[-0.5em]

\hspace{1.2em}When combined with existing work, our results yield the first constant-round quantum-secure multiparty computation for both classical and quantum functionalities {\em in the plain model}, under the {\em polynomial} hardness of quantum fully-homomorphic encryption and quantum learning with errors.

\end{abstract}

% We provide the first $\mathit{constant}$-$\mathit{round}$ construction of post-quantum non-malleable commitments under the minimal assumption that $\mathit{post}$-$\mathit{quantum}$ $\mathit{one}$-$\mathit{way}$ $\mathit{functions}$ exist. We achieve the standard notion of non-malleability with respect to commitments. Prior constructions required $\Omega(\log^*\lambda)$ rounds under the same assumption.

% We achieve our results through a new technique for constant-round non-malleable commitments which is easier to use in the post-quantum setting. The technique also yields an almost elementary proof of security for constant-round non-malleable commitments in the classical setting, which may be of independent interest.

% As an application, when combined with existing work, our results yield the first constant-round post-quantum secure multiparty computation under the $\mathit{polynomial}$ hardness of quantum fully-homomorphic encryption and quantum learning with errors.

% \para{For FOCS'23 Reviewers:} We have included the comments from the STOC'23 reviewers, along with our corresponding responses, in \Cref{sec:STOC-comments}. We hope that this information proves helpful during the reviewing process.

%-- Starting numbering pages with Roman letters --------
	\pagenumbering{roman}
% -- Creat ToC
\tableofcontents
\addcontentsline{toc}{section}{Table of Contents}
\clearpage

%-- Starting numbering pages with arabic --------
\pagenumbering{arabic}
	
%!TEX root = ../main.tex
\section{Introduction}

Commitments are one of the central primitives in modern cryptography. They are two-party protocols that enable a sender (or committer) to commit to a message to a receiver. It is required that an efficient receiver should learn nothing about the committed message until later the committer chooses to open (or decommit to) it. However, this vanilla promise of ``information hiding'' does not rule out the following mauling or man-in-the-middle (MIM) attack:
An adversary $\mcal{M}$ could play the role of a receiver in one instance of the commitment (referred to as the {\em left session}), while simultaneously playing as a committer in another (referred to as the {\em right session}). In this situation, $\mcal{M}$ can potentially make the value committed in the right session depend on the value in the left session, in a malicious manner that is to her advantage. Notice that this is not breaking the hiding property of the commitment scheme, as $\mcal{M}$ could conduct the above attack without explicitly learning the value committed in the left session.	

To protect against such attacks, Dolev, Dwork, and Naor \cite{STOC:DolDwoNao91} introduced the concept of {\em non-malleable commitments}. Such commitments capture the MIM attack by requiring that in the MIM execution, the {\em joint distribution} of $\mcal{M}$'s final output {\em and} the value committed in the right session is computationally indistinguishable for any values committed by the honest committer in the left session. Of course, $\mcal{M}$ can always relay without any modifications the messages between the left honest committer and the right honest receiver. This is handled by augmenting the commitment scheme with a {\em tag}, and $\mcal{M}$ is considered winning the MIM game only if she does not use the same tag on both sides.\footnote{Copying the tag can be shown equivalent to copying the entire interaction assuming one-way functions.}

%\xiao{@Omkant: I choose to keep the above two paragraph as it is informative for general audience (FOCS reviewers).}

Non-malleable commitments have found several applications in cryptography. They turn out to be a critical ingredient in resolving the exact round complexity of secure multiparty computation in the standard (plain) model \cite{EC:KatOstSmi03,C:PanPasVai08,FOCS:Wee10,STOC:Goyal11,EC:GMPP16,C:BGJKKS18,EPRINT:CCGJO19}, as well as protecting such protocols against concurrent attacks \cite{JC:Canetti00,FOCS:Canetti01,STOC:CLOS02,EC:Pass03,STOC:PraSah04,FOCS:MicPasRos06,EC:BDHMN17,ICALP:ChaLiaPan20}.

Since their introduction, the central question in this area has been the construction of {\em constant-round} non-malleable commitments under the minimal assumption that one-way functions (OWFs) exist. The original work \cite{STOC:DolDwoNao91} presented a logarithmic-round construction assuming only OWFs. The works of Barak \cite{FOCS:Barak02}, and Pass and Rosen \cite{STOC:PasRos05} succeeded in obtaining constant-round constructions under the (stronger but standard) assumption of  (polynomially-hard) collision-resistant hash functions; this assumption is inherited due to the use of non-black-box simulation techniques \cite{FOCS:Barak01}. After a long line of follow-up works \cite{C:PanPasVai08,TCC:LinPasVen08,STOC:LinPas09,FOCS:Wee10}, constant-round non-malleable commitments assuming only OWFs were first constructed in independent and concurrent works of Goyal \cite{STOC:Goyal11} and Lin and Pass \cite{STOC:LinPas11}. Since then, several constructions optimizing various aspects of this primitive, such as the {\em exact} round complexity \cite{TCC:Pass13,FOCS:GRRV14,STOC:GoyPanRic16,C:COSV17,FOCS:LinPasSon17,TCC:Khurana17,FOCS:KhuSah17,TCC:BitLin18,FOCS:GoyRic19,EC:Khurana21} and black-box usage of primitives \cite{FOCS:Wee10,STOC:Goyal11,FOCS:GLOV12}, have been proposed, thus achieving an almost complete understanding of this primitive in the {\em classical} setting where all parties, including the adversary, as well as the communication are classical.
% \xiao{@Omkant: there are some papers that I cited by you omitted. Should I add it to your version as well? Or you do this on purpose because they are of secondary importance?\om{It was not on purpose -- please add them back but do not repeat the ones we have already cited.}}

% Besides serving as an immediate solution to real-life tasks like auction bidding, they have also proven useful for other major cryptographic tasks such as coin-tossing, non-malleable zero-knowledge, multi-party computational, and more.

% In the classical setting, we now have a good understanding of this notion after thorough investigations retarding, e.g., its round complexity \cite{FOCS:Barak02,STOC:PasRos05,C:PanPasVai08,STOC:LinPas09,STOC:LinPas11,TCC:Pass13,FOCS:GRRV14,STOC:GoyPanRic16,C:COSV17,TCC:Khurana17,FOCS:KhuSah17,FOCS:LinPasSon17,TCC:BitLin18,FOCS:GoyRic19,EC:Khurana21}, black-box usage of underlying primitives \cite{FOCS:Wee10,STOC:Goyal11,FOCS:GLOV12}, composability \cite{FOCS:PasRos05,TCC:LinPasVen08,C:COSV17}, practical implementation \cite{CCS:BGRRV15}, and the combination of these features. 
% In particular, constant-round constructions of non-malleable commitments from the minimal assumption of the existence of one-way functions (OWFs) are known in the plain model~\cite{STOC:LinPas11,STOC:Goyal11,FOCS:GLOV12,STOC:GoyPanRic16,FOCS:GoyRic19}.  
% \takashi{added this sentence. please correct the citation if inappropriate.}
% There are also improved constructions with trusted setups like a CRS or Random Oracle \cite{STOC:DiCIshOst98,EC:DKOS01,C:Pass03,INDOCRYPT:FKMV12}.

Unfortunately, the results in the classical setting do not usually translate to the quantum setting where one or more parties may be quantum machines. Existing classical techniques often require the ability to rewind and copy the adversaries' code, both of which are not possible in the quantum setting due to the no-cloning theorem \cite{wootters1982single} and quantum state disturbance \cite{fuchs1996quantum}.

These issues, together with the fact that quantum computers might be possible one day, have resulted in a significant push toward developing tools and techniques to reason about security in the presence of quantum adversaries. Some examples include zero-knowledge \cite{STOC:Watrous06,STOC:BitShm20}, signatures \cite{C:BonZha13}, pseudo-random functions \cite{FOCS:Zhandry12}, secure computation \cite{C:BCKM21b,EC:ABGKM21}, and so on.

\para{Post-Quantum Non-Malleable Commitments.} A particularly appealing goal in this direction is the construction of the so-called {\em post-quantum} secure protocols, where the honest parties and their communication channels are entirely {\em classical} but the adversary is allowed to be a {\em quantum} (polynomial-time) algorithm. Such protocols have the feature that even if the adversary somehow gains early access to quantum computing capabilities, the honest parties are not required to catch up to remain protected against it.

Agarwal, Bartusek, Goyal, Khurana, and Malavolta \cite{EC:ABGKM21}---in their pursuit of constant-round post-quantum secure multiparty computation---construct a constant-round post-quantum non-malleable commitments w.r.t. the special case of {\em synchronous schedules} where, upon receiving a message in the left session, $\mcal{M}$ must respond with the corresponding message of the right session immediately, in reach round of the protocol. They achieve this assuming super-polynomial quantum hardness of the learning with errors (QLWE) problem.%---an assumption that carries over to their final result on secure multiparty computation.

%Assuming the (slightly) super-polynomial quantum hardness of Learning with Errors (QLWE), they present the first constant-round commitments that are non-malleable w.r.t.\ {\em synchronous schedules}; This refers to the requirement that in the MIM execution, whenever $\mcal{M}$ receives a message in the right (resp.\ left) session, she should immediately send the corresponding left-session (resp.\ right-session) message. While there exist {\em constant-round} compilers that lift synchronous non-malleability to the full-fledged (i.e., asynchronous) one in the classical setting (e.g., \cite{FOCS:Wee10}), it was not known how to go from the synchronous to the asynchronous in the post-quantum setting, without incurring a super-constant overhead on round complexity. 

In the general (i.e., asynchronous) setting, the first positive result was recently obtained by Bitansky, Lin, and Shmueli \cite{BLS21} who, in the  post-quantum setting, construct a $\log^*(\secpar)$-round protocol assuming only post-quantum OWFs. They present a general compiler to convert any $k$-round  ($\epsilon$-simulatable) post-quantum extractable commitment to $k^{O(1)}\cdot \log^*(\secpar)$-round post-quantum non-malleable commitments, and then rely on the very recent work of Chia, Chung, Liang, and Yamakawa \cite{cryptoeprint:2021:1516} where such an extractable commitment protocol is constructed in constant rounds from only post-quantum OWFs (see also \cite[Section 1.2]{cryptoeprint:2021:1516}). This brings us tantalizingly close to constant rounds. However, the techniques in \cite{BLS21} rely on scheduling and amplification techniques, which inherently require non-constant rounds to support (the standard requirement of) large tags or identities. It is unclear if these techniques can yield constant rounds.

The central question in this area thus still remains open:
\begin{quote}
{\bf Question 1:} {\em Do constant-round post-quantum non-malleable commitments, assuming only post-quantum one-way functions, exist?}
\end{quote}

\subsection{Challenge: Robust Simulatable Extraction}\label{sec:challenge}
Toward answering {\bf Question 1}, let us first discuss about the challenges.

As mentioned above, proving non-malleability of a commitment scheme requires one to show that the {\em joint distribution} of the final state of the MIM $\mcal{M}$ (denoted by $\msf{st}_\mcal{M}$) {and} the value committed in the right session (denoted by $\tilde{m}$) are computationally indistinguishable when the left-session committed value changes from any $m_0$ to any $m_1 \ne m_0$. Typically, this is done by a careful design of a sequence of hybrids, where the left value is gradually changed from $m_0$ to $m_1$; And non-malleability will then be established by showing that the joint distribution of $(\msf{st}_\mcal{M}, \tilde{m})$ is indistinguishable between each pair of adjacent hybrids. As in many other cryptographic proofs, one usually needs to reduce the indistinguishability between adjacent hybrids to some computational hardness assumptions. But this step is especially hard for non-malleability proofs due to the following ``inefficient testability'' issue---The value $\tilde{m}$ is hidden in the transcript of the right interaction, and no efficient machine could obtain it. Thus, the event that ``the joint distribution of $(\msf{st}_\mcal{M}, \tilde{m})$ changes'' is not efficiently testable, thus forbidding an efficient reduction to the underlying hardness assumptions.

The most common template to address the above issue is to (efficiently) extract the value $\tilde{m}$ from some ``extractable gadget'' (e.g., extractable commitments or proofs of knowledge) in the right session. The hope is: if the extracted value, denoted by $\tilde{m}'$, is equal to $\tilde{m}$ with good-enough probability, then one can conduct the above reduction using $(\msf{st}_{\mcal{M}}, \tilde{m}')$ in place of $(\msf{st}_{\mcal{M}}, \tilde{m})$. That is, since the extraction of $\tilde{m}'$ is efficient, it saves the reduction from the aforementioned inefficient testability issue.

To properly implement this template, it is crucial to maintain the following conditions (the combination of which we call {\em Robust Simulatable Extractability}.)
\begin{itemize}
\item
{\bf Extractability:} One can extract the committed message $\tilde{m}$ in the right from some extractable gadget; %\dlt{session by rewinding};
\item
{\bf Simulatability:} One can simulate $\mcal{M}$'s final (i.e., post-extraction) state while extracting $\tilde{m}$;
\item
{\bf Robustness:} %\dlt{The rewinding for the extraction}
The extraction of $\tilde{m}$ (in the right session) does not harm the hiding property of the left session (or some left-session gadget on whose hiding property the security reduction relies). 
\end{itemize}
Roughly speaking, previous designs of non-malleable protocols (in the classical setting) can be thought of as developing different (and better-and-better) techniques that enable the above robust simulatable extractability.

However, robust simulatable extractability turns out to be hard to obtain {\em in the post-quantum setting}.   First, as mentioned earlier, special techniques are needed to perform simulation for quantum adversaries due to the no-cloning theorem; Extracting a desired value while simultaneously simulating the quantum adversaries' internal state is even harder. Fortunately, the recent works \cite{cryptoeprint:2021:1516,cryptoeprint:2021:1543} did provide a constant-round solution to simulatable extraction based solely on post-quantum OWFs. 

However, the picture becomes unclear when we additionally require robustness. Known simulatable extraction techniques treat the adversary as a single reversible operation (i.e., unitary) and ``rewind'' it coherently. However, if the adversary talks in straight-line with the external left committer $C$ (that cannot be rewound), those techniques seem  inapplicable. Actually, this robustness issue already appeared in the classical setting. But we would like to point out that the quantum power of adversaries complicates it further: It is reasonable to expect that the extractor (to be constructed) may need to ``read'' the messages exchanged between the MIM adversary $\mcal{M}$ and the left committer $C$. However, known quantum rewinding strategies need to treat the adversary as a reversible operation. When we view $(C,\mcal{M})$ as a joint adversary to perform quantum rewinding, it is unclear if the simulator can ``read'' (technically, {\em measure}) the messages exchanged between $C$ and $\mcal{M}$, because this may irreversibly collapse the internal quantum state of the joint $(C,\mcal{M})$ adversary. Therefore, it is unclear if existing post-quantum rewinding techniques could be used in the MIM setting when robustness is a concern. This indeed represents the major difficulty when one tries to build constant-round post-quantum non-malleable commitments by quantizing the security proofs of the classical ones (e.g., \cite{STOC:Goyal11,STOC:LinPas11,FOCS:GLOV12,FOCS:GRRV14,STOC:GoyPanRic16,C:COSV17,FOCS:GoyRic19}).

\para{Existing Techniques.} The two recent works mentioned earlier
demonstrated possible solutions to the robust simulatable extractability issue, if one is willing to make stronger hardness assumptions, {\em or} does not insist on constant rounds:
\begin{itemize}
\item
\cite{BLS21} took a similar approach as in \cite{STOC:DolDwoNao91}. %drew inspirations from the classical work of \cite{STOC:DolDwoNao91}. %\xiao{``DDN trick''? @Omkant: I remember that you know a better name for this. What is it? ``The LOG-N trick''?}
Roughly, the idea is to introduce enough rewinding opportunities for extraction (usually referred to as ``slots'') in the construction such that there is always a ``free slot'', namely, a slot that does not interleave with any messages exchange between $\mcal{M}$ and the left committer $C$ (who will become the external challenger of the hardness-providing gadget in the security proof); Then, one can extract $\tilde{m}$ from this free slot using known post-quantum rewinding strategies. This approach is unlikely to give a constant-round construction (even in the classical setting). The constructions from \cite{BLS21} require at least $\log^*(\secpar)$ rounds.
%\footnote{Herein, $\log^*(\secpar)$ denotes the {\em iterated logarithm of $\secpar$}, i.e., the number of times the logarithm function must be iteratively applied before the result is less than or equal to 1.}. 
%The improvement by  \cite{cryptoeprint:2021:1516} inherits the same limitation. 
\item
\cite{EC:ABGKM21} took a similar approach as in \cite{EC:PasWee10,FOCS:KhuSah17}, which essentially obtained robust simulatable extractability via {\em complexity leveraging}. At a high level, the idea is to hide the committed value in some computationally hard gadget, whose hardness depends on the tag. This hard gadget is cleverly designed to admit the following strategy: In the MIM execution, because of the asymmetry of the left and right tags (i.e., $t \ne \tilde{t}$), the reduction can extract $\tilde{m}$ by breaking the {\em right} hard gadget via brute force in some slightly super-polynomial time $T(\secpar)$; However, the MIM adversary cannot break the {\em left} hard gadget in time $\poly\big(T(\secpar)\big)$. Since the extraction in this template is conducted via brute force, the robust simulatable extraction issue can be circumvented. However, this approach is unlikely to work without super-polynomial hardness assumptions due to the use of complexity leveraging.
\end{itemize}

\subsection{Our Results}
In this work, we answer {\bf Question 1} affirmatively by providing the first constant-round construction of post-quantum non-malleable commitments assuming only post-quantum one-way functions.

As discussed above, major classical approaches to OWF-based constant-round non-malleable commitments seem not to be ``quantum-friendly.'' Therefore, we first propose a new classical construction whose security proof is quantum-friendly. That is, when designing the new protocol, we restrict ourselves to techniques that avoid state-cloning and the above robust simulatable extractability issue. We find this result already interesting as it improves the diversity of the approaches known in the classical setting.

Next, we show that our new classical construction can be made quantum-secure once its building blocks are replaced by their post-quantum counterparts. This is possible because our classical construction is deliberately designed to be quantum-friendly. 
\begin{theorem}[Informal]\label{main:thm:intro}
Assuming the existence of post-quantum one-way functions, there exists a constant-round construction of post-quantum non-malleable commitments.
\end{theorem}

\Cref{main:thm:intro} yields interesting corollaries w.r.t.\ {\em multi-party computation} (MPC) in the quantum era. For classical functionalities, the recent work \cite{EC:ABGKM21} presents the first post-quantum MPC in constant rounds, from a mildly super-polynomial {\em quantum hardness of Learning with Errors} (QLWE) assumption and a QLWE-based circular security assumption. They need the super-polynomial hardness of QLWE (only) to build constant-round post-quantum non-malleable commitments, which serve as a building block to their MPC. Plugging our non-malleable commitment into their framework yields the following result:
\begin{corollary}[Informal]\label{main:thm:cor-1}
Assuming (polynomial) QLWE and the QLWE-based circular security assumption (as in~\cite{EC:ABGKM21}), there exists a constant-round construction of post-quantum MPC for classical functionalities, i.e., an MPC protocol secure against QPT adversaries where honest parties only need to perform classical computation.
\end{corollary}
For quantum functionalities,  the recent work \cite{C:BCKM21a} presents a constant-round quantum-secure MPC for quantum functionalities {\em in the CRS model}, based on the hardness of QLWE. It is easy to see that \Cref{main:thm:cor-1} provides a constant-round implementation for the CRS required by the \cite{C:BCKM21a} protocol. This observation leads to the first constant-round quantum-secure MPC for quantum functionalities from polynomial hardness assumptions {\em without any trusted setup}.
\begin{corollary}[Informal]\label{main:thm:cor-2}
Assuming (polynomial) QLWE and the QLWE-based circular security assumption (as in~\cite{EC:ABGKM21}), there exists a constant-round construction of quantum-secure MPC for quantum functionalities.
\end{corollary}

% Finally, we show an application to post-quantum {\em multi-party computation} (MPC). The recent work of \cite{EC:ABGKM21} presents the first post-quantum MPC in constant rounds, from a mildly super-polynomial {\em quantum hardness of Learning with Errors} (QLWE) assumption and a QLWE-based circular security assumption. They need the super-polynomial hardness of QLWE (only) to build constant-round post-quantum non-malleable commitments, which serve as a building block to their MPC. Plugging our non-malleable commitment into their framework yields the following result:
% \begin{corollary}[Informal]
% Assuming (polynomial) QLWE and the QLWE-based circular security assumption (as in~\cite{EC:ABGKM21}), there exists a constant-round construction of post-quantum MPC.
% \end{corollary}

%!TEX root = ../main.tex
\section{Technical Overview}
\label{sec:tech-overview}
% We first describe the main idea behind our construction to achieve non-malleability in the classical setting. Then, it is easy to see why this proof of non-malleability extends to the post-quantum setting (which we discuss in \Cref{sec:tech-overview:post-quantum}).
We will first present a construction $\langle C, R \rangle^{\msf{OneSided}}_{\msf{tg}}$ that is non-malleable {\em in the classical setting} with the following restrictions:
\begin{itemize}
\item
{\bf Small-Tag:} It only supports tags from the polynomial-size space $[n] \coloneqq \Set{1, \ldots, n(\secpar)}$, where $n(\secpar)$ is a polynomial on the security parameter $\secpar$;
\item
{\bf One-Sided:} Its non-malleability holds only if, in the MIM game, the left-session tag $t$ is smaller than the right-session tag $\tilde{t}$.
\item
{\bf Synchronous:} Its non-malleability holds only in the synchronous setting. This refers to the setting where upon receiving a message in the right (resp.\ left) session, the MIM adversary immediately responds with the corresponding message in the left (resp.\ right) session.
\end{itemize} 
It is a common approach in the literature to first obtain a construction under the above conditions, and then convert it to a full-fledged non-malleable commitment. While there exist standard techniques that take care of the latter step, the former step (i.e., constructing $\langle C, R\rangle^{\msf{OneSided}}_{\msf{tg}}$) is typically where difficulty lies. %Thus, this technical overview is mostly devoted to the ideas behind $\langle C, R\rangle^{\msf{OneSided}}_{\msf{tg}}$. 

In the classical setting, once we obtain $\langle C, R\rangle^{\msf{OneSided}}_{\msf{tg}}$, we can apply known techniques to remove the {\bf One-Sided} restriction, yielding a small-tag, synchronous protocol, which we denote as $\langle C, R\rangle^{\msf{sync}}_{\msf{tg}}$. Then, known compilers (e.g., \cite{FOCS:Wee10}) can be used to remove the {\bf Small-Tag} and {\bf Synchronous} restrictions {\em at one stroke}.  This leads to a full-fledged non-malleable commitment $\langle C, R\rangle^{\msf{async}}_{\msf{TG}}$ in the classical setting. 

We emphasize that all the above protocols are non-malleable {\em only in the classical setting}. However, they are designed deliberately using quantum-friendly techniques, which makes it possible to quantize their security proofs. We elaborate on that in the sequel.

\para{Post-Quantum Tag Amplification.} Recall that our eventual goal is to achieve {\em post-quantum} non-malleability. The use of quantum-friendly techniques will allow us to prove {\em post-quantum} non-malleability of $\langle C, R\rangle^{\msf{OneSided}}_{\msf{tg}}$. Also, the aforementioned (classical) conversion from $\langle C, R\rangle^{\msf{OneSided}}_{\msf{tg}}$ to $\langle C, R\rangle^{\msf{sync}}_{\msf{tg}}$ extends naturally to the post-quantum setting as well. However, the classical compiler from $\langle C, R\rangle^{\msf{sync}}_{\msf{tg}}$ to $\langle C, R\rangle^{\msf{async}}_{\msf{TG}}$ seems not to extend to the post-quantum setting. 

This has already been observed in \cite{BLS21}. The authors of \cite{BLS21} addressed this issue by constructing a new tag amplifier that converts a small-tag, asynchronous post-quantum non-malleable commitment to a large-tag (i.e, $t \in [2^\secpar]$), asynchronous post-quantum non-malleable commitment. But it is worth noting that this tag amplifier requires the small-tag protocol to be {\em asynchronously} secure;  This is in contrast to the aforementioned classical compiler, which handles asynchronicity and tag-size amplification at one stroke. 

To overcome this problem, we will first show that our post-quantum version of $\langle C, R\rangle^{\msf{sync}}_{\msf{tg}}$ can be modified to achieve non-malleability in the asynchronous setting, yielding a protocol $\langle C, R\rangle^{\msf{async}}_{\msf{tg,PQ}}$, which is exactly a small-tag, asynchronously non-malleable commitment in the post-quantum setting. Now, the \cite{BLS21} tag amplifier can be applied to $\langle C, R\rangle^{\msf{async}}_{\msf{tg,PQ}}$, leading to the full-fledged post-quantum non-malleable commitment $\langle C, R\rangle^{\msf{async}}_{\msf{TG,PQ}}$ we want.

\para{Organization of Technical Overview.} In \Cref{sec:tech-overview:small-tag:one-sided,sec:tech-overview:proving-NM}, we overview the main idea behind our construction of  $\langle C, R\rangle^{\msf{OneSided}}_{\msf{tg}}$ in the classical setting. This is the most technically involved part where the main difficulty lies. 
%Then, we provide a comparison with \cite{STOC:Goyal11} in \Cref{sec:tech-overview:compare:Goyal11}.

As mentioned earlier, protocol $\langle C, R\rangle^{\msf{OneSided}}_{\msf{tg}}$ is designed using only quantum-friendly techniques. Thus, the proof of its non-malleability extends to the post-quantum setting once we replace its building blocks by their post-quantum counterparts. (We provide a high-level explanation for that on \Cpageref{pageref:why-quantum-friendly}.) We then show how this can be done in \Cref{sec:tech-overview:small-tag:one-sided:sync:PQ}, where we obtain the one-sided, small-tag, synchronous, {\em post-quantum} non-malleable commitment $\langle C, R\rangle^{\msf{OneSided}}_{\msf{tg,PQ}}$.

Finally, we provide in \Cref{sec:tech-overview:full-fledged} a brief overview on how to convert $\langle C, R\rangle^{\msf{OneSided}}_{\msf{tg,PQ}}$ to the full-fledged post-quantum non-malleable commitment  $\langle C, R\rangle^{\msf{async}}_{\msf{TG,PQ}}$, achieving our eventual goal.

% Next, we show that the proof of non-malleability of $\langle C, R\rangle^{\msf{sync}}_{\msf{tg}}$ actually extends to the post-quantum setting (\Cref{sec:tech-overview:small-tag:sync:PQ}).

% We then show how to obtain $\langle C, R\rangle^{\msf{sync}}_{\msf{tg}}$ again in the classical setting (\Cref{sec:tech-overview:small-tag}). Next, we show that the proof of non-malleability of $\langle C, R\rangle^{\msf{sync}}_{\msf{tg}}$ actually extends to the post-quantum setting (\Cref{sec:tech-overview:small-tag:sync:PQ}). After that, we show how to upgrade $\langle C, R\rangle^{\msf{sync}}_{\msf{tg}}$ to $\langle C, R\rangle^{\msf{async}}_{\msf{tg}}$, achieving asynchronous security in the post-quantum setting (\Cref{sec:tech-overview:small-tag:async:PQ}). Finally, we apply the \cite{BLS21} tag amplifier to obtain the full-fledge post-quantum non-malleable commitment (\Cref{sec:tech-overview:post-quantum:tag-amp}).

\subsection{Small-Tag, One-Sided, Synchronous, Classical Setting: Construction}
\label{sec:tech-overview:small-tag:one-sided}

Our construction of $\langle C, R\rangle^{\msf{OneSided}}_{\msf{tg}}$ is easy to describe. It supports tags from the space $[n(\secpar)]$, where $n(\secpar)$ is any (fixed) polynomial on the security parameter $\secpar$. 

To proceed with a tag $t \in [n]$, we first ask the committer $C$ to commit to $m$ using a statistically binding scheme $\msf{com}=\Com(m;r)$ (e.g., Naor's commitment). Then, the receiver $R$ sends a {\em non-interactive hard puzzle} that has exactly $t$ distinct solutions; $R$ also gives a witness-indistinguishable proof of knowledge\footnote{Indeed, a WI {\em argument} of knowledge suffices (see \Cref{sec:tech-overview:small-tag:one-sided:sync:PQ}).} (referred to as {\bf WIPoK-1}) to prove that he knows one of the $t$ solutions.  Finally, $C$ is required to prove using another WIPoK (referred to as {\bf WIPoK-2}) that he knows {\em either} the value committed in $\msf{com}$ {\em or} one solution to $R$'s hard puzzle. 

\begin{figure}
 \centering
         \fbox{
         \includegraphics[width=0.95\textwidth,page=1]{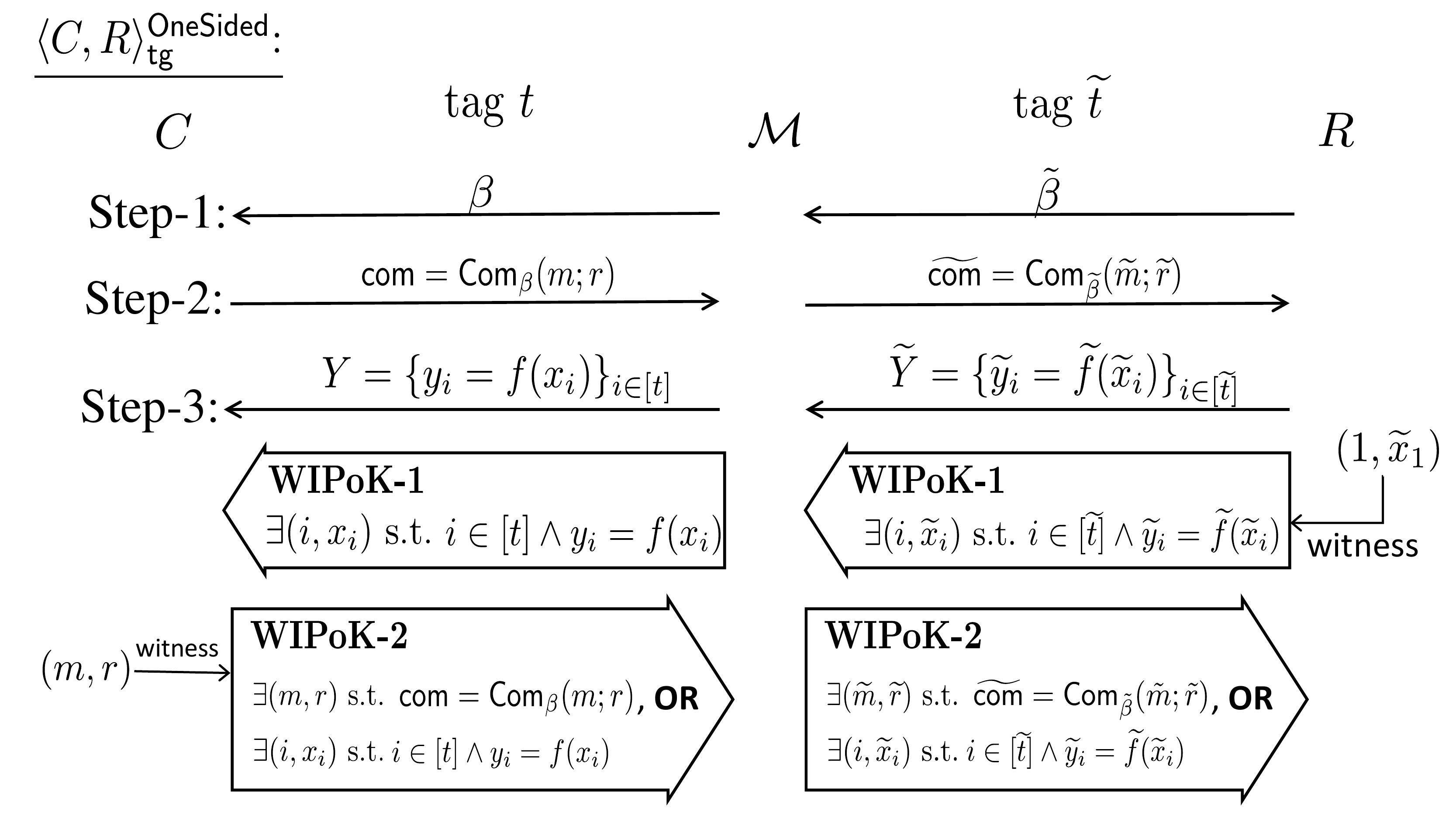}
         }
         \caption{Man-in-the-Middle Execution of $\langle C, R\rangle^{\msf{OneSided}}_{\msf{tg}}$}
         \label{figure:one-sided:tech-overview:real}
\end{figure}

We depict this construction (in the MIM setting) in \Cref{figure:one-sided:tech-overview:real}, where the ``hard puzzle'' is implemented with the problem of ``finding one of the preimages of the $t$ OWF images $\Set{y_i}_{i \in [t]}$''. Throughout this overview, we assume that the OWF $f$ is injective (otherwise, this ``hard puzzle'' may have more than $t$ solutions). But this is only to ease the presentation---We will describe in \Cref{sec:removing-injectivity} a simple trick to remove this injectivity requirement. 
%We will talk about how to remove this injectivity requirement later.

The main intuition underlying this construction is best illustrated by the proof of its non-malleability, which we show next in \Cref{sec:tech-overview:proving-NM}.

\subsection{Proving Non-Malleability}
\label{sec:tech-overview:proving-NM}

Instead of giving out the security proof directly, we start with a naive attempt that will fail. It will reveal the main difficulty in the proof of non-malleability and how our design addresses it.

% \red{Recall that the essence of non-malleability lies in the design a mechanism such that in the MIM game, the MIM adversary $\mcal{M}$ cannot change the right committed value $\tilde{m}$, while the left committed value changes from $m_0$ to $m_1$. Our idea is to achieve this effect using the following pigeonhole-principle style argument over the tags.}

\para{The First Attempt.}  In the MIM execution shown in \Cref{figure:one-sided:tech-overview:real}, imagine that we invoke the knowledge extractor\footnote{We remark that there exist different definitions for {\em Proofs of Knowledge}. Throughout this paper, we use Lindell's formalism called {\em Witness Extended Emulation} \cite{JC:Lindell03}. This definition ensures the existence of a knowledge extractor $\mcal{WE}$ such that for any $P^*$ that convinces $V$ with probability $p$, $\mcal{WE}^{P^*}$ will extract a valid witness with probability $\ge p - \negl(\secpar)$. See \Cref{sec:prelim:WIPoK} for more details.} for the right {\bf WIPoK-2} to extract a witness $w'$. It is easy to see that the only possible values for $w'$ is $(\tilde{m}, \tilde{r})$ or $(1, \tilde{x}_1)$, because they are the only information that $\mcal{M}$ could potentially possess assuming it cannot break the one-wayness of the OWF $\tilde{f}$ in the right {\bf Step-3}. Furthermore, we know for sure that $w'$ can only be $(\tilde{m}, \tilde{r})$ because of the WI property of the right {\bf WIPoK-1} and the following (standard) argument: Imagine that we switch to $(2, \tilde{x}_2)$ as the witness to finish the right {\bf WIPoK-1}. By the WI property of the right {\bf WIPoK-1}, we know that the distribution of the extracted $w'$ cannot change (up to negligible probability). That is, {\em if $w'$ could take the value $(1, \tilde{x}_1)$ before we switch to $(2, \tilde{x}_2)$ in the right {\bf WIPoK-1}, it would keep taking the value $(1, \tilde{x}_1)$ when we are using $(2, \tilde{x}_2)$ in the right {\bf WIPoK-1}}. But this cannot happen as it breaks the one-wayness of $\tilde{f}$ on the image $\tilde{y}_1$ using a standard reduction---Consider an external challenger for the one-wayness of $\tilde{f}$; The challenger sends us a challenge $y^*$; We run the game shown in \Cref{figure:one-sided:tech-overview:real}, while using $y^*$ in place of $\tilde{y}_1$ and using $(2, \tilde{x}_2)$ as the witness to perform the right {\bf WIPoK-1}. If we finally extract $w' = (1, x_1)$, we find a preimage for $y^*$.

With the above observation, we {\em hope to} prove non-malleability in the following way: We change the left committed value from $m_0$ to $m_1$, while extracting the right committed value $w' = (\tilde{m},\tilde{r})$ using the knowledge extractor for the right {\bf WIPoK-2}. If $\mcal{M}$ really makes $\tilde{m}$ depend on the left committed value, then the reduction can detect this change by checking the extracted $w'$. This breaks the hiding property\footnote{It is easy to see that our construction, in the stand-alone setting, is a computationally hiding commitment.} of the left execution when the left committed value changes from $m_0$ to $m_1$.  However, this naive idea will not work: When we invoke the knowledge extractor for the right {\bf WIPoK-2}, $\mcal{M}$ also rewinds the left {\bf WIPoK-2}. So, the left interaction is not ``in straight-line'' anymore. Thus, we could not build the reduction to the hiding property of the left interaction.

\para{Reduction to Naor's Commitment?} Though the above attempt does not work, it inspires the following thoughts: What if we reduce non-malleability to the hiding property of Naor's commitment in the left {\bf Step-1} and {\bf Step-2}? There is at least some hope as the left Naor's commitment is not interleaved with the right {\bf WIPoK-2} (recall that we are in the synchronous setting). That is, we can generate the {\bf Step-1} and {\bf Step-2} messages in the left interaction by forwarding messages between $\mcal{M}$ and an (external) challenger for the hiding property of Naor's commitment; Meanwhile, we extract $w' = (\tilde{m}, \tilde{r})$ from the right {\bf WIPoK-2} as before. Then, if $\mcal{M}$ makes the value $\tilde{m}$ depend on the value committed in the first two steps in the left, we win the hiding game by checking the extracted $w'$. 

\begin{figure}[!tb]
     \begin{subfigure}[t]{0.47\textwidth}
         \centering
         \fbox{
         \includegraphics[width=\textwidth,page=4]{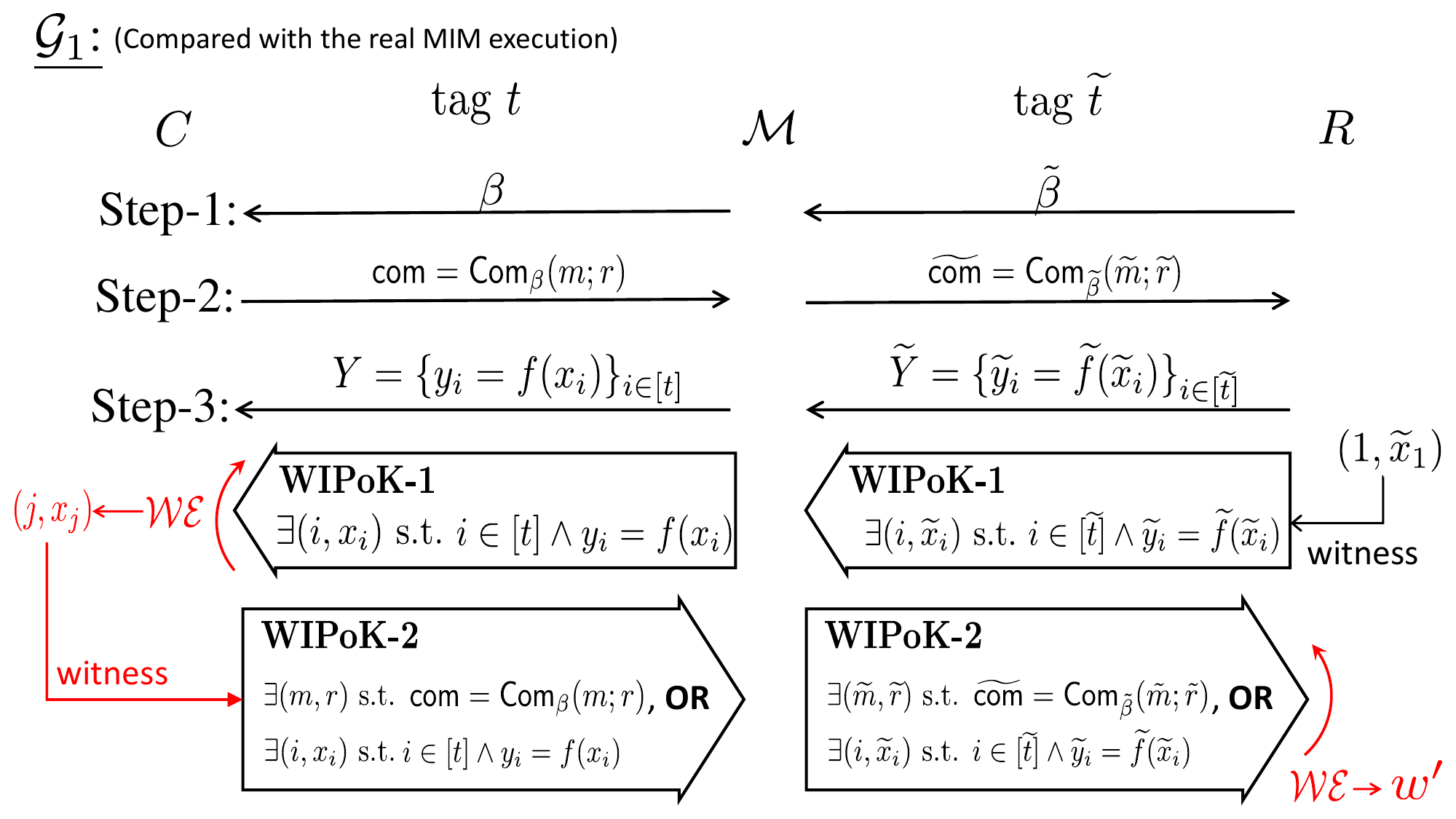}
         }
         \caption{}
         \label{figure:tech-overview:one-sided:G1}
     \end{subfigure}
     \hspace{6.5pt}
     \begin{subfigure}[t]{0.47\textwidth}
         \centering
         \fbox{
         \includegraphics[width=\textwidth,page=5]{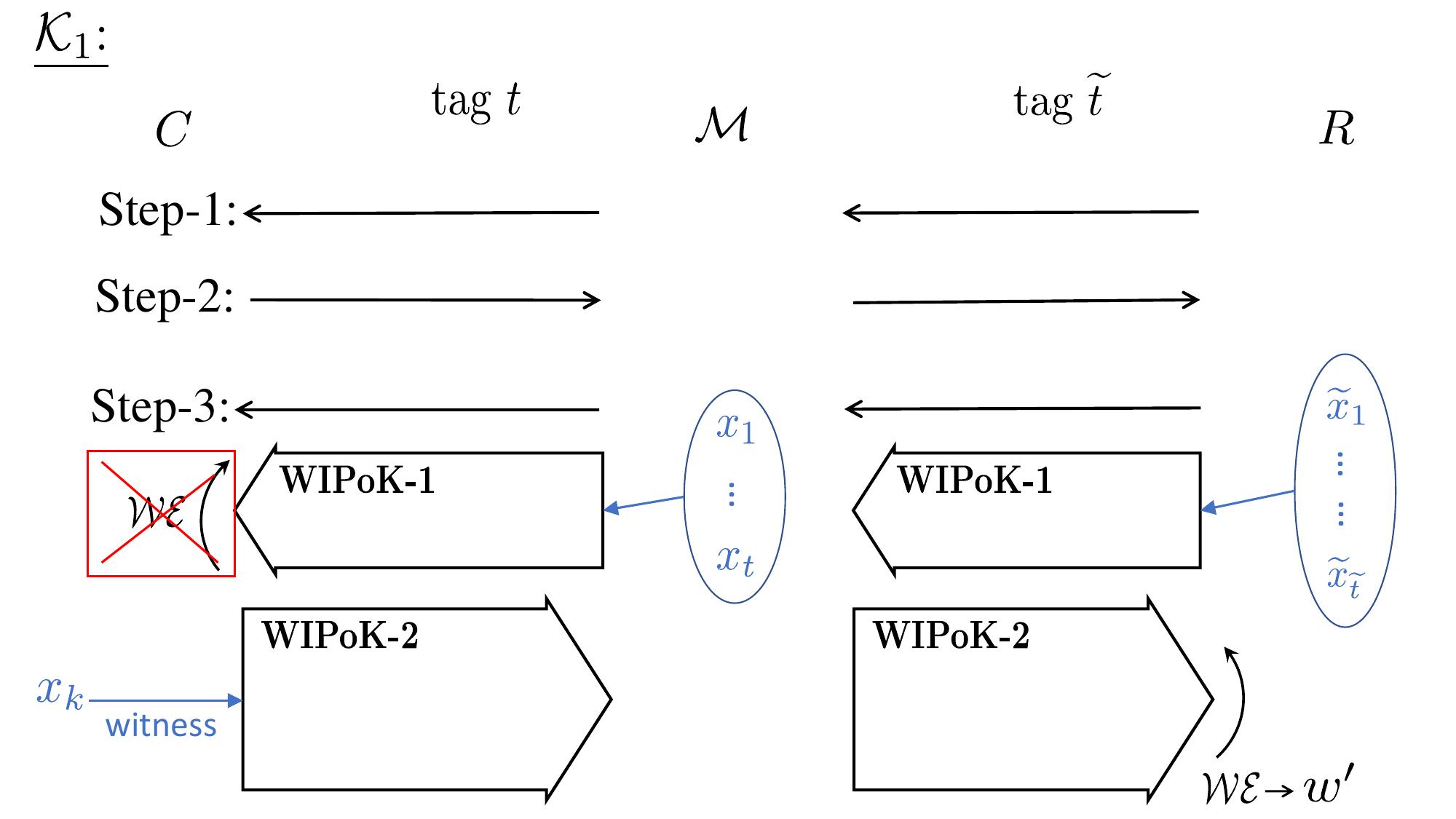}
         }
         \caption{}
         \label{figure:tech-overview:one-sided:K1}
     \end{subfigure}
     \caption{Machines $\mcal{G}_1$ and $\mcal{K}_1$}
     \label{figure:tech-overview:one-sided:G1-K1}
\end{figure}

\label{para:rebutal}To implement this idea, we first need to make the interaction happening after the left {\bf Step-2} {\em independent of $m$}; Otherwise, the above reduction will not work---Because the left Naor's commitment is now coming from an external challenger so that the reduction does not posses the value $m$ anymore, which is required to finish the remaining steps (in particular, the left {\bf WIPoK-2}) of the left interaction. To address this issue, we consider an intermediate execution $\mcal{G}_1$ shown in \Cref{figure:tech-overview:one-sided:G1}---The only difference (shown in red) between $\mcal{G}_1$ and the real MIM game (\Cref{figure:one-sided:tech-overview:real}) is that $\mcal{G}_1$ uses the knowledge extractor $\mcal{WE}$ in the left {\bf WIPoK-1}, and uses the extracted witness $(j, x_j)$ as the witness to go through the left {\bf WIPoK-2}. In this way, the interaction after the left {\bf Step-2} does not depend on $m$ anymore. However, this introduces new problems: We can no longer make use of the WI property of the right {\bf WIPoK-1} as before, because this $\mcal{WE}$ needs to rewind the left {\bf WIPoK-1}, resulting in rewindings of the right {\bf WIPoK-1} as well (since we are in the {\bf Synchronous} setting). Thus, we can no longer argue that the extracted $w'$ must be $(\tilde{m},\tilde{r})$ as in {\bf The First Attempt} part.

% \dlt{Fortunately, we can still show that $w' = (\tilde{m},\tilde{r})$ with good-enough probability by an alternative argument that leverages the asymmetry of tags, i.e.\ $t <\tilde{t}$ (recall that this is exactly our {\bf One-Sided} condition). In the following, we elaborate on this argument.}

Actually, there is a deeper reason why this naive attempt is bound to fail in establishing $w' = (\tilde{m},\tilde{r})$. When we run $\mcal{WE}$ for the left {\bf WIPoK-1}, $\mcal{M}$ could also learn the witness $(1, \tilde{x}_1)$ used by $R$ in the right {\bf WIPoK-1}. When we switch the witness from $(m,r)$ to the extracted $(j, x_j)$ in the left {\bf WIPoK-2}, nothing stops $\mcal{M}$ from switching her witness from $(\tilde{m}, \tilde{r})$ to $(1, \tilde{x}_1)$. Therefore, we can no longer argue that the extracted $w'$ must be $(\tilde{m}, \tilde{r})$. We emphasize that the WI property of the left {\bf WIPoK-2} does not help in ruling out this possibility---{\em WI does not protect against a man-in-the-middle adversary who is trying to make the right WIPoK instance depend on the left WIPoK instance.} Instead, this is a ``non-malleability'' type of requirement rather than (plain) witness indistinguishability. In that sense, the above argument did not really address the issue of non-malleability; Rather, it simply ``pushed'' the non-malleability requirement to the WIPoK used in the final {\bf WIPoK-2} stage.

Thus, new ideas seem necessary to enforce non-malleability. This leads us to the real virtue of our design---In the sequel, we will demonstrate that the $t$-solution hard puzzle introduces ``asymmetry'' between the right and the left sessions as $t<\tilde{t}$; This will allow us to perform a pigeon-hole-style argument, which can be used to  address the heart of the non-malleability problem. 

\para{A Pigeon-Hole Argument.} To address the above issue, we consider another hybrid $\mcal{K}_1$ shown in \Cref{figure:tech-overview:one-sided:K1}. Hybrid $\mcal{K}_1$ is similar to $\mcal{G}_{1}$ but it does not rewind the left {\bf WIPoK-1} for witness extraction (we will explain shortly which witness $\mcal{K}_1$ will use to perform the left {\bf WIPoK-2}).

We first make a crucial observation: There are $\tilde{t}$ preimages $\Set{\tilde{x}_1, \ldots, \tilde{x}_{\tilde{t}}}$ that $R$ could potentially use as the witness to perform the right {\bf WIPoK-1}. In contrast, there are at most $t$ preimages $\Set{x_1, \ldots, x_t}$ that $\mcal{M}$ could potentially use as the witness to perform the left {\bf WIPoK-1}. Since the {\bf WIPoK-1} stage itself is not ``non-malleable'', it is possible that $\mcal{M}$'s witness used there depends on $R$'s witness used in the right {\bf WIPoK-1}.\footnote{Again, standard witness-indistinguishability does not rule out such dependency.} For example, if $R$ uses $\tilde{x}_1$ to perform the right {\bf WIPoK-1}, $\mcal{M}$ may use, say, $x_3$ in the left {\bf WIPoK-1}; If $R$ instead uses $\tilde{x}_2$, $\mcal{M}$ may switch to $x_5$ in the left {\bf WIPoK-1}. 
% \xiao{@Takashi: I know that I should write $(1, x_1)$ and $(\tilde{m}, \tilde{r})$ instead of $x_1$ and $\tilde{m}$. But I believe people can understand what we mean. So, I choose to use the succinct form.}

 Now, recall that $t\le \tilde{t}-1$ (because we have the {\bf One-Sided} condition). It then follows from the pigeon-hole principle that there must exist a distinct pair $\tilde{x}_i, \tilde{x}_j \in \Set{\tilde{x}_1, \ldots, \tilde{x}_{\tilde{t}}}$ and an $x_k\in\Set{x_1, \ldots, x_t}$ such that $(\tilde{x}_i, \tilde{x}_j, x_k)$ form the following correspondence: {\em No matter $R$ uses $\tilde{x}_i$ or $\tilde{x_j}$ in the right {\bf WIPoK-1}, $\mcal{M}$ always uses (the same) $x_k$ in the left {\bf WIPoK-1}}. 

 If we assume that $\mcal{K}_1$ somehow ``magically'' knows this pigeon-hole tuple $(\tilde{x}_i, \tilde{x}_j, x_k)$, then we can prove $w' = \tilde{m}$ using the following argument. Consider two scenarios:
 \begin{enumerate}
 \item
 When $R$ uses $\tilde{x}_i$ in the right {\bf WIPoK-1} and $C$ (or $\mcal{K}_1$) uses $x_k$ in the left {\bf WIPoK-2}, we can prove that the extracted $w'$ can only take the values of $\tilde{m}$ or $\tilde{x}_i$. Intuitively, this is because $\tilde{m}$ and $\tilde{x}_i$ are the only witnesses that $\mcal{M}$ could potentially use for the right {\bf WIPoK-2}, assuming it cannot break the one-wayness of the right OWF $\tilde{f}$. We suppress the full proof in this informal discussion, as we will show a formal (and slightly different) argument in the next subsection.
 \item
 When $R$ uses $\tilde{x}_j$ in the right {\bf WIPoK-1} and $C$ uses (the same) $x_k$ in the left {\bf WIPoK-2}, we can prove that the extracted $w'$ can only take the values of $\tilde{m}$ or $\tilde{x}_j$. This follows from a similar argument as in the above bullet. 
  \end{enumerate}  

Next, observe that $\mcal{K}_1$ does {\em not} rewind the {\bf WIPoK-1} stage. By the witness indistinguishability of the right {\bf WIPoK-1}, it follows that the witness used by $\mcal{M}$ in the right {\bf WIPoK-2} cannot change when $R$ switches between $\tilde{x}_i$ and $\tilde{x}_j$ in the right {\bf WIPoK-1} (otherwise, we can invoke the knowledge extractor of the right {\bf WIPoK-2} to extract the used witness to detect this change). This, together with the above two bullets, implies that the extracted $w'$ can only take the value $\tilde{m}$ if $R$ uses $\tilde{x}_i$ (or $\tilde{x}_j$) in the right {\bf WIPoK-1}. Notice that in this argument, we do not rely on the ``non-malleability'' (or even WI) of the final {\bf WIPoK-2} stage, because in both scenarios described above, $C$ uses the same $x_k$ in the left {\bf WIPoK-2}. 

Although the above reasoning looks promising, it suffers from the following obstacles:
\begin{enumerate}
\item \label[Obstacle]{item:tech-overview:K1-obstacle:1}
In the above, we assumed that $\mcal{K}_1$ ``magically'' knows the pigeon-hole tuple $(\tilde{x}_i, \tilde{x}_j, x_k)$. However, it is unclear how $\mcal{K}_1$ could (efficiently) learn this tuple. Note that $\mcal{K}_1$ cannot try to learn $x_k$ by rewinding (i.e., running the knowledge extractor for) the left {\bf WIPoK-1}; Otherwise, we will be back to hybrid $\mcal{G}_1$ and cannot rely on the WI property of the right {\bf WIPoK-1} (as we did in the previous paragraph).
\item \label[Obstacle]{item:tech-overview:K1-obstacle:2}
The above pigeon-hole correspondence is not accurate. In fact, it is possible that $\mcal{M}$ changes the {\em distributions} (instead of {\em concrete values}, as in the previous discussion) of the witness used in the left {\bf WIPoK-1}. For example, when $R$ uses $\tilde{x}_1$ in the right {\bf WIPoK-1}, $\mcal{M}$ samples a witness for the left {\bf WIPoK-1} uniformly at random from $\Set{x_1, \ldots, x_t}$; While $R$ switches to $\tilde{x}_2$, $\mcal{M}$ may decide to sample a witness according to the Gaussian distribution instead. In this case, even with unbounded computational power, it is unclear how to identify (or define) the tuple $(\tilde{x}_i, \tilde{x}_j, x_k)$.
\end{enumerate} 

Nevertheless, this hybrid $\mcal{K}_1$ divulges the usefulness of the $t$-solution hard puzzle. It provides an approach to argue $w' = \tilde{m}$, thus having a potential to help us establish non-malleability. In the following, we show our main technical lemma which overcomes the above two obstacles, allowing us to formalize the above argument properly.

\para{The Main Technical Lemma.} Let us summarize what we have so far. We described two hybrids: Hybrid $\mcal{G}_1$ seems to help us reduce non-malleability to the computational-hiding property of the left Naor's commitment. But it actually pushes the non-malleability requirement to the final {\bf WIPoK-2} stage (or more technically, we are not able to prove that the extracted $w'$ is equal to the right-side committed value $\tilde{m}$). In contrast, hybrid $\mcal{K}_1$ does not suffer from this issue. It instead relies on a pigeon-hole-style argument, which employs the structure of the $t$-solution hard puzzle to enforce non-malleability. However, to formalize this argument, one has to (i) figure out how $\mcal{K}_1$ can learn the pigeon-hole tuple (\Cref{item:tech-overview:K1-obstacle:1}), and (ii) perform the pigeon-hole argument in a ``distributional'' manner (\Cref{item:tech-overview:K1-obstacle:2}).

\begin{figure}[!h]
     \begin{subfigure}[t]{0.47\textwidth}
         \centering
         \fbox{
         \includegraphics[width=\textwidth,page=2]{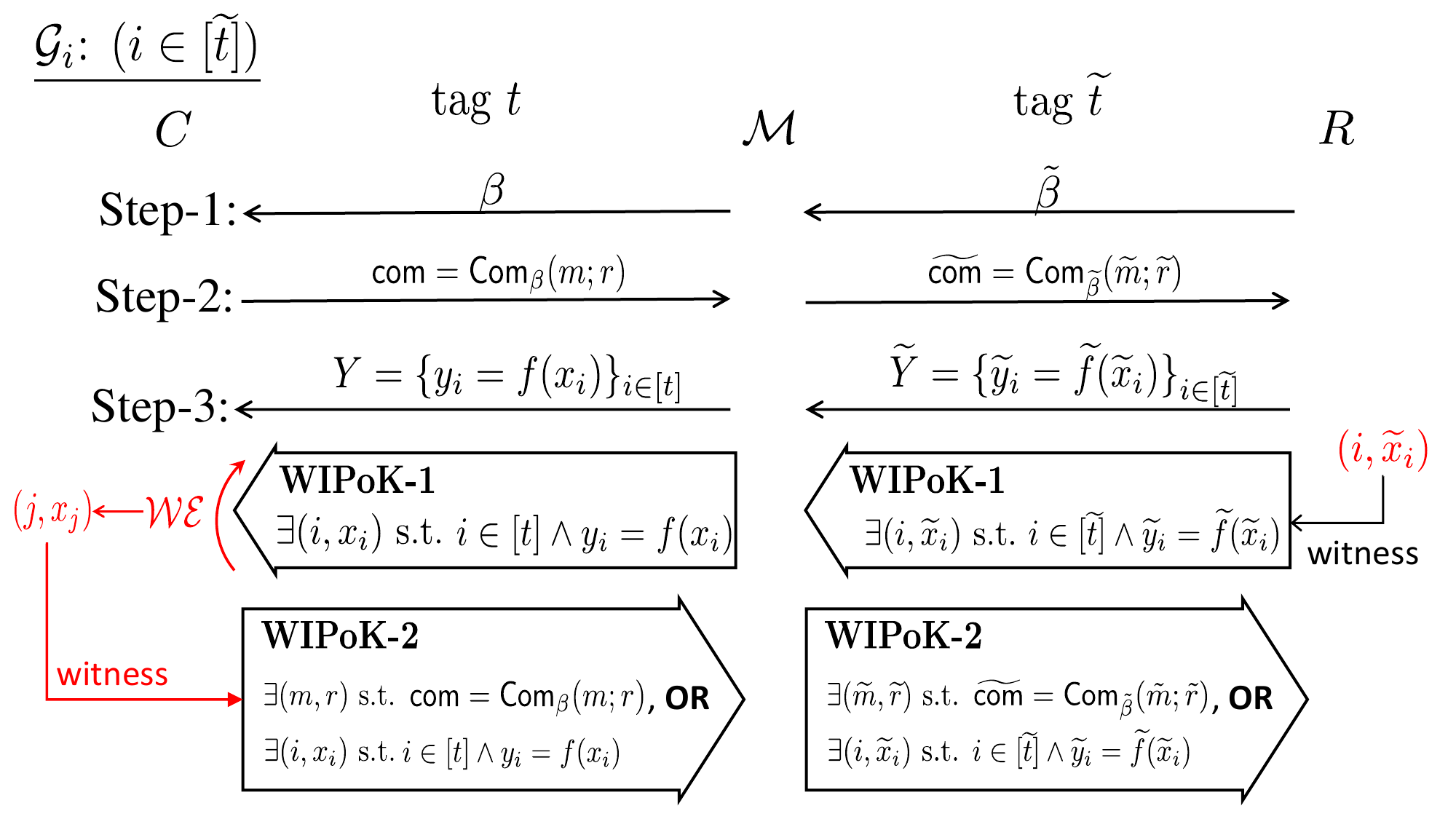}
         }
         \caption{}
         \label{figure:one-sided:tech-overview:Gi}
     \end{subfigure}
     \hspace{6.5pt}
     \begin{subfigure}[t]{0.47\textwidth}
         \centering
         \fbox{
         \includegraphics[width=\textwidth,page=3]{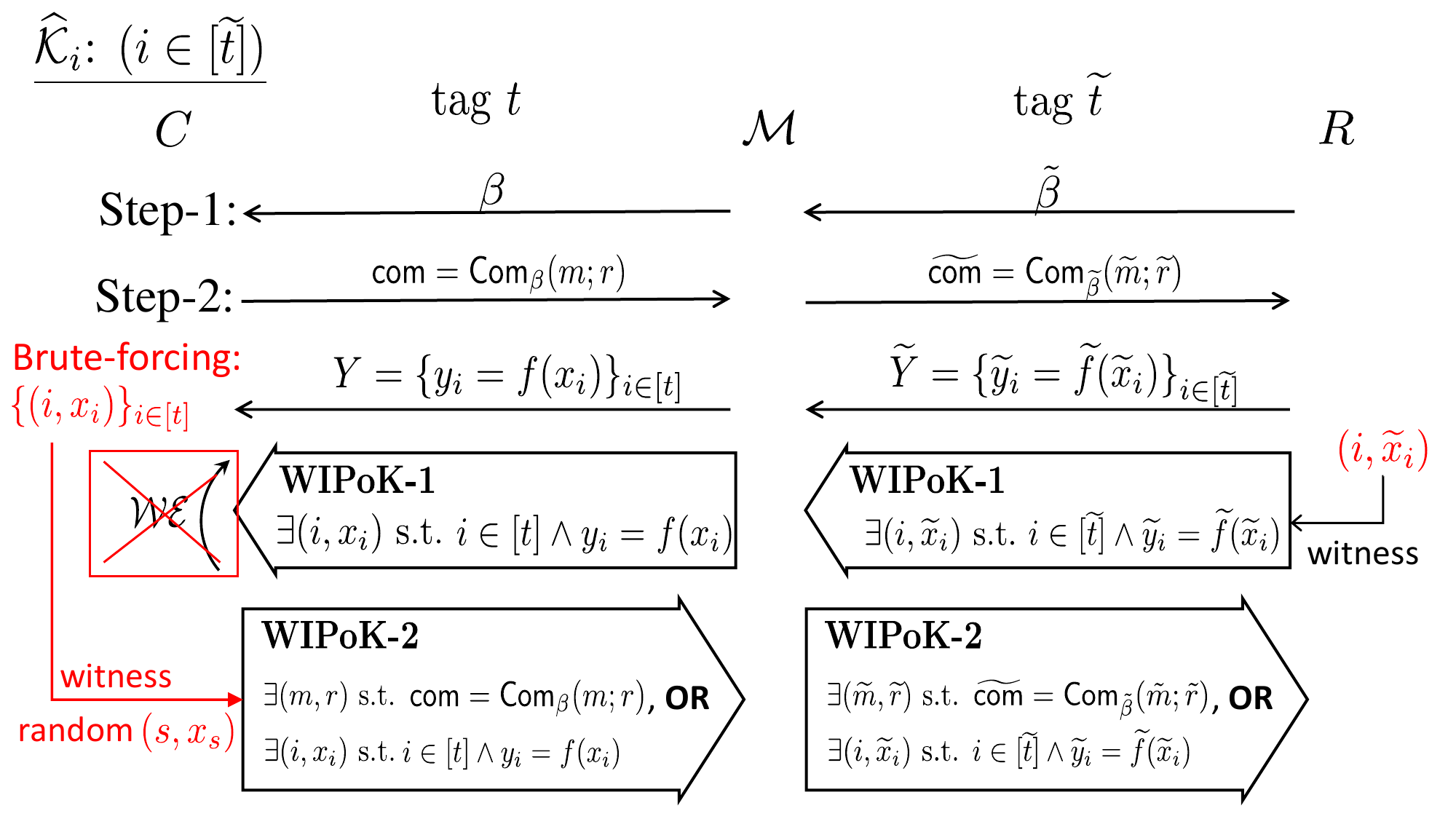}
         }
         \caption{}
         \label{figure:one-sided:tech-overview:Ki}
     \end{subfigure}
     \caption{Machines $\mcal{G}_i$ and $\hat{\mcal{K}}_i$ {\scriptsize (Difference is highlighted in red color)}}
     \label{figure:one-sided:tech-overview:Gi-Ki}
\end{figure}

In the following, we present our main technical lemma. This lemma can be understood as a combination of $\mcal{G}_1$ and $\mcal{K}_1$---In its proof, we will define analogs of these two hybrids (i.e., the $\Set{\mcal{G}_i}_{i\in [\tilde{t}]}$ and $\Set{\hat{\mcal{K}}_i}_{i\in [\tilde{t}]}$ in \Cref{figure:one-sided:tech-overview:Gi-Ki}) and ``jump between'' them to take advantage of both hybrids, while performing the pigeon-hole argument in a ``distributional'' fashion.

First, we define  machines $\mcal{G}_i$ for each $i\in [\tilde{t}]$ (as shown in \Cref{figure:one-sided:tech-overview:Gi}). Machine $\mcal{G}_i$ for $i \ne 1$ behaves identically to the $\mcal{G}_1$  described earlier (\Cref{figure:tech-overview:one-sided:G1}), except that  $\mcal{G}_i$ uses $(i, \tilde{x}_i)$ as the witness when performing the right {\bf WIPoK-1}. Again, we define the {\em extracted witness} $w'$ to be the witness extracted by running the knowledge extractor for the right {\bf WIPoK-2} in $\mcal{G}_i$.
Then, we show the main technical lemma, which we will use to establish non-malleability later.
\begin{lemma}[Informal Version of \Cref{lem:small-tag:proof:se:proof:K}]\label{lem:tech-overview:main}
Assume that {\em in the real MIM execution}, $\mcal{M}$ convinces the honest right receiver $R$ with some noticeable probability $p(\secpar)$. Then, there exists some $i \in [\tilde{t}]$ such that the extracted witness $w'$ in $\mcal{G}_i$ must be a valid opening $(\tilde{m},\tilde{r})$ for $\tilde{\msf{com}}$ with another noticeable probability $p'(\secpar)$. 
%\takashi{For simplicity, I modified "extracted value" to "extracted witness". The issue that $\tilde{r}$ is not unique can be mentioned later in the overview where we consider the post-quantum security.}
%\takashi{I replaced "non-negligible" with "noticeable" for convenience in the rewinding step and post-quantum version.}
\end{lemma}
\subpara{Proof of \Cref{lem:tech-overview:main}.} We prove \Cref{lem:tech-overview:main} by contradiction. In the following, we assume for contradiction that for all $ i \in [\tilde{t}]$, the extracted $w'$ contains a valid opening to $\tilde{\msf{com}}$ with at most {\em non-noticeable} probability. (For simplicity, we ignore the difference between ``non-noticeable" and ``negligible" and use the term ``negligible'' in place of ``non-noticeable'' in this proof.)

First, we claim that in game $\mcal{G}_i$ (for any $i \in [\tilde{t}]$), $R$ must also be convinced with probability $p \pm \negl(\secpar)$. This claim follows from the PoK property of the left {\bf WIPoK-1} and the WI property of the right {\bf WIPoK-1} and the left {\bf WIPoK-2}. Since this proof is rather standard, we do not elaborate on it in this overview. More details can be found in \Cref{sec:lem:small-tag:proof:se:proof:K:proof}.

%\xiao{Add the argument.}

%Now, imagine that we use the knowledge extractor for the right {\bf WIPoK-2} in $\mcal{G}_i$, to extract a witness which we denote again as $w'$. 
Next, we observe that in $\mcal{G}_i$, the extracted witness $w'$ can only take the values $(\tilde{m},\tilde{r})$ or $(i, \tilde{x}_i)$; Otherwise, we break the one-wayness of the right OWF $\tilde{f}$ (this follows from the same argument we made earlier). However, our assumption (for contradiction) says that in each $\mcal{G}_i$, $w' \ne (\tilde{m},\tilde{r})$ except for with negligible probability. Therefore, we obtain the following inequality:
\begin{equation}\label[Inequality]{eq:tech-overview:Gi}
\forall i \in [\tilde{t}],~ \big|\Pr[w' = (i, \tilde{x}_i) ~\text{in}~\mcal{G}_i] - p\big| \le \negl(\secpar).
\end{equation}

Next, we define $\Set{\hat{\mcal{K}}_i}_{i \in [\tilde{t}]}$ (shown in \Cref{figure:one-sided:tech-overview:Ki})\footnote{For readers trying to find a correspondence between this technical overview and its main-body counterpart (i.e., \Cref{sec:small-tag-one-sided-sync-classical}), we would like to point out that the $\hat{\mcal{K}}_i$ defined here does not match any machine in \Cref{sec:small-tag-one-sided-sync-classical}. In the current overview, $\hat{\mcal{K}}_i$ actually serves the functionality of several machines defined in \Cref{sec:lem:small-tag:proof:se:proof:K:proof,sec:lem:bound:Ki:proof,sec:proof:claim:K'':non-abort}. But $\hat{\mcal{K}}_i$ is most similar to the $\mcal{K}''_i$ on \Cpageref{pageref:one-sided:K''i} (and depicted in \Cref{figure:one-sided:K''i}), which is essentially $\hat{\mcal{K}}_i$ but additionally runs a knowledge extractor for the right {\bf WIPoK-2}.}.  For each $i \in [\tilde{t}]$, $\hat{\mcal{K}}_{i}$ behaves identically as $\mcal{G}_i$ except that $\hat{\mcal{K}}_{i}$ does not invoke the knowledge extractor for the left {\bf WIPoK-1}; Instead, it obtains all the preimages $\Set{x_i}_{i \in [t]}$ for $\Set{y_i}_{i \in [t]}$ in the left {\bf Step-3} by {\em brute force}; It then picks a random index $s \pick [t]$ and uses preimage $(s, x_s)$ as the witness to conduct the left {\bf WIPoK-2}. First, observe that if the $(s, x_s)$ picked by $\hat{\mcal{K}}_i$ hits the $(j, x_j)$ extracted from the left {\bf WIPoK-1} in $\mcal{G}_i$ (see \Cref{figure:one-sided:tech-overview:Gi}), then the games $\hat{\mcal{K}}_i$ and $\mcal{G}_i$ are identical (modulo that $\hat{\mcal{K}}_i$ is inefficient). Since $\hat{\mcal{K}}_i$ picks $s$ uniformly at random from $[t]$, we know that $\hat{\mcal{K}}_i$ will be identical to $\mcal{G}_i$ with probability $\ge 1/t$. It then follows from \Cref{eq:tech-overview:Gi} that
\begin{equation}\label[Inequality]{eq:tech-overview:Ki}
\forall i \in [\tilde{t}],~ \Pr[w' = (i, \tilde{x}_i) ~\text{in}~\hat{\mcal{K}}_i] \ge \frac{1}{t} \cdot p - \negl(\secpar).
\end{equation}

Next, observe that in $\hat{\mcal{K}}_i$, {\bf WIPoK-1} is not rewound anymore. Therefore, by the (non-uniform, to compensate for the brute-forcing step) WI property of the right {\bf WIPoK-1}, it follows that 
\begin{equation}\label[Inequality]{eq:tech-overview:Ki-K1}
\forall i \in [\tilde{t}],~ {\bigg|}\Pr[w' = (i, \tilde{x}_i) ~\text{in}~\hat{\mcal{K}}_i] - \Pr[w' = (i, \tilde{x}_i) ~\text{in}~\hat{\mcal{K}}_1] \bigg|\le \negl(\secpar).
\end{equation}
Combining \Cref{eq:tech-overview:Ki,eq:tech-overview:Ki-K1}, we have
\begin{equation}\label[Inequality]{eq:tech-overview:K1}
\forall i \in [\tilde{t}],~ \Pr[w' = (i, \tilde{x}_i) ~\text{in}~\hat{\mcal{K}}_1] \ge \frac{1}{t} \cdot p - \negl(\secpar).
\end{equation}
\Cref{eq:tech-overview:K1} implies the following
\begin{equation}\label[Inequality]{eq:tech-overview:K1:lower}
\Pr[\text{$w'$ is a valid witness in}~\hat{\mcal{K}}_1] \ge \tilde{t}\cdot\frac{1}{t} \cdot p - \negl(\secpar) \ge p + \frac{p}{t} -\negl(\secpar),
\end{equation}
where ``$w'$ is a valid witness'' refers to the event that $w'=(\tilde{m},\tilde{r}) \vee w'=(1, \tilde{x}_1) \vee \ldots \vee w' = (\tilde{t}, \tilde{x}_{\tilde{t}})$, and the last ``$\ge$''  follows from the requirement that $t < \tilde{t}$ (i.e., we are in the {\bf One-Sided} setting).

On the other hand, we claim that 
\begin{equation}\label[Inequality]{eq:tech-overview:K1:upper}
\Pr[\text{$w'$ is a valid witness in}~\hat{\mcal{K}}_1] \le  p + \negl(\secpar).
\end{equation}
This can be seen by comparing $\hat{\mcal{K}}_1$ with the real MIM execution shown in \Cref{figure:one-sided:tech-overview:real}: Recall that $p$ is the probability of $R$ being convinced {\em in the real MIM execution}. The only difference between $\hat{\mcal{K}}_1$ and the real MIM execution is the witness used in the left {\bf WIPoK-2}. By the (non-uniform) WI property of the left {\bf WIPoK-2}, we know that $R$ must be convinced in $\hat{\mcal{K}}_1$ with probability $\le p +\negl(\secpar)$. Then, by the proof of knowledge property, we must extract a valid witness in the right {\bf WIPoK-2} in $\hat{\mcal{K}}_1$ with probability upper-bounded by $p +\negl(\secpar)$ as well (this is exactly \Cref{eq:tech-overview:K1:upper}).

Observe that \Cref{eq:tech-overview:K1:upper} contradicts \Cref{eq:tech-overview:K1:lower} because ${p}/{t}$ is non-negligible. This gives us the desired contradiction, thus finishing the proof of \Cref{lem:tech-overview:main}.

\para{Completing the Proof of Non-Malleability.}\label{para:rebuttal-3} With \Cref{lem:tech-overview:main}, non-malleability can be reduced to the computational-hiding property of Naor's commitment as follows. Assuming that $\mcal{M}$ breaks the non-malleability of our scheme w.r.t. a distinguisher $D$, we construct an adversary $\Adv_{\msf{hiding}}$ against the computational-hiding property of Naor's commitment as follows:
 %that picks a random $i \in [\tilde{t}]$, and runs $\mcal{G}_i$ with the MIM adversary $\mcal{M}$ in the following way:
\begin{itemize}
\item {\bf Main-Thread Simulation:}
$\Adv_{\msf{hiding}}$ runs $\mcal{G}_1$ where it embeds the instance of the hiding game of Naor's commitment in the left {\bf Step-1} and {\bf Step-2}, and finishes other steps just as $\mcal{G}_1$.
If $R$ rejects, it sets $\tilde{m}\defeq \bot$. (Notice that at the end of this step, $\mcal{M}$ will give an output that is  computationally indistinguishable with the one from the real MIM execution. This is due to the similarity between $\mcal{G}_1$ and the real MIM execution.) 
\item {\bf Rewinding:} \label{item:tech-overview:page-ref:rewinding} 
Unless $R$ rejects in the above, 
$\Adv_{\msf{hiding}}$ repeats the following $N$ (to be specified later) times to extract the message $\tilde{m}$ committed in the right session:
\begin{itemize}
    \item It rewinds $\mcal{M}$ to the point right after the completion of {\bf Step-2}. 
    \item It picks a random index $i\pick [\tilde{t}]$, and finishes the execution in the same manner as $\mcal{G}_i$ (depicted in \Cref{figure:one-sided:tech-overview:Gi}). 
    \item Extract a witness $\tilde{w}$ from the right {\bf WIPoK-2} of the simulated execution. 
    \item If $\tilde{w}=(\tilde{m},\tilde{r})$ is a valid opening to $\tilde{\msf{com}}$, it breaks the loop. Otherwise, it continues.
\end{itemize}
If it fails to extract $\tilde{m}$ within $N$ trials, it aborts and outputs a random guess.
\item {\bf Decision:}
It runs the distinguisher $D$ on $\mcal{M}$'s final output at the end of the {\bf Main-Thread Simulation} step and $\tilde{m}$, and outputs whatever $D$ outputs. 
\end{itemize} 
It is easy to see that the above reduction works unless it fails to extract $\tilde{m}$. We show that the failure probability can be made an arbitrarily small noticeable $\epsilon(\secpar)$ if we take sufficiently large $N=\poly(\secpar)$ depending on $\epsilon$.
We call the snapshot (including the transcript and $\mcal{M}$'s internal state) at the end of {\bf Step-2} of the main thread a \emph{prefix}. 
For each prefix $\msf{pref}$, let $p_\msf{pref}$ be the probability that the right receiver accepts in $\mcal{G}_1$ starting from $\msf{pref}$. 
Then, \Cref{lem:tech-overview:main} ensures that\footnote{Strictly speaking, there are two minor differences from \Cref{lem:tech-overview:main}. First, we are considering experiments for each fixed $\msf{pref}$ whereas \Cref{lem:tech-overview:main} considers the whole experiment. This is not an issue since the proof of \Cref{lem:tech-overview:main} does not touch messages before {\bf Step-3} and thus works for any fixed $\msf{pref}$. Another difference is that we define $p_\msf{pref}$ w.r.t. $\mcal{G}_1$ whereas $p$ is defined w.r.t. the real MIM experiment in \Cref{lem:tech-overview:main}. This is not an issue since $\mcal{G}_1$ is computationally indistinguishable from the real MIM experiment.} for any fixed prefix $\msf{pref}$ such that $p_{\msf{pref}}\geq \epsilon(\secpar)$, 
if we run $\mcal{G}_i$ for random $i\pick [\tilde{t}]$ and extract $\tilde{w}$ from the right {\bf {WIPoK-2}} (as is done in each repetition described in the above {\bf Rewinding} step), then we will extract $\tilde{w}=(\tilde{m},\tilde{r})$ with at least another noticeable probability $\epsilon'(\secpar)$ (More accurately, it should be the $\epsilon'(\secpar)$ guaranteed by \Cref{lem:tech-overview:main} {\em divided by $\tilde{t}$} as we pick an $i$ randomly from $[\tilde{t}]$. But this does not affect the argument here as $\tilde{t}$ is bounded by a polynomial on $\secpar$---recall that we are in the {\bf Small-Tag} setting.) Now, we consider the following two cases:%\footnote{Strictly speaking, we should require $\epsilon$ to be noticeable rather than non-negligible to make sure that $N$ is polynonmial. In the actual proof, this is done by proving }
\begin{itemize}
    \item \underline{The Case of $p_\msf{pref}\geq \epsilon(\secpar)$:} In this case, if we set $N=\Omega(\epsilon'(\secpar)^{-1} \cdot \secpar)$, the reduction algorithm fails with at most a negligible probability (i.e., $(1-\epsilon'(\secpar))^{\Omega(\epsilon'(\secpar)^{-1} \cdot \secpar)}$).%\footnote{A keen reader may notice that $\tilde{p}_\msf{pref}^{-1}$ is not necessarily a polynomial because we only assume $\tilde{p}_\msf{pref}$ to be non-negligible. In the main body, we require it to be noticeable, so this issue does not occur.}
    \item \underline{The Case of $p_\msf{pref}<\epsilon(\secpar)$:}
    In this case, the right $R$ rejects in the main-thread except for a probability upper-bounded by $\epsilon(\secpar)$. Thus, the reduction  fails with probability at most $\epsilon(\secpar)$. 
\end{itemize}
Overall, the reduction algorithm works with an additive loss of $\epsilon(\secpar)$. %This copmpletes the proof of non-malleability. 
Since we have the freedom to set $\epsilon(\secpar)$ to an arbitrarily small noticeable function, the reduction from non-malleability to the computational hiding of Naor's commitment can be done properly. We omit further details.

\para{Why Our Proof Is ``Quantum-Friendly''.\label{pageref:why-quantum-friendly}} Before going on, let us explain why the above proof of non-malleability in the classical setting could extend to the post-quantum setting. At a high-level, our security proof enjoys the following properties:
\begin{itemize}
\item
The rewinding-extraction procedure described on \Cpageref{item:tech-overview:page-ref:rewinding} never goes to touch the left Naor's commitment, which is the ``hiding gadget'' that provides computational hardness for the reduction. This property saves us from the robust simulatable extractability issue described in \Cref{sec:challenge}.
 \item
 Moreover, this extraction procedure is to first perform the ``main-thread'' execution as hybrid $\mcal{G}_1$; If this main thread is accepted, then it starts $N$ ``rewinding threads'' for extraction. This structure is very similar to the classical extractor for the canonical three-round extractable commitments in \cite{TCC:PasWee09}. Thus, it allows us to invoke (a generalization of) the quantum simulatable-extraction lemma from \cite{cryptoeprint:2021:1516} (developed originally to quantize the \cite{TCC:PasWee09} extractable commitments) to quantize the our proof of non-malleability. More details are provided in \Cref{sec:tech-overview:small-tag:one-sided:sync:PQ}.
\end{itemize}
In contrast, no previous constructions of (classical) constant-round non-malleable commitments achieved the above two properties {\em simultaneously}. This is why it is hard to quantize their security proofs {\em even with the \cite{cryptoeprint:2021:1516} lemma in hand}.

\para{On the Necessity of $N$ Rewindings.} Given that \Cref{lem:tech-overview:main} already guarantees a noticeable probability $p'(\secpar)$ for successful extraction, one may wonder why we need to perform $N$ rewindings in the above reduction to the computational hiding of Naor's commitment. That is, machine $\Adv_{\msf{hiding}}$ seems to work as desired even if we set $N = 1$---If it extracts $\tilde{m}$, then the reduction is done; Otherwise, simply ask the reduction to guess at random in the Naor's hiding game. At the first glance, this seems to give a $\frac{1}{2} + p'(\secpar)$ advantage, which is sufficient for the security reduction.

Though this strategy looks appealing due to its simplicity, it does not work. To explain the reason, we need to refer to \Cref{lem:small-tag:proof:se:proof:K} on page \Cpageref{lem:small-tag:proof:se:proof:K} (i.e., the formal version of \Cref{lem:tech-overview:main}). The explanation (provided below) is rather technical and orthogonal to the current discussion. The reader may feel free to skip it and continue at \Cref{sec:tech-overview:small-tag:one-sided:sync:PQ} directly.

Notice that \Cref{lem:small-tag:proof:se:proof:K} refers to the prefix $\msf{pref}$ of the MIM execution, which consists of the first two rounds of the protocol. In particular, $\msf{pref}$ includes the left Naor's commitment from the hiding challenger. \Cref{lem:small-tag:proof:se:proof:K} says that for any noticeable probability $\epsilon$, if a $\msf{pref}$ leads to an accepting MIM execution with probability greater than $\epsilon$, then the extraction succeeds with a probability {\em lower-bounded} by $\epsilon'/\tilde{t}$, a value polynomially related to (but smaller than) $\epsilon$.

At a high level, \Cref{lem:small-tag:proof:se:proof:K} is not directly applicable in our reduction because there may be a correlation between $\mcal{M}$'s view and the success of extraction. That is, conditioned on the success of  extraction, the {\em joint distribution} of $\mcal{M}$'s view and the extracted right value may be distinguishable from those in the real MIM experiment. Consider the following distinguishing example: there are two prefixes $\msf{pref}_1$ and $\msf{pref}_2$ that lead to an accepting execution with the same probability $\epsilon$, and they appear with the same probability in the real experiment. Suppose that the extraction success probability is $\epsilon'/\tilde{t}$ for $\msf{pref}_1$, but $\epsilon$ ($> \epsilon'/\tilde{t}$) for $\msf{pref}_2$. Then, if we run the reduction  with \Cref{lem:small-tag:proof:se:proof:K} directly, the distinguisher may get to see $\msf{pref}_2$ much more often than $\msf{pref}_1$ (conditioned on successful extraction). But they should happen with the same probability in the real experiment.

The $N$ rewindings in our construction essentially ``amplify'' \Cref{lem:small-tag:proof:se:proof:K} to the following stronger version: For any noticeable $\epsilon$, if $\mcal{M}$ convinces $R$ with probability greater than $\epsilon$, then the extraction succeeds with overwhelming probability (i.e., $1-\negl(\secpar)$).  This effectively ensures the following: Conditioned on a prefix that leads to acceptance with probability greater than $\epsilon$ {\em and} that $R$ is indeed convinced, the joint distribution of $\mcal{M}$'s view and the extracted $\tilde{m}$ are negligibly close to their distribution in the real MIM execution. (Other cases are easy to handle: If $R$ is not convinced, we simply use $\bot$ as the extracted value; Also, by an averaging argument, no more than $\epsilon$ fraction of prefixes can lead to acceptance with probability $< \epsilon$.) Thus, our strategy eventually makes the joint distribution from simulation to be $\epsilon$-close to the real one. Since we have the freedom to pick any noticeable $\epsilon$, our reduction can achieve a non-negligible advantage in Naor's hiding game. 

We refer the reader to \Cref{lem:small-tag:proof:se:proof} for a formal treatment of this issue.

\takashi{Commented out the paragraphs Removing Injectivity and Comparison with \cite{STOC:Goyal11}.}

\subsection{Small-Tag, One-Sided, Synchronous, Post-Quantum Setting}
\label{sec:tech-overview:small-tag:one-sided:sync:PQ}
Next, we explain how to make the $\langle C, R\rangle^{\msf{OneSided}}_{\msf{tg}}$ described in \Cref{sec:tech-overview:small-tag:one-sided} non-malleable in the {\em post-quantum} setting.

For that, we simply instantiate $\langle C, R\rangle^{\msf{OneSided}}_{\msf{tg}}$ with post-quantum building blocks. (We denote this post-quantum version by $\langle C, R\rangle^{\msf{OneSided}}_{\msf{tg,PQ}}$.)
Here, we have to be careful about what security notion we should require for WIPoK. In the security proof in the classical setting, we often extract a witness from WIPoK while continuing the rest of the experiment. %This is not possible in the quantum setting if we only require the original post-quantum PoK property defined by Unruh \cite{EC:Unruh12} where the extractor many disturb prover's internal state during extraction. To avoid this issue, 
For this to work, we implicitly require the knowledge extractor to have a \emph{simulation} property \cite{C:HalSmiSon11,AFRICACRYPT:LunNie11}, i.e., it can simulate the prover's internal state as in a real execution, while performing the extraction task. While this is almost trivial in the classical setting, it becomes hard in the post-quantum setting due to the no-cloning theorem. In particular, constant-round constructions for such ``simulatable'' post-quantum WIPoKs are not known from (polynomially-hard) post-quantum OWFs.\footnote{A recent work by Lombardi, Ma, and Spooner~\cite{cryptoeprint:2021:1543} gave such WIPoKs from super-polynomial hardness of post-quantum OWFs.} 

Fortunately, a recent work~\cite{cryptoeprint:2021:1516} shows that a constant-round construction from post-quantum OWFs is possible if we (i) relax the PoK to \emph{argument} of knowledge (AoK), which only requires extractability against (quantum) polynomial-time adversaries,  and (ii) relax the simulation requirement to the \emph{$\epsilon$-close} simulation property, where an (arbitrarily small) noticeable simulation error is allowed. Next, we show that such a $\epsilon$-simulatable WIAoK suffices for our purpose.
First, the relaxation from PoK to AoK is totally fine since we only apply the knowledge extractors for polynomial-time adversaries (possibly with non-uniform advice). 
Second, a noticeable error coming from extraction does not affect the proof of non-malleability if we take the noticeable error to be much smaller than the assumed MIM adversary's advantage. Thus, we ignore the noticeable errors, and assume that the error is negligible in the rest of this overview. 
 
By using such a post-quantum WIAoK  (as well as the post-quantum version of Naor's commitment $\Com$ and a post-quantum injective OWF $f$), we can see that the main technical part of the security proof in the classical setting (the proof of \Cref{lem:tech-overview:main}) can be migrated to the post-quantum setting almost immediately.

The only non-trivial issue is how to complete the reduction to the computational hiding of Naor's commitment, assuming (a post-quantum analog of) \Cref{lem:tech-overview:main}. In the classical setting, we rely on a rewinding argument that is not applicable anymore when $\mcal{M}$ is a quantum machine (again, due to the no-cloning theorem). Thus, we use a different argument here---We observe that the following lemma, which is a generalization of \cite[Lemma 4]{cryptoeprint:2021:1516}, suffices for this step.
\begin{lemma}[Extract-and-Simulate Lemma (Informal)]
\label{lem:tech:extract_and_simulate}
Let $\mcal{G}$ and $\mcal{K}$ be QPT algorithms that satisfy the following:\footnote{In the formal version of this lemma~(\Cref{lem:extract_and_simulate}), $\mcal{G}$ and $\mcal{K}$ additionally take $1^{\gamma^{-1}}$ as input so that they can run extractors with simulation errors  depending on $\gamma$. We ignore this in this overview since we assume negligibly-close simulation for simplicity.}  %that takes a quantum state $\rho$ as input and outputs  $b\in \{\top,\bot\}$ and a quantum state $\rho_\out$. 
%Suppose that there exists a QPT algorithm $\hat{\mcal{K}}$ that satisfies the the following for some classical string $s^*$.  
\begin{enumerate}
 \item  \label[Property]{item:tech:s_star_or_bot}
    $\mcal{K}$ takes $1^\secpar$ and a quantum state $\rho$ as input and outputs some unique $s^*$ or otherwise $\bot$. 
    \item \label[Property]{item:tech:gamma_delta}
 For any noticeable function $\gamma(\secpar)$, there exists a noticeable function $\delta(\secpar)$ such that for any polynomial-size quantum state $\rho$, %\footnote{Similarly to \Cref{footnote:sequence_s}, we consider a sequence $\{\rho_\secpar\}_{\secpar\in \mathbb{N}}$ and denote by $\rho$ to mean $\rho_\secpar$.}    
 if 
$$
\Pr[b=\top : (b,\rho_\out) \leftarrow \mcal{G}(1^\secpar,\rho)]\geq  \gamma(\secpar),
$$  
then 
$$
\Pr[\mcal{K}(1^\secpar,\rho)=s^*]\geq   \delta(\secpar).
$$
\end{enumerate}
Then, there exists a QPT algorithm $\SimExt$ such that for any polynomial-size quantum state $\rho$ and 
noticeable function $\epsilon=\epsilon(\secpar)$, 
\begin{align}\label{eq:tech:ind}
\{\SimExt(1^\secpar,1^{\epsilon^{-1}},\rho)\}_{\secpar}
\statind_{\epsilon} 
\{(\rho_\out,\Gamma_b(s^*)):(b,\rho_\out)\leftarrow \mcal{G}(1^\secpar,\rho)\}_{\secpar}
\end{align}
where 
$$
\Gamma_b(s^*)\defeq 
\begin{cases}
s^* & \text{if}~ b=\top \\
\bot & \text{otherwise}
\end{cases}.
$$
\end{lemma}
The above lemma is a generalization of \cite[Lemma 4]{cryptoeprint:2021:1516} and can be proven in a similar manner.\footnote{Indeed, such a generalized version was mentioned in the technical overview of \cite{cryptoeprint:2021:1516}.} We will explain the main intuition for this proof at the end of this subsection.

Assuming a post-quantum analog of \Cref{lem:tech-overview:main}, we can complete the reduction to the computational hiding of Naor's commitment by using \Cref{lem:tech:extract_and_simulate} as follows:
We let $\rho$ be the state of the experiment $\mathcal{G}_1$ (as defined in \Cref{figure:one-sided:tech-overview:Gi}, \Cref{sec:tech-overview:small-tag:one-sided}) right after finishing {\bf Step-2}; Let 
$\mathcal{G}$ be the experiment that runs the rest of  $\mathcal{G}_1$  starting from the state $\rho$;\footnote{Similar to the classical setting in \Cref{sec:tech-overview:small-tag:one-sided}, we stipulate that $\rho$ does {\em not} include the left committer $C$'s internal state $\msf{ST}_C$ at the end of {\bf Step-2} (otherwise, we cannot reduce non-malleability to the hiding of the left commitment). This requirement is valid as the remaining steps of $\mcal{G}_1$, which will be executed by $\mcal{G}$, do not depend on $\msf{ST}_C$.}  At the end, $\mcal{G}$ outputs the final state of $\mcal{M}$ as $\rho_\msf{out}$, and outputs the final decision of the right receiver $R$ as $b$. We define $\mcal{K}$ as follows:
\begin{itemize}
    \item Upon receiving a state $\rho$, simulate $\mcal{G}_i$  for $i\pick [\tilde{t}]$. 
    \item Extract a witness $\tilde{w}$ from the right {\bf WIPoK-2} of the simulated execution. 
    \item If $\tilde{w}=(\tilde{m},\tilde{r})$ is a valid opening to $\tilde{\msf{com}}$, output $\tilde{m}$. Otherwise, output $\bot$.
\end{itemize}
It is worth noting that $\mcal{K}$ corresponds to the repeated steps in the rewinding argument in \Cref{sec:tech-overview:small-tag:one-sided}. 
By construction, it is clear that $\mcal{K}$ outputs $\tilde{m}$ whenever it does not output $\bot$ (for an overwhelming fraction of $\tilde{\beta}$ by the statistical binding property of Naor's commitment). Therefore, it satisfies \Cref{item:tech:s_star_or_bot} in \Cref{lem:tech:extract_and_simulate} with $s^*\defeq \tilde{m}$ (for each fixed prefix). (We stress that we set $s^*=\tilde{m}$ instead of $s^*=(\tilde{m},\tilde{r})$ for ensuring the uniqueness of $s^*$ by the statistical binding property of Naor's commitment.) 
Moreover, noting that $\mathcal{G}_1$ is computationally indistinguishable from the real MIM experiment, \Cref{lem:tech-overview:main} implies that $\mcal{K}$ also satisfies \Cref{item:tech:gamma_delta} in \Cref{lem:tech:extract_and_simulate}. 
Thus, there exists a machine $\mcal{SE}$ that satisfies 
\Cref{eq:tech:ind} for $s^*=\tilde{m}$. Using this $\mcal{SE}$, we can construct an adversary $\Adv_{\msf{hiding}}$ against the computational-hiding property of Naor's commitment as follows.

Assuming that $\mcal{M}$ breaks the non-malleability of our scheme w.r.t. a distinguisher $D$, we construct an adversary $\Adv_{\msf{hiding}}$ against the computational-hiding property of Naor's commitment as follows. 
\begin{itemize}
\item {\bf Prefix Generation:}
$\Adv_{\msf{hiding}}$ runs the real MIM experiment\footnote{Remark that the real MIM experiment and $\mcal{G}_1$ are identical until {\bf Step-2}.} for an adversary $\mcal{M}$ until {\bf Step-2} finishes, where it embeds the instance of the hiding game of Naor's commitment in the left {\bf Step-2} while simulating other steps of the left $C$ and the right $R$ honestly. 
Let $\rho$ be the state of the experiment at this point. 
\item {\bf Extract-and-Simulate:} It runs $\SimExt(1^\secpar,1^{\epsilon^{-1}},\rho)$ to extract $\tilde{m}^*$ while simulating the output $\rho_\out$ of the experiment $\mcal{G}_1$. We remark that it succeeds in extracting $\tilde{m}^*$ whenever $R$ accepts in the simulated experiment except for probability $\epsilon$. 
\item {\bf Decision:}
It runs the distinguisher $D$ on $\rho_\out$ and $\tilde{m}$, and outputs whatever $D$ outputs. 
\end{itemize}

By \Cref{lem:tech:extract_and_simulate}, the $(\rho_\out,\tilde{m})$ given to $D$ is indistinguishable from that in $\mcal{G}_1$ (up to an error $\epsilon$), which in turn is indistinguishable from that of the real MIM experiment. Since we have the freedom to set $\epsilon$ to any arbitrarily small noticeable function, this completes the reduction. 
%Using the computational indistinguishability of  $\mcal{G}_1$ and the real man-in-the-middel experiment, we can see that the above reduction works. 

%\xiao{@Takashi: Can I ask you to fill this part? I think you can summarize this part much bettern than me. We just need to pointing out what are the possible quantum weirdness, and how to handle them. This is the place to mention your quantum Extract-Simulate lemma.}

%\xiao{Let's denote this construction by $\langle C, R\rangle^{\msf{OneSided}}_{\msf{tg,PQ}}$. I'll change the notations in the main-body as-well.}

\para{Proof Idea of \Cref{lem:tech:extract_and_simulate}.} 
We assume that $\rho$ is a pure state w.l.o.g., and denote it by $\ket{\psi}$. By a similar usage of Jordan's lemma as in \cite{C:ChiChuYam21,cryptoeprint:2021:1516}, for any noticeable function $\delta(\secpar)$, we can decompose $\ket{\psi}$ as $\ket{\psi}=\ket{\psi_{< \delta}}+\ket{\psi_{\geq \delta}}$ such that 
\begin{align}\label{eq:small_part}
\Pr[\mcal{K}(1^\secpar,\ket{\psi_{< \delta}})=s^*]<  \delta(\secpar)    
\end{align}
and 
\begin{align}\label{eq:large_part}
\Pr[\mcal{K}(1^\secpar,\ket{\psi_{\geq \delta}})=s^*]\geq   \delta(\secpar).
\end{align}

By the contraposition of \Cref{item:tech:gamma_delta} of \Cref{lem:tech:extract_and_simulate} and \Cref{eq:small_part}, it holds that
$$
\Pr[b=\top : (b,\rho_\out) \leftarrow \mcal{G}(1^\secpar,\ket{\psi_{< \delta}})]<  \gamma(\secpar),
$$  
where $\gamma(\secpar)$ is a noticeable function that can be made arbitrarily small by taking sufficiently small $\delta(\secpar)$.  That is, $\mcal{G}$ simply ``rejects" (i.e., outputs $b=\bot$) except for a small  probability $\gamma(\secpar)$, in which case $\Gamma_b(s^*)=\bot$. The simulation of this case is easy because we do not need to extract $s^*$ except for probability $\gamma(\secpar)$, which can be made arbitrarily small. 
This shows \Cref{lem:tech:extract_and_simulate} for the case where $\ket{\psi}$ only has $\ket{\psi_{< \delta}}$ component. 

On the other hand, by \Cref{eq:large_part}, we can extract $s^*$ from $\ket{\psi_{\geq \delta}}$ with an overwhelming probability if we can repeat $\mcal{K}(1^\secpar,\ket{\psi_{\geq \delta}})$ for $\Theta(\delta(\secpar)^{-1}\secpar)$ times. Though such a repetition is not possible in general, the way of defining $\ket{\psi_{\geq \delta}}$ (as explained in \cite{cryptoeprint:2021:1516}) enables us to do so. As a result, we can extract $s^*$ from $\ket{\psi_{\geq \delta}}$ almost without disturbing the state by  the Almost-as-Good-as-New
lemma \cite[Lemma 2.2]{Aar05} (a.k.a.\ the gentle measurement lemma), which shows \Cref{lem:tech:extract_and_simulate} for the case where $\ket{\psi}$ only has $\ket{\psi_{\geq \delta}}$ component. 

In the above, we separately analyze 
$\ket{\psi_{< \delta}}$ and
$\ket{\psi_{\geq \delta}}$. In fact, we can see from the way we define them that they do not interfere with each other during the above described extraction procedure. 
As a result, the case of a superposition of $\ket{\psi_{< \delta}}$ and
$\ket{\psi_{\geq \delta}}$ can be reduced to the above two cases.

See \Cref{lem:extract_and_simulate} for the formal statement of the lemma and 
\Cref{sec:proof_extract_and_simulate} for the full proof.

\subsection{Full-Fledged Post-Quantum Non-Malleable Commitments}
\label{sec:tech-overview:full-fledged}

To make $\langle C, R\rangle^{\msf{OneSided}}_{\msf{tg,PQ}}$ a full-fledged post-quantum non-malleable commitments, we need to remove the one-side, small-tag, and synchronous restrictions. Due to space constraints, we will only briefly describe how to do that in this overview. We will provide more detailed overviews in the corresponding sections in the main body.

\para{Removing the One-Sided Restriction.} Notice that the non-malleability proof of $\langle C, R\rangle^{\msf{OneSided}}_{\msf{tg}}$ (and $\langle C, R\rangle^{\msf{OneSided}}_{\msf{tg,PQ}}$) makes use of the one-sided condition in both {\bf Step-3} and {\bf WIPoK-1}. Let us refer to these two steps as a {\bf Slot}. As demonstrated in \Cref{sec:tech-overview:small-tag:one-sided}
, it is the asymmetry condition $t < \tilde{t}$ for this {\bf Slot} that allows us to argue that the extracted $w'$ must contain the right committed value $\tilde{m}$ with good-enough probability.

We can use the following (standard) approach to remove the one-sided restriction. We simply repeat the {\bf Slot} twice, sequentially. The first repetition, referred to as {\bf Slot-A}, is identical to the original {\bf Slot}; In particular, both parties use the tag $t$. The second repetition, referred to as {\bf Slot-B}, will instead use $(n - t)$ as the tag. 

Now, in a MIM execution, if $t < \tilde{t}$, we can perform the same non-malleable proof as in \Cref{sec:tech-overview:small-tag:one-sided} (and \Cref{sec:tech-overview:small-tag:one-sided:sync:PQ}) with {\bf Slot-A} playing the role of {\bf Slot}. Otherwise (i.e., $t > \tilde{t}$), it must hold that $(n - t) < (n - \tilde{t})$; In this case, we can perform the non-malleable proof again, with {\bf Slot-B} playing the role of {\bf Slot}. (Note that by the definition of non-malleability, we do not need to consider the case $t = \tilde{t}$.)

\para{Handling Asynchronous Schedules.} 
%There are mainly three types of schedules in the asynchronous setting that may cause problems for our proof strategy shown in \Cref{sec:tech-overview:small-tag:one-sided} (which is  for the synchronous setting): 
First, we observe that the proof strategy used in \Cref{sec:tech-overview:small-tag:one-sided}, which is  for the synchronous setting, relies crucially on the following three conditions:% (In other words, the asynchronous schedules can be handled using the same proof strategy {\em if we can somehow maintain these conditions}.)   
\begin{enumerate}
\item
Both {\bf Step-1} and {\bf Step-2} in both the left and right sessions should finish before {\bf Step-3} in either of the two sessions starts. This is because our security proof considers each fixed prefix, which is the state right after the end of {\bf Step-2} (and before the start of {\bf Step-3}). 
%The right {\bf WIPoK-2} should not be interleaved with the left {\bf Step-1} and {\bf Step-2}. This is because we need to rewind the right {\bf WIPoK-2} to extract $\tilde{m}$ in $\mcal{G}_i$, while not rewinding the left Naor's commitment as we need to reduce non-malleability to its computationally-hiding property.\takashi{I don't think that is the real reason because rewinding Naor's commitment does not harm its security. The real reason should be because we fix each prefix in the security proof.}
\item
The right {\bf WIPoK-2} should not be interleaved with the left {\bf WIPoK-1}. This is because $\mcal{G}_i$ rewinds the left {\bf WIPoK-1}; Meanwhile, we need to rewind the right {\bf WIPoK-2} to extract $\tilde{m}$. We do not want these two parts to rewind each other recursively.  
\item
The left {\bf Step-3} should happen before the right {\bf WIPoK-2}. This is because in machine $\hat{\mcal{K}}_i$, brute-forcing is performed for the left {\bf Step-3}; At the same time, we need to rely on the WI property of the right {\bf WIPoK-2} to argue the similarity among all the $\hat{\mcal{K}}_i$'s for different $i \in [\tilde{t}]$ (i.e., \Cref{eq:tech-overview:Ki-K1}).
\end{enumerate}
To prove non-malleability in the asynchronous setting, our strategy is to introduce more gadgets into the protocol that enforce the above three conditions. As long as these conditions are satisfied, we can then rely on essentially the same proof of non-malleability as in the synchronous case. 

For example, to enforce the third condition above, we can add an extractable commitment $\ExtCom$ between {\bf Step-3} and {\bf WIPoK-1}, where $C$ commits to $m$ again (and additionally proves in {\bf WIPoK-2} that he did it as required). Then, in the MIM execution, if the left {\bf Step-3} happens after (the first message of) the right {\bf WIPoK-2}, then the right $\ExtCom$ must happen before the  left {\bf Step-3}. In this case, we can extract $\tilde{m}$ from the right $\ExtCom$ to finish the proof of non-malleability, instead of performing the same proof as in \Cref{sec:tech-overview:small-tag:one-sided}. We will use similar ideas to enforce  all the conditions mentioned above.

Of course, in the real proof, we need to do the above argument for the two slot-version described above in the {\bf Removing the One-Sided Restriction} part. Moreover, whenever we add one more gadget, cautions are needed to resolve new problems that it may cause by interacting with the ``already-introduced'' gadgets in some undesirable way. We will elaborate on this in \Cref{sec:small-tag:async:classical}, where we also provide a more detailed overview (in \Cref{sec:small-tag:classical:async:high-level}).

\para{Tag Amplification.} After the above two steps, we have already obtained a small-tag, asynchronous post-quantum non-malleable commitment $\langle C, R \rangle^{\msf{async}}_{\msf{tg,PQ}}$. As mentioned earlier, now we can apply the \cite{BLS21} tag amplifier. %which leads to the final full-fledged post-quantum non-malleable commitment. 
Here is one more caveat---The tag amplifier requires $\langle C, R \rangle^{\msf{async}}_{\msf{tg,PQ}}$ to additionally satisfy a condition called {\em robust extractability}, which means that the extractability holds even if the cheating committer interacts with an external bounded-round machine that cannot be rewound. We show that our $\langle C, R \rangle^{\msf{async}}_{\msf{tg,PQ}}$ already satisfies it, thanks to the added instances of $\ExtCom$ (which was introduced to handle asynchronous schedules originally).
Thus, we can apply their tag amplifier to our $\langle C, R \rangle^{\msf{async}}_{\msf{tg,PQ}}$ protocol to obtain the final full-fledged post-quantum non-malleable commitment scheme. 

\subsection{Organization}
In \Cref{sec:prelim}, we present necessary notations and preliminaries.

In \Cref{sec:small-tag-one-sided-sync-classical}, we show the small-tag, one-sided, synchronous non-malleable commitment $\langle C, R \rangle^{\msf{OneSided}}_{\msf{tg}}$ in the classical setting. 

In \Cref{sec:small-tag:classical:sync}, we remove the one-sided restriction, obtaining the small-tag, synchronous non-malleable commitment $\langle C, R \rangle^{\msf{sync}}_{\msf{tg}}$ (still in the classical setting).

In \Cref{sec:small-tag:async:classical}, we show how to handle asynchronous schedules {\em in the classical setting}. This leads to the small-tag, synchronous non-malleable commitment $\langle C, R \rangle^{\msf{async}}_{\msf{tg}}$. As mentioned earlier, the way we handle asynchronous schedules is quantum-friendly. We could have done this part directly for {\em the quantum version of} $\langle C, R \rangle^{\msf{sync}}_{\msf{tg}}$; But we find that it conveys our ideas better to first show the techniques in the classical setting, because it can avoid confusing the reader with quantum phenomena that are orthogonal to the discussion of asynchronous schedules. Once one understands our techniques for asynchronous schedules in the classical setting, it is easy to see why it extends to the post-quantum setting.

In \Cref{sec:extract_and_simulate}, we show a post-quantum extract-and-simulate lemma, which plays an important role later when we quantize our classical proof of non-malleability.

%In \Cref{pq:sec:small-tag-one-sided-sync-pq}, we show how to quantize the non-malleable proof of $\langle C, R \rangle^{\msf{OneSided}}_{\msf{tg}}$. This gives the post-quantum version $\langle C, R \rangle^{\msf{OneSided}}_{\msf{tg,PQ}}$ (i.e., a small-tag, one-sided, synchronous, polst-qunatum non-malleable commitment).

In \Cref{sec:full:pq:nmcom}, we describe  the full-fledged post-quantum non-malleable commitment. To do that, we first show in \Cref{summary:pq:sec:small-tag-one-sided-sync-pq} how to quantize the non-malleable proof of $\langle C, R \rangle^{\msf{OneSided}}_{\msf{tg}}$. This gives its post-quantum analog $\langle C, R \rangle^{\msf{OneSided}}_{\msf{tg,PQ}}$ (i.e., a small-tag, one-sided, synchronous, post-quantum non-malleable commitment).
Next, we describe in \Cref{sec:full:pq:nmcom:async} how to remove the one-sided and synchronous restrictions from $\langle C, R \rangle^{\msf{OneSided}}_{\msf{tg,PQ}}$, to obtain the protocol $\langle C, R \rangle^{\msf{async}}_{\msf{tg,PQ}}$; This step mimics the techniques we used in the classical setting converting $\langle C, R \rangle^{\msf{OneSided}}_{\msf{tg}}$ to $\langle C, R \rangle^{\msf{async}}_{\msf{tg}}$. Finally, in 
\Cref{sec:full:pq:nmcom:tag-amp}, we obtain the final full-fledged post-quantum non-malleable commitment by applying the \cite{BLS21}
tag amplifier to $\langle C, R \rangle^{\msf{async}}_{\msf{tg,PQ}}$.

In \Cref{sec:app-to-MPC}, we show the application of our post-quantum commitments to post-quantum MPC.

%!TEX root = ../main.tex
\section{Preliminaries}
\label{sec:prelim}
\subsection{Basic Notations} 
Let $\secpar \in \Naturals$ denote security parameter. 
For a positive integer $n$, let $[n]$ denote the set $\{1,2,...,n\}$.
For a finite set $\calX$, $x\sample \calX$ means that $x$ is uniformly chosen from $\calX$.

A function $f:\mathbb{N}\ra [0,1]$ is said to be \emph{negligible} if for all polynomial $p$ and sufficiently large $\secpar \in \mathbb{N}$, we have $f(\secpar)< 1/p(\secpar)$; it is said to be \emph{overwhelming} if $1-f$ is negligible, and said to be \emph{noticeable} if there is a polynomial $p$ such that $f(\secpar)\geq  1/p(\secpar)$ for sufficiently large $\secpar\in \mathbb{N}$.
We denote by $\poly$ an unspecified polynomial and by $\negl$ an unspecified negligible function.

Honest (classical) parties are modeled as interactive Turing machines (ITMs). We use PPT and QPT to denote (classical) probabilistic polynomial time and quantum polynomial time, respectively.
For a classical probabilistic or quantum algorithm $\A$, $y\sample \A(x)$ means that $\A$ is run on input $x$ and outputs $y$.
An adversary (or malicious party) is modeled as a non-uniform PPT (or QPT in the case of post-quantum security) algorithm.
When we consider a non-uniform QPT adversary, we often specify it by a sequence of polynomial-size quantum circuits with quantum advice $\{\A_\secpar, \rho_\secpar\}_{\secpar\in\mathbb{N}}$. 
In an execution with the security parameter $\secpar$, $\A$ runs $\A_{\secpar}$ taking $\rho_\secpar$ as the advice.  
For simplicity, we often omit the index  $\secpar$ and just write $\A(\rho)$ to mean a non-uniform QPT algorithm specified by  $\{\A_\secpar, \rho_\secpar\}_{\secpar\in \mathbb{N}}$.

For an $\NP$ language $\Lang$ and a true statement in this language $ x \in \Lang$, we use $\Relation_\Lang(x)$ ($\Relation$ stands for ``relation'') to denote the set of all witnesses for $x$.
We will refer to the OR-composition of $\NP$ languages, which are defined in \Cref{def:OR-Comp}.
\begin{definition}[OR-Composition of $\NP$ Languages]\label{def:OR-Comp}
Let $\Lang_1$ and $\Lang_2$ be two $\NP$ languages. The {\em OR-composition} of them (dubbed $\Lang_1 \vee \Lang_2$) is the new $\NP$ language defined as follows:
$$\Lang_1 \vee \Lang_2 \coloneqq \Set{(x_1, x_2) ~|~ x_1 \in \Lang_1 \vee x_2 \in \Lang_2}.$$
\end{definition}

\para{Notations for Indistinguishability.}
%We define computational and statistical indistinguishability of quantum states. % similarly to \cite{STOC:BitShm20,C:ChiChuYam21,cryptoeprint:2021:1516}.
%except that we denote by $\qcind$ insteaf of $\cind$ to mean computational indistinguishability against quantum distinguishers to distinct it from the classical computational indistinguishability.
We may consider random variables over bit strings or over quantum states. 
This will be clear from the context.
We use the same notations for classical and quantum computational indistinguishability, but there should be no fear of confusion; It means computational indistinguishability against PPT (resp.\ QPT) distinguishers whenever we consider classical (resp.\ post-quantum) security.
For ensembles of random variables $\mathcal{X}=\{X_i\}_{\secpar\in \mathbb{N},i\in I_\secpar}$ and $\mathcal{Y}=\{Y_i\}_{\secpar\in \mathbb{N},i\in I_\secpar}$ over the same set of indices $I=\bigcup_{\secpar\in\mathbb{N}}I_\secpar$ and a function $\delta$,       
we use $\mathcal{X}\compind_{\delta}\mathcal{Y}$ to mean that for any non-uniform PPT (resp.\ QPT) algorithm $\A$, there exists a negligible function $\negl(\cdot)$ such that for all $\secpar\in\mathbb{N}$, $i\in I_\secpar$, we have
\[
|\Pr[\A(X_i)]-\Pr[\A(Y_i)]|\leq \delta(\secpar) + \negl(\secpar).
\]
We say that $\mathcal{X}$ and $\mathcal{Y}$ are $\delta$-computationally indistinguishable if the above holds. 
In particular, when the above holds for $\delta=0$, we say that $\calX$ and $\calY$ are computationally indistinguishable, and simply write $\calX\compind \calY$.

Similarly, we use $\calX\statind_{\delta}\calY$ to mean that for any unbounded time  algorithm $\A$, there exists a negligible function $\negl(\cdot)$ such that for all $\secpar\in\mathbb{N}$, $i\in I_\secpar$, we have 
\[
|\Pr[\A(X_i)]-\Pr[\A(Y_i)]|\leq \delta(\secpar) + \negl(\secpar).
\]
In particular, when the above hold for $\delta=0$, we say that $\calX$ and $\calY$ are statistically indistinguishable, and simply write $\calX\statind \calY$.
Moreover,  
we write $\calX \equiv \calY$ to mean
that $X_i$ and $Y_i$ are distributed identically for all $i\in I$. 

When we consider an ensemble $\calX$ that is only indexed by $\secpar$ (i.e., $I_\secpar=\{\secpar\}$), we write $\calX=\{X_\secpar\}_\secpar$ for simplicity.

\if0
\paragraph{Interactive Quantum Machines and Oracle-Aided Quantum Machines.} 
The following explanation is almost taken verbatim from \cite{cryptoeprint:2021:1516}. 
We rely on the definition of interactive quantum machines and oracle-aided quantum machines that are given oracle access to an interactive quantum machine, following \cite{EC:Unruh12}.
Roughly, an interactive quantum machine $\A$ is formalized by a unitary over registers $\regM$ for receiving and sending messages,  and $\regA$ for maintaining $\A$'s internal state.
%We say that $\A$ is (non-uniform) QPT if the unitary can be implemented by a (non-uniform) QPT algorithm.
For two interactive quantum machines $\A$ and $\B$ that share the same message register $\regM$, an interaction between $\A$ and $\B$ proceeds by alternating invocations of $\A$ and $\B$ while exchanging messages over  $\regM$.

An oracle-aided quantum machine $\mathcal{S}$ given oracle access to an interactive quantum machine $\A$ with an initial internal state $\rho$ (denoted by $\mathcal{S}^{\A(\rho)}$) is allowed to apply the unitary part of $\A$ (the unitary obtained by deferring all measurements by $\A$ and omitting these measurements)  and its inverse in a black-box manner. $\mathcal{S}$ is only allowed to act on $\A$'s internal register $\regA$ through oracle access.
We refer to \cite{EC:Unruh12} for formal definitions of interactive quantum machines and black-box access to them.
\fi

\subsection{Commitment Schemes}
We define (classically-secure and post-quantum) commitments. The following definitions are based on those in \cite{cryptoeprint:2021:1516}. 
\begin{definition}[Commitment]\label{def:commitment}
A {\em commitment scheme} is a pair of PPT ITMs $\langle C, R \rangle$ that satisfies the following. Let $m\in \bits^{\ell(\secpar)}$ (where $\ell(\cdot)$ is some polynomial) is a message that $C$ wants to commit to. The protocol consists of the following stages:
\begin{itemize}
\item
{\bf Commit Stage:} $C(m)$ and $R$ interact with each other to generate a transcript $\tau$, a decommitment $\decom$ as $C$'s private output, and a decision value $b_{\mathrm{com}}\in\{\top,\bot\}$ as $R$'s private output.  
%$C$'s state $\ST_{C}$, and 
%$R$'s output $b_{\mathrm{com}}\in\bit$ indicating acceptance $(i.e., b_{\mathrm{com}}=1)$ 
%or rejection $(i.e., b_{\mathrm{com}}=0)$.
 We denote this execution by $(\tau,\decom,b_{\mathrm{com}}) \gets \langle C(m), R \rangle(1^\secpar)$.
 %When $C$ is honest, $\ST_C$ is classical, but when we consider a malicious quantum committer $C^*(\rho)$, we allow it to generate any quantum state $\ST_{C^*}$.  Similarly, a malicious quantum receiver $R^*(\rho)$ can output any quantum state, which we denote by $\OUT_{R^*}$ instead of $b_{\mathrm{com}}$. 
\item
{\bf Decommit Stage:}
$C$ sends the committed message $m$ and the decommitment $\decom$ to $R$, 
and $R$ outputs decision 
$b_{\mathrm{dec}}\in\{\top,\bot\}$. 
We denote this execution by $b_{\mathrm{dec}}\gets \Verify(\tau,m,\decom)$.\footnote{We assume that there is no state information for $R$ kept from the commit stage.}
W.l.o.g., we assume that whenever $R$ rejects at the end of the commit stage (i.e., $b_\mathrm{com} = \bot$), $R$ will reject at the end of decommit stage (i.e., $b_\mathrm{dec} = \bot$). (Note that w.l.o.g., one can think that $b_\mathrm{com}$ is included in $\tau$, because we can always modify the protocol to ask $R$ to send $b_\mathrm{com}$ as the last-round message of the commit stage.)
\end{itemize}
The scheme satisfies the following completeness requirement:
\begin{enumerate}
\item
{\bf Completeness.} For any polynomial $\ell:\mathbb{N} \rightarrow \mathbb{N}$ and any $m \in \bits^{\ell(\secpar)}$, it holds that
\begin{equation*}
\Pr[b_{\mathrm{com}}=b_{\mathrm{dec}}=\top : 
\begin{array}{l}
(\tau, \decom, b_{\mathrm{com}}) \gets \langle C(m),R \rangle(1^\secpar) \\
b_{\mathrm{dec}}\gets \Verify(\com,m,\decom)
\end{array}
] = 1.
\end{equation*}
\end{enumerate}
\end{definition}

\begin{definition}[Computationally Hiding]\label{def:comp_hiding}
A commitment scheme $\langle C, R \rangle$ is {\em computationally hiding} if for any 
non-uniform PPT receiver $R^*$ and any polynomial $\ell : \mathbb{N} \rightarrow \mathbb{N}$, the following holds:
% \begin{equation*}
% \left|\Pr[\OUT_{R^*}\langle C(m_0),R^* \rangle(1^\secpar)=1]-  \Pr[\{\OUT_{R^*}\langle C(m_1),R^* \rangle(1^\secpar)=1]\right|=\negl(\secpar)
% \end{equation*}
$$\big\{ \OUT_{R^*}\langle C(m_0),R^* \rangle(1^\secpar)\big\}_{\secpar \in \mathbb{N}, ~m_0, m_1 \in \bits^{\ell(\secpar)}} ~\cind~ \big\{ \OUT_{R^*}\langle C(m_1),R^* \rangle(1^\secpar)\big\}_{\secpar \in \mathbb{N}, ~m_0, m_1 \in \bits^{\ell(\secpar)}},$$
where $\OUT_{R^*}\langle C(m_b),R^* \rangle(1^\secpar)$ $(b \in \bits)$ denotes the output of $R^*$ at the end of the commit stage. 
%\begin{equation*}
%\{\OUT_{R^*}\langle C(m_0),R^*(\rho) \rangle(1^\secpar) \}_{\secpar} \cind \{\OUT_{R^*}\langle C(m_1),R^*(\rho) \rangle(1^\secpar) \}_{\secpar}
%\end{equation*}
%where $\OUT_{R^*}$ denotes the output of $R^*$, which can be quantum.

We say that $\langle C, R \rangle$ is \emph{post-quantum} computationally hiding if the above holds for all non-uniform \emph{QPT} $R^*$. 
\end{definition}

\begin{definition}[Statistically Binding]\label{def:stat-binding} 
A commitment scheme $\langle C, R \rangle$ is {\em statistically binding} if for any unbounded-time committer $C^*$, the following holds: 
\begin{align*}
    \Pr[
    \begin{array}{l}
    \exists~m_0,m_1,\decom_0,\decom_1,~s.t.~m_0\neq m_1 ~\land\\
     \Verify(\tau,m_0,\decom_0)=\Verify(\tau,m_1,\decom_1)=\top
    \end{array}
    :(\tau,\decom,b_{\mathrm{com}}) \gets \langle C^*, R \rangle(1^\secpar)]=\negl(\secpar).
\end{align*}
\end{definition}

\begin{definition}[Committed Values]\label{def:val}
For a commitment scheme $\langle C, R \rangle$, 
we define the value function as follows:
\begin{equation*}
    \val(\tau)\defeq 
    \begin{cases}
    m&\text{~if~}\exists\text{~unique~}m\text{~s.t.~}\exists~\decom, \Verify(\tau,m,\decom)=1\\
    \bot &\text{otherwise}
    \end{cases},
\end{equation*}
where $\tau$ and $\decom$ are defined in the commit stage in \Cref{def:commitment}.
%We say that $\tau$ is valid if $\val_{\Prot}(\tau)\neq \bot$ and invalid if $\val_{\Prot}(\tau)=\bot$.
\end{definition}

\subsection{Non-Malleable Commitments}\label{sec:nmcom}
\para{Classically Non-Malleable Commitments.} We first define non-malleable commitments in the classical setting. This definition follows the formalization in \cite{TCC:LinPasVen08,STOC:GoyPanRic16}. We consider a man-in-the-middle adversary $\mcal{M}$ interacting with a committer $C$ in the {\em left}, and a receiver $R$ in the {\em right}. We denote the relevant entities used in the right interaction as the ``tilde'd'' version of the corresponding entities on the left. In particular, suppose that $C$ commits to $m$ in the left interaction, and $\mcal{M}$ commits to $\tilde{m}$ on the right, i.e., we set $\tilde{m}=\val(\tilde{\tau})$ where $\tilde{\tau}$ is the transcript of the right session. Let $\msf{mim}^\mcal{M}_{\langle C, R \rangle}(\secpar, m, z)$ denote the random variable that is the pair $(\OUT_{\mcal{M}},\tilde{m})$, consisting of $\mcal{M}$'s output as well as the value committed to by $\mcal{M}$ on the right (assuming $C$ commits to $m$ on the left), where $z$ is $\mcal{M}$'s non-uniform advice. We use a {\em tag-based} (or ``identity-based'') specification, and ensure that $\mcal{M}$ uses a distinct tag $\tilde{t}$ on the right from the tag $t$ it uses on the left. This is done by stipulating that $\msf{mim}^\mcal{M}_{\langle C, R \rangle}(\secpar, m, z)$ outputs a special value $\bot_{tag}$ when $\mcal{M}$ uses the same tag in both the left and right executions. The reasoning is that this corresponds to the uninteresting case when $\mcal{M}$ is simply acting as a channel, forwarding messages from $C$ on the left to $R$ on the right and vice versa. 
\begin{definition}[Classical Non-Malleable Commitments]\label{def:NMCom:classical}
A commitment scheme $\langle C, R \rangle$ is said to be {\em (classically) non-malleable} if for every (non-uniform) PPT man-in-the-middle adversary $\mcal{M}$ and every polynomial $\ell:\Naturals \rightarrow \Naturals$, it holds that
$$\big\{\msf{mim}^\mcal{M}_{\langle C, R \rangle}(\secpar, m_0, z)\big\}_{\secpar \in \Naturals, m_0, m_1 \in \bits^{\ell(\secpar)}, z\in \bits^*} \cind \big\{\msf{mim}^\mcal{M}_{\langle C, R \rangle}(\secpar, m_1, z)\big\}_{\secpar \in \Naturals, m_0, m_1 \in \bits^{\ell(\secpar)}, z\in \bits^*}.$$
\end{definition}

\para{Post-Quantum Non-Malleable Commitments.} In the post-quantum setting, non-malleability can be defined similarly, except that the man-in-the-middle adversary can be QPT, instead of being PPT. (In particular, the final output of the adversary could be a quantum state.) To avoid using confusing notations, we put an over-line on top of the variable $\msf{mim}$ in the post-quantum setting.

\begin{definition}[Post-Quantum Non-Malleable Commitments]\label{def:NMCom:pq}
A commitment scheme $\langle C, R \rangle$ is said to be {\em post-quantumly non-malleable} if for every (non-uniform) QPT man-in-the-middle adversary $\mcal{M} = \Set{\mcal{M}_\secpar, \rho_\secpar}_{\secpar \in \Naturals}$ and every polynomial $\ell:\Naturals \rightarrow \Naturals$, it holds that
$$\big\{\bar{\msf{mim}}^{\mcal{M}_\secpar}_{\langle C, R \rangle}(\secpar,m_0, \rho_\secpar)\big\}_{\secpar \in \Naturals, m_0, m_1 \in \bits^{\ell(\secpar)}} \cind \big\{\bar{\msf{mim}}^{\mcal{M}_\secpar}_{\langle C, R \rangle}(\secpar,m_1, \rho_\secpar)\big\}_{\secpar \in \Naturals, m_0, m_1 \in \bits^{\ell(\secpar)}}.$$
where ``$\cind$'' refers to computational indistinguishability against QPT distinguishers.
\end{definition}

\begin{remark}[On Entangled Auxiliary Information.]
The above definition does not consider entanglement between $\mcal{M}$'s auxiliary input and distinguisher's auxiliary input. 
However, \cite[Claim 3.1]{BLS21} shows that the above definition implies the version that considers such entanglement. 
%\xiao{make a remark about the entanglement between $\mcal{M}$'s auxiliary input and the distinguisher's. Refer the reader to \cite[Claim 3.1]{BLS21} for more details.}
\end{remark}

\para{Synchronous Adversaries:} This notion refers to man-in-the-middle adversaries who upon receiving a message in the left (reps.\ right) session, immediately respond with the corresponding message in the right (resp.\ left) session. An adversary is said to be {\em asynchronous} if it is not synchronous.

\subsection{Extractable Commitments}
\para{Classically Extractable Commitments.} A commitment scheme is extractable if there exists an efficient extractor such that, the committed value can be extracted. We present the definition in \Cref{definition:ext-com}, which is taken from \cite{TCC:PasWee09}.  Constant-round constructions of (classically) extractable commitments are known from OWFs \cite{TCC:PasWee09}.
% Constructions for such commitment already existed implicitly in the implementation of concurrent zero-knowledge protocols in \cite{FOCS:PraRosSah02,TCC:Rosen04}. This concept and construction was made explicit in \cite{TCC:MOSV06}, which also inherited the concurrent extractability from \cite{FOCS:PraRosSah02}. The standalone version was later formalized and used in other works \cite{TCC:PasWee09,EC:GGJS12,EC:GarKiyPan17}. It suffices our purpose once the extractability property holds in a standalone setting. We now present the definition (in \Cref{definition:ext-com}) and construction (in \Cref{protocol:ext-com}) used in \cite{TCC:PasWee09}.
\begin{definition}[Classically Extractable Commitments]\label{definition:ext-com} A commitment $\ExtCom = \langle C,R\rangle$ is {\em extractable} if there exists an expected polynomial-time probabilistic oracle machine (the extractor) $\mathcal{E}$ that given oracle access to any PPT cheating committer $C^*$ outputs a pair $(\tau, \sigma^*)$ such that:
\begin{itemize}
\item
{\bf Simulation:} $\tau$ is identically distributed to the view of $C^*$ at the end of interacting with an honest receiver $R$ in commitment phase.

\item
{\bf Extraction:} the probability that $\tau$ is accepting and $\sigma^* =\bot$ is negligible.
\item
{\bf Binding:} if $\sigma^* \ne \bot$, then it is statistically impossible to open $\tau$ to any value other than $\sigma^*$.
\end{itemize}
\end{definition}

\para{Post-Quantum (Robust) Extractable Commitments.}
We define the post-quantum analog of extractable commitments. In the post-quantum setting, we often need an extractor that (almost) does not disturb the (potentially malicious) committer's state during the extraction. 
However, it is not known that such a post-quantum extractable commitments exist from (polynomially hard) post-quantum OWFs.\footnote{A recent work of \cite{cryptoeprint:2021:1543} gave a constant-round construction of such extractable commitments from super-polynomial hardness of post-quantum OWFs.} 
Fortunately, a recent work \cite{cryptoeprint:2021:1516} showed that a constant-round construction from post-quantum OWFs is possible if we relax the extractability to allow an (arbitrarily small) noticeable simulation error.  
The following definitions are taken from \cite{BLS21} with some notational adaptations.

Let $\langle C,R \rangle$ be a (possibly tag-based) commitment scheme. A sequential committed-value oracle $O^\infty[\langle C,R \rangle]$ acts as follows in interaction with a committer $C^*$: 
it interacts with $C^*$ in
many sequential sessions; in each session,
\begin{itemize}
    \item it participates with $C^*$
in the commit phase of $\langle C,R \rangle$ as the honest receiver $R$ (using a tag
chosen adaptively by $C^*$), obtaining a transcript $\tau$, {\bf and}
\item if $C^*$ is non-aborting in the commit phase and sends request $\mathsf{break}$, it returns $\val(\tau)$.
\end{itemize}
The single-session oracle $O^1[\langle C,R \rangle]$ is similar to $O^\infty[\langle C,R \rangle]$, except that it interacts with the adversary in a single session. When the commitment scheme is clear from the context, we write $O^\infty$ (or $O^1$)
for simplicity. 

%A formal definition of such post-quantum extractable commitments is given below.\footnote{The following definition is referred to as \emph{strong} extractability with $\epsilon$-simulation in \cite{cryptoeprint:2021:1516}.}

Then, the definition of post-quantum $\epsilon$-simulatable commitment is given below. 

\begin{definition}[Post-Quantum $\epsilon$-Simulatable Extractable Commitment]\label{def:epsilon-sim-ext-com:pq}
A commitment scheme $\langle C, R \rangle$ is said to be a {\em post-quantum $\epsilon$-simulatable extractable commitment} if it satisfies the following in addition to post-quantum computational hiding and statistical binding. 
There exists a QPT algorithm $\SimExt$ such that for any noticeable $\epsilon(\secpar)$ and any non-uniform QPT committer $\{C^*_\secpar,\rho_\secpar\}_{\secpar\in \mathbb{N}}$,
\begin{equation*}
\big\{ \SimExt(1^\secpar,1^{\epsilon^{-1}},C^*_\secpar,\rho_\secpar) \big\}_{\secpar \in \mathbb{N}}
\cind_\epsilon
\big\{{C^*_\secpar}^{O^{\infty}}(\rho_\secpar)\big\}_{\secpar \in \mathbb{N}},
\end{equation*}
where ``$\cind_\epsilon$'' refers to $\epsilon$-close computational indistinguishability against QPT distinguishers.
\end{definition}

\if0
\begin{definition}[Post-Quantum $\epsilon$-Simulatable Extractable Commitment]\label{def:epsilon-sim-ext-com:pq}
A commitment scheme $\langle C, R \rangle$ is said to be a {\em post-quantum $\epsilon$-simulatable extractable commitment} if it satisfies the following in addition to post-quantum computational hiding and statistical binding. 
There exists a QPT algorithm $\SimExt$ such that for any noticeable $\epsilon(\secpar)$, any non-uniform QPT $\A(\rho)$, and $r$-round interactive machine $\B$,  
\begin{equation*}
\big\{ \OUT_{\SimExt}\langle \B,\SimExt^{\A(\rho)}(1^\secpar,1^{\epsilon^{-1}}) \big\}_\secpar
\cind_\epsilon
\big\{\OUT_{\A}\langle \B,\A^{O^{\infty}}(\rho)\rangle \big\}_\secpar
\end{equation*}
where ``$\cind_\epsilon$'' refers to $\epsilon$-close computational indistinguishability against QPT distinguishers.
\end{definition}
\fi

\begin{lemma}[{\cite[Lemma 10]{cryptoeprint:2021:1516}}]
Assuming the existence of post-quantum OWFs, there exist constant-round post-quantum $\epsilon$-simulatable extractable commitments.   
\end{lemma}
\begin{remark}
The $\epsilon$-simulatable extractability as defined in \Cref{def:epsilon-sim-ext-com:pq} is called post-quantum \emph{strong} $\epsilon$-simulatable extractability in \cite{cryptoeprint:2021:1516}. Though their definition looks different from \Cref{def:epsilon-sim-ext-com:pq}, they are actually equivalent---First, it is easy to see that \Cref{def:epsilon-sim-ext-com:pq} is equivalent to a modified version of \Cref{def:epsilon-sim-ext-com:pq} where we give $O^1$ instead of $O^{\infty}$. 
Second, \cite[Lemma 3.2]{BLS21} shows that the $O^1$ and $O^{\infty}$ versions of \Cref{def:epsilon-sim-ext-com:pq} are equivalent.
\end{remark}

We also define the \emph{robust} version of the above definition following \cite{BLS21}. 
Roughly speaking, $r$-robust $\epsilon$-simulatable extractability means that the $\epsilon$-simulatable extractability holds even if the adversary interacts with an external $r$-round interactive machine that cannot be rewound by the extractor. 
The formal definition is given below. 

\begin{definition}[Post-Quantum $r$-Robust $\epsilon$-Simulatable Extractable Commitment]\label{def:robust-epsilon-sim-ext-com:pq}
A commitment scheme $\langle C, R \rangle$ is said to be a {\em post-quantum $r$-robust $\epsilon$-simulatable extractable commitment} if it satisfies the following in addition to post-quantum computational hiding and statistical binding. 
There exists a QPT algorithm $\SimExt$ such that for any noticeable $\epsilon(\secpar)$, non-uniform QPT committer $\{C^*_\secpar,\rho_\secpar\}_{\secpar\in \mathbb{N}}$, and $r$-round interactive machine $B$, 
\begin{equation*}
\big\{ \OUT_{\SimExt}\langle B(1^\secpar,z), \SimExt(1^\secpar,1^{\epsilon^{-1}},C^*_\secpar,\rho_\secpar)\rangle \big\}_{\secpar\in \mathbb{N}, z\in \bit^*}
\cind_\epsilon
\big\{ \OUT_{C^*_\secpar}\langle B(1^\secpar,z),{C^*_\secpar}^{O^{\infty}}(\rho_\secpar)\rangle\big\}_{\secpar\in \mathbb{N}, z\in \bit^*}
\end{equation*}
where ``$\cind_\epsilon$'' refers to $\epsilon$-close computational indistinguishability against QPT distinguishers.
\end{definition}

\subsection{Witness-Indistinguishable Proofs of Knowledge}
\label{sec:prelim:WIPoK}
\para{Classically Witness-Indistinguishable Proofs of Knowledge.} First, we provide a definition from \cite{Goldreich01,JC:Lindell03}.
\begin{definition}[Proofs of Knowledge \cite{Goldreich01,JC:Lindell03}]\label{def:PoK}
Let $\Lang$ be an $\NP$ language and $\kappa: \Naturals \rightarrow [0,1]$. We say that a pair of PPT ITMs $\langle P, V \rangle$ is an interactive proof of knowledge for $\Lang$ with knowledge error $\kappa$ if the following two conditions hold:
\begin{enumerate}
\item {\bf Completeness.} For every true statement $x\in\Lang$ and every witness $w \in \Relation_{\Lang}(x)$, it holds that
$$\Pr[\OUT_V\langle P(w), V \rangle(x)= \top] = 1.$$
\item {\bf Validity (with error $\kappa$).} There exists a polynomial $q(\cdot)$ and a probabilistic oracle machine $K$ such that for every $P^*$, and every $x,z, r \in \bits^*$, machine $K$ satisfies the following condition: 
\begin{quote}
Denote by $p(x,z,r)$ the probability that $V(x)$ accepts when interacting with the prover specified by $P^*(x,z;r)$. If $p(x,z,r)>\kappa(|x|)$, then, on input $x$ and with oracle access to oracle $P^*(x,z;r)$, $K$ outputs a solution $w \in \Relation_\Lang(x)$ within an expected number of steps bounded by $\frac{q(|x|)}{p(x, z, r) - \kappa(|x|)}$.
\end{quote}
\end{enumerate}
\end{definition}

We present an alternative formalism of knowledge soundness called {\em witness-extended emulation} due to Lindell \cite{JC:Lindell03}. This formalism turns out to be more suitable for our application. 
%\takashi{low priority comment: I'm wondering if we should unify ``emulation" and ``simulation". I believe they are the same meaning in the current manuscript.}

\begin{definition}[PoK via Witness-Extended Emulation {\cite{JC:Lindell03}}]\label{def:WEE}
Let $\Lang$ be an $\NP$ language and $\kappa: \Naturals \rightarrow [0,1]$. We say that a pair of ITMs $\langle P, V \rangle$ is an interactive proof of knowledge for $\Lang$ with {\em witness-extended emulation} if it satisfies the same completeness requirement as in \Cref{def:PoK}, and the following requirement: There exists an expected PPT oracle machine $\msf{WE}$, called the {\em emulation-extractor},  such that for any machine $P^*$ the following conditions hold 
%\xiao{Modify this definition: change $\mcal{T}$ to the final of $P^*$. Find where Lindell did this and cite it.}:
\begin{enumerate}
\item\label[Condition]{item:WEE:1}
$\Set{(\ST'_{P^*},b'): (\ST'_{P^*},b', w') \gets \mcal{WE}^{P^*(x, z;r)}(x)}_{\secpar,x, z, r} \equiv \Set{(\ST_{P^*},b):(\ST_{P^*},b) \gets \langle P^*(x, z;r), V(x)\rangle }_{\secpar,x, z, r}$,
where in the RHS above, $\ST_{P^*}$ and $b$ denote the state of $P^*$ and output of $V$ respectively, at the end of the execution $\langle P^*(x, z;r), V(x)\rangle$.
\item\label[Condition]{item:WEE:2}
$\Pr[b'=1~\wedge~w'\notin \Relation_{\Lang}(x) : (\ST'_{P^*},b', w') \gets \mcal{WE}^{P^*(x, z;r)}(x)] = \negl(|x|)$.
\end{enumerate}
% \takashi{To match the post-quantum definition, I slightly modified the definition}
\if0
\begin{enumerate}
\item\label[Condition]{item:WEE:1}
$\Set{\mcal{T}': (\mcal{T}', w') \gets \mcal{WE}^{P^*(x, z;r)}(x)}_{x, z, r} \equiv \Set{\mcal{T}:\mcal{T} \gets \langle P^*(x, z;r), V(x)\rangle }_{x, z, r}$;
\item\label[Condition]{item:WEE:2}
$\Pr[\mcal{T}'~\text{is accepting}~\wedge~w'\notin \Relation_{\Lang}(x) : (\mcal{T}', w') \gets \mcal{WE}^{P^*(x, z;r)}(x)] = \negl(|x|)$.
\end{enumerate}
\fi
\end{definition}
The following lemma connects \Cref{def:WEE} with \Cref{def:PoK}.
\begin{lemma}[{\cite[Lemma 3.1]{JC:Lindell03}}]
Let $\langle P, V\rangle$ be a proof of knowledge for $\Lang \in \NP$ with negligible knowledge error. Then, $\langle P, V\rangle$ is also a proof of knowledge for $\Lang$ with witness-extended emulation.
\end{lemma}

% \xiao{Add the claim I showed to Omkant as a corollary. Also, ask Takshi if that argument extends to $\epsilon$-quantum setting easily.}
Next, we define witness indistinguishable proof
of knowledge:
\begin{definition}[WIPoKs]\label{def:WIPoK}
A witness-indistinguishable proof
of knowledge for an $\NP$ language $\Lang$ is a proof of knowledge for $\Lang$ with witness extended emulation (as per \Cref{def:WEE}) which additionally satisfies the following property
\begin{itemize}
\item
{\bf Witness-Indistinguishability.} For any non-uniform PPT machine $V^*$, we have
$$\Set{\OUT_{V^*}\langle P(w_0), V^*\rangle(x)}_{\secpar, x, w_0, w_1} \cind \Set{\OUT_{V^*}\langle P(w_1), V^*\rangle(x)}_{\secpar, x, w_0, w_1},$$
where $\secpar \in \Naturals$, $x \in \Lang \cap \bits^\secpar$, and $w_0, w_1 \in \Relation_\Lang(x)$.
\end{itemize}
\end{definition}

It is well-known that constant-round witness-indistinguishable proofs of knowledge (as per \Cref{def:WIPoK}) can be built from one-way functions.

\para{Post-Quantum Witness-Indistinguishable Arguments of Knowledge}
We define a post-quantum analog of the WIPoKs described in \Cref{def:WIPoK}. A natural way to do so is to just replace all adversaries with any QPT algorithms.
More precisely, we require that the witness indistinguishability holds against QPT distinguishers and that the knowledge extractor extracts a witness without disturbing the prover's state.  
However, a constant-round construction for such WIPoKs is not known based on (polynomially hard) post-quantum OWFs.\footnote{A recent work of \cite{cryptoeprint:2021:1543} gave such WIPoKs from super-polynomial hardness of post-quantum OWFs.} 
Fortunately, a recent work \cite{cryptoeprint:2021:1516} showed that a constant-round construction from post-quantum OWFs is possible if we relax the proof of knowledge via witness-extended emulation to \emph{argument} of knowledge via witness-extended \emph{$\epsilon$-close} emulation, where a cheating prover is limited to QPT and a (arbitrarily small) noticeable simulation error is allowed. 
A formal definition of such post-quantum WIAoKs is given below.
\begin{definition}[Post-Quantum WIAoK with $\epsilon$-Close Emulation]\label{def:PQWIAoK}
Let $\Lang$ be an $\NP$ language. We say that a pair of PPT ITMs $\langle P, V \rangle$ is a post-quantum witness-distinguishable arguments of knowledge with $\epsilon$-close emulation
for $\Lang$ if the following three conditions hold:
\begin{enumerate}
\item {\bf Completeness.} For every $(x,w)\in\Lang$, it holds that
$$\Pr[\OUT_V\langle P(w), V \rangle(x)= \top] = 1.$$
\item {\bf AoK via Witness-Extended $\epsilon$-Close Emulation.}\label[Property]{item:PQWEE} 
There exists a QPT oracle machine $\mcal{WE}$, called the {\em witness-extended emulator},  such that for any non-uniform QPT machine $\{P^*_{\secpar},\rho_\secpar\}_{\secpar\in \Naturals}$ the following conditions hold: for any noticeable function $\epsilon(\cdot)$ (referred to as the error parameter),  \takashi{modified the interface of $\mcal{WE}$ so that it takes $P^*_\secpar$ and $\rho_\secpar$ as an input instead of an oracle. Please fix it if the oracle version is left.}
\begin{enumerate}
\item\label[Condition]{item:PQWEE:1}
$\Set{(\ST'_{\P^*},b'): (\ST'_{\P^*},b', w') \gets \mcal{WE}(1^{\epsilon^{-1}},P^*_{\secpar},\rho_{\secpar},x)}_{\secpar,x} \cind_{\epsilon} \Set{(\ST_{\P^*},b):(\ST_{P^*},b) \gets \langle P^*_{\secpar}(\rho_{\secpar}), V(x)\rangle }_{\secpar,x}$
where $\secpar \in \Naturals$, $x\in \bit^\secpar$, and ``$\cind_{\epsilon}$'' refers to $\epsilon$-computational indistinguishability against QPT distinguishers;
\item\label[Condition]{item:PQWEE:2}
$\Pr[b'=1~\wedge~w'\notin \Relation_{\Lang}(x) : (\ST'_{\P^*},b', w') \gets \mcal{WE}(1^{\epsilon^{-1}},P^*_{\secpar},\rho_{\secpar},x)] =0$,\footnote{We can assume the probability is $0$ instead of $\negl(|x|)$ without loss of generality. If the probability is $\negl(|x|)$, we can modify $\mcal{WE}$ to output $b'=0$ whenever $w'\notin \Relation_{\Lang}(x)$. This only negligibly affects the distribution in the LHS of \Cref{item:PQWEE:1}, which can be absorbed into $\cind_{\epsilon}$.}
\end{enumerate}
where $(\ST_{P^*},b) \gets \langle P^*_{\secpar}(\rho^*_{\secpar}), V(x)\rangle$ means that $\ST_{P^*}$ is the final state of $P^*_{\secpar}$ and $b$ is the decision output by $V$.  
 \item
{\bf Witness-Indistinguishability.} For any non-uniform QPT machine $\Set{V^*_\secpar,\rho_\secpar}_{\secpar \in \Naturals}$, we have
$$\Set{\OUT_{V^*_\secpar}\langle P(w_0), V^*_\secpar(\rho_\secpar)\rangle(x)}_{\secpar, x, w_0, w_1} \cind \Set{\OUT_{V^*_\secpar}\langle P(w_1), V^*_\secpar(\rho_\secpar)\rangle(x)}_{\secpar, x, w_0, w_1},$$
where $\secpar \in \Naturals$, $x \in \Lang \cap \bits^\secpar$, $w_0, w_1 \in \Relation_\Lang(x)$, and ``$\cind$'' refers to computational indistinguishability against QPT distinguishers.
\end{enumerate}
\end{definition}

The work of \cite{cryptoeprint:2021:1516} gives a constant-round construction of post-quantum \emph{$\epsilon$-zero-knowledge} arguments of knowledge from post-quantum OWFs. Since $\epsilon$-zero-knowledge implies witness indistinguishability, we have the following lemma.
\begin{lemma}[{\cite[Corollary 2]{cryptoeprint:2021:1516}}]
Assuming the existence of post-quantum OWFs, there exist constant-round post-quantum witness-indistinguishable arguments of knowledge as per \Cref{def:PQWIAoK}.
\end{lemma}

%!TEX root = ../main.tex
\section{Small-Tag, One-Sided, Synchronous, Classical Setting}
\label{sec:small-tag-one-sided-sync-classical}

In this section, we show the small-tag, one-sided, synchronous non-malleable commitment scheme $\langle C, R \rangle^{\msf{OneSided}}_{\msf{tg}}$ in the classical setting. 
Though our final goal is to construct \emph{post-quantum} non-malleable commitments, we give a classically-secure construction here since the security proof for its post-quantum version $\langle C, R \rangle^{\msf{OneSided}}_{\msf{tg,PQ}}$ (in \Cref{pq:sec:small-tag-one-sided-sync-pq}) is very similar to that in the classical case except for one step where we rely on the extract-and-simulate lemma (\Cref{lem:extract_and_simulate}).

\subsection{Construction of \textnormal{$\langle C, R \rangle^{\msf{OneSided}}_{\msf{tg}}$}}

An intuitive explanation of this construction is already provided in \Cref{sec:tech-overview:small-tag:one-sided}. We now present the formal description in \Cref{prot:one-sided:classical}. 
This construction is based on the following building blocks:
%We need the following building blocks:
\begin{itemize}
\item
An {\em injective} OWF $f$; To convey the main idea better, we first present the construction assuming the existence of {\em injective} OWFs.  
We explain how to relax the assumption to the existence to {\em any} OWFs in \Cref{sec:removing-injectivity}. 
\item
Naor's commitment; We use $\beta$ to denote the first message for Naor's commitment, and $\Com_\beta(m;r)$ to denote the second message where a string $m$ is committed using randomness $r$.
\item
A witness-indistinguishable proof of knowledge $\WIPoK$. %with knowledge error $\kappa(\secpar) = \negl(\secpar)$ 
\end{itemize}

\begin{ProtocolBox}[label={prot:one-sided:classical}]{Small-Tag One-Sided Synchronous NMCom \textnormal{$\langle C, R\rangle^{\msf{OneSided}}_{\msf{tg}}$}}
The tag space is defined to be $[n]$ where $n$ is a polynomial on $\secpar$. Let $t \in [n]$ be the tag for the following interaction. Let $m$ be the message to be committed to. 

\para{Commit Stage:}
\begin{enumerate}
\item\label[Step]{item:one-sided:step:Naor-rho}
Receiver $R$ samples and sends the first message $\beta$ for Naor's commitment; 
\item\label[Step]{item:one-sided:step:committing}
Committer $C$ commits to $m$ using the second message of Naor's commitment. Formally, $C$ samples a random tape $r$ and sends $\msf{com} = \Com_\beta(m;r)$;
\item\label[Step]{item:one-sided:step:OWFs}
$R$ computes $\Set{y_i = f(x_i)}_{i \in [t]}$ with $x_i \pick \bits^\secpar$ for each $i \in [t]$. $R$ sends $Y = (y_1, \ldots, y_t)$ to $C$;
\item\label[Step]{item:one-sided:step:WIPoK:1}
{\bf (WIPoK-1.)}\footnote{We assume that the first round of this step goes to the opposite direction to \Cref{item:one-sided:step:OWFs}. This is true for typical WIPoK constructions. See \Cref{rmk:optional-ACK} for more details.} $R$ and $C$ execute an instance of $\WIPoK$ where $R$ proves to $C$ that he ``knows'' the pre-image of some $y_i$ contained in $Y$ (defined in \Cref{item:one-sided:step:OWFs}). Formally, $R$ proves that $Y \in \Lang^t_f$, where \begin{equation}\label[Language]{eq:one-sided:Lang:OWF}
\Lang^t_f \coloneqq \Set{(y_1, \ldots, y_t)~|~\exists (i, x_i) ~s.t.~ i \in [t] \wedge y_i = f(x_i)}.
\end{equation}
Note that $R$ uses $(1,x_1)$ as the witness when executing this $\WIPoK$.
\item\label[Step]{item:one-sided:step:WIPoK:2}
{\bf (WIPoK-2.)} $C$ and $R$ execute an instance of $\WIPoK$ where $C$ proves to $R$ that he ``knows'' {\em either} the message committed in $\msf{com}$ (defined in \Cref{item:one-sided:step:committing}), {\em or} the pre-image of some $y_i$ contained in $Y$ (defined in \Cref{item:one-sided:step:OWFs}). Formally, $C$ proves that $(\msf{com}, Y) \in \Lang_\beta \vee \Lang^t_f$, where $\Lang_\beta \vee \Lang^t_f$ denotes the  OR-composed language (as per \Cref{def:OR-Comp}), $\Lang^t_f$ was defined in \Cref{eq:one-sided:Lang:OWF} and 
\begin{equation}\label[Language]{eq:one-sided:Lang:Com}
\Lang_\beta\coloneqq \Set{\msf{com} ~|~ \exists (m, r)~s.t.~ \msf{com} = \Com_\beta(m;r)}.
\end{equation}
Note that $C$ uses the $(m, r)$ defined in \Cref{item:one-sided:step:committing} as the witness when executing this $\WIPoK$.
\end{enumerate}
\para{Decommit Stage:} 
$C$ sends $(m, r)$. $R$ accepts if $\msf{com} = \Com_\beta(m;r)$, and rejects otherwise.
\end{ProtocolBox}

\begin{remark}[Inserting an ACK round if necessary]\label{rmk:optional-ACK}
In \Cref{prot:one-sided:classical}, we emphasize that \Cref{item:one-sided:step:OWFs} and the first message of \Cref{item:one-sided:step:WIPoK:1} should be {\em separated}, i.e., they cannot be combined into one round even if the first message of \Cref{item:one-sided:step:WIPoK:1} happens to go from $R$ to $C$ as well, which may be possible depending on the choice of WIPoK protocols instantiating \Cref{item:one-sided:step:WIPoK:1}. In that case, one can simply insert a dummy round right after \Cref{item:one-sided:step:OWFs}, where $C$ sends an ``ACK'' symbol to acknowledge the reception of \Cref{item:one-sided:step:OWFs}; Only then will $R$ starts the execution of \Cref{item:one-sided:step:WIPoK:1}. This is necessary because some steps in our security proof need to non-uniformly fix the execution up to the closing of \Cref{item:one-sided:step:OWFs}, and make use of the (non-uniform) WI property of \Cref{item:one-sided:step:WIPoK:1}.
\end{remark}

\para{Security.} Completeness is straightforward from the description of \Cref{prot:one-sided:classical}. 
The statistical binding property follows from that of Naor's commitment. Computational-hiding property of any non-malleable commitment scheme follows directly from its non-malleability. So, we only need to prove the non-malleability of our protocol, which is established by the following theorem.
\begin{theorem}\label{thm:one-sided:non-malleability}
The commitment scheme $\langle C, R\rangle^{\msf{OneSided}}_{\msf{tg}}$ in \Cref{prot:one-sided:classical} is non-malleable against  one-sided synchronous PPT adversaries with tag space $[n]$, with $n$ being any polynomial on $\secpar$.
\end{theorem}	
We prove  \Cref{thm:one-sided:non-malleability} in subsequent subsections.

\subpara{Structure of the Proof of \Cref{thm:one-sided:non-malleability}.}
Since the proof of \Cref{thm:one-sided:non-malleability} is lengthy, we provide a road map here.
We prove \Cref{thm:one-sided:non-malleability} in \Cref{one-sided:non-malleability:proof:classical,sec:one-sided:core-lemma:proof,lem:small-tag:proof:se:proof,sec:lem:small-tag:proof:se:proof:K:proof,sec:lem:bound:Ki:proof,sec:proof:claim:K'':non-abort}.
In each subsection, we introduce a lemma or claim whose proof is deferred to the next subsection. Specifically,
\begin{itemize}
    \item In the rest of \Cref{one-sided:non-malleability:proof:classical}, we prove \Cref{thm:one-sided:non-malleability} assuming that \Cref{lem:one-sided:proof:core} is correct.
     \item In \Cref{sec:one-sided:core-lemma:proof}, we prove \Cref{lem:one-sided:proof:core} assuming that \Cref{lem:small-tag:proof:se} is correct.
      \item In  \Cref{lem:small-tag:proof:se:proof}, we prove \Cref{lem:small-tag:proof:se} assuming that \Cref{lem:small-tag:proof:se:proof:K} is correct.
        \item In  \Cref{sec:lem:small-tag:proof:se:proof:K:proof}, we prove \Cref{lem:small-tag:proof:se:proof:K} assuming that \Cref{lem:bound:Ki} is correct.
        \item  In  \Cref{sec:lem:bound:Ki:proof}, we prove \Cref{lem:bound:Ki} assuming that \Cref{claim:K'':non-abort} is correct.
        \item In \Cref{sec:proof:claim:K'':non-abort}, we prove \Cref{claim:K'':non-abort}. 
\end{itemize}
Putting everything together, we complete the proof of  \Cref{thm:one-sided:non-malleability}.
\begin{remark}
We remark that the proof of \Cref{lem:small-tag:proof:se} is the only step that cannot be directly translated to the post-quantum setting. Looking ahead, the post-quantum counterpart of this step (i.e., \Cref{pq:lem:small-tag:proof:se}) will make use of a new \emph{post-quantum extract-and-simulate} lemma, which we state and prove in \Cref{sec:extract_and_simulate}.
\end{remark}
%\takashi{I think this information should be useful for readers. Is it fine to put it here?} \xiao{I agree.}
%, which enables us to do the following rewinding argument in the post-quantum setting.  
Moreover, we provide a moderately-detailed summary of this proof in \Cref{sec:proof:summar:one-sided}, with the hope to further elucidate the logic flow behind the proof of \Cref{thm:one-sided:non-malleability}. It is recommended that the reader start reading \Cref{sec:proof:summar:one-sided} after he/she has read  
\Cref{one-sided:non-malleability:proof:classical,sec:one-sided:core-lemma:proof,lem:small-tag:proof:se:proof,sec:lem:small-tag:proof:se:proof:K:proof,sec:lem:bound:Ki:proof,sec:proof:claim:K'':non-abort,sec:removing-injectivity} (at least once), because \Cref{sec:proof:summar:one-sided} uses several notations defined in preceding subsections.

\subsection{Proving Non-Malleability (Proof of \Cref{thm:one-sided:non-malleability})}
\label{one-sided:non-malleability:proof:classical}

% To prove \Cref{thm:one-sided:non-malleability}, we show that
% \begin{equation}\label[Expression]{eq:one-sided:goal}
% \Set{\msf{mim}^\mcal{M}_{\langle C, R \rangle}(\secpar, m_0, z)}_{\secpar, m_0, m_1, z} \cind \Set{\msf{mim}^\mcal{M}_{\langle C, R \rangle}(\secpar, m_1, z)}_{\secpar, m_0, m_1, z},
% \end{equation}
% where $\secpar \in \Naturals$, $m_0, m_1 \in \bits^{\secpar}$, and $z \in \bits^*$.

In the rest of this section, we write $\langle C, R \rangle$ to mean $\langle C, R\rangle^{\msf{OneSided}}_{\msf{tg}}$ for notational simplicity. 
To prove \Cref{thm:one-sided:non-malleability}, let us first define the game that captures the man-in-the-middle execution corresponding to $\msf{mim}^\mcal{M}_{\langle C, R \rangle}(\secpar, m, z)$ w.r.t.\ the $\langle C, R \rangle$ defined in \Cref{prot:one-sided:classical}.

\para{Game $H^{\mcal{M}}(\secpar,m,z)$:\label{gameH:description}} This is the man-in-the-middle execution of the commit stage of the $\langle C, R \rangle$ defined in \Cref{prot:one-sided:classical}, where the left committer commits to $m$ and $\mcal{M}$'s non-uniform advice is $z$. The output of this game consists of the following three parts:
\begin{enumerate}
\item
$\OUT_{\mcal{M}}$: this is the output of $\mcal{M}$ at the end of this game;
\item
$\tilde{\tau}$: this is defined to be $\tilde{\tau} \coloneqq (\tilde{\beta}, \tilde{\msf{com}})$, where $\tilde{\beta}$ and $\tilde{\msf{com}}$ are the \Cref{item:one-sided:step:Naor-rho} and \Cref{item:one-sided:step:committing} messages exchanged between $\mcal{M}$ and the honest receiver $R$ (i.e., in the right session);
\item
$b \in \Set{\top, \bot}$: this is the output of the honest receiver $R$, indicating if the man-in-the-middle's commitment (i.e., the right session) is accepted ($b = \top$) or not ($b = \bot$).
\end{enumerate}

\para{Notation.} For any $(\OUT_{\mcal{M}}, \tilde{\tau}, b)$ in the support of $H^{\mcal{M}}(\secpar,m,z)$, the following $\msf{val}_b(\tilde{\tau})$ defines the value committed in the man-in-the-middle's commitment (i.e.,  the right session):
\begin{equation}\label{eq:def:val-b-tau}
\msf{val}_b(\tilde{\tau}) \coloneqq 
\begin{cases}
\msf{val}(\tilde{\tau}) & b = \top\\
\bot & b = \bot
\end{cases},
\end{equation}
where $\msf{val}(\tilde{\tau})$ denote the value statistically-bound in \Cref{item:one-sided:step:Naor-rho,item:one-sided:step:committing} of the right session. (Recall that these two steps constitute a Naor's commitment.) Also, throughout this paper,  we assume for simplicity that Naor's commitment is perfectly binding (instead of only statistically binding). This assumption only suppresses an {\em additive error of negligible amount} in relevant lemmas, which will not affect any of our results.

By definition, it is easy to see that:
\begin{equation}\label{eq:classical:mim-H}
\big\{\msf{mim}^\mcal{M}_{\langle C, R \rangle}(\secpar, m, z)\big\} \idind \big\{\big(\OUT_{\mcal{M}}, \val_b(\tilde{\tau})\big): (\OUT_{\mcal{M}}, \tilde{\tau}, b) \gets H^{\mcal{M}}(\secpar,m,z) \big\},
\end{equation}
where both ensembles are indexed by $\secpar \in \Naturals$, $m \in \bits^{\ell(\secpar)}$, and $z \in \bits^*$.

Therefore, to prove that \Cref{prot:one-sided:classical} satisfies \Cref{def:NMCom:classical}, it suffices to show the following:
\begin{align*}
& \big\{\big(\OUT^0_{\mcal{M}}, \val_{b^0}(\tilde{\tau}^0)\big): (\OUT^0_{\mcal{M}}, \tilde{\tau}^0, b^0) \gets H^{\mcal{M}}(\secpar,m_0,z) \big\} \\
\cind ~&
\big\{\big(\OUT^1_{\mcal{M}}, \val_{b^1}(\tilde{\tau}^1)\big): (\OUT^1_{\mcal{M}}, \tilde{\tau}^1, b^1) \gets H^{\mcal{M}}(\secpar,m_1,z) \big\}, \numberthis \label{eq:classical:H:m-0:m-1}
\end{align*}
where both ensembles are indexed by $\secpar \in \Naturals$, $(m_0, m_1) \in \bits^{\ell(\secpar)} \times \bits^{\ell(\secpar)}$, and $z \in \bits^*$.

\para{Proof by Contradiction.} Our high-level strategy is proof by contradiction. We assume for contradiction that there are a (possibly non-uniform) PPT machine $\mcal{D}$ and a function $\delta(\secpar) = 1/\poly(\secpar)$ such that for infinitely many $\secpar \in \Naturals$,
\begin{equation}\label[Inequality]{eq:one-sided:proof:contra-assump}
\bigg|\Pr[\mcal{D}\big(1^\secpar, \OUT^0_{\mcal{M}}, \val_{b^0}(\tilde{\tau}^0)\big)=1] - \Pr[\mcal{D}\big(1^\secpar, \OUT^1_{\mcal{M}}, \val_{b^1}(\tilde{\tau}^1)\big)=1] \bigg|\ge 3 \cdot \delta(\secpar),
\end{equation}
where the first probability is taken over the random procedure $(\OUT^0_{\mcal{M}}, \tilde{\tau}^0, b^0) \gets H^{\mcal{M}}(\secpar,m_0,z)$ and the second probability is taken over the random procedure $(\OUT^1_{\mcal{M}}, \tilde{\tau}^1, b^1) \gets H^{\mcal{M}}(\secpar,m_1,z)$.

We then show the following lemma:
\begin{lemma}\label{lem:one-sided:proof:core}
There exists a hybrid $G$ such that for any PPT $\mcal{M}$ and the $\delta(\secpar)$ defined above, the following holds
\begin{enumerate}
\item\label[Property]{item:lem:one-sided:proof:core:1}
$\big\{(\OUT^0, \Val^0) : (\OUT^0, \Val^0) \gets G^{\mcal{M}}(\secpar,m_0,z) \big\} \cind \big\{(\OUT^1, \Val^1) : (\OUT^1, \Val^1) \gets G^{\mcal{M}}(\secpar,m_1,z) \big\}$, where both ensembles are indexed by $\secpar \in \Naturals$, $(m_0, m_1) \in \bits^{\ell(\secpar)} \times \bits^{\ell(\secpar)}$, and $z \in \bits^*$.
\item\label[Property]{item:lem:one-sided:proof:core:2}
\begingroup\fontsize{9.5pt}{0pt}\selectfont
$\big\{(\OUT^G, \Val^G):(\OUT^G, \Val^G) \gets G^{\mcal{M}}(\secpar,m,z) \big\} \cind_{\delta(\secpar)} \big\{ \big(\OUT^H, \val_{b^H}(\tilde{\tau}^H)\big): (\OUT^H, \tilde{\tau}^H, b^H) \gets H^{\mcal{M}}(\secpar,m,z) \big\}$,
\endgroup  
where both ensembles are indexed by $\secpar \in \Naturals$, $m \in \bits^{\ell(\secpar)}$, and $z \in \bits^*$.
\end{enumerate}
\end{lemma}
It is easy to see that if \Cref{lem:one-sided:proof:core} is true, it contradicts our assumption in \Cref{eq:one-sided:proof:contra-assump}. 
 Thus, it will finish the proof of non-malleability. Indeed, this lemma is the most technically involved part. We prove it in \Cref{sec:one-sided:core-lemma:proof}. 

\subsection{Proof of \Cref{lem:one-sided:proof:core}}
\label{sec:one-sided:core-lemma:proof}

First, we provide a new but equivalent interpretation of the game $H^{\mcal{M}}(\secpar, m, z)$ in \Cref{hybrid:H:reinterpretation},. (We also provide a picture in \Cref{figure:one-sided:H:re-interpretation} to illustrate it.)
\begin{AlgorithmBox}[label={hybrid:H:reinterpretation}]{Re-interpretation of Game \textnormal{$H^{\mcal{M}}(\secpar, m, z)$}}
Game $H^{\mcal{M}}(\secpar, m, z)$ can be split into the following stages:
\begin{enumerate}
\item \label[Stage]{hybrid:H:reinterpretation:1}
{\bf Prefix Generation:}
First, execute \Cref{item:one-sided:step:Naor-rho,item:one-sided:step:committing} of the man-in-the-middle game of \Cref{prot:one-sided:classical}. That is, it plays as the left honest committer committing to $m$ and the right honest receiver, with $\mcal{M}(1^\secpar,z)$ being the man-in-the-middle adversary.

\subpara{Notation:} Let $\msf{st}_{\mcal{M}}$ denote the state of $\mcal{M}$ at the end of \Cref{item:one-sided:step:committing}; Let $\msf{st}_C$ (resp.\ $\msf{st}_R$) denote the state of the honest committer (resp.\ receiver) at the end of \Cref{item:one-sided:step:committing}; Let $\tilde{\tau}=(\tilde{\beta}, \tilde{\msf{com}})$\footnote{Recall that $\tilde{\beta}$ and  $\tilde{\msf{com}}$ are the \Cref{item:one-sided:step:Naor-rho,item:one-sided:step:committing} messages in the right session; They constitute an execution of Naor's commitment.} and ${\tau}=({\beta}, {\msf{com}})$. In terms of notation, we denote the execution of this stage by 
\begin{equation}\label[Expression]{expression:hybrid:H:prefix}
(\msf{st}_{\mcal{M}},  \msf{st}_C, \msf{st}_R, \tau,\tilde{\tau})\gets H^{\mcal{M}}_{\msf{pre}}(\secpar, m, z).
\end{equation}
We will call the tuple $(\msf{st}_{\mcal{M}},  \msf{st}_C, \msf{st}_R, \tau,\tilde{\tau})$ the {\em prefix} and denote it by $\msf{pref}$. It is worth noting that this $\msf{pref}$ contains all the information such that a PPT machine can ``complete'' the remaining execution of $H^{\mcal{M}}(\secpar, m, z)$ starting from $\msf{pref}$.

\item \label[Stage]{stage:hybrid:H:remainder}
{\bf The Remainder:}
Next, it simply resumes from where the {\bf Prefix Generation} stage stops, to finish the remaining steps of the man-in-the-middle execution $H^{\mcal{M}}(\secpar, m, z)$.

\subpara{Notation:} We introduce the following notations to describe this stage. Define a PPT machine $\Adv$ that takes as input $(\msf{st}_{\mcal{M}}, \tilde{\tau})$; Machine $\mcal{A}$ is supposed to run the residual strategy of $\mcal{M}$ starting from $\msf{st}_{\mcal{M}}$. Also, define a PPT machine $\mcal{B}$ that takes as input $(\msf{st}_C, \msf{st}_R, \tilde{\tau})$; Machine $\mcal{B}$ is supposed to run the residual strategies of the honest committer $C$ and receiver $R$, starting from 
$\msf{st}_C$ and $\msf{st}_R$ respectively\footnote{Note that it is actually redundant to give $\tilde{\tau}$ as common input to these parties---It can be included in $\msf{st}_{\mcal{M}}$ and $\msf{st}_R$. We choose to make $\tilde{\tau}$ explicit for notational convenience.}. With the above notations, we can denote the execution of the remaining steps of $H^{\mcal{M}}(\secpar, m,z)$ by
\begin{equation}\label[Expression]{expression:hybrid:H:remainder}
(\OUT_\mcal{A}, b)\gets \langle \mcal{A}(\msf{st}_{\mcal{M}}), \mcal{B}(\msf{st}_C, \msf{st}_R) \rangle(1^\secpar, \tilde{\tau}),
\end{equation}
where $\OUT_\mcal{A}$ is the output of $\mcal{A}$, and $b \in \Set{\bot, \top}$ is the output of the honest receiver $R$ (in the right), indicating if the man-in-the-middle's commitment (i.e., the right session) is accepted ($b = \top$) or not ($b = \bot$). (We remark that $\OUT_\Adv$ is nothing but the man-in-the-middle $\mcal{M}$'s final output.)

\item
{\bf Output:} It outputs the tuple $(\OUT_\Adv, \tilde{\tau}, b)$.  
\end{enumerate}
\end{AlgorithmBox}

%\xiao{Say that this is the major lemma for this proof. }
We claim the following lemma regarding \Cref{hybrid:H:reinterpretation}.
\begin{lemma}\label{lem:small-tag:proof:se}
Let $H^{\mcal{M}}_{\msf{pre}}(\secpar, m, z)$, $\mcal{A}$, $\mcal{B}$ be as defined in \Cref{hybrid:H:reinterpretation}.
There exists a PPT machine $\mcal{SE}$ (the simulation-extractor) such that for any $(\msf{st}_{\mcal{M}},  \msf{st}_C, \msf{st}_R, \tau,\tilde{\tau})$ in the support of $H^{\mcal{M}}_{\msf{pre}}(\secpar, m, z)$, any noticeable $\epsilon(\secpar)$, it holds that
\begin{align*}
& \big\{ (\OUT_{\mcal{SE}}, \msf{Val}_{\mcal{SE}}) : (\OUT_{\mcal{SE}}, \msf{Val}_{\mcal{SE}}, b_{\mcal{SE}}) \gets \SimExt^{\A(\msf{st}_{\mcal{M}})}(1^\secpar,1^{\epsilon^{-1}},\msf{st}_{R},\tau, \tilde{\tau})\big\}_{\secpar\in \Naturals} \\
\cind_{\epsilon(\secpar)} ~&
\big\{\big(\OUT_\mcal{A}, \msf{val}_b(\tilde{\tau})\big) : (\OUT_\mcal{A}, b) \gets \langle \mcal{A}(\msf{st}_{\mcal{M}}), \mcal{B}(\msf{st}_C, \msf{st}_R) \rangle(1^\secpar, \tilde{\tau})\big\}_{\secpar\in \Naturals} 
\end{align*}
%\takashi{$b_{\mcal{SE}}$ seems redundant. It will make more sense to just remove it or otherwise include $b_{\mcal{SE}}$ and $b$ in the LHS and RHS respectively. } \xiao{Actually, I do need $b_{\mcal{SE}}$ for notational convenience (e.g., in \Cref{eq:one-sided:proof:se:itself}, \Cref{claim:bounding-E1E2,bound:Val-ne-m}). Without this $b_{\mcal{SE}}$, I need to explain several points in words. We can discuss if there is a better solution.}\takashi{I still prefer omitting $b_{\mcal{SE}}$ from the statement of this lemma and introducing it in the proof of this lemma for consistency to the extract-and-simulate lemma. We may explain that  we use the notation $(\OUT_{\mcal{SE}}, \msf{Val}_{\mcal{SE}}, b_{\mcal{SE}}) \gets \SimExt^{\A(\msf{st}_{\mcal{M}})}(1^\secpar,1^{\epsilon^{-1}})$ for convenience even though $b_{\mcal{SE}}$ is not part of the output.} \xiao{After I fixed the bug, I feel it is convenient to have the $b_{\mcal{SE}}$ here, because I used it in several places (due to my way to fix the bug). So, let's talk about what to do with this $b_{\mcal{SE}}$.}
\end{lemma}
\begin{remark}
The third output $b_{\mcal{SE}}$ of $\mcal{SE}$ is redundant in the above lemma. We include it in the output of $\mcal{SE}$ for convenience in the proof of this lemma in \Cref{lem:small-tag:proof:se:proof}.
\end{remark}

\Cref{lem:small-tag:proof:se} is the main technical lemma for the current proof of \Cref{lem:one-sided:proof:core}. We present its proof in \Cref{lem:small-tag:proof:se:proof}. In the following, we finish the proof of \Cref{lem:one-sided:proof:core}  assuming that \Cref{lem:small-tag:proof:se} is true.

%\xiao{Say that \Cref{lem:small-tag:proof:se} is the main technical lemma for the current proof of \Cref{item:lem:one-sided:proof:core:2}. We present its proof in \Cref{lem:small-tag:proof:se:proof}. In the following, we finish the proof of \Cref{lem:one-sided:proof:core}  assuming that \Cref{lem:small-tag:proof:se} is true.}

Now, we are ready to present the description of $G$ as required by \Cref{lem:one-sided:proof:core}.
\begin{AlgorithmBox}[label={hybrid:G}]{Hybrid \textnormal{${G}^{\mcal{M}}(\secpar, m, z)$}}
This hybrid proceeds as follows:
\begin{enumerate}
\item
{\bf Prefix Generation:}
This stage is identical to \Cref{hybrid:H:reinterpretation:1} of $H^{\mcal{M}}(\secpar, m, z)$. Formally, it executes 
$$(\msf{st}_{\mcal{M}},  \msf{st}_C, \msf{st}_R, \tau,\tilde{\tau}) \gets H^{\mcal{M}}_{\msf{pre}}(\secpar, m, z),$$
where $H^{\mcal{M}}_{\msf{pre}}(\secpar, m, z)$ is defined in \Cref{hybrid:H:reinterpretation:1} of \Cref{hybrid:H:reinterpretation}.
\item
{\bf The Remainder:}
Define $\mcal{A}$ in the same way as in \Cref{stage:hybrid:H:remainder} of $H^{\mcal{M}}(\secpar, m, z)$. With this $\mcal{A}$  and the $(\msf{st}_{\mcal{M}},  \msf{st}_R, \tau,\tilde{\tau})$ from the previous stage, ${G}^{\mcal{M}}(\secpar, m, z)$ invokes the $\mcal{SE}$ prescribed in \Cref{lem:small-tag:proof:se}. Formally, it executes the following procedure:
$$(\OUT_{\mcal{SE}}, \msf{Val}_{\mcal{SE}}, b_{\mcal{SE}}) \gets \SimExt^{\A(\msf{st}_{\mcal{M}})}(1^\secpar,1^{\delta^{-1}},\msf{st}_R,\tau,\tilde{\tau}),$$
where the $\delta$ is the statistical distance that we want to show for \Cref{item:lem:one-sided:proof:core:2} of \Cref{lem:one-sided:proof:core}.

\begin{remark}
We emphasize that in this stage, ${G}^{\mcal{M}}(\secpar, m, z)$ does {\em not} make use of $\msf{st}_C$.
\end{remark}
\item
{\bf Output:}
% $\bar{G}^{\mcal{M}_\secpar}(m,\beta_\secpar)$ sets $\OUT_{\bar{G}}\coloneqq \OUT_{\mcal{SE}}$ and $ \msf{Val}_{\bar{G}}\coloneqq \msf{Val}_{\mcal{SE}}$, and 
It outputs $(\OUT_{\mcal{SE}}, \msf{Val}_{\mcal{SE}})$.\footnote{Note that $b_{\mcal{SE}}$ is not included in the output of $G$. Indeed, $b_{\mcal{SE}}$ is not important for the current machine $G^{\mcal{M}}(\secpar, m, z)$. We choose to include it in the output of $\SimExt^{\A(\msf{st}_{\mcal{M}})}(1^\secpar,1^{\delta^{-1}},\msf{st}_R,\tilde{\tau})$ only for notational convenience when we define/prove properties about $\mcal{SE}$ itself.}
\end{enumerate}
\end{AlgorithmBox}

\para{Proving \Cref{item:lem:one-sided:proof:core:1} of \Cref{lem:one-sided:proof:core}.} Observe that hybrid $G^{\mcal{M}}(\secpar, m, z)$ is an efficient machine because both $H^{\mcal{M}}_{\msf{pre}}(\secpar, m, z)$ and $\SimExt^{\A(\msf{st}_{\mcal{M}})}(1^\secpar,1^{\delta^{-1}},\msf{st}_R,\tau,\tilde{\tau})$ are efficient. Moreover, it does not rewind \Cref{item:one-sided:step:Naor-rho,item:one-sided:step:committing} of the man-in-the-middle execution, and $\SimExt^{\A(\msf{st}_{\mcal{M}})}(1^\secpar,1^{\delta^{-1}},\msf{st}_R,\tau,\tilde{\tau})$ does {\em not} need to know $\msf{st}_C$. Therefore, \Cref{item:lem:one-sided:proof:core:1} of \Cref{lem:one-sided:proof:core} follows immediately from the computational-hiding property of the left Naor's commitment (i.e., \Cref{item:one-sided:step:Naor-rho,item:one-sided:step:committing} in the left session). 

\para{Proving \Cref{item:lem:one-sided:proof:core:2} of \Cref{lem:one-sided:proof:core}.} First, observe that the distribution of the prefix $(\msf{st}_{\mcal{M}},  \msf{st}_C, \msf{st}_R, \tau,\tilde{\tau})$ is identical in $G^{\mcal{M}}(\secpar, m, z)$ and $H^{\mcal{M}}(\secpar, m, z)$.
For each fixed prefix, \Cref{lem:small-tag:proof:se} implies that $G^{\mcal{M}}(\secpar, m, z)$ and $H^{\mcal{M}}(\secpar, m, z)$ are $\delta(\secpar)$-computationally indistinguishable where we remark that $G^{\mcal{M}}(\secpar, m, z)$ runs $\mcal{SE}$ with the second input $1^{\delta^{-1}}$. 
This immediately implies \Cref{item:lem:one-sided:proof:core:2} of \Cref{lem:one-sided:proof:core}. 
%$G^{\mcal{M}}(\secpar, m, z)$ and $H^{\mcal{M}}(\secpar, m, z)$ (as per \Cref{hybrid:H:reinterpretation}) are identical up to the end of \Cref{item:one-sided:step:committing} of the man-in-the-middle game of \Cref{prot:one-sided:classical}.

\subsection{Proof of \Cref{lem:small-tag:proof:se}}
\label{lem:small-tag:proof:se:proof}

\begin{figure}[!h]
     \begin{subfigure}[t]{0.47\textwidth}
         \centering
         \fbox{
         \includegraphics[width=\textwidth,page=1]{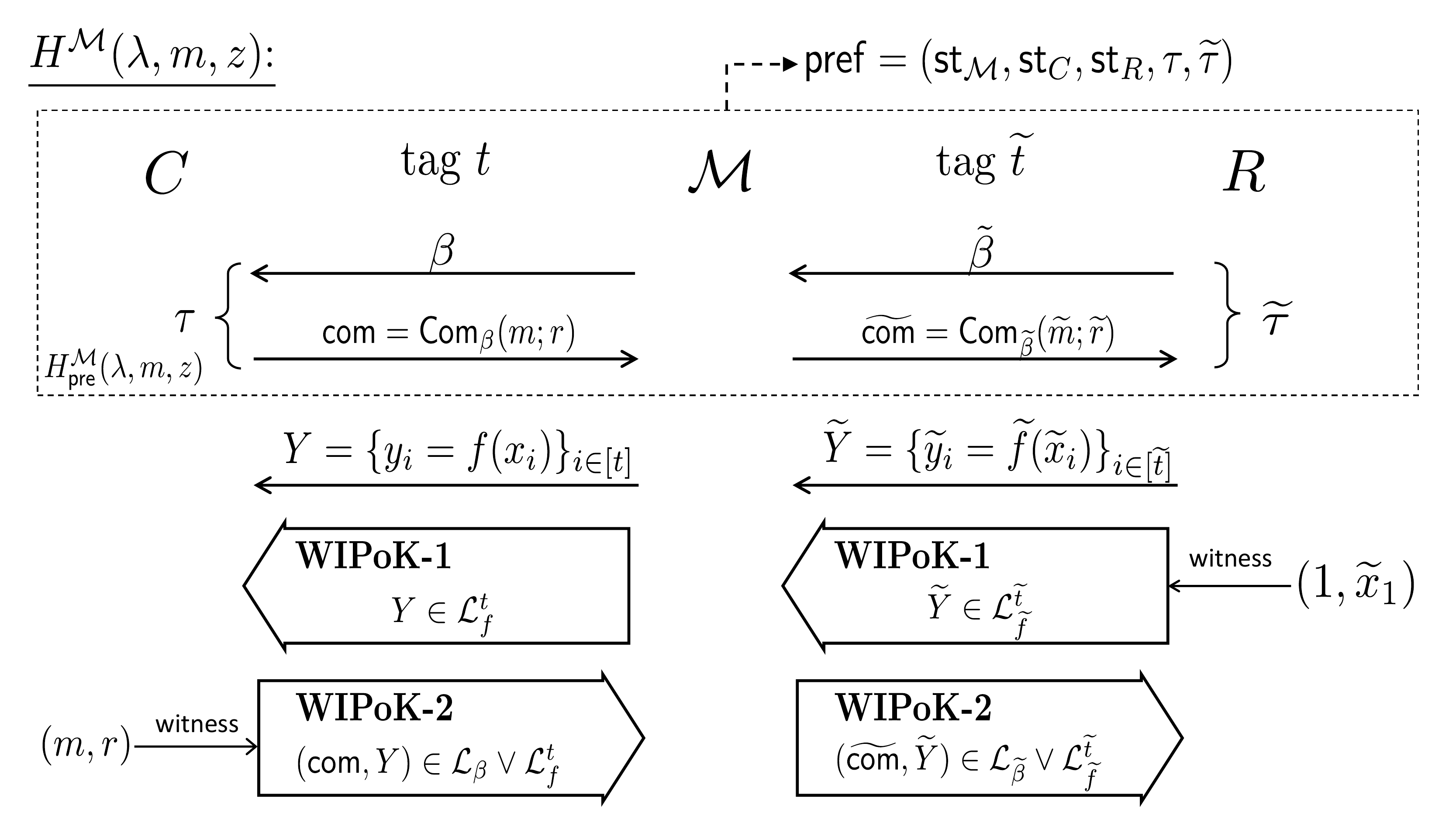}
         }
         \caption{}
         \label{figure:one-sided:H:re-interpretation}
     \end{subfigure}
     \hspace{6.5pt}
     \begin{subfigure}[t]{0.47\textwidth}
         \centering
         \fbox{
         \includegraphics[width=\textwidth,page=2]{figures/figures-new.pdf}
         }
         \caption{}
         \label{figure:one-sided:G1}
     \end{subfigure}
     \caption{Machines $H^{\mcal{M}}$ and $\mcal{G}_1$ {\scriptsize (Difference is highlighted in red color)}}
     \label{figure:one-sided:games:M:G1}
\end{figure}

%\xiao{Provide some intuition. Say that our machine $\mcal{SE}$ will make use of two machines $\mcal{G}_1$ that simualtes the main-thread, and an extractor $\mcal{K}$ that extracts the value committed in $\tau$. Thus, in the following, we will first define these two machines, and the present the description of $\mcal{SE}$ (in \Cref{machine:se}).}

Now, we prove \Cref{lem:small-tag:proof:se} by constructing the required machine $\mcal{SE}$. Our machine $\mcal{SE}$ (for \Cref{lem:small-tag:proof:se}) will make use of two machines: (1) a $\mcal{G}_1$ that simulates the main-thread, and (2) an extractor $\mcal{K}$ that extracts the value committed in $\tilde{\tau}$. 
Thus, in the following, we will first define these two machines, and then present the description of $\mcal{SE}$ in \Cref{machine:se}.

First, we describe a machine $\mcal{G}_1$ that simulates the real execution {\em without using $\msf{st}_C$}.   

\para{Machine $\mcal{G}_1$:} (Illustrated in \Cref{figure:one-sided:G1}\label{pageref:one-sided:proof:G1}) For any prefix $\msf{pref}$, $\mcal{G}_1(\msf{st}_{\mcal{M}}, \msf{st}_R, \tau,\tilde{\tau})$ behaves identically to \Cref{stage:hybrid:H:remainder} of $H^{\mcal{M}}(\secpar, m, z)$ shown in \Cref{hybrid:H:reinterpretation} (and depicted in \Cref{figure:one-sided:H:re-interpretation}), except for the following difference: Instead of executing the left {\bf WIPoK-1} honestly, it uses the witness-extended emulator $\mcal{WE}$ (as per \Cref{def:WEE}) of the left {\bf WIPoK-1} to extract a witness, %(the existence of this $\mcal{SE}$ is guaranteed by the proof of knowledge property of the left {\bf WIPoK-1} as per \Cref{def:WEE}), 
and

\begin{itemize}
\item
 If the left committer accepts the left {\bf WIPoK-1} and the extracted witness is valid (i.e., it is a $(j, x_j)$ pair such that $y_j = f(x_j)$ for some $j \in [t]$), $\mcal{G}_1$ uses $(j, x_j)$ to finish the left {\bf WIPoK-2}. Similarly to $H^{\mcal{M}}(\secpar, m, z)$, $\mcal{G}_1$ eventually outputs $\mcal{M}$'s final state and the right receiver's decision bit $b$; 
 \item
 If the left committer accepts the left {\bf WIPoK-1}  but the extracted witness is invalid, it aborts immediately and outputs $(\bot, \bot)$.
 \item If the left committer rejects the left {\bf WIPoK-1}, it runs the rest of execution of $H^{\mcal{M}}(\secpar, m, z)$ to output $\mcal{M}$'s final state and the right receiver's decision bit $b$.  Note that it does not need to run the left {\bf WIPoK-2} in this case since the left committer aborts after the left {\bf WIPoK-1}. %\takashi{Previously, it is not explained what happens if the left committer rejects the left {\bf WIPoK-1}. I believe we need to mention this case.}
\end{itemize}
We denote the above procedure by $(\OUT, b) \gets \mcal{G}_1(\msf{st}_{\mcal{M}}, \msf{st}_R, \tau,\tilde{\tau})$. 
It is worth noting that $\mcal{G}_1$ does {\em not} need to know $\msf{st}_C$.

\begin{remark}[Precise Meaning of Extraction]\label{rem:meaning_extraction}
In the above description of $\mcal{G}_1$, we say that ``it uses the witness-extended emulator of the left {\bf WIPoK-1} to extract a witness". 
More accurately, this means the following: We consider a cheating prover against the left {\bf WIPoK-1} that takes as advice the states of $\mcal{M}$ and the right receiver right before the start of the left {\bf WIPoK-1}, interacts with the left committer by simulating $\mcal{M}$ and the right receiver, and outputs the states of $\mcal{M}$ and the right receiver at the end of the left {\bf WIPoK-1}. Machine $\mcal{G}_1$ runs the witness-extended emulator of the left {\bf WIPoK-1} (as per \Cref{def:WEE}) w.r.t. the this cheating prover to simulate the states of $\mcal{M}$ and the right receiver right at the end of the left {\bf WIPoK-1}, the decision bit of the left {\bf WIPoK-1} of the left committer, while extracting a (candidate of) witness of $\Lang_f^t$.  
We will use similar convention many times throughout this paper. 
%\takashi{I believe this remark is useful especially for the post-quantum case. Please feel free to improve the explanation.}
\end{remark}

%\xiao{Next, prove a lemma relates the simulated main-thread with the real main-thread.} 
Next, we prove a lemma that shows that  $\mcal{G}_1$ simulates the real MIM execution. 
\begin{lemma}\label{lem:main-thread:sim:real}
For the $H^{\mcal{M}}_{\msf{pre}}(\secpar, m, z)$, $\mcal{A}$, and $\mcal{B}$ defined in \Cref{hybrid:H:reinterpretation}, for any $\msf{pref}=(\msf{st}_{\mcal{M}},  \msf{st}_C, \msf{st}_R, \tau,\tilde{\tau})$ in the support of $H^{\mcal{M}}_{\msf{pre}}(\secpar, m, z)$, it holds that
\begin{align*}
& \big\{\big(\OUT, ~b \big): (\OUT, b) \gets \mcal{G}_1(\msf{st}_{\mcal{M}}, \msf{st}_R, \tau,\tilde{\tau}) \big\}_{\secpar \in \Naturals} \\
\cind ~&
\big\{\big(\OUT_\mcal{A}, ~b\big) : (\OUT_\mcal{A}, b) \gets \langle \mcal{A}(\msf{st}_{\mcal{M}}), \mcal{B}(\msf{st}_C, \msf{st}_R) \rangle(1^\secpar, \tilde{\tau})\big\}_{\secpar\in \Naturals}. 
\end{align*}
% \takashi{I removed $\val(\tilde{\tau})$ from both sides. That's redundant and even confusing to me.} \xiao{Without $\val(\tilde{\tau}$, I'm not sure if you can prove \Cref{claim:proof:SE:simulated-main-thread} later---adding (even the same) extra information to two computationally indistinguishable ensembles may suddenly make the distinguishable, as the added information could be correlated to the variable sin the original ensembles.}
\end{lemma}
\begin{proof}
We first define a hybrid machine $\mcal{G}'_1$ below.
\begin{figure}[!h]
\centering
\fbox{
     \includegraphics[width=0.55\textwidth,page=3]{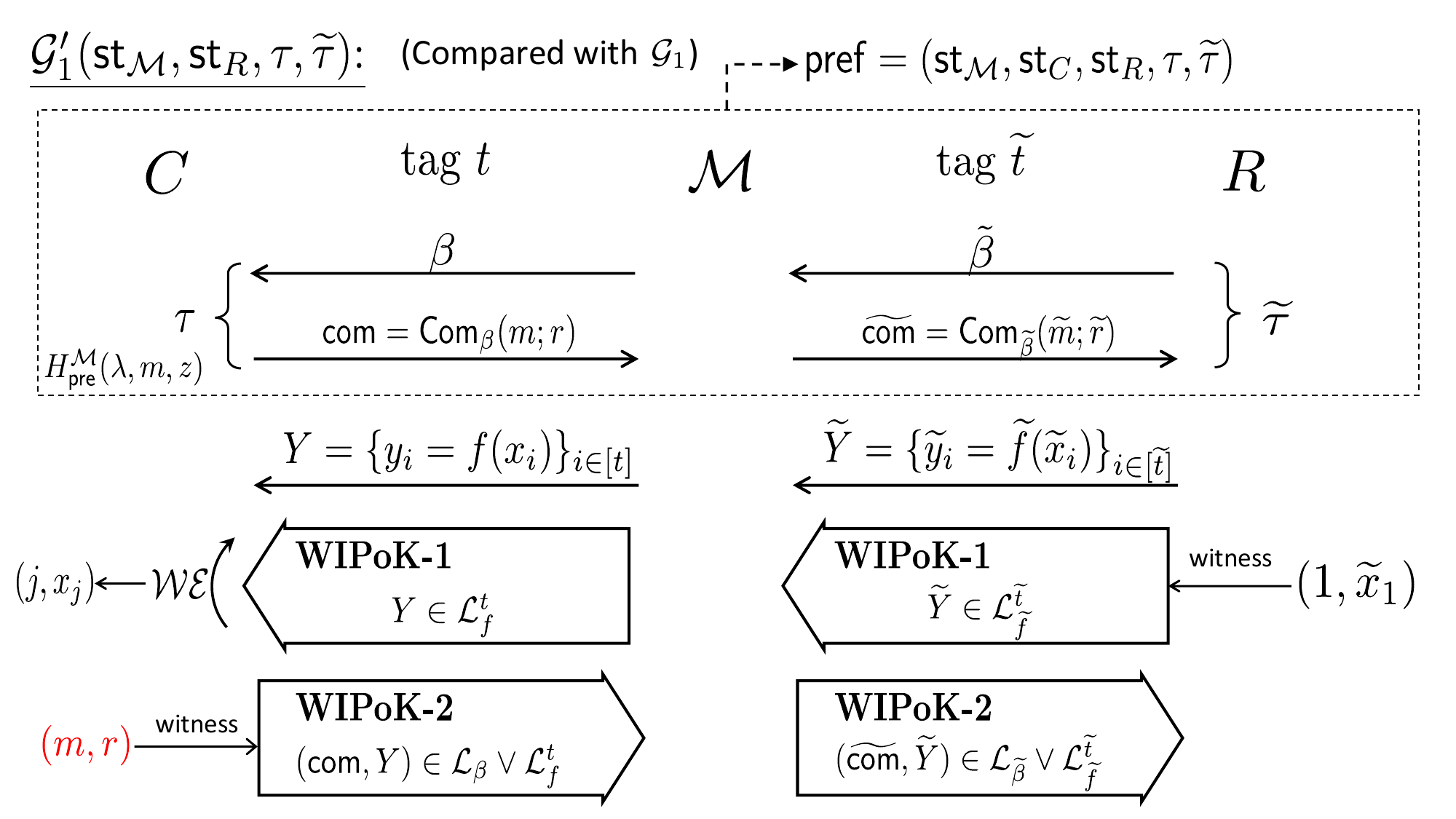}
     }
     \caption{Machine $\mcal{G}'_1$ {\scriptsize (Difference with $\mcal{G}_1$ is highlighted in red color)}}
     \label{figure:one-sided:G'1}
\end{figure}

\para{Machine $\mcal{G}'_1$:} (Illustrated in \Cref{figure:one-sided:G'1}\label{pageref:one-sided:proof:G'1}) For any prefix $\msf{pref}$, $\mcal{G}'_1(\msf{st}_{\mcal{M}}, \msf{st}_R, \tau,\tilde{\tau})$ behaves identically to $\mcal{G}_1(\msf{st}_{\mcal{M}}, \msf{st}_R, \tau,\tilde{\tau})$ except that it uses $(m,r)$ as the witness in the left {\bf WIPoK-2} even if it succeeds in extracting a valid witness from the left {\bf WIPoK-1}.

By the WI property of the left {\bf WIPoK-2}, it holds that 
\begin{align} 
\begin{split}\label{eq:G1_to_G'1}
& \big\{\big(\OUT, ~b\big): (\OUT, b) \gets \mcal{G}_1(\msf{st}_{\mcal{M}}, \msf{st}_R, \tau,\tilde{\tau}) \big\}_{\secpar \in \Naturals} \\
\cind ~&
\big\{\big(\OUT, ~b\big): (\OUT, b) \gets \mcal{G}'_1(\msf{st}_{\mcal{M}}, \msf{st}_R, \tau,\tilde{\tau}) \big\}_{\secpar \in \Naturals}
\end{split}.
\end{align}
%where we remark that $\val(\tilde{\tau})$ can be treated as an advice in the reduction. 

Note that the only difference between $\mcal{G}'_1(\msf{st}_{\mcal{M}}, \msf{st}_R, \tau,\tilde{\tau})$ and $\langle \mcal{A}(\msf{st}_{\mcal{M}}), \mcal{B}(\msf{st}_C, \msf{st}_R) \rangle(1^\secpar, \tilde{\tau})$ is that the former runs the witness-extended emulator of the left {\bf WIPoK-1} (but does not use the extracted witness at all). Thus, by the PoK property of the left {\bf WIPoK-1}, it holds that 
\begin{align} 
\begin{split}\label{eq:G'1_to_real}
& \big\{\big(\OUT, ~b\big): (\OUT, b) \gets \mcal{G}'_1(\msf{st}_{\mcal{M}}, \msf{st}_R, \tau,\tilde{\tau}) \big\}_{\secpar \in \Naturals} \\
\cind ~&
\big\{\big(\OUT_\mcal{A}, ~b\big) : (\OUT_\mcal{A}, b) \gets \langle \mcal{A}(\msf{st}_{\mcal{M}}), \mcal{B}(\msf{st}_C, \msf{st}_R) \rangle(1^\secpar, \tilde{\tau})\big\}_{\secpar\in \Naturals}
\end{split}.
\end{align}

By combining \Cref{eq:G1_to_G'1,eq:G'1_to_real}, we obtain \Cref{lem:main-thread:sim:real}. 
%\xiao{TODO: prove this lemma. It follows from the PoK property of the left {\bf WIPoK-1} and the WI property of the left {\bf WIPoK-2}}

\end{proof}

Next, we define the probability of $R$ being convinced in the execution of $\mcal{G}_1$. This value plays an important role later in our proof.
\begin{definition}\label{def:p-sim-pref}
For any $\msf{pref}=(\msf{st}_{\mcal{M}},  \msf{st}_C, \msf{st}_R, \tau,\tilde{\tau})$ in the support of $H^{\mcal{M}}_{\msf{pre}}(\secpar, m, z)$, we define the following value $p^{\msf{Sim}}_{\msf{pref}}$:
$$p^{\msf{Sim}}_{\msf{pref}} \coloneqq \Pr[b = \top : (\OUT, b) \gets \mcal{G}_1(\msf{st}_\mcal{M}, \msf{st}_R, \tau,\tilde{\tau})].$$
\end{definition}

Next, we show a technical lemma that gives an extractor $\mcal{K}$ {\em without the simulation property}. 
\begin{lemma}\label{lem:small-tag:proof:se:proof:K}
Let $H^{\mcal{M}}_{\msf{pre}}(\secpar, m, z)$, $\mcal{A}$, and $\mcal{B}$ be as defined in \Cref{hybrid:H:reinterpretation}. There exists an expected PPT machine $\mcal{K}$ such that for any $\msf{pref}=(\msf{st}_{\mcal{M}},  \msf{st}_C, \msf{st}_R, \tau,\tilde{\tau})$ in the support of $H^{\mcal{M}}_{\msf{pre}}(\secpar, m, z)$ and any noticeable $\epsilon(\secpar)$, the following holds:
\begin{enumerate}
\item \label[Property]{property:small-tag:proof:se:proof:K:syntax}
{\bf (Syntax.)} $\mcal{K}$ takes as input $(1^\secpar, \msf{st}_R, \tau,\tilde{\tau})$ and makes oracle access to $\Adv(\msf{st}_{\mcal{M}})$. It outputs a value $\msf{Val}_{\mcal{K}} \in \bits^{\ell(\secpar)} \cup \Set{\bot}$ such that $\msf{Val}_{\mcal{K}} = \msf{val}(\tilde{\tau})$ whenever $\msf{Val}_{\mcal{K}} \ne \bot$.
(Also see \Cref{rmk:small-tag:proof:se:proof:K:output} for an intuitive explanation.)

\item \label[Property]{property:small-tag:proof:se:proof:K}
If $p^{\msf{Sim}}_{\msf{pref}} \ge \epsilon(\secpar)$, then it holds that
$$\Pr[\Val_{\mcal{K}} = \msf{val}(\tilde{\tau}) : \Val_{\mcal{K}} \gets \mcal{K}^{\A(\msf{st}_{\mcal{M}})}(1^\secpar, \msf{st}_R, \tau,\tilde{\tau})] \ge \frac{\epsilon'(\secpar)}{\tilde{t}},$$
where  $\epsilon'(\secpar) \coloneqq \frac{\epsilon(\secpar)}{10t^2}$.
\end{enumerate}
\end{lemma}
\begin{remark}[On the Output of $\mcal{K}$]
\label{rmk:small-tag:proof:se:proof:K:output}
% \xiao{Say it will become clear in the ``\underline{$\mcal{K}_i$'s Output}'' part on \Cpageref{subpara:K-i:output}.}
% The output $\msf{Val}_{\mcal{K}}$ can take three types of values: 
% \begin{enumerate}
% \item
% $\msf{Val}_{\mcal{K}}$ is a witness for $\tilde{Y} \in \Lang^{\tilde{t}}_{\tilde{f}}$ (see \Cref{eq:one-sided:Lang:OWF}). As we will show later, this case will not happen except for with negligible probability. 
% \item
% $\msf{Val}_{\mcal{K}}$ is a witness for $\tilde{\msf{com}} \in \Lang_{\tilde{\beta}}$ (see \Cref{eq:one-sided:Lang:Com}). In this case, $\msf{Val}_{\mcal{K}}$ is exactly the value committed in 
% This is the case we really expect to happen.
% \item
% $\msf{Val}=\bot$. See the discussion below.
% \end{enumerate}
The output $\msf{Val}_{\mcal{K}}$ is expected to be the value committed in $\tilde{\tau}$, i.e., $\msf{val}(\tilde{\tau})$. If $\msf{Val}_{\mcal{K}} = \bot$, it indicates that $\mcal{K}$ did not extract the correct $\msf{val}(\tilde{\tau})$. We explicitly include $\bot$ in the range of $\mcal{K}$ for the following purpose: Looking ahead, the simulation-extractor $\mcal{SE}$ that we are going to build will invoke $\mcal{K}$ for several times, until the value $\msf{val}(\tilde{\tau})$ is extracted. However, $\mcal{SE}$ cannot tell if the extracted value is indeed $\msf{val}(\tilde{\tau})$. Thus, the case $\msf{Val}_{\mcal{K}} = \bot$ serves as an indicator, telling $\mcal{SE}$ if the extraction by $\mcal{K}$ succeeds (more accurately, fails). 

As a vigilant reader may have already realized, there is an alternative formalism: simply ask $\mcal{K}$ to output both $\msf{val}(\tau)$ {\em and the decommitment information} so that $\mcal{SE}$ can test by itself whether $\mcal{K}$'s extraction is successful. We remark that this approach does work for the current proof {\em in classical setting}. However, it may not extend when we prove {\em post-quantum} non-malleability. In short, that is because to make this proof work in the quantum setting, our technique requires that the valid output of $\mcal{K}$ should be {\em unique}; Only in this way can we ask $\mcal{K}$ to ``forget'' other information, such that the extraction procedure can be ``un-computed'' to rewind $\mcal{M}$ back without much disturbance to its initial state. (This point will become clearer in the proof of \Cref{lem:extract_and_simulate}.) However, if we include the decommitment information in the output of $\mcal{K}$, then the valid output may not be unique, even if the commitment scheme is perfect-binding---There could exists different ways to decommit to the {\em unique} committed value.

To make the proof consistent in both classical and quantum settings, we choose the current formalism in \Cref{property:small-tag:proof:se:proof:K:syntax}.
\end{remark}

We will prove \Cref{lem:small-tag:proof:se:proof:K} in \Cref{sec:lem:small-tag:proof:se:proof:K:proof}.
In the rest of this subsection, we finish the proof of \Cref{lem:small-tag:proof:se} assuming \Cref{lem:small-tag:proof:se:proof:K} is true.

We present the description of $\mcal{SE}$ (for \Cref{lem:small-tag:proof:se}) in \Cref{machine:se}.
\begin{AlgorithmBox}[label={machine:se}]{Simulation-Extractor \textnormal{$\mcal{SE}^{\A(\msf{st}_{\mcal{M}})}(1^\secpar,1^{\epsilon^{-1}},\msf{st}_R, \tau, \tilde{\tau})$}}
Let $H^{\mcal{M}}_{\msf{pre}}(\secpar, m, z)$, $\mcal{A}$, and $\mcal{B}$ be as defined in \Cref{hybrid:H:reinterpretation}. Let $\epsilon(\secpar)$ be a noticeable function. For $(\msf{st}_{\mcal{M}},  \msf{st}_C, \msf{st}_R, \tau,\tilde{\tau})$ in the support of $H^{\mcal{M}}_{\msf{pre}}(\secpar, m, z)$, machine $\mcal{SE}^{\A(\msf{st}_{\mcal{M}})}(1^\secpar,1^{\epsilon^{-1}}, \msf{st}_R,\tau,\tilde{\tau})$ proceeds as follows:
\begin{enumerate}
\item\label[Stage]{stage:machine:se:main-thread-sim}
{\bf Main-Thread Simulation.} It first uses the machine $\mcal{G}_1$ to simulate \Cref{stage:hybrid:H:remainder} of $H^{\mcal{M}}(\secpar, m, z)$. That is, it computes $(\OUT, b) \gets \mcal{G}_1(\msf{st}_{\mcal{M}}, \msf{st}_{R}, \tau,\tilde{\tau})$. It sets $\msf{OUT}_{\mcal{SE}} \coloneqq \OUT$ and $b_{\mcal{SE}} \coloneqq b$. Then, 
\begin{itemize}
\item
if $b_{\mcal{SE}} = \bot$, it sets $\msf{Val}_{\mcal{SE}} \coloneqq \bot$ and jumps directly to \Cref{machine:se:output}; 
\item
otherwise, it goes to the next step.
\end{itemize}

\item \label[Stage]{stage:machine:se:rewinding}
{\bf Rewinding:} $\mcal{SE}^{\A(\msf{st}_{\mcal{M}})}(1^\secpar,1^{\epsilon^{-1}},\msf{st}_R, \tau, \tilde{\tau})$ then loops the following procedure for $\frac{\tilde{t}}{\epsilon'(\secpar)} \cdot \secpar$ times, where $\epsilon'(\secpar) \coloneqq \frac{\epsilon(\secpar)}{10t^2}$:  
\begin{itemize}
\item
Execute $\Val_{\mcal{K}} \gets \mcal{K}^{\A(\msf{st}_{\mcal{M}})}(1^\secpar, \msf{st}_R, \tau,\tilde{\tau})$, where $\mcal{K}$ is provided by  \Cref{lem:small-tag:proof:se:proof:K}.  If $\Val_{\mcal{K}} \ne \bot$,\footnote{Note that by \Cref{property:small-tag:proof:se:proof:K:syntax} in \Cref{lem:small-tag:proof:se:proof:K}, this means $\Val_{\mcal{K}} = \msf{val}(\tilde{\tau})$.} set $\Val_{\mcal{SE}} \coloneqq \Val_{\mcal{K}}$ and jump to \Cref{machine:se:output}; Otherwise, go to the next loop.
%\red{Takashi: If we want to closely follow the post-quantum security proof, the rewinding should be run before the main thread.But this may be fine since running the main thread first is more intuitive and this step will be absorbed into the simulate-and-extract lemma anyway. The counterpart of this algorithm will not explicitly appear in the post-quantum security proof.}
\end{itemize}

\item
It sets $\msf{Val}_{\mcal{SE}} = \bot$. (Note that if this stage is reached, it means that the number of the loop in the previous stage reached its upper bound $\frac{\tilde{t}}{\epsilon'(\secpar)} \cdot \secpar$, but $\mcal{K}$ did not extract the message committed in $\tilde{\tau}$ yet.)

\item \label[Stage]{machine:se:output}
{\bf Output:} it outputs $(\msf{OUT}_{\mcal{SE}}, \msf{Val}_{\mcal{SE}}, b_{\mcal{SE}})$.
\end{enumerate}
\end{AlgorithmBox}

%\xiao{We first prove that the simulated main-thread is computationally indistinguisahble as the real one.}
We first prove that the simulated main-thread is computationally indistinguishable from the real one. 
\begin{MyClaim}\label{claim:proof:SE:simulated-main-thread}
Let $H^{\mcal{M}}_{\msf{pre}}(\secpar, m, z)$, $\mcal{A}$, $\mcal{B}$ be as defined in \Cref{hybrid:H:reinterpretation}. For any $(\msf{st}_{\mcal{M}},  \msf{st}_C, \msf{st}_R, \tau,\tilde{\tau})$ in the support of $H^{\mcal{M}}_{\msf{pre}}(\secpar, m, z)$ and any noticeable $\epsilon(\secpar)$, it holds that
\begin{align*}
& \big\{\big(\OUT_{\mcal{SE}}, \msf{val}_{b_{\mcal{SE}}}(\tilde{\tau}) \big): (\OUT_{\mcal{SE}}, \Val_{\mcal{SE}}, b_{\mcal{SE}}) \gets \mcal{SE}^{\A(\msf{st}_{\mcal{M}})}(1^\secpar,1^{\epsilon^{-1}}, \msf{st}_R, \tau,\tilde{\tau})\big\}_{\secpar \in \Naturals} \\
\cind ~&
\big\{\big(\OUT_\mcal{A}, \msf{val}_b(\tilde{\tau})\big) : (\OUT_\mcal{A}, b) \gets \langle \mcal{A}(\msf{st}_{\mcal{M}}), \mcal{B}(\msf{st}_C, \msf{st}_R) \rangle(1^\secpar, \tilde{\tau})\big\}_{\secpar\in \Naturals}. \numberthis \label{claim:proof:SE:simulated-main-thread:eq}
\end{align*}
\end{MyClaim}
\begin{proof}
Since we consider each fixed prefix, we can efficiently compute $\msf{val}_{b_{\mcal{SE}}}(\tilde{\tau})$ and $\msf{val}_{b}(\tilde{\tau})$ from $b_{\mcal{SE}}$ and $b$, respectively, by using $\val(\tilde{\tau})$ as a non-uniform advice\footnote{Note that this is possible because both the LHS and the RHS of \Cref{claim:proof:SE:simulated-main-thread:eq} use the same {\em fixed} $\tilde{\tau}$, which does not depend on the randomness of the ensembles.}. 
Then, it suffices to prove 
\begin{align*}
& \big\{\big(\OUT_{\mcal{SE}}, b_{\mcal{SE}}\big): (\OUT_{\mcal{SE}}, \Val_{\mcal{SE}}, b_{\mcal{SE}}) \gets \mcal{SE}^{\A(\msf{st}_{\mcal{M}})}(1^\secpar,1^{\epsilon^{-1}}, \msf{st}_R, \tau,\tilde{\tau})\big\}_{\secpar \in \Naturals} \\
\cind ~&
\big\{\big(\OUT_\mcal{A}, b\big) : (\OUT_\mcal{A}, b) \gets \langle \mcal{A}(\msf{st}_{\mcal{M}}), \mcal{B}(\msf{st}_C, \msf{st}_R) \rangle(1^\secpar, \tilde{\tau})\big\}_{\secpar\in \Naturals}. 
\end{align*}
Since $\mcal{SE}$ just runs $\mcal{G}_1$ in the main thread, the above follows directly from \Cref{lem:main-thread:sim:real}. 

\if0
\begin{xiaoenv}{Proving \Cref{claim:proof:SE:simulated-main-thread}}
Todo: This is simple to prove: 
\begin{itemize}
\item
First, note that it follows from \Cref{lem:main-thread:sim:real} that $\big\{\big(\OUT_\mcal{SE}, ~b_{\mcal{SE}}, ~\msf{val}(\tilde{\tau})\big) \big\} \cind \big\{\big(\OUT_\mcal{A}, ~b, ~\msf{val}(\tilde{\tau})\big) \big\}$;
\item
Then, it follows straightforwardly that $\big\{\big(\OUT_\mcal{SE}, \msf{val}_{b_{\mcal{SE}}}(\tilde{\tau})\big) \big\} \cind \big\{\big(\OUT_\mcal{A}, \msf{val}_b(\tilde{\tau})\big) \big\}$.
\end{itemize}
\end{xiaoenv}
\fi
\end{proof}

Given \Cref{claim:proof:SE:simulated-main-thread}, to prove \Cref{lem:small-tag:proof:se}, it suffices to show the following inequality:
\begin{equation}\label[Inequality]{eq:one-sided:proof:se:itself}
\Pr[\Val_{\mcal{SE}} \ne  \msf{val}_{b_{\mcal{SE}}}(\tilde{\tau}): (\OUT_{\mcal{SE}}, \Val_{\mcal{SE}}, b_{\mcal{SE}}) \gets \mcal{SE}^{\A(\msf{st}_{\mcal{M}})}(1^\secpar,1^{\epsilon^{-1}},\msf{st}_R,\tau,\tilde{\tau})] \le \epsilon(\secpar) + \negl(\secpar).
\end{equation}
To prove \Cref{eq:one-sided:proof:se:itself}, let us first define two events:
\begin{itemize}
\item
{\bf Event $E_{\le \epsilon}$:}\footnote{We remark that this event actually does not depend on the random procedure $\mcal{SE}^{\A(\msf{st}_{\mcal{M}})}(1^\secpar,1^{\epsilon^{-1}},\msf{st}_R, \tau, \tilde{\tau})$. It is only about the property of the prefix $(\msf{st}_{\mcal{M}},  \msf{st}_C, \msf{st}_R, \tau,\tilde{\tau})$ given as input to machine $\mcal{SE}$.} This is the event that the prefix $(\msf{st}_{\mcal{M}},  \msf{st}_C, \msf{st}_R, \tau,\tilde{\tau})$ lead to an accepting execution of $\mcal{SE}$ with probability at most $\epsilon$. Formally, it denotes the event that $\msf{pref}=(\msf{st}_{\mcal{M}},  \msf{st}_C, \msf{st}_R, \tau,\tilde{\tau})$ satisfies the following inequality:
% $$\Pr[b = \top : (\OUT_{\mcal{A}}, b)\gets \langle \mcal{A}(\msf{st}_{\mcal{M}}), \mcal{B}(\msf{st}_C, \msf{st}_R) \rangle(1^\secpar, \tilde{\tau})] \le \epsilon(\secpar).$$
$$\Pr[b_{\mcal{SE}} = \top : (\OUT_{\mcal{SE}}, \Val_{\mcal{SE}}, b_{\mcal{SE}}) \gets \mcal{SE}^{\A(\msf{st}_{\mcal{M}})}(1^\secpar,1^{\epsilon^{-1}},\msf{st}_R,\tau,\tilde{\tau})] \le \epsilon(\secpar).$$

\begin{remark}\label{rmk:p-sim-pref:alternative}
We emphasize that the above probability (on the LHS) is exactly the $p^{\msf{Sim}}_{\msf{pref}}$ defined in \Cref{def:p-sim-pref}, because $\mcal{SE}$ obtains $b_{\mcal{SE}}$ by running $\mcal{G}_1$ (see \Cref{stage:machine:se:main-thread-sim} of \Cref{machine:se}).
\end{remark}

\item
{\bf Event $E_{\msf{acc}}$:} This is the event that $b_{\mcal{SE}} = \top$ at the end of the execution of $\mcal{SE}^{\A(\msf{st}_{\mcal{M}})}(1^\secpar,1^{\epsilon^{-1}},\msf{st}_R,\tau,\tilde{\tau})$. Namely, it means that the honest receiver $R$  accepts in the execution simulated by $\mcal{SE}$. (The subscript ``$\msf{acc}$'' stands for ``accept'').
\end{itemize}
We first show that the the probability of $(E_{\msf{acc}} \wedge E_{\le \epsilon})$ is upper-bounded by $\epsilon(\secpar)$. 

\begin{MyClaim}\label{claim:bounding-E1E2}
It holds that
$$\Pr[E_{\msf{acc}} \wedge E_{\le \epsilon}:(\OUT_{\mcal{SE}}, \Val_{\mcal{SE}}, b_{\mcal{SE}}) \gets \mcal{SE}^{\A(\msf{st}_{\mcal{M}})}(1^\secpar,1^{\epsilon^{-1}},\msf{st}_R,\tau,\tilde{\tau})]  \le \epsilon(\secpar).$$
\end{MyClaim}
\begin{proof}
This is an immediate result of the definitions of these two events. In more details,
\begin{equation}\label[Inequality]{eq:bounding-E1E2}
\Pr[E_{\msf{acc}} \wedge E_{\le \epsilon}] 
= \Pr[E_{\msf{acc}} ~|~ E_{\le \epsilon}] \cdot \Pr[E_{\le \epsilon}] 
 \le \Pr[E_{\msf{acc}} ~|~ E_{\le \epsilon}] \cdot 1 = \epsilon(\secpar) \cdot 1,	
\end{equation}
where all the probabilities are taken over $(\OUT_{\mcal{SE}}, \Val_{\mcal{SE}}, b_{\mcal{SE}}) \gets \mcal{SE}^{\A(\msf{st}_{\mcal{M}})}(1^\secpar,1^{\epsilon^{-1}},\msf{st}_R,\tau,\tilde{\tau})$.

\end{proof}

Next, we upper-bound the LHS of \Cref{eq:one-sided:proof:se:itself} by $\epsilon(\secpar)+\Pr[\Val_{\mcal{SE}} \ne  \msf{val}_{b_{\mcal{SE}}}(\tilde{\tau})~|~E_{\msf{acc}} \wedge \neg E_{\le \epsilon}]$:

\begin{MyClaim}\label{bound:Val-ne-m}
It holds that	
$$\Pr[\Val_{\mcal{SE}} \ne  \msf{val}_{b_{\mcal{SE}}}(\tilde{\tau})] \le \epsilon(\secpar) + \Pr[\Val_{\mcal{SE}} \ne  \msf{val}_{b_{\mcal{SE}}}(\tilde{\tau})~|~E_{\msf{acc}} \wedge \neg E_{\le \epsilon}],$$ where both  probabilities are taken over  $(\OUT_{\mcal{SE}}, \Val_{\mcal{SE}}, b_{\mcal{SE}}) \gets \mcal{SE}^{\A(\msf{st}_{\mcal{M}})}(1^\secpar,1^{\epsilon^{-1}},\msf{st}_R,\tau,\tilde{\tau})$.
\end{MyClaim}
\begin{proof}
All the probabilities below are taken over  $(\OUT_{\mcal{SE}}, \Val_{\mcal{SE}}, b_{\mcal{SE}}) \gets \mcal{SE}^{\A(\msf{st}_{\mcal{M}})}(1^\secpar,1^{\epsilon^{-1}},\msf{st}_R,\tau,\tilde{\tau})$: 
\begin{align*}
\Pr[\Val_{\mcal{SE}} \ne  \msf{val}_{b_{\mcal{SE}}}(\tilde{\tau})] 
& = \Pr[\Val_{\mcal{SE}} \ne  \msf{val}_{b_{\mcal{SE}}}(\tilde{\tau}) \wedge E_{\msf{acc}}] + \Pr[\Val_{\mcal{SE}} \ne  \msf{val}_{b_{\mcal{SE}}}(\tilde{\tau}) \wedge \neg E_{\msf{acc}}]	\\
& = \Pr[\Val_{\mcal{SE}} \ne  \msf{val}_{b_{\mcal{SE}}}(\tilde{\tau}) \wedge E_{\msf{acc}}] \numberthis \label{bound:Val-ne-m:step:1}	\\
& = \Pr[\Val_{\mcal{SE}} \ne  \msf{val}_{b_{\mcal{SE}}}(\tilde{\tau}) \wedge E_{\msf{acc}} \wedge E_{\le \epsilon}] +\Pr[\Val_{\mcal{SE}} \ne  \msf{val}_{b_{\mcal{SE}}}(\tilde{\tau}) \wedge E_{\msf{acc}} \wedge \neg E_{\le \epsilon}]	\\
& \le \Pr[ E_{\msf{acc}} \wedge E_{\le \epsilon}] +  \Pr[\Val_{\mcal{SE}} \ne  \msf{val}_{b_{\mcal{SE}}}(\tilde{\tau}) \wedge E_{\msf{acc}} \wedge \neg E_{\le \epsilon}]\\
& \le \epsilon(\secpar) +  \Pr[\Val_{\mcal{SE}} \ne  \msf{val}_{b_{\mcal{SE}}}(\tilde{\tau}) \wedge E_{\msf{acc}} \wedge \neg E_{\le \epsilon}] \numberthis \label[Inequality]{bound:Val-ne-m:step:2}\\
&\le \epsilon(\secpar) +  \Pr[\Val_{\mcal{SE}} \ne  \msf{val}_{b_{\mcal{SE}}}(\tilde{\tau}) ~|
~ E_{\msf{acc}} \wedge \neg E_{\le \epsilon}],
\end{align*}
where \Cref{bound:Val-ne-m:step:1} follows from the fact that if $b_{\mcal{SE}} = \bot$ then $\Val_{\mcal{SE}} =\msf{val}_{b_{\mcal{SE}}}(\tilde{\tau}) = \bot$, and \Cref{bound:Val-ne-m:step:2} follows from \Cref{claim:bounding-E1E2}.

\end{proof}

Recall that our goal is to establish \Cref{eq:one-sided:proof:se:itself}. Due to \Cref{bound:Val-ne-m}, it  now suffices to show 
\begin{equation}\label{bad:conditioned:E1-nE2}
\Pr[\Val_{\mcal{SE}} \ne  \msf{val}_{b_{\mcal{SE}}}(\tilde{\tau}) ~|
~ E_{\msf{acc}} \wedge \neg E_{\le \epsilon}] = \negl(\secpar),
\end{equation}
where the probability is taken over ${(\OUT_{\mcal{SE}}, \Val_{\mcal{SE}}, b_{\mcal{SE}}) \gets \mcal{SE}^{\A(\msf{st}_{\mcal{M}})}(1^\secpar,1^{\epsilon^{-1}},\msf{st}_R,\tau,\tilde{\tau})}$.

\begin{proof}[Proof of \Cref{bad:conditioned:E1-nE2}]
Given $E_{\msf{acc}}$ (i.e., $b_{\mcal{SE}} = \top$), we know that $\mcal{SE}^{\A(\msf{st}_{\mcal{M}})}(1^\secpar,1^{\epsilon^{-1}},\msf{st}_R,\tau,\tilde{\tau})$ must have entered \Cref{stage:machine:se:rewinding} to execute the machine $\mcal{K}^{\A(\msf{st}_{\mcal{M}})}(1^\secpar, \msf{st}_R, \tau, \tilde{\tau})$ for at most $\frac{\tilde{t}}{\epsilon'(\secpar)} \cdot \secpar$ times, with early termination only if the event $\Val_{\mcal{K}} = \msf{val}(\tilde{\tau})$ happens (recall that $\msf{val}_{b_{\mcal{SE}}}(\tilde{\tau}) = \msf{val}(\tilde{\tau})$ when $b_{\mcal{SE}} = \top$).

Given $\neg E_{\le \epsilon}$ (thus, $p^{\msf{Sim}}_{\msf{pref}} > \epsilon(\secpar)$, see \Cref{rmk:p-sim-pref:alternative}), it follows from \Cref{property:small-tag:proof:se:proof:K} in \Cref{lem:small-tag:proof:se:proof:K} that each time $\mcal{K}$ is invoked, it outputs $\Val_{\mcal{K}} = \msf{val}(\tilde{\tau})$ with probability $\ge \epsilon'(\secpar)/\tilde{t}$.

Therefore, we have 
\begin{equation*}
\Pr[\Val_{\mcal{SE}} \ne  \msf{val}_{b_{\mcal{SE}}}(\tilde{\tau}) ~|
~ E_{\msf{acc}} \wedge \neg E_{\le \epsilon}] = \bigg(1 - \frac{\epsilon'(\secpar)}{\tilde{t}}\bigg)^{\frac{\tilde{t}}{\epsilon'(\secpar)}\cdot \secpar} = 2^{-O(\secpar)} = \negl(\secpar),
\end{equation*}
where the probability is taken over ${(\OUT_{\mcal{SE}}, \Val_{\mcal{SE}}, b_{\mcal{SE}}) \gets \mcal{SE}^{\A(\msf{st}_{\mcal{M}})}(1^\secpar,1^{\epsilon^{-1}},\msf{st}_R,\tau,\tilde{\tau})}$.

This finishes the proof of \Cref{bad:conditioned:E1-nE2}.

\end{proof} 
This eventually finishes the proof of \Cref{lem:small-tag:proof:se} (modulo the proof of \Cref{lem:small-tag:proof:se:proof:K}, which we present in \Cref{sec:lem:small-tag:proof:se:proof:K:proof}
).

\subsection{Extractor $\mcal{K}$ (Proof of \Cref{lem:small-tag:proof:se:proof:K})}
\label{sec:lem:small-tag:proof:se:proof:K:proof}

\begin{figure}[!tb]
\centering
\fbox{
     \includegraphics[width=0.55\textwidth,page=4]{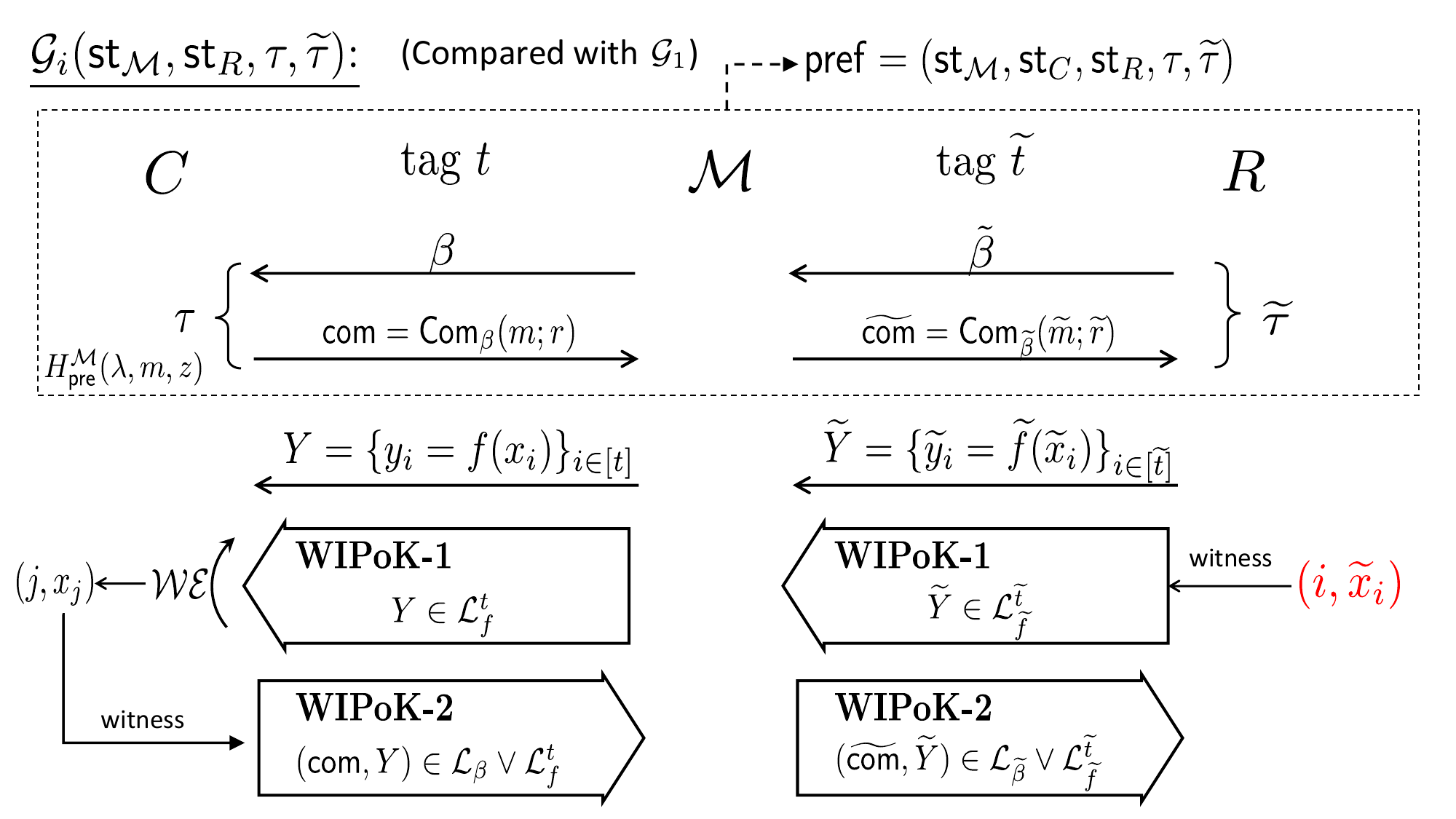}
     }
     \caption{Machine $\mcal{G}_i$ {\scriptsize (Difference with $\mcal{G}_1$ is highlighted in red color)}}
     \label{figure:one-sided:Gi}
\end{figure}

\para{Machine $\mcal{G}_{i}$ ($i \in [\tilde{t}]$):\label{machine:Gi:new}} (Illustrated in \Cref{figure:one-sided:Gi}.)
Recall that we have already defined the machine $\mcal{G}_1(\msf{st}_{\mcal{M}}, \msf{st}_R, \tau,\tilde{\tau})$ on \Cpageref{pageref:one-sided:proof:G1}. Now, for $i \in [\tilde{t}] \setminus \Set{1}$, $\mcal{G}_i(\msf{st}_{\mcal{M}}, \msf{st}_R, \tau,\tilde{\tau})$ behaves identically to $\mcal{G}_1(\msf{st}_{\mcal{M}}, \msf{st}_R, \tau,\tilde{\tau})$ except that it uses $(i, \tilde{x}_i)$ as the witness in the right {\bf WIPoK-1}.

\begin{figure}[!tb]
     \begin{subfigure}[t]{0.47\textwidth}
         \centering
         \fbox{
         \includegraphics[width=\textwidth,page=5]{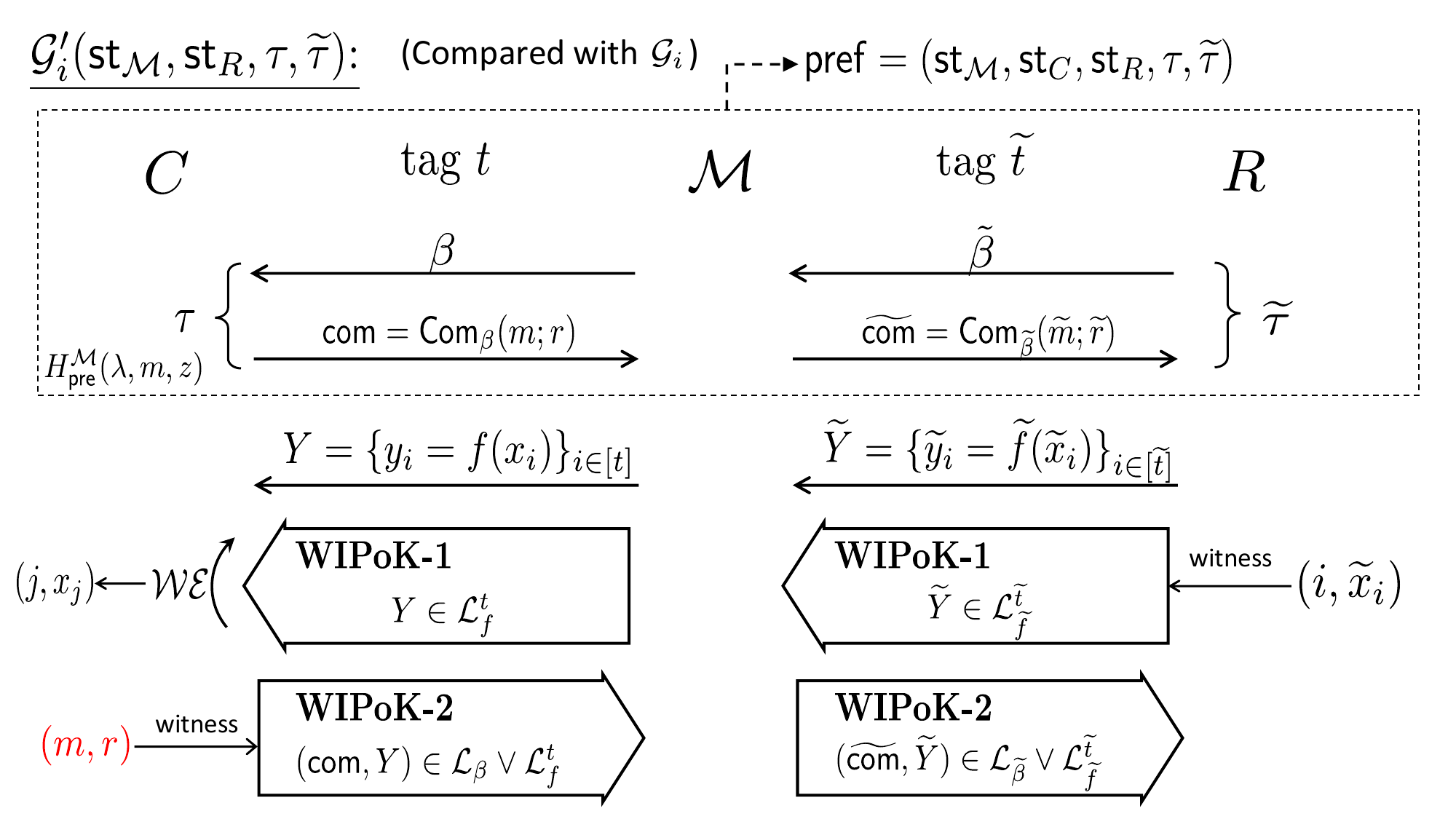}
         }
         \caption{}
         \label{figure:one-sided:G'i}
     \end{subfigure}
     \hspace{6.5pt}
     \begin{subfigure}[t]{0.47\textwidth}
         \centering
         \fbox{
         \includegraphics[width=\textwidth,page=6]{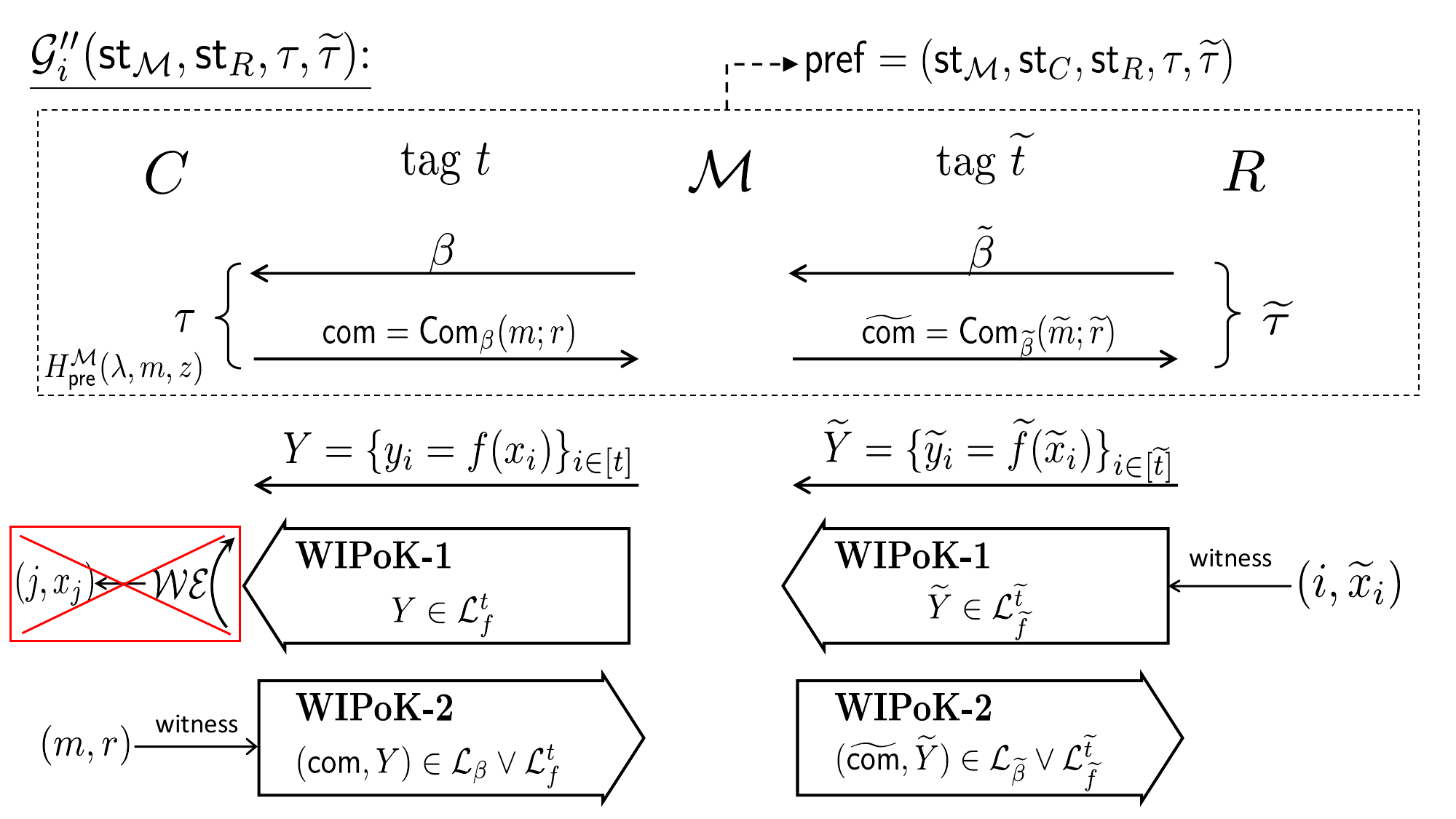}
         }
         \caption{}
         \label{figure:one-sided:G''i}
     \end{subfigure}
     \caption{Machines $\mcal{G}'_i$ and $\mcal{G}''_i$ {\scriptsize (Difference is highlighted in red color)}}
     \label{figure:one-sided:hybrid:G'i:G''i}
\end{figure}

\begin{MyClaim}\label{claim:bound:Gi:new}  
$\forall i \in [\tilde{t}], ~\Pr[b = \top : (\OUT, b) \gets \mcal{G}_i(\msf{st}_{\mcal{M}}, \msf{st}_R, \tau,\tilde{\tau})] \ge p^{\msf{Sim}}_{\msf{pref}} - \negl(\secpar)$.
\end{MyClaim}
\begin{proof}
We define hybrid machines $\mcal{G}'_i$ and $\mcal{G}''_i$ as follows.

\para{Machine $\mcal{G}'_{i}$ ($i \in [\tilde{t}]$):\label{machine:G'i:new}} (Illustrated in \Cref{figure:one-sided:G'i}.)
Recall that we have already defined the machine $\mcal{G}'_1(\msf{st}_{\mcal{M}}, \msf{st}_R, \tau,\tilde{\tau})$ on \Cpageref{pageref:one-sided:proof:G'1}. Now, for $i \in [\tilde{t}] \setminus \Set{1}$, $\mcal{G}'_i(\msf{st}_{\mcal{M}}, \msf{st}_R, \tau,\tilde{\tau})$ behaves identically to $\mcal{G}'_1(\msf{st}_{\mcal{M}}, \msf{st}_R, \tau,\tilde{\tau})$ except that it uses $(i, \tilde{x}_i)$ as the witness in the right {\bf WIPoK-1}.

\para{Machine $\mcal{G}''_i$ ($i \in [\tilde{t}]$):} (Illustrated in \Cref{figure:one-sided:G''i}\label{pageref:one-sided:proof:G''1}) For any prefix $\msf{pref}$, $\mcal{G}''_i(\msf{st}_{\mcal{M}}, \msf{st}_R, \tau,\tilde{\tau})$ behaves identically to $\mcal{G}'_i(\msf{st}_{\mcal{M}}, \msf{st}_R, \tau,\tilde{\tau})$ except that it honestly runs the left {\bf WIPoK-1}, instead of running the witness-extended emulator $\mcal{WE}$. In other words, $\mcal{G}''_i(\msf{st}_{\mcal{M}}, \msf{st}_R, \tau,\tilde{\tau})$ behaves identically to $\langle \mcal{A}(\msf{st}_{\mcal{M}}), \mcal{B}(\msf{st}_C, \msf{st}_R) \rangle(1^\secpar, \tilde{\tau})$ except that it uses $(i,\tilde{x}_i)$ as the witness in the right {\bf WIPoK-1}. 
In particular, $\mcal{G}''_i(\msf{st}_{\mcal{M}}, \msf{st}_R, \tau,\tilde{\tau})$ is identical to $\langle \mcal{A}(\msf{st}_{\mcal{M}}), \mcal{B}(\msf{st}_C, \msf{st}_R) \rangle(1^\secpar, \tilde{\tau})$. 

Then, \Cref{claim:bound:Gi:new} follows from the following sequence of inequalities. 
\begin{itemize}
\item
By the  WI property of the left {\bf WIPoK-2} and the definition of $p^{\msf{Sim}}_{\msf{pref}}$ (\Cref{def:p-sim-pref}), it holds that:
$$\Pr[b = \top : (\OUT, b) \gets \mcal{G}'_1(\msf{st}_{\mcal{M}}, \msf{st}_R, \tau,\tilde{\tau})] \ge p^{\msf{Sim}}_{\msf{pref}} - \negl(\secpar).$$
\item
By the PoK property (per \Cref{def:WEE}) of the left {\bf WIPoK-1} and the above inequality, it holds that:
$$\Pr[b = \top : (\OUT, b) \gets \mcal{G}''_1(\msf{st}_{\mcal{M}}, \msf{st}_R, \tau,\tilde{\tau})] \ge p^{\msf{Sim}}_{\msf{pref}} - \negl(\secpar).$$
\item
By the WI property of the right {\bf WIPoK-1} and the above inequality, it holds that:
$$\forall i \in [\tilde{t}],~\Pr[b = \top : (\OUT, b) \gets \mcal{G}''_i(\msf{st}_{\mcal{M}}, \msf{st}_R, \tau,\tilde{\tau})] \ge p^{\msf{Sim}}_{\msf{pref}} - \negl(\secpar).$$
\item
By the PoK property of the left {\bf WIPoK-1} and the above inequality, it holds that:
$$\forall i \in [\tilde{t}],~\Pr[b = \top : (\OUT, b) \gets \mcal{G}'_i(\msf{st}_{\mcal{M}}, \msf{st}_R, \tau,\tilde{\tau})] \ge p^{\msf{Sim}}_{\msf{pref}} - \negl(\secpar).$$
\item
By the  WI property of the left {\bf WIPoK-2} and the above inequality, it holds that:
$$\forall i \in [\tilde{t}],~\Pr[b = \top : (\OUT, b) \gets \mcal{G}_i(\msf{st}_{\mcal{M}}, \msf{st}_R, \tau,\tilde{\tau})] \ge p^{\msf{Sim}}_{\msf{pref}} - \negl(\secpar).$$
\end{itemize}
This finishes the proof of \Cref{claim:bound:Gi:new}.
% \begin{xiaoenv}{}This proof is simple:
% \begin{itemize}
% \item
% Define the machines $\mcal{G}'_i$ and $\mcal{G}''_i$ shown in \Cref{figure:one-sided:G'i,figure:one-sided:G''i} respectively.
% \item
% By the  WI property of the left {\bf WIPoK-2}, it holds that:
% $$\Pr[b = \top : (\OUT, b) \gets \mcal{G}'_1(\msf{st}_{\mcal{M}}, \msf{st}_R, \tau,\tilde{\tau})] \ge p^{\msf{Sim}}_{\msf{pref}} - \negl(\secpar).$$
% \item
% By the PoK property of the left {\bf WIPoK-1}, it holds that:
% $$\Pr[b = \top : (\OUT, b) \gets \mcal{G}''_1(\msf{st}_{\mcal{M}}, \msf{st}_R, \tau,\tilde{\tau})] \ge p^{\msf{Sim}}_{\msf{pref}} - \negl(\secpar).$$
% \item
% By the WI property of the right {\bf WIPoK-1}, it holds that:
% $$\forall i \in [\tilde{t}],~\Pr[b = \top : (\OUT, b) \gets \mcal{G}''_i(\msf{st}_{\mcal{M}}, \msf{st}_R, \tau,\tilde{\tau})] \ge p^{\msf{Sim}}_{\msf{pref}} - \negl(\secpar).$$
% \item
% By the PoK property of the left {\bf WIPoK-1}, it holds that:
% $$\forall i \in [\tilde{t}],~\Pr[b = \top : (\OUT, b) \gets \mcal{G}'_i(\msf{st}_{\mcal{M}}, \msf{st}_R, \tau,\tilde{\tau})] \ge p^{\msf{Sim}}_{\msf{pref}} - \negl(\secpar).$$
% \item
% By the  WI property of the left {\bf WIPoK-2}, it holds that:
% $$\forall i \in [\tilde{t}],~\Pr[b = \top : (\OUT, b) \gets \mcal{G}_i(\msf{st}_{\mcal{M}}, \msf{st}_R, \tau,\tilde{\tau})] \ge p^{\msf{Sim}}_{\msf{pref}} - \negl(\secpar).$$
% \end{itemize}
% \end{xiaoenv}

\end{proof}

\begin{figure}[!tb]
     \begin{subfigure}[t]{0.47\textwidth}
         \centering
         \fbox{
         \includegraphics[width=\textwidth,page=7]{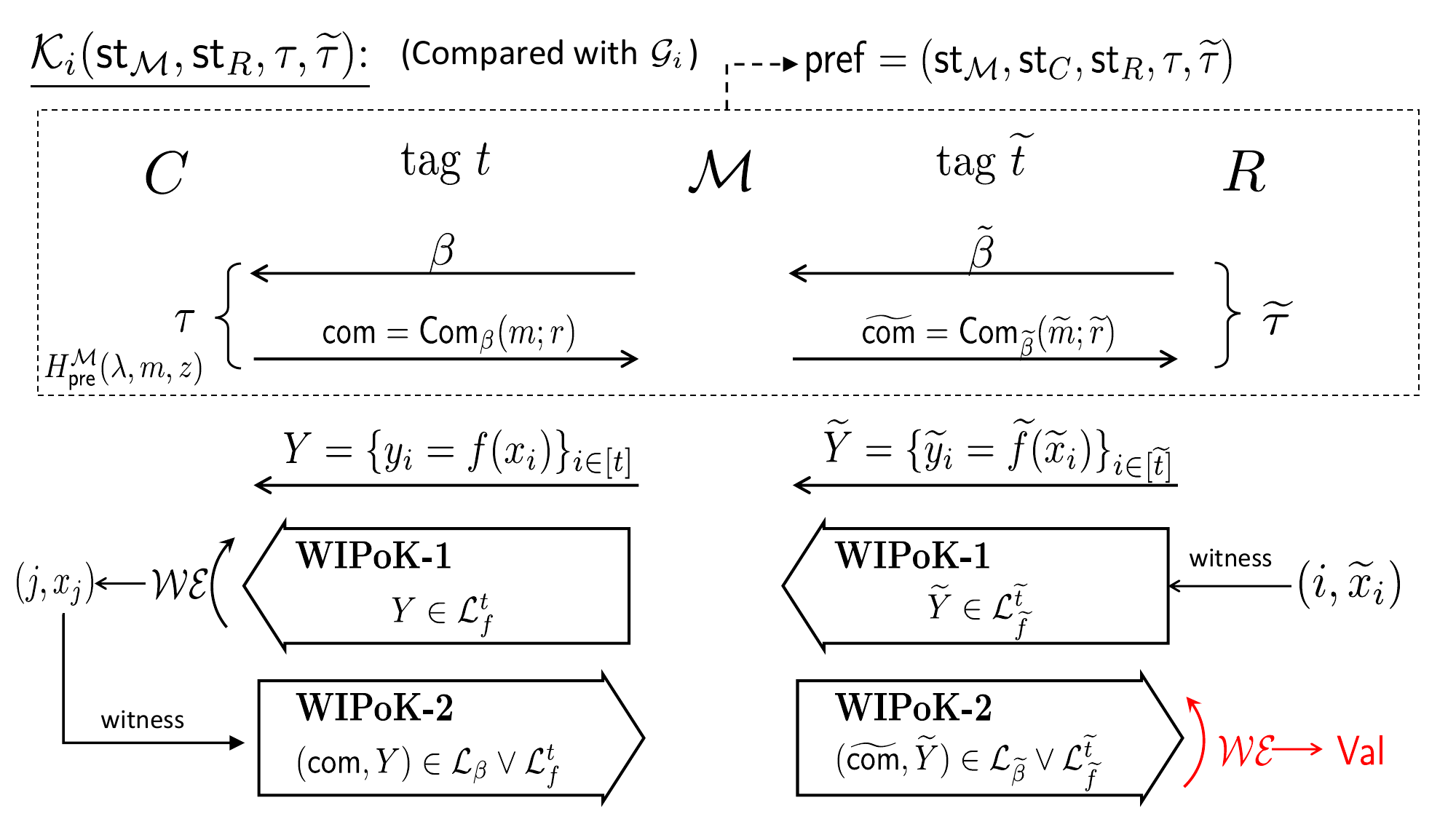}
         }
         \caption{}
         \label{figure:one-sided:Ki}
     \end{subfigure}
     \hspace{6.5pt}
     \begin{subfigure}[t]{0.47\textwidth}
         \centering
         \fbox{
         \includegraphics[width=\textwidth,page=8]{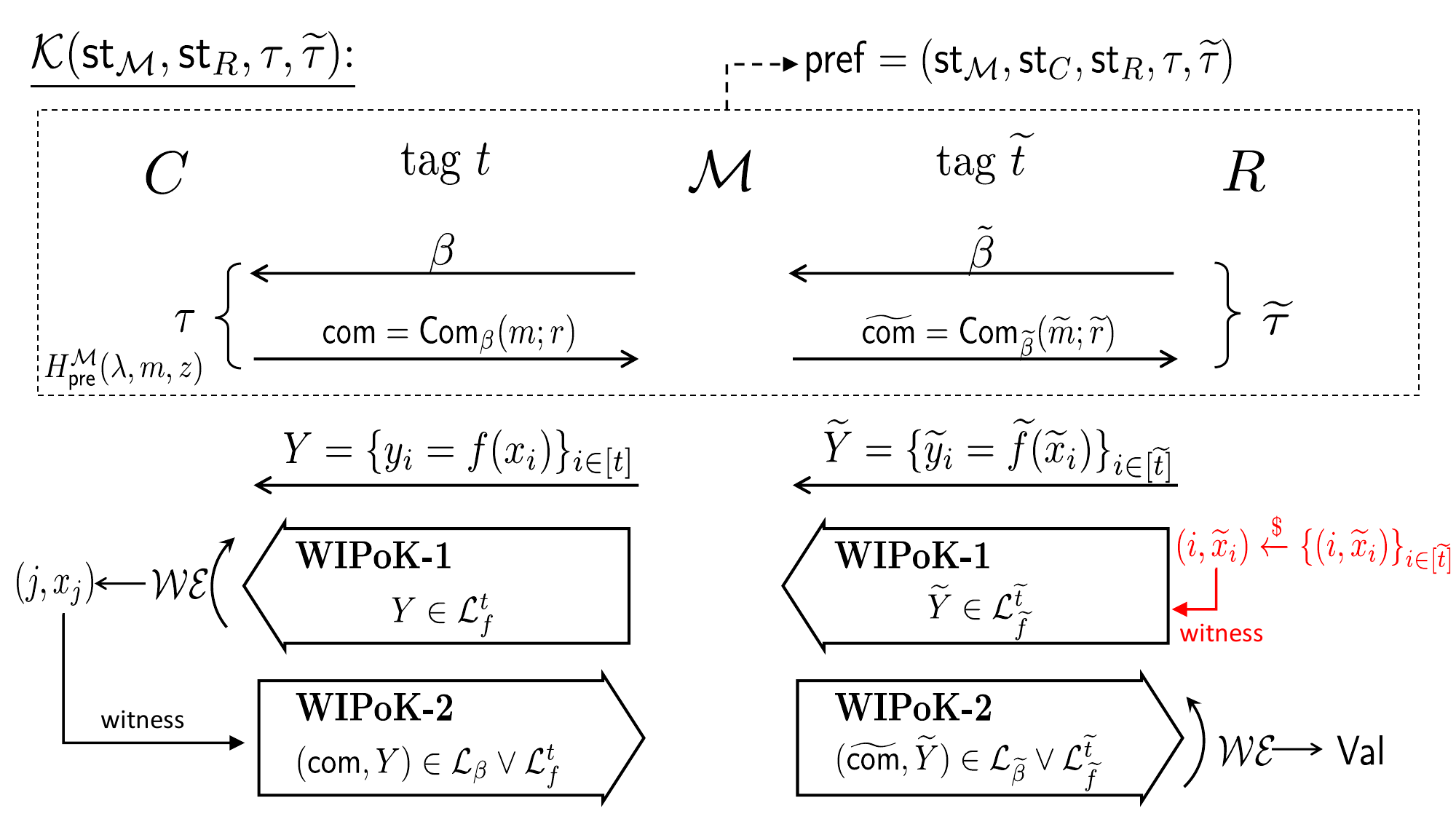}
         }
         \caption{}
         \label{figure:one-sided:K}
     \end{subfigure}
     \caption{Machines $\mcal{K}_i$ and $\mcal{K}$ {\scriptsize (Difference is highlighted in red color)}}
     \label{figure:one-sided:hybrid:Ki:K}
\end{figure}

\para{Machine $\mcal{K}_i$ ($i\in [\tilde{t}]$):} (Illustrated in \Cref{figure:one-sided:Ki}.) For a prefix $\msf{pref}$, $\mcal{K}_i(\msf{st}_{\mcal{M}}, \msf{st}_R, \tau,\tilde{\tau})$ 
behaves identically to the $\mcal{G}_i(\msf{st}_{\mcal{M}}, \msf{st}_R, \tau,\tilde{\tau})$ depicted in \Cref{figure:one-sided:Gi}, except for the following difference. Machine $\mcal{K}_i(\msf{st}_{\mcal{M}}, \msf{st}_R, \tau,\tilde{\tau})$ uses the witness-extended emulator $\mcal{WE}$ (as per \Cref{def:WEE}) to finish the right {\bf WIPoK-2}, instead of playing the role of the honest receiver. 

\subpara{$\mcal{K}_i$'s Output:\label{subpara:K-i:output}} Let $w'$ denote the third output of the witness-extended emulator $\mcal{WE}$ (see \Cref{def:WEE}), which is supposed to be the witness used by $\mcal{M}$ in the right {\bf WIPoK-2} (for the statement $(\tilde{\msf{com}}, \tilde{Y}$) w.r.t.\ the language $\Lang_{\tilde{\beta}}\vee \Lang^{\tilde{t}}_{\tilde{f}}$). Depending on the value of $w'$, we define a value $\msf{Val} \in \bits^{\ell(\secpar)} \cup \Set{\bot_{\tilde{Y}}, \bot_{\msf{invalid}}}$ as follows:
\begin{enumerate}
\item \label[Case]{K-i:output:case:1}
If $w'$ is a valid witness for $(\tilde{\msf{com}}, \tilde{Y}) \in \Lang_{\tilde{\beta}}\vee \Lang^{\tilde{t}}_{\tilde{f}}$, then there are tow sub-cases:
\begin{enumerate}
\item \label[Case]{K-i:output:case:1a}
$w'$ is a valid witness for $\tilde{\msf{com}} \in \Lang_{\tilde{\beta}}$. In this case\footnote{Technically, we should argue that $\mcal{K}_i$ is able to detect if this case is happening. This is easy to do as we explicitly give $\tilde{\tau}$ as an input to $\mcal{K}_i$---Upon obtaining $w'$, it can just re-execute Naor's commitment by itself to see if $w'$ is consistent with $\tilde{\tau}$.}, $w'$ consists of the value $\msf{val}(\tilde{\tau})$, i.e., the value committed in $\tilde{\tau}=(\tilde{\beta}, \tilde{\msf{com}})$, {\em and} the randomness $\tilde{r}$. We set $\msf{Val} \coloneqq \msf{val}(\tilde{\tau})$. Importantly, notice that we do {\em not} include the randomness $\tilde{r}$ in $\msf{Val}$ (as explained in \Cref{rmk:small-tag:proof:se:proof:K:output}.). 
\item \label[Case]{K-i:output:case:1b}
$w'$ is a valid witness for $\tilde{Y} \in \Lang^{\tilde{t}}_{\tilde{f}}$: In this case, we set $\msf{Val} \coloneqq \bot_{\tilde{Y}}$.
\end{enumerate}

\item \label[Case]{K-i:output:case:2}
Otherwise, set $\msf{Val} \coloneqq \bot_{\msf{invalid}}$.
\end{enumerate}
% \xiao{It seems that we can define the output of $\mcal{K}_i$ just as $w'$. We only need to use the notion $\msf{Val}$ for $\mcal{K}$'s output. If we really did this change, we need to modify the picture for $\mcal{K}_i$. It should be $\mcal{SE} \rightarrow w'$.} 
% \takashi{I don't think any change is needed here.}\xiao{Oh, what I wanted to do is only for notational clarity. I didn't mean to say that this step is not accurate. I just want to improve the presentation.}\takashi{I thought that the output cannot be $w'$ for \Cref{rmk:small-tag:proof:se:proof:K:output}.}
The output of $\mcal{K}_i$ is defined to be the above $\msf{Val}$. Notice that this is in contrast to all previous machines, for which the output is defined to be the man-in-the-middle $\mcal{M}$'s output and the honest receiver's decision bit. We emphasize that such a $\msf{Val}$ satisfies the syntax requirement in \Cref{property:small-tag:proof:se:proof:K:syntax} of \Cref{lem:small-tag:proof:se:proof:K}. In particular, {\em $\msf{Val} = \msf{val}(\tilde{\tau})$ whenever $\msf{Val} \ne \bot$}.\footnote{Note that here we defined two types of abortion: $\bot_{\tilde{Y}}$ and $\bot_{\msf{invalid}}$, while \Cref{property:small-tag:proof:se:proof:K:syntax} of \Cref{lem:small-tag:proof:se:proof:K} only allows a single abortion symbol $\bot$. We remark that this is only a cosmetic difference---It can be made consistent using the following rules: $\bot = \bot_{\tilde{Y}}$ {\em and} $\bot = \bot_{\msf{invalid}}$ (i.e., $\msf{Val} \ne \bot  \Leftrightarrow (\msf{Val} \ne \bot_{\tilde{Y}} \wedge \msf{Val} \ne \bot_{\msf{invalid}})$).}

\begin{MyClaim}\label{claim:bound:Ki}
$\forall i \in [\tilde{t}], ~\Pr[\Val \ne \bot_{\msf{invalid}} :\Val \gets \mcal{K}_i(\msf{st}_{\mcal{M}}, \msf{st}_R, \tau,\tilde{\tau})] \ge p^{\msf{Sim}}_{\msf{pref}} - \negl(\secpar)$.
\end{MyClaim}
\begin{proof}
This claim follows immediately from \Cref{claim:bound:Gi:new} and the PoK property (as per \Cref{def:WEE}) of the right {\bf WIPoK-2}. 
%witness-extended emulation property (as per \Cref{def:WEE}) of the right {\bf WIPoK-2}.

\end{proof}

Finally, we are ready to define the extractor $\mcal{K}$ as required by \Cref{lem:small-tag:proof:se:proof:K}. Intuitively, $\mcal{K}$ can be conceived as an average-case version of $\Set{\mcal{K}_i}_{i\in[\tilde{t}]}$:

\para{Extractor $\mcal{K}$:\label{extractorK:description}} (Illustrated in \Cref{figure:one-sided:K}.)  For a prefix $\msf{pref}$, $\mcal{K}(\msf{st}_{\mcal{M}}, \msf{st}_R, \tau,\tilde{\tau})$ samples uniformly at random an index $i \pick [\tilde{t}]$, executes $\mcal{K}_i(\msf{st}_{\mcal{M}}, \msf{st}_R, \tau,\tilde{\tau})$, and outputs whatever $\mcal{K}_i(\msf{st}_{\mcal{M}}, \msf{st}_R, \tau,\tilde{\tau})$ outputs.

\begin{remark}[On the Notation of $\mcal{K}$]\label{rmk:notation-K}
It is worth noting that in the above, we write machine $\mcal{K}$ as $\mcal{K}(\msf{st}_{\mcal{M}}, \msf{st}_R, \tau,\tilde{\tau})$, while it was written in \Cref{lem:small-tag:proof:se:proof:K} as $\mcal{K}^{\A(\msf{st}_{\mcal{M}})}(1^\secpar, \msf{st}_R, \tau, \tilde{\tau})$. This is only a cosmetic difference. 
\end{remark}

Next, we show that the extractor $\mcal{K}$ satisfies the requirements in \Cref{lem:small-tag:proof:se:proof:K}.

\para{Running Time of $\mcal{K}$.} Observe that for each $i \in [\tilde{t}]$, $\mcal{K}_i(\msf{st}_{\mcal{M}}, \msf{st}_R, \tau,\tilde{\tau})$ differs from the real man-in-the-middle game only in the following places: 
\begin{itemize}
\item
$(i,\tilde{x}_i)$ is used in the right {\bf WIPoK-1};
\item
the witness-extended emulator $\mcal{WE}$ is used in the right {\bf WIPoK-2} and the left {\bf WIPoK-1}; 
\item
the left {\bf WIPoK-2} is done using the  $(j, x_j)$ extracted by $\mcal{WE}$ from the left {\bf WIPoK-1}.
\end{itemize}
Since the witness-extended emulator $\mcal{WE}$ (as per \Cref{def:WEE}) runs in expected PPT, so does $\mcal{K}_i$. Thus, $\mcal{K}$ is expected PPT.

\para{Proving \Cref{property:small-tag:proof:se:proof:K:syntax} of \Cref{lem:small-tag:proof:se:proof:K}.} It is straightforward to see that the $\mcal{K}$ defined above satisfies the syntax requirement in \Cref{lem:small-tag:proof:se:proof:K} (see also \Cref{rmk:notation-K}). In particular, we have $\msf{Val}_{\mcal{K}} = \msf{val}(\tilde{\tau})$ whenever $\msf{Val}_{\mcal{K}} \ne \bot$, because this is true for each $\mcal{K}_i$ by definition (see the paragraph for ``\underline{$\mcal{K}_i$'s Output}'' on \Cpageref{subpara:K-i:output}).

\para{Proving \Cref{property:small-tag:proof:se:proof:K} of \Cref{lem:small-tag:proof:se:proof:K}.} First, recall that \Cref{property:small-tag:proof:se:proof:K} requires us to show that 
for any $\msf{pref}$ in the support of $H^{\mcal{M}}_{\msf{pre}}(\secpar, m, z)$ , if $p^{\msf{Sim}}_{\msf{pref}} \ge \epsilon(\secpar)$, then it holds that
\begin{equation}\label[Inequality]{eq:bound:averageK}
\Pr[\Val_{\mcal{K}} = \msf{val}(\tilde{\tau}) : \Val_{\mcal{K}} \gets \mcal{K}(\msf{st}_{\mcal{M}}, \msf{st}_R, \tau, \tilde{\tau})] \ge \frac{\epsilon'(\secpar)}{\tilde{t}}.
\end{equation} 
Also recall that $\mcal{K}(\msf{st}_{\mcal{M}}, \msf{st}_R, \tau, \tilde{\tau})$ is defined to execute the machine $\mcal{K}_i(\msf{st}_{\mcal{M}}, \msf{st}_R, \tau, \tilde{\tau})$  with  $i$ uniformly sampled from $[\tilde{t}]$. Therefore, \Cref{eq:bound:averageK} can be reduced to the following \Cref{lem:bound:Ki}. We will prove \Cref{lem:bound:Ki} in \Cref{sec:lem:bound:Ki:proof}, which will eventually finish the current proof of \Cref{lem:small-tag:proof:se:proof:K}.
\begin{lemma}\label{lem:bound:Ki}
Let $\epsilon(\secpar) = \frac{1}{\poly(\secpar)}$ and $\epsilon'(\secpar) = \frac{\epsilon(\secpar)}{t^2}$.
For any $\msf{pref} = (\msf{st}_{\mcal{M}},\msf{st}_C, \msf{st}_R, \tau, \tilde{\tau})$, if $p^{\msf{Sim}}_{\msf{pref}} \ge \epsilon(\secpar)$, then there exists an $i \in [\tilde{t}]$ such that 
$$\Pr[\Val = \msf{val}(\tilde{\tau}) :\Val \gets \mcal{K}_i(\msf{st}_{\mcal{M}}, \msf{st}_R, \tau, \tilde{\tau})] \ge \epsilon'(\secpar).$$
\end{lemma}

\subsection{Proof of \Cref{lem:bound:Ki}}
\label{sec:lem:bound:Ki:proof}

\para{Notation.} We highly recommend reviewing the ``\underline{$\mcal{K}_i$'s Output}'' part on \Cpageref{subpara:K-i:output} (in particular, the meanings of $\msf{Val}$, $\bot_{\tilde{Y}}$, and $\bot_{\msf{invalid}}$) before starting to read this subsection.  Recall that $\mcal{K}_i$'s output $\msf{Val}$ is determined by the $w'$ output by the witness-extended emulator of the right {\bf WIPoK-2}.  %simulation-extractor $\mcal{SE}$.
In this subsection, we will need to refer to this $w'$, though it is not explicitly included as a part of $\mcal{K}_i$'s output. Particularly, we will make use of the following notation: whenever we write an expression of the form 
 $$\Pr[\text{Some Event $E_{w'}$ about $w'$}: \msf{Val} \gets \mcal{K}_i(\msf{st}_{\mcal{M}}, \msf{st}_R, \tau, \tilde{\tau})],$$
it should be understood as the probability of $E_{w'}$, where $w'$ refers to the $w'$ generated during the random procedure $\msf{Val} \gets\mcal{K}_i(\msf{st}_{\mcal{M}}, \msf{st}_R, \tau, \tilde{\tau})$, over which the probability is taken. 

Using these notations, we can partition the event $\msf{Val} = \bot_{\tilde{Y}}$ as the following {\em mutually exclusive} events: $w' = (1, \tilde{x}_1)$, $\ldots$, $w' =(\tilde{t}, \tilde{x}_{\tilde{t}})$, where $\tilde{y}_i = \tilde{f}(\tilde{x}_i)$ for each $i \in [\tilde{t}]$. Formally, we express this relation by 
\begin{equation}\label{eq:bot-tilde-Y:partition}
\Pr[\msf{Val} = \bot_{\tilde{Y}}: \msf{Val} \gets \mcal{K}_i(\msf{st}_{\mcal{M}}, \msf{st}_R, \tau, \tilde{\tau})] = \sum_{i = 1}^{\tilde{t}} \Pr[w' = (i,\tilde{x}_i): \msf{Val} \gets \mcal{K}_i(\msf{st}_{\mcal{M}}, \msf{st}_R, \tau, \tilde{\tau})]
\end{equation}

With the above notations, we prove \Cref{lem:bound:Ki} in the following.

\para{Proof for \Cref{lem:bound:Ki}.} We assume for contradiction that for some $\msf{pref}$ satisfying $p^{\msf{Sim}}_{\msf{pref}} \ge \epsilon(\secpar)$, it holds that 
\begin{equation}\label[Inequality]{eq:proof:averageK:contra-assump}
\forall i \in [\tilde{t}], ~\Pr[\Val = \msf{val}(\tilde{\tau}) :\Val \gets \mcal{K}_i(\msf{st}_{\mcal{M}}, \msf{st}_R, \tau, \tilde{\tau})] < \epsilon'(\secpar).
\end{equation}
\begin{MyClaim}\label{cliam:bound:Ki-xi}
Under the assumption in \Cref{eq:proof:averageK:contra-assump}, it holds that 
$$\forall i \in [\tilde{t}], ~\Pr[w' = (i, \tilde{x}_i) :\Val \gets \mcal{K}_i(\msf{st}_{\mcal{M}}, \msf{st}_R, \tau, \tilde{\tau})] \ge p^{\msf{Sim}}_{\msf{pref}} - \epsilon'(\secpar) - \negl(\secpar).$$
\end{MyClaim}
\begin{proof}
% First, recall that the event $\Val \ne \bot$ (i.e., $\Val$ is a valid witness) can be partitioned into the following {\em disjoint} events: $\Val = \msf{val}(\tilde{\tau})$, $\Val = \tilde{x}_1$, \ldots, $\Val = \tilde{x}_{\tilde{t}}$. Therefore,
In this proof, all the probabilities are taken over the random procedure $\Val \gets \mcal{K}_i(\msf{st}_{\mcal{M}}, \msf{st}_R, \tau, \tilde{\tau})$.

 First, notice that
\begin{align*}
\forall i \in [\tilde{t}], ~\Pr[\Val \ne \bot_{\msf{invlid}}] 
& = 
\Pr[\Val = \msf{val}(\tilde{\tau})] +  \Pr[\Val = \bot_{\tilde{Y}}] \\
\text{(by \Cref{eq:bot-tilde-Y:partition})} ~& = 
\Pr[\Val = \msf{val}(\tilde{\tau})] + \Pr[w' = (i, \tilde{x}_i)] + \sum_{j \in [\tilde{t}]\setminus \Set{i}} \Pr[w' = (j, \tilde{x}_j)]. \numberthis \label{eq:bound:Ki-xi:1}
\end{align*}
Then, the following holds:
{\begingroup\fontsize{10.5pt}{0pt}\selectfont
\begin{align}
\forall i \in [\tilde{t}], ~\Pr[w' = (i, \tilde{x}_i)] 
& = 
\Pr[\Val \ne \bot_{\msf{invalid}}] - \Pr[\Val = \msf{val}(\tilde{\tau})] -  \bigg(\sum_{j \in [\tilde{t}]\setminus \Set{i}} \Pr[w' = (j, \tilde{x}_j)] \bigg)  \label{eq:bound:Ki-xi:derive:1} \\
& \ge p^{\msf{Sim}}_{\msf{pref}} - \negl(\secpar) - \Pr[\Val = \msf{val}(\tilde{\tau})] -  \bigg(\sum_{j \in [\tilde{t}]\setminus \Set{i}} \Pr[w' = (j,\tilde{x}_j)] \bigg)  \label[Inequality]{eq:bound:Ki-xi:derive:2}\\
& \ge p^{\msf{Sim}}_{\msf{pref}} - \negl(\secpar) - \epsilon'(\secpar) -  \bigg(\sum_{j \in [\tilde{t}]\setminus \Set{i}} \Pr[w' = (j,\tilde{x}_j)] \bigg),\label[Inequality]{eq:bound:Ki-xi:derive:3}
\end{align}\endgroup}where \Cref{eq:bound:Ki-xi:derive:1} follows from \Cref{eq:bound:Ki-xi:1}, \Cref{eq:bound:Ki-xi:derive:2} follows from \Cref{claim:bound:Ki}, and \Cref{eq:bound:Ki-xi:derive:3} follows from \Cref{eq:proof:averageK:contra-assump}.

Now, to prove \Cref{cliam:bound:Ki-xi}, it suffices to show that 
\begin{equation}
\forall i \in [\tilde{t}],~\forall j \in [\tilde{t}]\setminus \Set{i}, ~\Pr[w' = (j,\tilde{x}_j) : \Val \gets \mcal{K}_i(\msf{st}_{\mcal{M}}, \msf{st}_R, \tau, \tilde{\tau})] = \negl(\secpar).
\end{equation}
This can be reduced via standard techniques to the one-wayness of the OWF $\tilde{f}$ in \Cref{item:one-sided:step:OWFs} of the right execution. In more details, we assume for contradiction that there exist $i^*, j^* \in [\tilde{t}]$ such that $i^* \ne j^*$ and that the probability $\Pr[w' = (j^*, \tilde{x}_{j^*}) : \Val \gets \mcal{K}_{i^*}(\msf{st}_{\mcal{M}}, \msf{st}_R, \tau, \tilde{\tau})]$ is non-negligible, where, by definition, $\tilde{x}_{j^*}$ is the preimage of $\tilde{y}_{j^*}$ under the right OWF $\tilde{f}$. Then, we can build a PPT adversary $\Adv_\textsc{owf}$ breaking one-wayness in the following way: $\Adv_\textsc{owf}$ obtains the challenge $y^*$ from the external one-wayness challenger; it then runs the machine $\mcal{K}_{i^*}(\msf{pref})$ internally, for which $\Adv_\textsc{owf}$ uses $y^*$ in place of $\tilde{y}_{j^*}$ when executing \Cref{item:one-sided:step:OWFs} in the right. Note that the internal execution of $\mcal{K}_{i^*}(\msf{pref})$ is identically to the real execution of $\mcal{K}_{i^*}$, thus the extracted $w' =(j^*, \tilde{x}_{j^*})$ must satisfy $\tilde{f}(\tilde{x}_{j^*}) = \tilde{y}_j^*~(=y^*)$ with non-negligible probability, breaking one-wayness. 

This finishes the proof of \Cref{cliam:bound:Ki-xi}.

\end{proof}

\begin{figure}[!tb]
     \begin{subfigure}[t]{0.47\textwidth}
         \centering
         \fbox{
         \includegraphics[width=\textwidth,page=9]{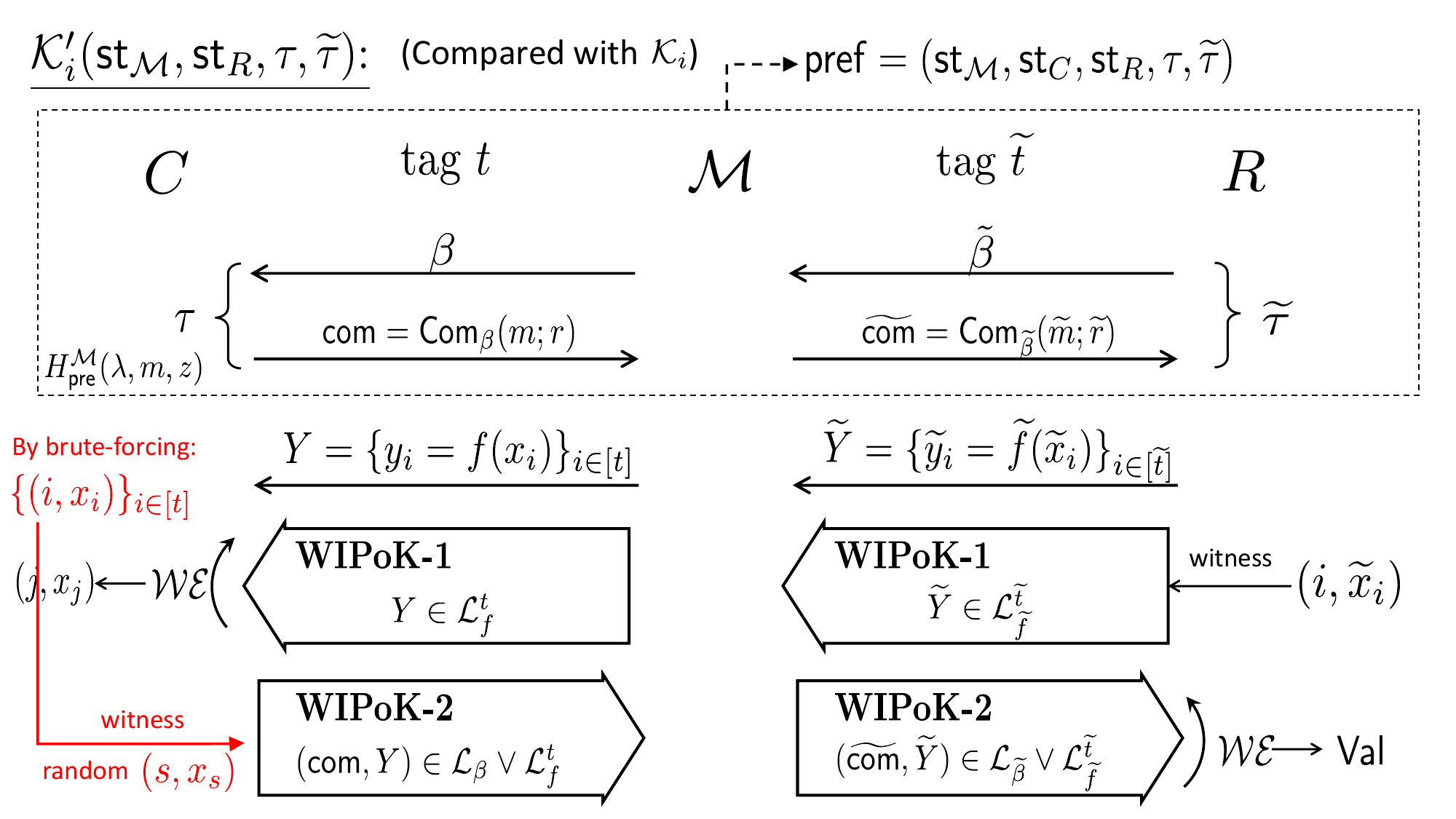}
         }
         \caption{}
         \label{figure:one-sided:K'i}
     \end{subfigure}
     \hspace{6.5pt}
     \begin{subfigure}[t]{0.47\textwidth}
         \centering
         \fbox{
         \includegraphics[width=\textwidth,page=10]{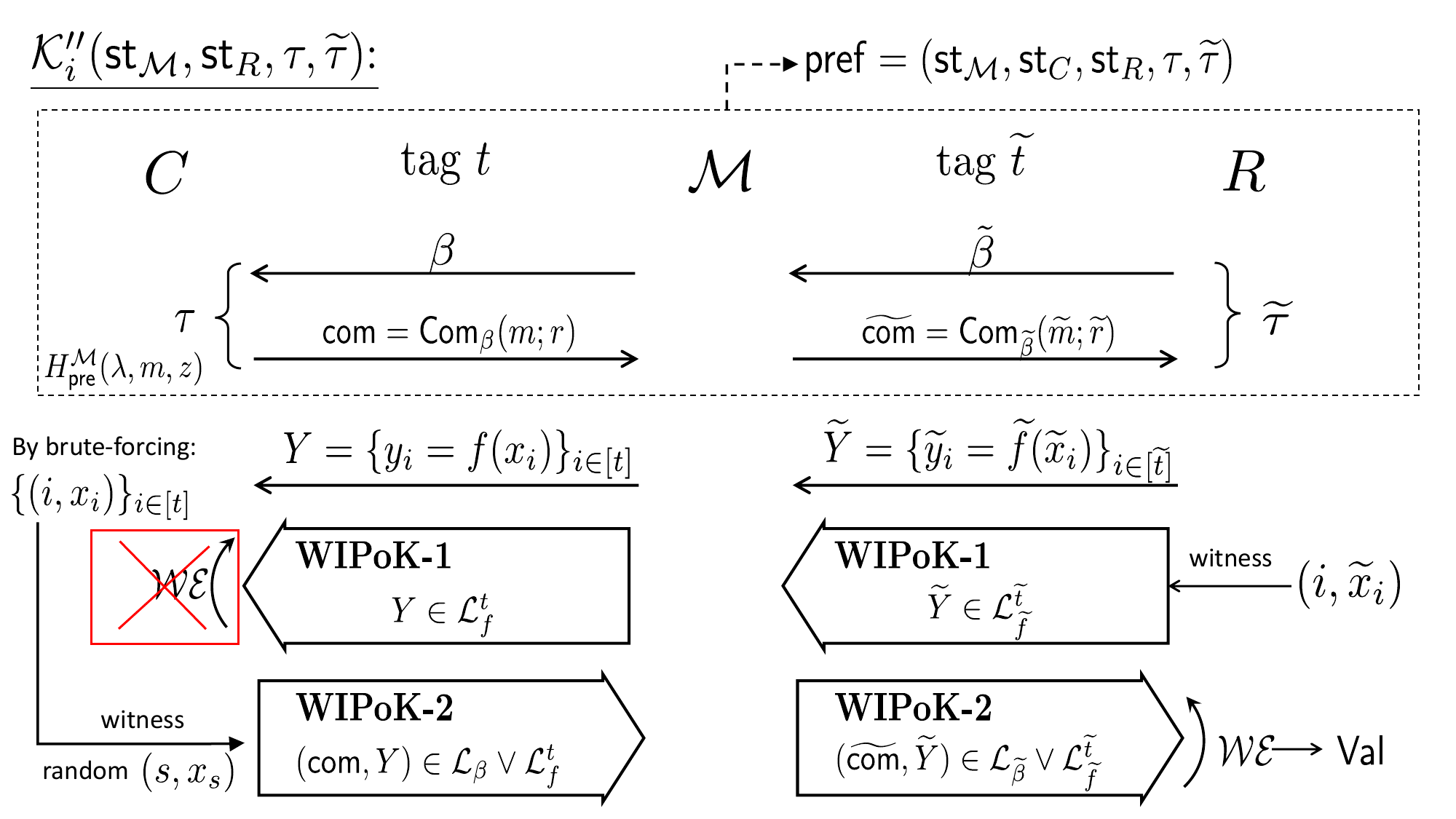}
         }
         \caption{}
         \label{figure:one-sided:K''i}
     \end{subfigure}
     \caption{Machines $\mcal{K}'_i$ and $\mcal{K}''_i$ {\scriptsize (Difference is highlighted in red color)}}
     \label{figure:one-sided:K'i-K''i}
\end{figure}

\para{Machine $\mcal{K}'_i$ ($i \in [\tilde{t}]$):\label{machineK':description}} (Illustrated in \Cref{figure:one-sided:K'i}.) For a prefix $\msf{pref}$, $\mcal{K}'_i(\msf{st}_{\mcal{M}}, \msf{st}_R, \tau, \tilde{\tau})$ proceeds as follows:
\begin{enumerate}
\item
It first finishes \Cref{item:one-sided:step:OWFs} of the man-in-the-middle execution in the same way as $\mcal{K}_i(\msf{st}_{\mcal{M}}, \msf{st}_R, \tau, \tilde{\tau})$. In particular, it will see in the left execution the values $Y = (y_1, \ldots, y_t)$ sent from $\mcal{M}$; 
\item
It then recovers $(x_1, \ldots, x_t)$ by brute-force: Namely, for each $i \in [t]$, it inverts the OWF $f$ to find $x_i$ s.t.\ $f(x_i) = y_i$. It is possible that there exist some ``bad'' $y_i$'s that are not in the range of $f$. For such bad $i$'s, it sets $x_i = \bot$. If all the $x_i$'s are bad, $\mcal{K}'_i(\msf{st}_{\mcal{M}}, \msf{st}_R, \tau, \tilde{\tau})$ halts and outputs $\msf{Fail}$;
\item \label[Step]{item:machineK':3}
If this step is reached, we know that $(x_1, \ldots, x_t)$ cannot be all-$\bot$. $\mcal{K}'_i(\msf{st}_{\mcal{M}}, \msf{st}_R, \tau, \tilde{\tau})$ then picks an $(s, x_s)$ uniformly at random from the good (i.e.\ non-$\bot$) $x_i$'s.
\item
Then, $\mcal{K}'_i(\msf{st}_{\mcal{M}}, \msf{st}_R, \tau, \tilde{\tau})$ continues to finish the execution in the same way as $\mcal{K}_i(\msf{st}_{\mcal{M}}, \msf{st}_R, \tau, \tilde{\tau})$, except that it uses $(s, x_s)$ as the witness when executing the left {\bf WIPoK-2}. 
\end{enumerate}
It is worth noting that the $(j, x_j)$ extracted by the witness-extended emulator for the left {\bf WIPoK-1} (inherited from $\mcal{K}_i(\msf{pref})$) is not used any more in $\mcal{K}'_i(\msf{st}_{\mcal{M}}, \msf{st}_R, \tau, \tilde{\tau})$. 

Obviously, if the $(s, x_s)$ picked by $\mcal{K}'_i(\msf{st}_{\mcal{M}}, \msf{st}_R, \tau, \tilde{\tau})$ is equal to the $(j, x_j)$ extracted in $\mcal{K}_i(\msf{st}_{\mcal{M}}, \msf{st}_R, \tau, \tilde{\tau})$ from its left {\bf WIPoK-1}, then $\mcal{K}'_i(\msf{st}_{\mcal{M}}, \msf{st}_R, \tau, \tilde{\tau})$ and $\mcal{K}_i(\msf{st}_{\mcal{M}}, \msf{st}_R, \tau, \tilde{\tau})$ are identical. Notice that this step relies on the injectivity of $f$ (see \Cref{rmk:injectivity-of-OWF}). Since $\mcal{K}'_i(\msf{st}_{\mcal{M}}, \msf{st}_R, \tau, \tilde{\tau})$ samples $(s,x_s)$ uniformly from all the good $(i, x_i)$'s, it must hold with probability at least $1/t$ that $(s, x_s) = (j,x_j)$.
Therefore, the following must hold:
{\begingroup\fontsize{10.5pt}{0pt}\selectfont
\begin{equation}\label[Inequality]{eq:bound:K'i}
\forall i \in [\tilde{t}], ~\Pr[w' = (i, \tilde{x}_i) : \Val \gets \mcal{K}'_i(\msf{st}_{\mcal{M}}, \msf{st}_R, \tau, \tilde{\tau})] \ge \frac{1}{t}\cdot \Pr[w' = (i, \tilde{x}_i) : \Val \gets \mcal{K}_i(\msf{st}_{\mcal{M}}, \msf{st}_R, \tau, \tilde{\tau})].
\end{equation}
\endgroup}
\begin{remark}[On Injectivity of the OWF]\label{rmk:injectivity-of-OWF}
We emphasize that throughout the proof of non-malleability, this is {\em the only} place where we rely on the injectivity of the OWF. In particular, we rely on the injectivity of the $f$ in the left session to ensure that there is a unique preimage for each $\Set{{y}_i}_{i \in [t]}$. Thus, if both the $x_s$ with $s = j$ (sampled by $\mcal{K}'_i$) and the extracted $x_j$ (in $\mcal{K}_i$) are a valid preimage for the same $y_j$, then the injectivity of $f$ implies that $x_s=x_j$. We will show how to remove injectivity in \Cref{sec:removing-injectivity}.
\end{remark}

\para{Machine $\mcal{K}''_i$ ($i\in [\tilde{t}]$):} (Illustrated in \Cref{figure:one-sided:K''i}.\label{pageref:one-sided:K''i}) For a prefix $\msf{pref}$,  $\mcal{K}''_i(\msf{st}_{\mcal{M}}, \msf{st}_R, \tau, \tilde{\tau})$ behaves identically to $\mcal{K}'_i(\msf{st}_{\mcal{M}}, \msf{st}_R, \tau, \tilde{\tau})$ except that it plays as the honest committer in the left {\bf WIPoK-1}, instead of running the witness-extended emulator. Recall that starting from $\mcal{K}'_i$, the witness $(j, x_j)$ extracted by the witness-extended emulator from the left {\bf WIPoK-1} is not used any more; Thus, machine $\mcal{K}''_i$ does not need to perform this witness-extended emulation.

By the PoK property %witness-extended emulation property (as per \Cref{def:WEE}) 
of the left {\bf WIPoK-1}, it holds that
\begin{equation}\label[Inequality]{eq:relation:K'-K''}
\forall i \in [\tilde{t}], ~\big|\Pr[w' = (i,\tilde{x}_i):\Val \gets \mcal{K}''_i(\msf{st}_{\mcal{M}}, \msf{st}_R, \tau, \tilde{\tau})] - \Pr[w' = (i,\tilde{x}_i):\Val \gets \mcal{K}'_i(\msf{st}_{\mcal{M}}, \msf{st}_R, \tau, \tilde{\tau})]\big| \le \negl(\secpar).
\end{equation}
Next, by the (non-uniform, see \Cref{rmk:non-uniform_reduction}) WI property of the right {\bf WIPoK-1}, it holds that
\begin{equation}\label[Inequality]{eq:relation:K:WI}
\forall i \in [\tilde{t}], ~\big|\Pr[w' = (i,\tilde{x}_i):\Val \gets \mcal{K}''_i(\msf{st}_{\mcal{M}}, \msf{st}_R, \tau, \tilde{\tau})] - \Pr[w' = (i,\tilde{x}_i):\Val \gets \mcal{K}''_1(\msf{st}_{\mcal{M}}, \msf{st}_R, \tau, \tilde{\tau})]\big|  \le \negl(\secpar).
\end{equation}
\begin{remark}[Power of Non-Uniform Reductions]\label{rmk:non-uniform_reduction}
Note that we can rely on the PoK and WI properties, although $\mcal{K}'_i$ and $\mcal{K}''_i$ perform brute-force to recover $(x_1,...,x_t)$. This is because the brute-forcing step is done before {\bf WIPoK-1} or {\bf WIPoK-2} starts; Thus, $(x_1,...,x_t)$ can be treated as a non-uniform advice in the reductions. This non-uniform type of argument will be used again in this section later. 
%A similar remark applies to many steps in the rest of this section.
\end{remark}

Then, we have the following claim:
\begin{MyClaim}\label{claim:bound:K''1}
% For any $\mcal{\tau}$, if $p_{\tilde{\tau}} \ge \epsilon(\secpar)$, then it holds that
$\forall i \in [\tilde{t}], ~\Pr[w' = (i,\tilde{x}_i):\Val \gets \mcal{K}''_1(\msf{st}_{\mcal{M}}, \msf{st}_R, \tau, \tilde{\tau})] \ge \frac{1}{t} \cdot \big(p^{\msf{Sim}}_{\msf{pref}} - \epsilon'(\secpar)\big) - \negl(\secpar)$.
\end{MyClaim}
\begin{proof} 
This claim follows from \Cref{cliam:bound:Ki-xi} and \Cref{eq:bound:K'i,eq:relation:K'-K'',eq:relation:K:WI}. Formally, (in the following, we omit the input $(\msf{st}_{\mcal{M}}, \msf{st}_R, \tau, \tilde{\tau})$ to $\mcal{K}_i$, $\mcal{K}'_i$, and $\mcal{K}''_i$)
\begin{align}
\forall i \in [\tilde{t}], ~\Pr[w' = (i,\tilde{x}_i):\Val \gets \mcal{K}''_1] 
 & \ge \Pr[w' = (i,\tilde{x}_i):\Val \gets \mcal{K}''_i]  - \negl(\secpar)
\label[Inequality]{claim:bound:K''1:proof:1}\\
 & \ge 
\Pr[w' = (i,\tilde{x}_i):\Val \gets \mcal{K}'_i]  - \negl(\secpar) 
\label[Inequality]{claim:bound:K''1:proof:2}\\
 & \ge 
\frac{1}{t}\cdot\Pr[w' = (i,\tilde{x}_i):\Val \gets \mcal{K}_i]  - \negl(\secpar) 
\label[Inequality]{claim:bound:K''1:proof:3}\\
& \ge 
\frac{1}{t} \cdot \big(p^{\msf{Sim}}_{\msf{pref}} - \epsilon'(\secpar)\big) - \negl(\secpar), \label[Inequality]{claim:bound:K''1:proof:4}
\end{align}
where \Cref{claim:bound:K''1:proof:1} follows from \Cref{eq:relation:K:WI}, \Cref{claim:bound:K''1:proof:2} follows from \Cref{eq:relation:K'-K''}, \Cref{claim:bound:K''1:proof:3} follows from \Cref{eq:bound:K'i}, and \Cref{claim:bound:K''1:proof:4} follows from \Cref{cliam:bound:Ki-xi}.

\end{proof}

Now, we make the last claim which, together with \Cref{claim:bound:K''1}, leads to the desired contradiction.
\begin{MyClaim}\label{claim:K'':non-abort}
$\Pr[\Val \ne \bot_{\msf{invalid}}:\Val \gets \mcal{K}''_1(\msf{st}_{\mcal{M}}, \msf{st}_R, \tau, \tilde{\tau})] \le p^{\msf{Sim}}_{\msf{pref}} + \negl(\secpar)$.
\end{MyClaim}
\para{Deriving the Contradiction.} Before proving \Cref{claim:K'':non-abort}, we first show why \Cref{claim:bound:K''1,claim:K'':non-abort} are contradictory (all the probabilities below are taken over $\Val \gets \mcal{K}''_1(\msf{st}_{\mcal{M}}, \msf{st}_R, \tau, \tilde{\tau})$):
\begin{align*}
\Pr[\Val \ne \bot_{\msf{invalid}}] 
& = \Pr[\Val = \msf{val}(\tilde{\tau})]  + \Pr[\Val = \bot_{\tilde{Y}}] \\
& = \Pr[\Val = \msf{val}(\tilde{\tau})] + \sum_{i =1 }^{\tilde{t}} \Pr[w' = (i,\tilde{x}_i)] \numberthis \label{eq:bound:Ki:final-contradiction:0}\\
& \ge  \sum_{i =1 }^{\tilde{t}} \Pr[w' = (i,\tilde{x}_i)] \\
&\ge \tilde{t} \cdot \frac{1}{t} \cdot \big(p^{\msf{Sim}}_{\msf{pref}} - \epsilon'(\secpar)\big) - \negl(\secpar) \numberthis \label[Inequality]{eq:bound:Ki:final-contradiction:1}\\
& \ge (1+\frac{1}{t}) \cdot (p^{\msf{Sim}}_{\msf{pref}} - \epsilon'(\secpar)) - \negl(\secpar) \numberthis \label[Inequality]{eq:bound:Ki:final-contradiction:2}\\
& = p^{\msf{Sim}}_{\msf{pref}} + \bigg(\frac{p^{\msf{Sim}}_{\msf{pref}}}{t} - \epsilon'(\secpar) - \frac{\epsilon'(\secpar)}{t}\bigg) - \negl(\secpar)\\
& \ge p^{\msf{Sim}}_{\msf{pref}} + \frac{10t^2 - t -1}{10t^3}\cdot \epsilon(\secpar) - \negl(\secpar), \numberthis \label[Inequality]{eq:bound:Ki:final-contradiction:3}\\
\end{align*}
where \Cref{eq:bound:Ki:final-contradiction:0} follows from \Cref{eq:bot-tilde-Y:partition}, \Cref{eq:bound:Ki:final-contradiction:1} follows from \Cref{claim:bound:K''1}, \Cref{eq:bound:Ki:final-contradiction:2} follows from the assumption that $\tilde{t} \ge t+1$, and \Cref{eq:bound:Ki:final-contradiction:3} follows from the assumption that $p^{\msf{Sim}}_{\msf{pref}} \ge \epsilon(\secpar)$ and our parameter setting $\epsilon'(\secpar) = \frac{\epsilon(\secpar)}{10t^2}$.

Recall that $t$ is the tag taking values from $[n]$ with $n$ being a polynomial of $\secpar$. Also recall that $\epsilon(\secpar)$ is an inverse polynomial on $\secpar$. Therefore, \Cref{eq:bound:Ki:final-contradiction:3} can be written as:
$$\Pr[\Val \ne \bot_{\msf{invalid}}:\Val \gets \mcal{K}''_1({\msf{pref}})] \ge p^{\msf{Sim}}_{\msf{pref}} + \frac{1}{\poly(\secpar)} - \negl(\secpar),$$
which contradicts \Cref{claim:K'':non-abort}.

This eventually finishes the proof of \Cref{lem:bound:Ki} (modulo the proof of \Cref{claim:K'':non-abort}, which we show in \Cref{sec:proof:claim:K'':non-abort}).

\subsection{Proof of \Cref{claim:K'':non-abort}}
\label{sec:proof:claim:K'':non-abort}
We first define two extra machines $\mcal{K}^{**}_1$ and $\mcal{K}^*_1$.

\begin{figure}[!tb]
     \begin{subfigure}[t]{0.47\textwidth}
         \centering
         \fbox{
         \includegraphics[width=\textwidth,page=11]{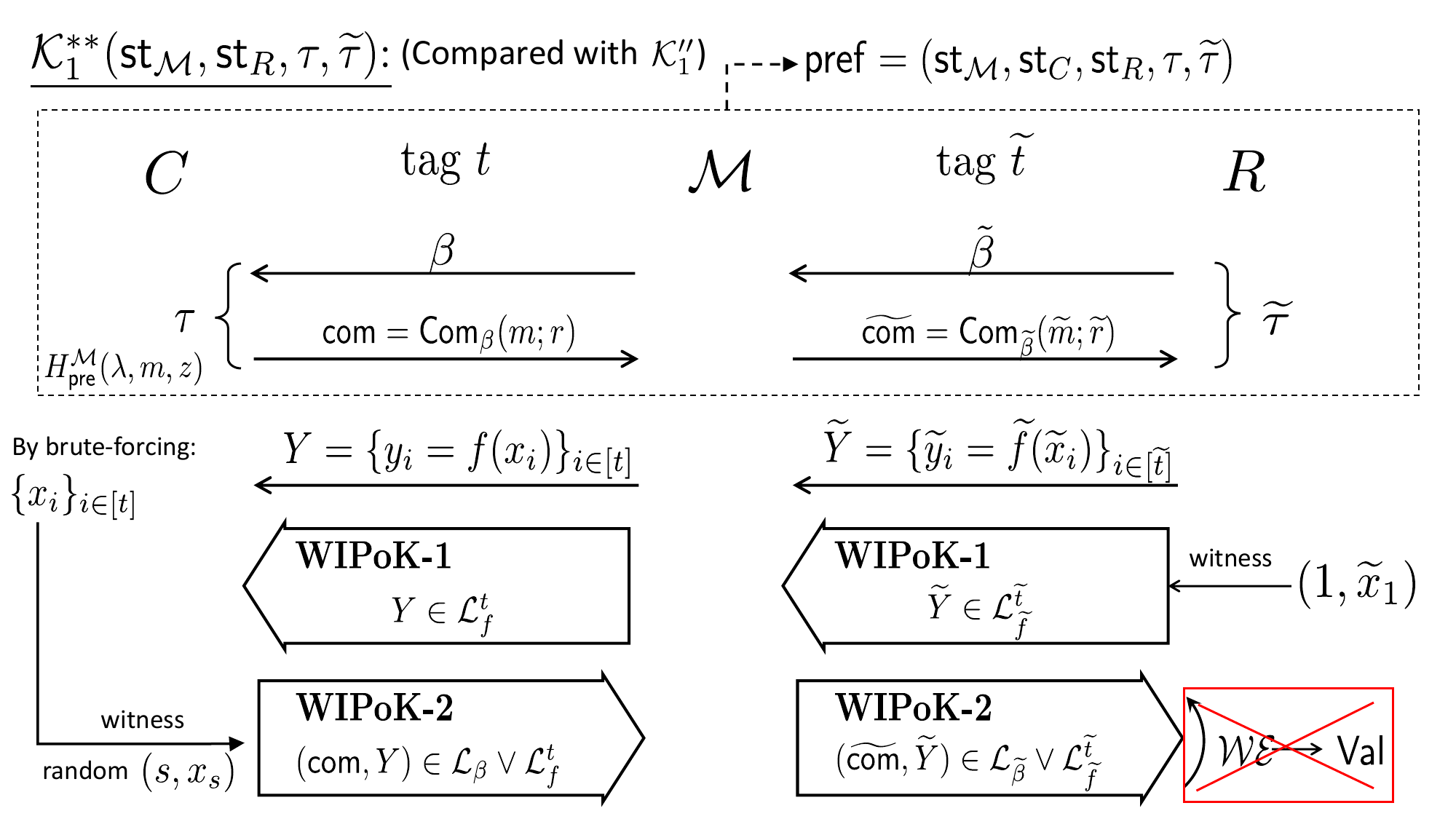}
         }
         \caption{}
         \label{figure:one-sided:K**i}
     \end{subfigure}
     \hspace{6.5pt}
     \begin{subfigure}[t]{0.47\textwidth}
         \centering
         \fbox{
         \includegraphics[width=\textwidth,page=12]{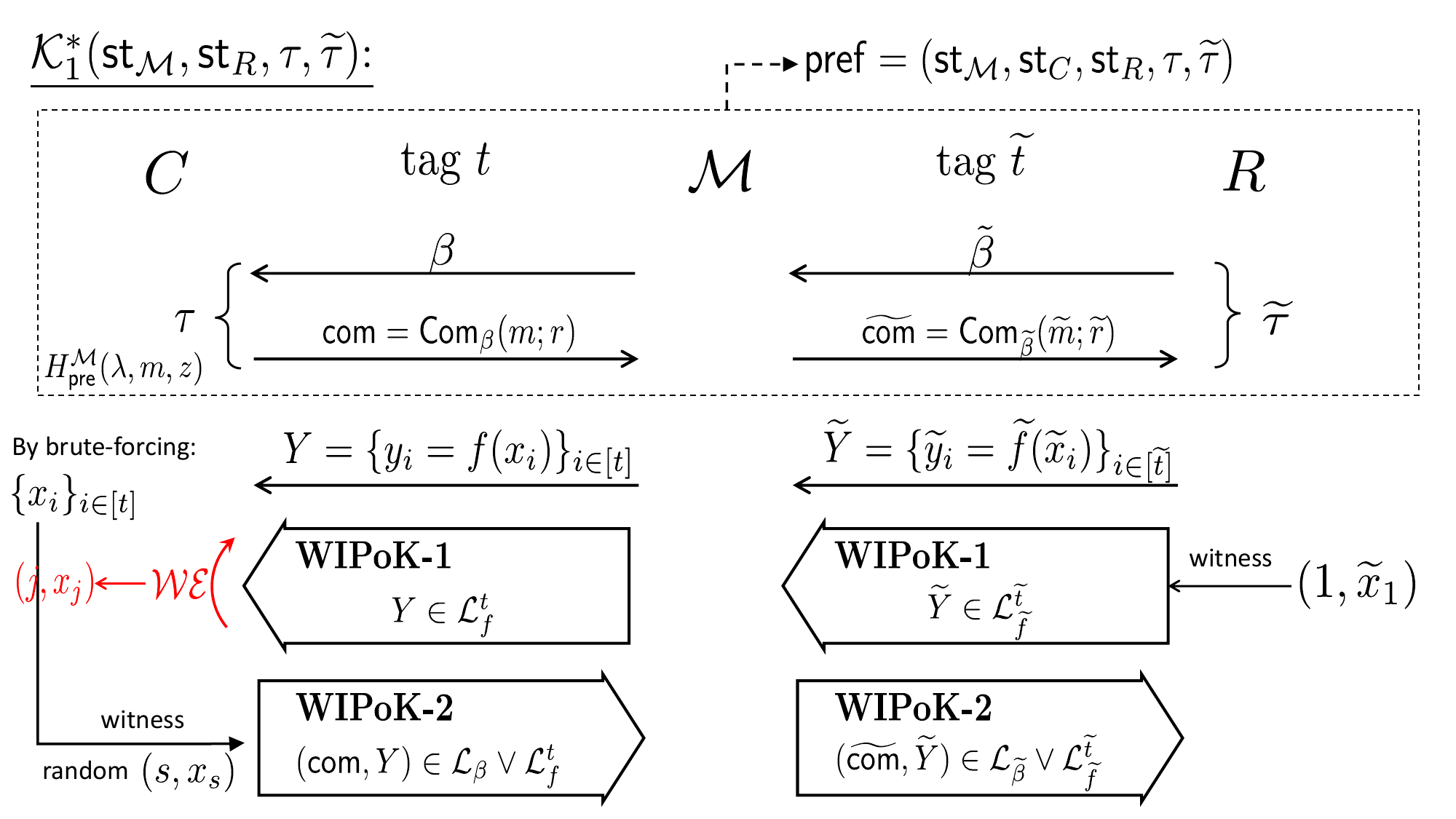}
         }
         \caption{}
         \label{figure:one-sided:K*i}
     \end{subfigure}
     \caption{Machines $\mcal{K}^{**}_1$ and $\mcal{K}^{*}_1$ {\scriptsize (Difference is highlighted in red color)}}
     \label{figure:one-sided:K**i-K*i}
\end{figure}
\para{Machine $\mcal{K}^{**}_1$:} (Illustrated in \Cref{figure:one-sided:K**i}.) For a prefix $\msf{pref}$, $\mcal{K}^{**}_1(\msf{st}_{\mcal{M}}, \msf{st}_R, \tau, \tilde{\tau})$ behaves identically as $\mcal{K}''_1(\msf{st}_{\mcal{M}}, \msf{st}_R, \tau, \tilde{\tau})$ except that $\mcal{K}^{**}_1(\msf{st}_{\mcal{M}}, \msf{st}_R, \tau, \tilde{\tau})$ finishes the right {\bf WIPoK-2} using the honest receiver's algorithm, instead of using the witness-extended emulator. 

\subpara{$\mcal{K}^{**}_1$'s Output.} We define the output of $\mcal{K}^{**}_1(\msf{st}_{\mcal{M}}, \msf{st}_R, \tau, \tilde{\tau})$ to be the honest receiver's decision $b \in \Set{\top, \bot}$. This is in contrast to previous hybrids $\mcal{K}'_i$, $\mcal{K}''_i$, and $\mcal{K}_i$, whose output is defined to be the value $\msf{Val}$ that depends on the value $w'$ extracted by the $\mcal{WE}$ for the right {\bf WIPoK-2}.

 By the (non-uniform) PoK property (as per \Cref{def:WEE})
 %witness-extended emulation property (as per \Cref{def:WEE}) 
 of the right {\bf WIPoK-2}, it holds that
\begin{equation}\label[Inequality]{eq:K''1-K**1}
\big|\Pr[\Val \ne \bot_{\msf{invalid}} :\Val\gets \mcal{K}''_1(\msf{st}_{\mcal{M}}, \msf{st}_R, \tau, \tilde{\tau})] - \Pr[b = \top : b\gets \mcal{K}^{**}_1(\msf{st}_{\mcal{M}}, \msf{st}_R, \tau, \tilde{\tau})] \big| \le \negl(\secpar).
\end{equation}

\para{Machine $\mcal{K}^{*}_1$:} (Illustrated in \Cref{figure:one-sided:K*i}.) For a prefix $\msf{pref}$, $\mcal{K}^{*}_1(\msf{st}_{\mcal{M}}, \msf{st}_R, \tau, \tilde{\tau})$ behaves identically as $\mcal{K}^{**}_1(\msf{st}_{\mcal{M}}, \msf{st}_R, \tau, \tilde{\tau})$ except the following difference: $\mcal{K}^{*}_1(\msf{st}_{\mcal{M}}, \msf{st}_R, \tau, \tilde{\tau})$ uses the witness-extended emulator to extract a witness $(j,x_j)$ from the left {\bf WIPoK-1}, and if $x_j$ is not a valid preimage for $y_j$,  $\mcal{K}^{*}_1$ aborts.

 By the (non-uniform) PoK property 
 %witness-extended emulation property (as per \Cref{def:WEE}) 
 of the left {\bf WIPoK-1}, it holds that
\begin{equation}\label[Inequality]{eq:K**1-K*1}
\big|\Pr[b = \top : b \gets \mcal{K}^{**}_1(\msf{st}_{\mcal{M}}, \msf{st}_R, \tau, \tilde{\tau})] - \Pr[b = \top : b\gets \mcal{K}^{*}_1(\msf{st}_{\mcal{M}}, \msf{st}_R, \tau, \tilde{\tau})] \big| \le \negl(\secpar).
\end{equation}

\para{Compare $\mcal{K}^{*}_1$ with $\mcal{G}_1$.} Now, let us compare $\mcal{K}^{*}_1$ with the $\mcal{G}_1$ depicted in \Cref{figure:one-sided:Gi}. They only differ in the witness used in the left {\bf WIPoK-2} (and that $\mcal{G}_1$ does not need to perform brute-forcing for $Y$, as it does not use those preimages). Therefore, by the (non-uniform) WI property of the left {\bf WIPoK-2}, it holds that
$$
\big|\Pr[b = \top : b \gets \mcal{K}^{*}_1(\msf{st}_{\mcal{M}}, \msf{st}_R, \tau, \tilde{\tau})] - \Pr[b = \top : (\OUT, b) \gets \mcal{G}_1(\msf{st}_{\mcal{M}}, \msf{st}_R, \tau, \tilde{\tau})] \big| \le \negl(\secpar).
$$
Also, recall (from \Cref{def:p-sim-pref}) that $\Pr[b = \top : (\OUT, b) \gets \mcal{G}_1(\msf{st}_{\mcal{M}}, \msf{st}_R, \tau, \tilde{\tau})]$ is exactly the definition of $p^{\msf{Sim}}_{\msf{pref}}$. Thus, the above implies:
\begin{equation}\label[Inequality]{eq:K*1-G1}
\big|\Pr[b = \top : b \gets \mcal{K}^{*}_1(\msf{st}_{\mcal{M}}, \msf{st}_R, \tau, \tilde{\tau})] - p^{\msf{Sim}}_{\msf{pref}} \big| \le \negl(\secpar).
\end{equation}

Therefore, the following holds:
\begin{align}
\Pr[\Val \ne \bot_{\msf{invalid}} :\Val\gets \mcal{K}''_1(\msf{st}_{\mcal{M}}, \msf{st}_R, \tau, \tilde{\tau})] & \le \Pr[b = \top : b\gets \mcal{K}^{**}_1(\msf{st}_{\mcal{M}}, \msf{st}_R, \tau, \tilde{\tau})] +  \negl(\secpar) \label[Inequality]{eq:K'':non-abort:final:1} \\
& \le \Pr[b = \top : b \gets \mcal{K}^{*}_1({\msf{pref}})] + \negl(\secpar) \label[Inequality]{eq:K'':non-abort:final:2} \\
& \le p^{\msf{Sim}}_{\msf{pref}} + \negl(\secpar), \label[Inequality]{eq:K'':non-abort:final:3}
\end{align}
where \Cref{eq:K'':non-abort:final:1} follows from \Cref{eq:K''1-K**1}, \Cref{eq:K'':non-abort:final:2} follows from \Cref{eq:K**1-K*1}, and \Cref{eq:K'':non-abort:final:3} follows from \Cref{eq:K*1-G1}. 

This finishes the proof of \Cref{claim:K'':non-abort}.

\subsection{Replacing the Injective OWF with Any OWF}
\label{sec:removing-injectivity}
In this part, we show how to replace the injective OWF with any OWF in \Cref{prot:one-sided:classical}.

As mentioned in \Cref{rmk:injectivity-of-OWF}, the injectivity is  used only when we switch from $\mcal{K}_i$ to $\mcal{K}'_i$ (on \Cpageref{machineK':description}). Let us first recall  why injectivity is important there: $\mcal{K}'_i$ learns the preimages of each $\Set{y_i}_{i\in[t]}$ via brute-forcing, while $\mcal{K}_i$ learns a preimage $x_j$ of $y_j$ {\em for some $j \in [t]$} by extraction (from the left {\bf WIPoK-1}). Our goal there is to show that if $\mcal{K}'_i$ picks a random preimage, then it will hit the $(j, x_j)$ extracted by $\mcal{K}_i$, except for a multiplicative $1/t$ loss on the probability (i.e., \Cref{eq:bound:K'i}). If the OWF $f$ is not injective, there is no guarantee that the $(j, x_j)$ extracted by $\mcal{K}_i$ is really contained in the set of all preimages learned by $\mcal{K}'_i$ via brute-forcing---Even if $\mcal{K}'_i$ guessed the index $j$ correctly, it could have brute-forced a $x'_j \ne x_j$ (though it holds that $y_j = f(x_j) = f(x'_j)$). In this case, \Cref{eq:bound:K'i} may not hold anymore.

From the above discussion, it is clear that if we want to remove the requirement of injectivity, we only need to find an alternative approach to ensure that the preimage extracted by $\mcal{K}_i$ will fall in the set of the preimages learned by $\mcal{K}'_i$ via brute-forcing. It turns out that this can be achieved by the following simple modification to \Cref{item:one-sided:step:OWFs,item:one-sided:step:WIPoK:1} of \Cref{prot:one-sided:classical}:
\begin{itemize}
\item
In \Cref{item:one-sided:step:OWFs}, ask $R$ to {\em additionally} commit (using a statistically-binding scheme) to the preimages $\Set{x_i}_{i\in [t]}$ that it used to generate $\Set{y_i}_{i\in [t]}$. Formally, we ask $R$ to send the following messages in  \Cref{item:one-sided:step:OWFs}:
$$Y = \Set{y_i = f(x_i)}_{i \in [t]},~\text{and}~\Set{\msf{com}_i = \Com_{\beta'}(x_i;r_i)}_{i \in [t]},$$ 
where $\Com_{\beta'}$ is Naor's commitment with $\beta'$ being the first Naor's message from $C$ (which could be sent in parallel with other messages in \Cref{item:one-sided:step:committing}).
\item
In \Cref{item:one-sided:step:WIPoK:1}, ask $R$ to additionally prove that the $Y$ and $\Set{\msf{com}_i}_{i \in [t]}$ are ``consistent''. Formally, instead of proving $Y \in \Lang^t_f$ (defined in \Cref{eq:one-sided:Lang:OWF}), we ask $R$ to prove $(Y,  \Set{\msf{com}_i}_{i \in [t]}) \in \Lang^t_{f, \beta'}$ using the $\msf{WIPoK}$, where 
{\begingroup\fontsize{10pt}{0pt}\selectfont
\begin{equation}\label[Language]{eq:one-sided:Lang:OWF:new}
\Lang^t_{f, \beta'}\coloneqq \Set{(y_1, \ldots, y_t), (\msf{com}_1, \ldots, \msf{com}_t)~|~\exists (i, x_i, r_i) ~s.t.~ i \in [t] \wedge y_i = f(x_i) \wedge \msf{com}_i = \Com_{\beta'}(x_i;r_i)}.
\end{equation}
\endgroup}
\end{itemize} 
It is worth noting that we do {\em not} need to modify \Cref{item:one-sided:step:WIPoK:2}. In particular, $C$ does {\em not} need to prove that it knows one of the $(x_i, r_i)$'s used by $R$ to generate $\Set{\msf{com}_i}_{i\in [t]}$; $C$ still only needs to prove $(\msf{com}, Y) \in \Lang_\beta \vee \Lang^t_f$.

Now, instead of performing brute-force over $\Set{y_i}_{i \in [t]}$, $\mcal{K}'_i$ does that over $\Set{\msf{com}_i}_{i \in [t]}$ to learn the committed values, and then picks a random $x_s$ satisfying $y_s = f(x_s)$ to use in the left {\bf WIPoK-2}.\footnote{Note that $\mcal{K}'_i$ (resp.\ $\mcal{K}_i$) will also learn the randomness used to generate the commitment via brute-forcing (resp.\ via extraction from {\bf WIPoK-1}). They simply ignore the randomness---indeed, both machines only need to know a preimage $(j,x_j)$ to finish {\bf WIPoK-2}; the corresponding randomness $r_i$ is irrelevant.} It is easy to see that the statistical-binding property of $\Com$ %and the knowledge soundness of the left {\bf WIPoK-1} 
ensures that \Cref{eq:bound:K'i} still holds for an overwhelming fraction of $\beta'$.

\begin{remark}[On One-Wayness]
Technically, we also need to argue that the above modification does not enable the man-in-the-middle  $\mcal{M}$ to invert the one-way function $\tilde{f}$ on $\tilde{Y} = \Set{\tilde{y}_i}_{i \in [\tilde{t}]}$. (Recall that the one-wayness of $\tilde{f}$ is used (only) in the proof of \Cref{cliam:bound:Ki-xi}.) This is straightforward---Any OWF remains one-way even if a commitment to its preimage is given.
\end{remark}

\subsection{Summarizing the Proof Strategy for \Cref{thm:one-sided:non-malleability}}
\label{sec:proof:summar:one-sided}

Since the proof of \Cref{thm:one-sided:non-malleability} is complicate, we summarize the logic flow in this subsection. We hope that it will help the reader obtain a better understanding of the whole picture, without being disturbed by details of secondary importance. But one could safely skip this subsection as it does not contain new information. In particular, it will not affect the understanding of succeeding sections.

At a high level, the proof of \Cref{thm:one-sided:non-malleability} can be divided into the following five steps:
\begin{enumerate}
\item \label[Step]{item:proof:summar:one-sided:1}
{\bf (Proof by Contradiction.)}
In the proof of \Cref{thm:one-sided:non-malleability}, we first denoted the MIM execution by $H^{\mcal{M}}(\secpar, m, z)$ (on \Cpageref{gameH:description}). 
We then assumed for contradiction that a (potentially non-uniform) PPT adversary can distinguish the outputs of $H^{\mcal{M}}(\secpar, m_0, z)$ and $H^{\mcal{M}}(\secpar, m_1, z)$ with advantage $3 \cdot \delta(\secpar)$, with $\delta(\secpar)$ being some polynomial (i.e., \Cref{eq:one-sided:proof:contra-assump}). Next, we required the existence of a machine $G^{\mcal{M}}$ (i.e., \Cref{lem:one-sided:proof:core}) such that 
\begin{enumerate}
\item \label[Property]{summary:G-M:property:1}
the outputs of $G^{\mcal{M}}(\secpar, m_0, z)$ and $G^{\mcal{M}}(\secpar, m_1, z)$ are computationally indistinguishable, and 
\item
$G^{\mcal{M}}(\secpar, m_b, z)$ and $H^{\mcal{M}}(\secpar, m_b, z)$ is (computationally) $\delta(\secpar)$-close for both $b = 0$ and $b =1$.
\end{enumerate}
The existence of such a $G^{\mcal{M}}$ finished the proof of \Cref{thm:one-sided:non-malleability}, because it contradicts the assumption that there exists some PPT distinguisher that tells the difference between $H^{\mcal{M}}(\secpar, m_0, z)$ and $H^{\mcal{M}}(\secpar, m_1, z)$ with advantage $3 \cdot \delta(\secpar)$. Thus, the remaining steps for the proof of \Cref{thm:one-sided:non-malleability} were then devoted to constructing such a $G^{\mcal{M}}$.

\item \label[Step]{item:proof:summar:one-sided:2}
{\bf (Constructing $G^{\mcal{M}}$.)} To build the $G^{\mcal{M}}$ required in \Cref{item:proof:summar:one-sided:1}, we first broke $H^{\mcal{M}}$ into two stages---the prefix part $H^{\mcal{M}}_{\msf{pre}}$ and the remainder $\langle \mcal{A}, \mcal{B}\rangle$ (i.e.,  \Cref{hybrid:H:reinterpretation}). Here, we defined an important variable called ``prefix'': 
\begin{itemize}
\item
An execution of $H^{\mcal{M}}_{\msf{pre}}$ fixes the prefix $\msf{pref}$, which contains the internal state $\msf{st}_C, \msf{st}_{\mcal{M}}, \msf{st}_R$ of the corresponding parties at the end of {\bf Step-2} in a MIM execution. Also, $\msf{pref}$ fixes the right Naor's commitment $\tilde{\tau}$.  Also note that the committed value in the right execution was defined to be $\msf{val}_b(\tilde{\tau})$ (i.e., \Cref{eq:def:val-b-tau}).
\end{itemize}
Then, we constructed a simulation-extractor $\mcal{SE}^{\mcal{M}}$ (in \Cref{lem:small-tag:proof:se}) such that, when continuing from any prefix $\msf{pref}$ (Importantly, $\mcal{SE}$ cannot use the intermediate state $\msf{st}_C$ for the committer; Otherwise, we cannot show \Cref{summary:G-M:property:1} by reducing it to the hiding property of Naor's commitment), $\mcal{SE}^{\mcal{M}}$ will $\epsilon$-simulate the remainder execution $\langle \mcal{A},\mcal{B} \rangle$, while also extracting the committed value $\msf{val}_b(\tilde{\tau})$ in the right execution.

With this $\mcal{SE}^{\mcal{M}}$, it is easy to construct $G^{\mcal{M}}$---$G^{\mcal{M}}$ simply runs $H^{\mcal{M}}_{\msf{pre}}$ to obtain the prefix, and then runs $\mcal{SE}$ to simulate the remainder execution and also extract $\msf{val}_b(\tilde{\tau})$. Thus, the remaining steps for the proof of \Cref{thm:one-sided:non-malleability} were then devoted to constructing such a $\mcal{SE}^{\mcal{M}}$.

\item \label[Step]{item:proof:summar:one-sided:3}
{\bf (Constructing $\mcal{SE}^{\mcal{M}}$.)} To construct $\mcal{SE}^{\mcal{M}}$, we first built two machines: 
\begin{enumerate}
\item
 a simulator $\mcal{G}_1$ (in \Cref{lem:main-thread:sim:real}) that can simulate the execution $\langle \mcal{A}, \mcal{B}\rangle$ without knowing $C$'s internal states $\msf{st}_C$. This is to address the point that $\mcal{SE}^{\mcal{M}}$  is not allowed to use $\msf{st}_C$. Here, we defined an important variable $p^{\msf{Sim}}_{\msf{pref}}$, which denotes the probability that the honest $R$ is convinced in the execution simulated by $\mcal{G}_1$.
 \item
an extractor $\mcal{K}$ (in \Cref{lem:small-tag:proof:se:proof:K}) such that 
for any prefix $\msf{pref}$ and noticeable $\epsilon(\secpar)$, if $p^{\msf{Sim}}_{\msf{pref}} \ge \epsilon(\secpar)$, then $\mcal{K}$ can extract the value committed in $\tilde{\tau}$ (contained in $\msf{pref}$) with probability polynomially related to $\epsilon(\secpar)$ (more accurately, the probability is $\frac{\epsilon'(\secpar)}{\tilde{t}}$ where $\epsilon'(\secpar) \coloneqq \frac{\epsilon(\secpar)}{10t^2}$).
\end{enumerate}
With $\mcal{G}_1$ and $\mcal{K}$, $\mcal{SE}^{\mcal{M}}$ can be constructed as follows: on input a $\msf{pref}$, it first runs $\mcal{G}_1$ to simulate the main-tread execution of $\langle \mcal{A}, \mcal{B}\rangle$; If the honest receiver is not convinced in the main-thread, $\mcal{SE}^{\mcal{M}}$ simply sets the committed value in the right execution as $\bot$; Otherwise, $\mcal{SE}^{\mcal{M}}$ will extract the value committed in $\tilde{\tau}$ (denoted as $\val(\tilde{\tau})$) by repeating $\mcal{K}$ for polynomially many times (more accurately, $\frac{\tilde{t}}{\epsilon'(\secpar)}\cdot \secpar$ times). As shown toward the end of \Cref{lem:small-tag:proof:se:proof}, such a $\mcal{SE}^{\mcal{M}}$ satisfied the requirements described in \Cref{item:proof:summar:one-sided:2}.
Thus, the remaining steps for the proof of \Cref{thm:one-sided:non-malleability} were then devoted to constructing the extractor $\mcal{K}$. (The construction and security of $\mcal{G}_1$ is relatively straightforward, so we do not repeat it in this summary.)

\item
{\bf (Constructing $\mcal{K}$.)} To construct $\mcal{K}$, we first defined a sequence of machines $\Set{\mcal{K}_i}_{i \in [\tilde{t}]}$ as shown in \Cref{figure:one-sided:Ki}. We then argued that there must exist an $i \in [\tilde{t}]$ (called ``the good $i$'') such that the $\msf{Val}$ output by $\mcal{K}_i$ must be $\val(\tau)$ with probability $\ge \epsilon'(\secpar)$. Then, machine $\mcal{K}$ was defined to pick a random $i \picks [\tilde{t}]$ and run machine $\mcal{K}_i$. Since  such a $\mcal{K}$ will hit the good $i$ with probability $1/\tilde{t}$, it will successfully extract $\val(\tilde{\tau})$ with probability $\frac{\epsilon'(\secpar)}{\tilde{t}}$, satisfying the requirement in \Cref{item:proof:summar:one-sided:3}. Now, to finish the proof of \Cref{thm:one-sided:non-malleability}, the only thing left is to prove that $\Set{\mcal{K}_i}_{i \in [\tilde{t}]}$ satisfies the above requirement.

\item
{\bf (Existence of Good $\mcal{K}_i$.)}
Informally, our current goal is to show the following (see \Cref{lem:bound:Ki} for the precise statement):
$$\exists i \in [\tilde{t}]~s.t.~\Pr[\msf{Val} = \val(\tilde{\tau}):\msf{Val} \gets \mcal{K}_i] \ge \epsilon'(\secpar).$$

We first assumed for contradiction that (i.e., \Cref{eq:proof:averageK:contra-assump}):
\begin{equation}\label[Inequality]{one-sided:proof:summary:assump-contra}
\forall  i \in  [\tilde{t}],~\Pr[\msf{Val} = \val(\tilde{\tau}) :\msf{Val} \gets \mcal{K}_i] \le \epsilon'(\secpar).
\end{equation}
We then derived the contradiction in the following way:
\begin{enumerate}
\item
We first showed that the $\msf{Val}$ output by $\mcal{K}_i$ must be a valid witness with probability $\ge p^{\msf{Sim}}_{\msf{pref}} - \negl(\secpar)$ (i.e., \Cref{claim:bound:Ki}). Next, observe that in $\mcal{K}_i$, $\msf{Val}$ cannot be $(j, \tilde{x}_j)$ for some $j \ne i$; Otherwise, it breaks the one-wayness of the right OWF $\tilde{f}$. Thus, when $\msf{Val}$ is a valid witness in $\mcal{K}_i$, it can only take the values $\val(\tilde{\tau})$ or $(i, \tilde{x}_i)$. This together with our assumption in \Cref{one-sided:proof:summary:assump-contra} implies that 
\begin{equation}\label[Inequality]{one-sided:proof:summary:Ki}
\forall i \in [\tilde{t}],~\Pr[\msf{Val} = (i, \tilde{x}_i): \msf{Val} \gets\mcal{K}_i] \ge p^{\msf{Sim}}_{\msf{pref}} - \epsilon'(\secpar) - \negl(\secpar).\end{equation}
(This is exactly \Cref{cliam:bound:Ki-xi}.)
\item
Next, we defined machine $\mcal{K}''_i$ (see \Cref{figure:one-sided:K''i}), which differs from $\mcal{K}_i$ in the following way: $\mcal{K}''_i$ does not rewind the left {\bf WIPoK-1}. Instead, it extracts the preimages of the right $\Set{y_i}_{i \in [t]}$ by brute force, and then picks a random preimage $(s, x_s)$ to use in the left {\bf WIPoK-2}. Since the $(s, x_s)$ guessed by $\mcal{K}''_i$ will hit the one used by $\mcal{K}_i$, the following holds:
\begin{equation}\label[Inequality]{one-sided:proof:summary:Ki-K''i}
\forall i \in [\tilde{t}], ~\Pr[\msf{Val} = (i, \tilde{x}_i) : \msf{Val} \gets\mcal{K}''_i] \ge \frac{1}{t}\cdot \Pr[\msf{Val} = (i, \tilde{x}_i) : \msf{Val} \gets\mcal{K}_i].
\end{equation}
(This is essentially \Cref{eq:bound:K'i}---Note that in \Cref{sec:lem:bound:Ki:proof}, we also defined a machine $\mcal{K}'_i$. But it is only a tool that helps us eventually connect $\mcal{K}''_i$ and $\mcal{K}_i$. So we ignore $\mcal{K}'_i$ in this relatively high-level summary.)
\item
Since in $\mcal{K}''_i$ we do not rewind the left {\bf WIPoK-1} anymore, we can now rely on the WI property of the right {\bf WIPoK-1} to connect all the $\Set{\mcal{K}''_i}_{i \in [\tilde{t}]}$ to $\mcal{K}''_1$. Namely, we have
\begin{equation}\label[Inequality]{one-sided:proof:summary:K''i-K''1}
\forall i \in [\tilde{t}], ~\big|\Pr[\msf{Val} = (i,\tilde{x}_i): \msf{Val} \gets\mcal{K}''_i] - \Pr[\msf{Val} = (i,\tilde{x}_i): \msf{Val} \gets\mcal{K}''_1]\big|  \le \negl(\secpar).
\end{equation}
(This is exactly \Cref{eq:relation:K:WI}.) We want to remark that in this step, we took advantage of the synchronous schedule: in the synchronous setting, it is guaranteed that the brute-forcing for $\Set{y_i}_{i\in [n]}$ happens {\em before} the right {\bf WIPoK-1}; Otherwise, we will not be able to reduce \Cref{one-sided:proof:summary:K''i-K''1} to the WI property of the right {\bf WIPoK-1}. 
\end{enumerate}

Now, by \Cref{one-sided:proof:summary:Ki,one-sided:proof:summary:Ki-K''i,one-sided:proof:summary:K''i-K''1}, the following holds (this is exactly \Cref{claim:bound:K''1})
\begin{equation}\label[Inequality]{one-sided:proof:summary:bound:K''1}
\forall i \in [\tilde{t}], ~\Pr[\msf{Val} = (i,\tilde{x}_i)~\text{in}~ \mcal{K}''_1] \ge \frac{1}{t} \cdot \big(p^{\msf{Sim}}_{\msf{pref}} - \epsilon'(\secpar)\big) - \negl(\secpar).
\end{equation}
This implies 
\begin{align*}
\Pr[\Val = (1, \tilde{x}_1)\vee \ldots \vee \Val = (\tilde{t}, \tilde{x}_{\tilde{t}}) : \msf{Val} \gets \mcal{K}''_1({\msf{pref}})] 
& \ge \tilde{t} \cdot \frac{1}{t} \cdot \big(p^{\msf{Sim}}_{\msf{pref}} - \epsilon'(\secpar) \big) + \negl(\secpar)\\
\text{(By our parameter setting. See \Cref{eq:bound:Ki:final-contradiction:3})} & \ge p^{\msf{Sim}}_{\msf{pref}} + \frac{10t^2 - t -1}{10t^3}\cdot \epsilon(\secpar) - \negl(\secpar)
\end{align*}
On the other hand, \Cref{claim:K'':non-abort} implies that (we do not repeat the proof of \Cref{claim:K'':non-abort} here as it is relatively straightforward)
$$
\Pr[\Val = (1, \tilde{x}_1)\vee \ldots \vee \Val = (\tilde{t}, \tilde{x}_{\tilde{t}}) : \msf{Val} \gets \mcal{K}''_1({\msf{pref}})] \le p_{\msf{pref}} + \negl(\secpar)
$$
This gives us the desired contradiction because $\frac{10t^2 - t -1}{10t^3}\cdot \epsilon(\secpar)$ is an (inverse) polynomial on $\secpar$. 
\end{enumerate}

%!TEX root = ../main.tex
\section{Small-Tag, Synchronous, Classical Setting}
\label{sec:small-tag:classical:sync}
In this section, we show how to make $\langle C, R \rangle^{\msf{OneSided}}_{\msf{tg}}$ (i.e., \Cref{prot:one-sided:classical}) secure without the ``one-sided'' restriction. We emphasize that this section still focuses on the small-tag, synchronous setting against classical (i.e., non-uniform PPT) adversaries. 

\subsection{High-Level Idea} 
\label{sec:small-tag:classical:sync:high-level}
Our proof for the non-malleability of $\langle C, R \rangle^{\msf{OneSided}}_{\msf{tg}}$ (shown in \Cref{prot:one-sided:classical}) works only if $t<\tilde{t}$. However, this is not guaranteed in the real main-in-the-middle attack---the adversary can of course use a smaller tag in the right session. Fortunately, this problem can be addressed by the so-called ``two-slot'' technique proposed by Pass and Rosen \cite{STOC:PasRos05}. The idea is to create a situation where no matter how the man-in-the-middle $\mcal{M}$ schedules the messages, there is always a ``slot'' for which the ``$t<\tilde{t}$'' condition holds; As long as this is true,  non-malleability can be proven using the same technique as in \Cref{sec:small-tag-one-sided-sync-classical}.

To do that, first observe that the only place where \Cref{prot:one-sided:classical} makes use of the tag $t$ is \Cref{item:one-sided:step:OWFs,item:one-sided:step:WIPoK:1}. Let us call these two messages a ``{\bf Slot}''. Our modification is to execute this {\bf Slot} twice sequentially, where
\begin{itemize}
\item
for the first execution (referred to as {\bf Slot-A}), we ask $R$ to use tag $t$ as in the original construction;
\item
for the second execution (referred to as {\bf Slot-B}), we ask $R$ to use $n-t$ as the tag.
\end{itemize}
We also modify the language instance proven by {\bf WIPoK-2} in the expected way---Now $C$ needs to prove that it ``knows'' the $(m,r)$ tuple, or the preimage of one of the $y_i$'s sent in {\bf Slot-A}, {\em or the preimage of one of the $y_i$'s sent in {\bf Slot-B}}.

By the above design, it is easy to see that one of the following case must happen for any man-in-the-middle execution (note that we are still in the synchronous setting):
\begin{enumerate}
\item
{$t = \tilde{t}$:} This is the trivial case that is already ruled out by the definition of non-malleability.
\item
{$t < \tilde{t}$:} In this case, non-malleability follows by applying the argument in \Cref{sec:small-tag-one-sided-sync-classical} to {\bf Slot-A}. We sometimes refer to this case by saying that {\em {\bf Slot-A} is the good slot}.
\item
{$t > \tilde{t}$:} In this case, it holds that $(n-t) < (n-\tilde{t})$. In other words, the tag for the left {\bf Slot-B} is smaller than the tag for the right {\bf Slot-B}. Therefore,  non-malleability follows by applying the argument in \Cref{sec:small-tag-one-sided-sync-classical} to {\bf Slot-B}. We sometimes refer to this case by saying that {\em {\bf Slot-B} is the good slot}.
\end{enumerate}
Therefore, the modified protocol is non-malleable without the ``one-sided'' restriction.

\subsection{Construction of $\langle C, R\rangle^{\msf{sync}}_{\msf{tg}}$}
 We refer to the construction described in \Cref{sec:small-tag:classical:sync:high-level} as $\langle C, R \rangle^{\msf{sync}}_{\msf{tg}}$, and formally present it in \Cref{prot:small-tag:sync:classical}. This protocol makes use of the same building blocks as for \Cref{prot:one-sided:classical}, namely:
 \begin{itemize}
 \item
An OWF $f$, (Note that we do not require injectivity any more, because \Cref{prot:small-tag:sync:classical} makes use of the technique discussed in \Cref{sec:removing-injectivity}.)
\item
Naor's commitment $\Com$;
\item
A witness-indistinguishable proof of knowledge $\WIPoK$ %with knowledge error $\kappa(\secpar) = \negl(\secpar)$ 
(as per \Cref{def:WIPoK}). 
%\xiao{WIAoK should already suffice(?)}\takashi{Yes. We may consider using WIAoK for the classical case too for consistency to the pq setting.}
 \end{itemize} 
We remark that we further optimize the ``two-slot'' approach from \Cref{sec:small-tag:classical:sync:high-level} by sending the first message of {\bf Slot-A} and the first message of {\bf Slot-B} together. 
%It is easy to see that this does not affect security.
\begin{ProtocolBox}[label={prot:small-tag:sync:classical}]{Small-Tag Synchronous NMCom \textnormal{$\langle C, R\rangle^{\msf{sync}}_{\msf{tg}}$}}
The tag space is defined to be $[n]$ where $n$ is a polynomial on $\secpar$. Let $t \in [n]$ be the tag for the following interaction. Let $m$ be the message to be committed to. 

\para{Commit Stage:}
\begin{enumerate}
\item\label[Step]{item:small-tag:sync:step:Naor-rho}
Receiver $R$ samples and sends the first message $\beta$ for Naor's commitment.
\item\label[Step]{item:small-tag:sync:step:committing}
Committer $C$ commits to $m$ using the second message of Naor's commitment. $C$ also sends a $\beta'$ that will be used as the first Naor's message for $R$ to generate a commitment in next step. Formally, $C$ samples $r$ and $\beta'$ and sends the tuple $\big(\msf{com} = \Com_\beta(m;r),~\beta'\big)$. 

\subpara{Comment:} The $\beta'$ is to address the injectivity issue as discussed in \Cref{sec:removing-injectivity}.
\item\label[Step]{item:small-tag:sync:step:OWFs}
$R$ performs the following computation:
\begin{enumerate}
\item \label[Step]{item:small-tag:sync:step:OWFs:A}
$R$ computes $\Set{y^A_i = f(x^A_i)}_{i \in [t]}$ with $x^A_i \pick \bits^\secpar$ for each $i \in [t]$, and sets $Y^A \coloneqq (y^A_1, \ldots, y^A_t)$. $R$ also computes $\Set{\msf{com}^A_i = \Com_{\beta'}(x^A_i;r^A_i)}_{i\in [t]}$, where $r^A_i$ is a random string for each $i\in [t]$.  
\item \label[Step]{item:small-tag:sync:step:OWFs:B}
$R$ computes $\Set{y^B_i = f(x^B_i)}_{i \in [n-t]}$ with $x^B_i \pick \bits^\secpar$ for each $i \in [n-t]$, and sets $Y^B \coloneqq (y^B_1, \ldots, y^B_{n-t})$. $R$ also computes $\Set{\msf{com}^B_i = \Com_{\beta'}(x^B_i;r^B_i)}_{i\in [n-t]}$, where $r^B_i$ is a random string for each $i\in [n-t]$.  
\end{enumerate}
$R$ sends the tuple $\big(Y^A,~\Set{\msf{com}^A_i = \Com_{\beta'}(x^A_i;r^A_i)}_{i\in [t]},~Y^B,~\Set{\msf{com}^B_i = \Com_{\beta'}(x^B_i;r^B_i)}_{i\in [n-t]}\big)$.

\subpara{Comment:} Compared with \Cref{item:one-sided:step:OWFs} of \Cref{prot:one-sided:classical}, the current \Cref{item:small-tag:sync:step:OWFs} has two differences: (1) $R$ additionally commits to the preimages of $y_i$'s. This is to address the injectivity issue as discussed in \Cref{sec:removing-injectivity}. (2) $R$ sends {\em two copies} of the $(Y, \Set{\msf{com}_i}_i)$ tuple, indexed by superscripts $A$ and $B$ respectively. This is corresponding to the ``two-slot'' approach discussed in \Cref{sec:small-tag:classical:sync:high-level} (with the first message of {\bf Slot-B} being sent in parallel with the first message of {\bf Slot-A}).

\item\label[Step]{item:small-tag:sync:step:WIPoK:1A}
{\bf (WIPoK-1-A.)} $R$ and $C$ execute an instance of $\WIPoK$ where $R$ proves to $C$ that he ``knows'' a pre-image of some $y^A_i$ contained in $Y^A$, and $y^A_i$ are consistent with $\msf{com}^A_i$ (as defined in \Cref{item:small-tag:sync:step:OWFs:A}). Formally, $R$ proves that $(Y^A, \Set{\msf{com}^A_i}_{i\in[t]}) \in \Lang^t_{f,\beta'}$, where
\begin{equation}\label[Language]{eq:small-tag:sync:Lang:OWF:A}
\Lang^t_{f,\beta'} \coloneqq \Bigg\{(y^A_1, \ldots, y^A_t), (\msf{com}^A_1, \ldots, \msf{com}^A_t)~\Bigg|~\exists (i, x^A_i, r^A_i) ~s.t.~~
\begin{array}{l}
i \in [t] ~\wedge \\
y^A_i = f(x^A_i) ~\wedge \\
\msf{com}^A_i = \Com_{\beta'}(x^A_i; r^A_i)
\end{array}\Bigg\}.
\end{equation}
Note that $R$ uses $(1, x^A_1, r^A_1)$ as the witness when executing this $\WIPoK$.

\item\label[Step]{item:small-tag:sync:step:WIPoK:1B}
{\bf (WIPoK-1-B.)} $R$ and $C$ execute an instance of $\WIPoK$ where $R$ proves to $C$ that he ``knows'' a pre-image of some $y^B_i$ contained in $Y^B$, and $y^B_i$ are consistent with $\msf{com}^B_i$ (as defined in \Cref{item:small-tag:sync:step:OWFs:B}). Formally, $R$ proves that $(Y^B, \Set{\msf{com}^B_i}_{i\in[n-t]}) \in \Lang^{n-t}_{f,\beta'}$, where
\begin{equation}\label[Language]{eq:small-tag:sync:Lang:OWF:B}
\Lang^{n-t}_{f,\beta'} \coloneqq \Bigg\{(y^B_1, \ldots, y^B_{n-t}), (\msf{com}^B_1, \ldots, \msf{com}^B_{n-t})~\Bigg|~\exists (i, x^B_i, r^B_i) ~s.t.~~ 
\begin{array}{l}
i \in [n-t] ~\wedge \\
y^B_i = f(x^B_i) ~\wedge \\
\msf{com}^B_i = \Com_{\beta'}(x^B_i; r^B_i)
\end{array}\Bigg\}.
\end{equation}
Note that $R$ uses $(1, x^B_1, r^B_1)$ as the witness when executing this $\WIPoK$.

\item\label[Step]{item:small-tag:sync:step:WIPoK:2}
{\bf (WIPoK-2.)} $C$ and $R$ execute an instance of $\WIPoK$ where $C$ proves to $R$ that he ``knows'' the message committed in $\msf{com}$ (defined in \Cref{item:small-tag:sync:step:committing}), {\em or} a pre-image of some $y^A_i$ contained in $Y^A$ (defined in \Cref{item:small-tag:sync:step:OWFs:A}), {\em or} a pre-image of some $y^B_i$ contained in $Y^B$ (defined in \Cref{item:small-tag:sync:step:OWFs:B}). Formally, $C$ proves that $(\msf{com}, Y^A, Y^B) \in \Lang_\beta \vee \Lang^t_f \vee \Lang^{n-t}_f$, where 
\begin{align}
& \Lang_\beta \coloneqq \Set{\msf{com} ~|~ \exists (m, r)~s.t.~ \msf{com} = \Com_\beta(m;r)}, \label[Language]{eq:small-tag:sync:Lang:Com}\\
& \Lang^t_f \coloneqq \Set{(y^A_1, \ldots, y^A_t) ~|~ \exists (i, x^A_i)~s.t.~ i \in [t] \wedge y^A_i = f(x^A_i)}, \\
& \Lang^{n-t}_f \coloneqq \Set{(y^B_1, \ldots, y^B_{n-t}) ~|~ \exists (i, x^B_i)~s.t.~ i \in [n-t] \wedge y^B_i = f(x^B_i)}.
\end{align}
Note that $C$ uses the $(m, r)$ defined in \Cref{item:small-tag:sync:step:committing} as the witness when executing this $\WIPoK$.

\subpara{Comment:} Compared with \Cref{item:one-sided:step:WIPoK:2} of \Cref{prot:one-sided:classical}, we add $\Lang^{n-t}_f$ as an additional OR part to the language to be proven. This is corresponding to the new {\bf Slot-B} as discussed in \Cref{sec:small-tag:classical:sync:high-level}.
\end{enumerate}
\para{Decommit Stage:} 
$C$ sends $(m, r)$. $R$ accepts if $\msf{com} = \Com_\beta(m;r)$, and rejects otherwise.
\end{ProtocolBox}

\para{Security.} Completeness is straightforward from the description of \Cref{prot:small-tag:sync:classical}. 
The statistical binding property follows from that of Naor's commitment. Computational hiding of any commitment scheme follows directly from non-malleability. So it remains for us to show that our commitment protocol is non-malleable, which we establish by the following theorem.

\begin{theorem}\label{thm:nm:small-tag:sync:classical}
The commitment scheme $\langle C, R\rangle^{\msf{sync}}_{\msf{tg}}$ in \Cref{prot:small-tag:sync:classical} is non-malleable against synchronous PPT adversaries with tag space $[n]$, with $n$ being any polynomial on $\secpar$.
\end{theorem}

\subsection{Proving Non-Malleability (Proof of \Cref{thm:nm:small-tag:sync:classical})}
\label{sec:proof:nm:small-tag:sync:classical}
%Due to \Cref{lem:adaptive-predetermined}, it suffices to prove \Cref{thm:nm:small-tag:sync:classical} assuming that the MIM $\mcal{M}$ only uses schedules of a fixed type $T \in \Set{T_{=}, T_{<}, T_{>}}$. 
Since we only consider synchronous adversaries in this section, we can assume that the adversary declares $t$ and $\tilde{t}$ at the beginning w.l.o.g. Thus, we can analyze the three cases of $t=\tilde{t}$, $t<\tilde{t}$, and $t>\tilde{t}$ {\em separately}.   
Since the case of $t=\tilde{t}$ is already ruled out by the definition of non-malleability, we prove non-malleability for the other two cases below. 
\if0
\begin{itemize}
\item {\bf Case 1: $t=\tilde{t}$.} This is the trivial case that is already ruled out by the definition of non-malleability.
\end{itemize}
We show the proof of \Cref{thm:nm:small-tag:sync:classical} for the other two cases in \Cref{sec:proof:nm:small-tag:sync:classical:case2,sec:proof:nm:small-tag:sync:classical:case3} respectively. 
\fi

\subsubsection{The Case of \textnormal{$t<\tilde{t}$}}
\label{sec:proof:nm:small-tag:sync:classical:case2}

The proof for this case is almost identical to the proof  of \Cref{prot:one-sided:classical} in \Cref{sec:small-tag-one-sided-sync-classical}. 
To avoid re-doing the same work, we only focus on the places where modifications are needed. In the following, we assume that the reader has already read and understood the proof in \Cref{sec:small-tag-one-sided-sync-classical}.

First, by repeating the similar arguments in \Cref{one-sided:non-malleability:proof:classical,sec:one-sided:core-lemma:proof,lem:small-tag:proof:se:proof},  we can see that it suffices to construct a machine $\mcal{K}$ that satisfies  similar properties as required in \Cref{lem:small-tag:proof:se:proof:K}. 
Roughly speaking, $\mcal{K}$ is required to extract the committed message $\tilde{m}$ in the right session with a noticeable probability if it is given a prefix that leads to acceptance with a noticeable probability.\footnote{Strictly speaking, the input of $\mcal{K}$ is not a prefix, but a prefix \emph{without $C$'s state $\ST_C$}. We also remark that the assumption is not that the prefix leads to $R$'s acceptance with a noticeable probability in the real experiment, but in the \emph{simulated} experiment defined similarly to $\mcal{G}_1$ in \Cref{lem:small-tag:proof:se:proof}.} 
The construction of $\mcal{K}$ is identical to that in \Cref{sec:lem:small-tag:proof:se:proof:K:proof} except that it extracts a witness $(j,x_j^A,r_j^A)$ from {\bf WIPoK-1-A} and 
always uses the same witness $(1,\tilde{x}^B_1,\tilde{r}^B_1)$ for the right {\bf WIPoK-1-B}.
%\footnote{In \Cref{sec:lem:small-tag:proof:se:proof:K:proof}, $\mcal{K}$ extracts $(j,x_j)$ from {\bf WIPoK-1}, which is the only slot in \Cref{prot:one-sided:classical}. We remark that $\mcal{K}$ in this section additionally extracts the randomness $r_j^A$ due to the modification to remove the injectivity requirement as explained in \Cref{sec:removing-injectivity}.} 
Specifically, $\mcal{K}$ honestly runs the man-in-the-middle experiment starting from the given prefix except for the following modifications:
\begin{enumerate}
    %\item Randomly choose $i\sample [\tilde{t}]$;
    \item \label{item:choose_i}
    Use the witness $(i,\tilde{x}^A_i,\tilde{r}^A_i)$ for the right {\bf WIPoK-1-A} for $i\pick [\tilde{t}]$;
    %\item Use the witness $(1,\tilde{x}^B_1,\tilde{r}^B_1)$ for the right {\bf WIPoK-1-B};
    \item Extract a witness $(j,x^A_j,r^A_j)$ from the left {\bf WIPoK-1-A};
    \item Use the witness $(j,x^A_j)$ in the left {\bf WIPoK-2};
    \item Extract a witness $\tilde{w}$ from the right {\bf WIPoK-2};
    \item If $\tilde{w}$ is a valid witness $(\tilde{m},\tilde{r})$ for $\tilde{\com}\in \Lang_{\tilde{\beta}}
    $, it outputs $\tilde{m}$. Otherwise, it outputs $\bot$. 
\end{enumerate}  
We stress again that $\mathcal{K}$ uses the same witness $(1,\tilde{x}^B_1,\tilde{r}^B_1)$ for the right {\bf WIPoK-1-B} regardless of the $i$ sampled in \Cref{item:choose_i}.\footnote{Changing the witness for {\bf Slot-B} to $(i,\tilde{x}^B_i,\tilde{r}^B_i)$ by using the WI property also works, but that is redundant.}
We also remark that $\mathcal{K}$ extracts a witness from the left {\bf WIPoK-1-A} but not from the left {\bf WIPoK-1-B}.

The proof that the above $\mcal{K}$ satisfies similar properties as in \Cref{lem:small-tag:proof:se:proof:K} is almost identical to that in \Cref{sec:lem:small-tag:proof:se:proof:K:proof,sec:lem:bound:Ki:proof,sec:proof:claim:K'':non-abort} in the one-sided setting.
The only difference is the following.
We define $\mathcal{K}_i$ similarly to that in \Cref{sec:lem:small-tag:proof:se:proof:K:proof}, i.e., it works similarly to $\mcal{K}$ with the random $i$ (in \Cref{item:choose_i}) being fixed. 
%to that in \Cref{sec:lem:small-tag:proof:se:proof:K:proof} except for the difference that $\mathcal{K}_i$ uses the witness $(i,\tilde{x}^A_i,\tilde{r}^A_i)$ for the right {\bf WIPoK-1-A} and the witness $(1,\tilde{x}^B_1,\tilde{r}^B_1)$ for the right {\bf WIPoK-1-B}.   
We need to ensure that the witness $\tilde{w}$ extracted from the right {\bf WIPoK-2} by $\mathcal{K}_i$ cannot be a witness for {\bf Slot-B}, i.e., a witness for $Y^B \in \Lang_f^{n-t}$ with a non-negligible probability. 
This is needed in the proof of a counterpart of \Cref{cliam:bound:Ki-xi} in \Cref{sec:lem:bound:Ki:proof}.
It can be shown as follows: 
\begin{itemize}
\item It is easy to see that $\tilde{w}=(j,\tilde{x}^B_j)$ for $j\neq 1$ with a negligible probability: Since $\mcal{K}_i$ uses $(1,\tilde{x}^B_1,\tilde{r}^B_1)$ for the right {\bf WIPoK-1-B}, extracting  $(j,\tilde{x}^B_j,\tilde{r}^B_j)$ for $j\neq 1$ directly breaks the one-wayness of $f$ or the computational hiding of $\Com$. 
\item Suppose that $\tilde{w}=(1,\tilde{x}^B_1)$  with a non-negligible probability. %By noting that $\mcal{K}_i$ does not rewind the right {\bf WIPoK-1-B}, 
The probability remains non-negligible even if we replace the witness used in the right {\bf WIPoK-1-B} with  $(2,\tilde{x}^B_2,\tilde{r}^B_2)$ by the WI property of the right {\bf WIPoK-1-B}. Then, due to a similar reason to the above case, this also breaks the one-wayness of $f$ or the computational hiding of $\Com$.  
It is worth mentioning that this step relies on the fact that the left {\bf WIPoK-1-A} and the right {\bf WIPoK-1-B} do not interleave with each other by the synchronicity assumption. (Otherwise we have to rewind the right {\bf WIPoK-1-B} when rewinding the left {\bf WIPoK-1-A}, in which case we cannot rely on the WI property of the right {\bf WIPoK-1-B}.) This point will become relevant when dealing with asynchronous adversaries in \Cref{sec:small-tag:async:classical}.
\end{itemize}
Except for the above, the rest of the proof follows from a straightforward adaptation of that in \Cref{sec:small-tag-one-sided-sync-classical}. 

\subsubsection{The Case of \textnormal{$t > \tilde{t}$}}
\label{sec:proof:nm:small-tag:sync:classical:case3}
In this case, we have $n-t < n-\tilde{t}$. Thus, we can simply do the same analysis as the case of $t < \tilde{t}$ with the roles of {\bf Slot-A} and {\bf Slot-B} being swapped. 
%\xiao{ToDo: only need to say the difference from \Cref{sec:proof:nm:small-tag:sync:classical:case2}.}

By combining the analysis for the above cases, we obtain \Cref{thm:nm:small-tag:sync:classical}.

\section{Small-Tag, Asynchronous, Classical Setting}
\label{sec:small-tag:async:classical}
In this section, we show how to make $\langle C, R \rangle^{\msf{sync}}_{\msf{tg}}$ (i.e., \Cref{prot:small-tag:sync:classical}) secure against asynchronous adversaries, to obtain the protocol $\langle C, R\rangle^{\msf{async}}_{\msf{tg}}$. We emphasize that this section still focuses on the small-tag setting against classical (i.e., non-uniform PPT) adversaries.

% In \Cref{sec:small-tag-one-sided-sync-classical}, we constructed a small-tag, one-sided, synchronous non-malleable commitment $\langle C, R \rangle^{\msf{OneSided}}_{\msf{tg}}$ secure in the classical setting. In this section, we show how to modify $\langle C, R \rangle^{\msf{OneSided}}_{\msf{tg}}$ to make it secure in the asynchronous (but still small-tag and classical) setting.

% We first show how to remove the ``one-sided'' restriction \Cref{sec:small-tag:classical:sync}, and then show how to deal with asynchronous adversaries in \Cref{sec:small-tag:classical:async}.

\subsection{High-Level Idea} 
\label{sec:small-tag:classical:async:high-level}
By a careful inspection, the proof of non-malleability of \Cref{prot:small-tag:sync:classical} (which is based on the proof of non-malleability of \Cref{prot:one-sided:classical}) relies on the following assumptions about the schedule. 
%We start by observing that 
\begin{itemize}
    \item 
    The prefix of both the left and right sessions should be generated before moving forward. That is, 
    the left (resp. right) message of \Cref{item:small-tag:sync:step:committing} is sent before the right (resp. left) message of \Cref{item:small-tag:sync:step:OWFs} is sent. 
    This is necessary because we need to analyze the protocol for each fixed prefix $\msf{pref}$ in the security proof for \Cref{prot:small-tag:sync:classical}.
    \item 
    The left message of \Cref{item:small-tag:sync:step:OWFs} is sent before the right {\bf WIPoK-1-A} starts. This is needed because we rely on brute-force search to break hiding of $\com_i^A$ or $\com_i^B$ in an intermediate hybrid, where we rely on (non-uniform) WI of the right {\bf WIPoK-1-A}. This is possible only if we can do brute-force before the right {\bf WIPoK-1-A} starts so that the result of the brute-force can be treated as a non-uniform advice in the reduction to WI. (See \Cref{rmk:non-uniform_reduction}.)
    \item 
    The left (resp. right) {\bf WIPoK-1-A}  and the right (resp. left) {\bf WIPoK-1-B} do not interleave with each other. This is needed because we need to ensure that when rewinding the left {\bf WIPoK-1-A} (resp. {\bf WIPoK-1-B}), we do not rewind the right {\bf WIPoK-1-B} (resp. {\bf WIPoK-1-A}) as explained in \Cref{sec:proof:nm:small-tag:sync:classical:case2}.
    \item The left {\bf WIPoK-1-B} finishes before the right {\bf WIPoK-2} starts. This is needed because we rewind both of them simultaneously in the construction of the extractor $\mcal{K}$.\footnote{In fact, it is also fine that the left {\bf WIPoK-1-A} \emph{starts} after the right {\bf WIPoK-2} \emph{finishes}, since they do not interleave with each other in this case either.} 
\end{itemize}
To deal with a schedule that does not satisfy some of the above conditions, we need to modify the protocol. Our modification is based on a combination of the following tricks.
\begin{enumerate}
    \item \label[Trick]{item:trick_extcom}
    Insert an extractable commitment to $m$ by $C$ between Steps $X$ and $X+1$. 
    We also modify the language for {\bf WIPoK-2} 
    to ensure that the committed message in the extractable commitment is the correct one.
    This 
    gives us an opportunity to extract the right message $\tilde{m}$ without extracting the left message $m$ in the man-in-the-middle game if the right Step $X+1$ starts before the left Step $X$ finishes. 
    \item \label[Trick]{item:trick_trapdoor}
    Insert a WIPoK proving some ``trapdoor statement" where $R$ plays the role of the prover between Steps $X$ and $X+1$. Here, the trapdoor is a witness for an NP language chosen by $R$. 
    We also modify the language for {\bf WIPoK-2} 
    to allow $C$ to use the trapdoor as a witness. Now, if the left Step $X+1$ starts before the right Step $X$ finishes, then we can  extract the left trapdoor without extracting the right trapdoor in the man-in-the-middle game. Then, we can use the extracted trapdoor as a witness for the left  {\bf WIPoK-2}. As a result, we can simulate the honest committer in the left session without using the message $m$. This enables us to extract the right message $\tilde{m}$ from the right {\bf WIPoK-2} without using the left message $m$, and eventually to prove non-malleability.  
    \item \label[Trick]{item:trick_repeat}
    Repeat Step $X$ as many times as the total round complexity of all steps before Step $X$ plus one. 
    %We call each repeated execution a \emph{slot}.  
    By the pigeonhole principle, 
    this ensures that there is at least one execution of the left (resp. right) Step $X$ that does not interleave with all steps before Step $X$ in the right (resp. left). In particular, when we rewind that execution, it does not affect the security of any primitive on the other side. 
\end{enumerate}

\subsection{Construction of $\langle C, R\rangle^{\msf{async}}_{\msf{tg}}$}
% We refer to the construction described in \Cref{sec:small-tag:classical:sync:high-level} as $\langle C, R \rangle^{\msf{sync}}_{\msf{tg}}$, and formally 
We present our construction in \Cref{prot:small-tag:classical} where constants $n_5,...,n_9$ are specified later.
%\begin{itemize}
%    \item $n_4$ is...
%    \item For $i=5,...$ $n_i=$...
%\end{itemize}
This protocol makes use of the same building blocks as for \Cref{prot:small-tag:sync:classical} plus an extractable commitment, namely:
 \begin{itemize}
 \item
An OWF $f$,  
\item
Naor's commitment $\Com$;
\item
A witness-indistinguishable proof of knowledge $\WIPoK$ 
(as per \Cref{def:WIPoK});
\item An extractable commitment $\ExtCom$ (as per \Cref{definition:ext-com}).
We denote by $\Verify_{\ExtCom}$ the verification algorithm of $\ExtCom$ in the decommit stage.  
\end{itemize} 

\begin{ProtocolBox}[label={prot:small-tag:classical}]{Small-Tag Asynchronous NMCom \textnormal{$\langle C, R\rangle^{\msf{async}}_{\msf{tg}}$}}
The tag space is defined to be $[n]$ where $n$ is a polynomial on $\secpar$. Let $t \in [n]$ be the tag for the following interaction. Let $m$ be the message to be committed to. 

\para{Commit Stage:}
\begin{enumerate}
\item\label[Step]{item:small-tag:async:step:Naor-rho}
Receiver $R$ samples and sends the first message $\beta$ for Naor's commitment.
\item\label[Step]{item:small-tag:async:step:committing}
Committer $C$ commits to $m$ using the second message of Naor's commitment. $C$ also sends a $\beta'$ that will be used as the first Naor's message for $R$ to generate a commitment in next step. Formally, $C$ samples $r$ and $\beta'$ and sends the tuple $\big(\msf{com} = \Com_\beta(m;r),~\beta'\big)$. 

\item\label[Step]{item:small-tag:async:step:W-td}  $R$ and $C$ do the following:
\begin{enumerate}
\item {\bf (TrapGen.)}\label[Step]{item:small-tag:async:step:W-td:Z}
$R$ computes $V_0=f(v_0)$ and $V_1=f(v_1)$ with $v_0,v_1\pick \bit^\secpar$ and sends $(V_0,V_1)$ to $C$ 
\item \label[Step]{item:small-tag:async:step:W-td:WIPoK} {\bf (WIPoK-Trap.)}
$R$ and $C$ execute an instance of $\WIPoK$ where $R$ proves to $C$ that he ``knows'' a pre-image of $V_0$ or $V_1$.
Formally, $R$ proves that $(V_0,V_1) \in \Lang_{f}^{\mathrm{OR}}$, where
\begin{equation}\label[Language]{eq:small-tag:async:Lang:TD}
\Lang_{f}^{\mathrm{OR}} \coloneqq \big\{(V_0,V_1)~\big|~\exists v ~s.t.~~f(v)=V_0~\vee~f(v)=V_1
\big\}.
\end{equation}
Note that $R$ uses $v_0$ as the witness when executing this $\WIPoK$.
\end{enumerate}
\subpara{Comment:}
This step is inserted for \Cref{item:trick_trapdoor} in \Cref{sec:small-tag:classical:async:high-level} where a witness for $(V_0,V_1) \in \Lang_{f}^{\mathrm{OR}}$ plays the role of a ``trapdoor". 

\item\label[Step]{item:small-tag:async:step:OWFs}
$R$ performs the following computation:
\begin{enumerate}
\item \label[Step]{item:small-tag:async:step:OWFs:A}
$R$ computes $\Set{y^A_i = f(x^A_i)}_{i \in [t]}$ with $x^A_i \pick \bits^\secpar$ for each $i \in [t]$, and sets $Y^A \coloneqq (y^A_1, \ldots, y^A_t)$. $R$ also computes $\Set{\msf{com}^A_i = \Com_{\beta'}(x^A_i;r^A_i)}_{i\in [t]}$, where $r^A_i$ is a random string for each $i\in [t]$.  
\item \label[Step]{item:small-tag:async:step:OWFs:B}
$R$ computes $\Set{y^B_i = f(x^B_i)}_{i \in [n-t]}$ with $x^B_i \pick \bits^\secpar$ for each $i \in [n-t]$, and sets $Y^B \coloneqq (y^B_1, \ldots, y^B_{n-t})$. $R$ also computes $\Set{\msf{com}^B_i = \Com_{\beta'}(x^B_i;r^B_i)}_{i\in [n-t]}$, where $r^B_i$ is a random string for each $i\in [n-t]$.  
\end{enumerate}
$R$ sends the tuple $\big(Y^A,~\Set{\msf{com}^A_i = \Com_{\beta'}(x^A_i;r^A_i)}_{i\in [t]},~Y^B,~\Set{\msf{com}^B_i = \Com_{\beta'}(x^B_i;r^B_i)}_{i\in [n-t]}\big)$.

\item\label[Step]{item:small-tag:async:step:ExtCom:1}
{\bf (ExtCom-1 $\times~n_5$.)}
$C$ and $R$ sequentially execute $n_5$ instances of $\ExtCom$ where $C$ commits to $m$. For $i\in [n_5]$, let $\tau^1_i$ and $\decom^1_i$ be the transcript and decommitment information (privately obtained by $C$) of the $i$-th execution. 

\subpara{Comment:}
This step is inserted for \Cref{item:trick_extcom} 
and repeated for \Cref{item:trick_repeat}  
in \Cref{sec:small-tag:classical:async:high-level}.

\item\label[Step]{item:small-tag:async:step:WIPoK:1A}
{\bf (WIPoK-1-A $\times~n_6$.)} $R$ and $C$ sequentially execute $n_6$ instances of $\WIPoK$ where $R$ proves to $C$ that he ``knows'' a pre-image of some $y^A_i$ contained in $Y^A$, and $y^A_i$ are consistent with $\msf{com}^A_i$ (as defined in \Cref{item:small-tag:async:step:OWFs:A}). Formally, $R$ proves that $(Y^A, \Set{\msf{com}^A_i}_{i\in[t]}) \in \Lang^t_{f,\beta'}$, where
\begin{equation}\label[Language]{eq:small-tag:async:Lang:OWF:A}
\Lang^t_{f,\beta'} \coloneqq \Bigg\{(y^A_1, \ldots, y^A_t), (\msf{com}^A_1, \ldots, \msf{com}^A_t)~\Bigg|~\exists (i, x^A_i, r^A_i) ~s.t.~~
\begin{array}{l}
i \in [t] ~\wedge \\
y^A_i = f(x^A_i) ~\wedge \\
\msf{com}^A_i = \Com_{\beta'}(x^A_i; r^A_i)
\end{array}\Bigg\}.
\end{equation}
Note that $R$ uses $(1, x^A_1, r^A_1)$ as the witness when executing this $\WIPoK$.

\subpara{Comment:}
This step is repeated for \Cref{item:trick_repeat}  
in \Cref{sec:small-tag:classical:async:high-level}.

\item\label[Step]{item:small-tag:async:step:ExtCom:2}
{\bf (ExtCom-2 $\times~n_7$.)}
$C$ and $R$ sequentially execute $n_7$ instances of $\ExtCom$ where $C$ commits to $m$. Let $\tau^2_i$ and $\decom^2_i$ be the transcript and decommitment (privately obtained by $C$) of the $i$-th execution. 

\subpara{Comment:}
This step is inserted for \Cref{item:trick_extcom} 
and repeated for \Cref{item:trick_repeat}  
in \Cref{sec:small-tag:classical:async:high-level}.

\item\label[Step]{item:small-tag:async:step:WIPoK:1B}
{\bf (WIPoK-1-B $\times~n_8$.)} $R$ and $C$ sequentially execute $n_8$ instances of $\WIPoK$ where $R$ proves to $C$ that he ``knows'' a pre-image of some $y^B_i$ contained in $Y^B$, and $y^B_i$ are consistent with $\msf{com}^B_i$ (as defined in \Cref{item:small-tag:async:step:OWFs:B}). Formally, $R$ proves that $(Y^B, \Set{\msf{com}^B_i}_{i\in[n-t]}) \in \Lang^{n-t}_{f,\beta'}$, where
\begin{equation}\label[Language]{eq:small-tag:async:Lang:OWF:B}
\Lang^{n-t}_{f,\beta'} \coloneqq \Bigg\{(y^B_1, \ldots, y^B_{n-t}), (\msf{com}^B_1, \ldots, \msf{com}^B_{n-t})~\Bigg|~\exists (i, x^B_i, r^B_i) ~s.t.~~ 
\begin{array}{l}
i \in [n-t] ~\wedge \\
y^B_i = f(x^B_i) ~\wedge \\
\msf{com}^B_i = \Com_{\beta'}(x^B_i; r^B_i)
\end{array}\Bigg\}.
\end{equation}
Note that $R$ uses $(1, x^B_1, r^B_1)$ as the witness when executing this $\WIPoK$.

\subpara{Comment:}
This step is repeated for \Cref{item:trick_repeat}  
in \Cref{sec:small-tag:classical:async:high-level}.

\item\label[Step]{item:small-tag:async:step:ExtCom:3}
{\bf (ExtCom-3 $\times~n_9$.)}
$C$ and $R$ sequentially execute $n_9$ instances of $\ExtCom$ where $C$ commits to $m$. Let $\tau^2_i$ and $\decom^2_i$ be the transcript and decommitment (privately obtained by $C$) of the $i$-th execution.

\subpara{Comment:}
This step is inserted for \Cref{item:trick_extcom} 
and repeated for \Cref{item:trick_repeat}  
in \Cref{sec:small-tag:classical:async:high-level}.

\item\label[Step]{item:small-tag:async:step:WIPoK:2}
{\bf (WIPoK-2.)} $C$ and $R$ execute an instance of $\WIPoK$ where $C$ proves to $R$ that he ``knows'' the (same) message committed in $\msf{com}$ (defined in \Cref{item:small-tag:async:step:committing}) and all commitments $\tau^c_i$ of $\ExtCom$ (in \Cref{item:small-tag:async:step:ExtCom:1,item:small-tag:async:step:ExtCom:2,item:small-tag:async:step:ExtCom:3}), {\em or} a pre-image of some $y^A_i$ contained in $Y^A$ (defined in \Cref{item:small-tag:async:step:OWFs:A}), {\em or} a pre-image of some $y^B_i$ contained in $Y^B$ (defined in \Cref{item:small-tag:async:step:OWFs:B})
{\em or} a preimage of either of $V_0$ or $V_1$ (defined in \Cref{item:small-tag:async:step:W-td:Z}). Formally, $C$ proves that $(\msf{com},\{\tau^c_i\}_{(c,i)\in \extcomindex}, Y^A, Y^B, V_0,V_1) \in \Lang_{\beta} \vee \Lang^t_f \vee \Lang^{n-t}_f \vee \Lang_{f}^{\mathrm{OR}}$,
where $\extcomindex=(\{1\}\times [n_5])\cup (\{2\}\times [n_7])\cup (\{3\}\times [n_9])$, 
\begin{align}
& \Lang_{\beta} \coloneqq \Bigg\{(\msf{com},\{\tau^c_i\}_{(c,i)\in \extcomindex}) ~\Bigg|
\begin{array}{l}
~ \exists (m, r,\{\decom^c_i\}_{(c,i)\in \extcomindex})~s.t.~\\ 
\begin{array}{l}
\msf{com} = \Com_\beta(m;r)~\wedge\\
\forall~(c,i)\in \extcomindex,~ \Verify_{\ExtCom}(m,\tau^c_i,\decom^c_i)=\top\\
\end{array}
\end{array}
\Bigg\}
\label[Language]{eq:small-tag:async:Lang:Com}\\
& \Lang^t_f \coloneqq \Set{(y^A_1, \ldots, y^A_t) ~|~ \exists (i, x^A_i)~s.t.~ i \in [t] \wedge y^A_i = f(x^A_i)}, \\
& \Lang^{n-t}_f \coloneqq \Set{(y^B_1, \ldots, y^B_{n-t}) ~|~ \exists (i, x^B_i)~s.t.~ i \in [n-t] \wedge y^B_i = f(x^B_i)},
\end{align}
and $\Lang_{f}^{\mathrm{OR}}$ is defined in \Cref{eq:small-tag:async:Lang:TD}. 
Note that $C$ uses the $(m, r, \{\decom^c_i\}_{(c,i)\in\extcomindex})$ defined in \Cref{item:small-tag:async:step:committing,item:small-tag:async:step:ExtCom:1,item:small-tag:async:step:ExtCom:2,item:small-tag:async:step:ExtCom:3} as the witness when executing this $\WIPoK$.

\subpara{Comment:} Compared with \Cref{item:small-tag:sync:step:WIPoK:2} of \Cref{prot:small-tag:sync:classical}, we add $\Lang_{f}^{\mathrm{OR}}$ as an additional OR part to the language to be proven for \Cref{item:trick_trapdoor} in \Cref{sec:small-tag:classical:async:high-level}. 
We also modify $\Lang_{\beta}$ to prove that all commitments $\tau^c_i$ of $\ExtCom$ commit to the same message as $\com$ for \Cref{item:trick_extcom} in \Cref{sec:small-tag:classical:async:high-level}. 
\end{enumerate}
\para{Decommit Stage:} 
$C$ sends $(m, r)$. $R$ accepts if $\msf{com} = \Com_\beta(m;r)$, and rejects otherwise.
\end{ProtocolBox}
\para{Choice of $n_i$.}
We recursively define $n_5,...,n_9$ so that 
$n_i$ is larger than the total round complexity of Steps $1$ to $i-1$. We also require that $n_5$ is larger than the round complexities of $\WIPoK$ and $\ExtCom$. It is easy to see that we can set $n_i$ to be constant if both $\WIPoK$ and $\ExtCom$ have constant rounds. In particular, \Cref{prot:small-tag:classical} runs in constant rounds.  
We remark that the above choice of $n_i$ is for simplifying the security analysis and we could significantly optimize the exact round complexity with  more careful analysis. But we choose not to press the issue further.

\para{Notation.}
For $c=1,2,3$, 
we refer to the $i$-th execution of $\ExtCom$ in {\bf ExtCom-$c$} as {\bf ExtCom-$c$-$i$}.
We also define {\bf WIPoK-1-A-$i$} and  {\bf WIPoK-1-B-$i$} similarly. 

\para{Security.}
The completeness is easy to see. Statistical binding follows from that of Naor's commitment. We prove computational hiding below.\footnote{Though computational hiding follows from non-malleability, we prove it here because we rely on the computational hiding in the proof of non-malleability.}
\begin{theorem}\label{thm:small-tag:async:classical:hiding}
The commitment scheme $\langle C, R\rangle^{\msf{async}}_{\msf{tg}}$ in \Cref{prot:small-tag:classical} is computationally hiding.
\end{theorem}
\begin{proof}
Let $R^*$ be a non-uniform PPT adversary against the computational hiding property of \Cref{prot:small-tag:classical}. 
We consider the following sequence of hybrids. 

\begin{itemize}
\item {Hybrid $H_1^{R^*}(\secpar,m)$:} This is the real experiment that simulates the interaction between $C(m)$ and $R^*$. That is, it runs $\langle C(m), R^*\rangle_{\msf{tg}}(1^\secpar)$ and outputs the final output of $R^*$. We have to prove 
\begin{align}\label{eq:break_hiding}
    \left|\Pr[H_1^{R^*}(\secpar,m_0)=1]-\Pr[H_1^{R^*}(\secpar,m_1)=1]\right|=\negl(\secpar).
\end{align}
for all $m_0,m_1$. 
\item {Hybrid $H_2^{R^*}(\secpar,m)$:} 
This is identical to $H_1^{R^*}(\secpar,m)$ except that it runs the emulation extractor for {\bf WIPoK-1-A-1} to extract a witness $(i,x^A_i,r^A_i)$ for  $(Y^A, \Set{\msf{com}^A_i}_{i\in[t]}) \in \Lang^t_{f,\beta'}$ where $Y^A$ and $\Set{\msf{com}^A_i}_{i\in[t]})$ are generated in \Cref{item:small-tag:async:step:OWFs:A}. We note that this experiment extracts $(i,x^A_i,r^A_i)$ but does not use it. 
By the PoK property (as per \Cref{def:WEE}) of {\bf WIPoK-1-A-1}, we have 
\begin{align}\label{eq:hiding_hybrid_one}
    \left|\Pr[H_1^{R^*}(\secpar,m)=1]-\Pr[H_2^{R^*}(\secpar,m)=1]\right|=\negl(\secpar)
\end{align}
for all $m$.

\item {Hybrid $H_3^{R^*}(\secpar,m)$:} 
This is identical to $H_2^{R^*}(\secpar,m)$ except that it uses $(i,x^A_i,r^A_i)$ extracted from {\bf WIPoK-1-A-1} in {\bf WIPoK-2}.
By the WI property of {\bf WIPoK-2}, we have 
\begin{align}\label{eq:hiding_hybrid_two}
    \left|\Pr[H_2^{R^*}(\secpar,m)=1]-\Pr[H_3^{R^*}(\secpar,m)=1]\right|=\negl(\secpar)
\end{align}
for all $m$.

We observe that $H_3^{R^*}(\secpar,m_b)$ uses $m$ only in the generation of $\com$ (in \Cref{item:small-tag:async:step:committing}) and $\tau^c_i$ for $(c,i)\in\extcomindex$  (in \Cref{item:small-tag:async:step:ExtCom:1,item:small-tag:async:step:ExtCom:2,item:small-tag:async:step:ExtCom:3}). Therefore, by the computational hiding property of Naor's commitment and $\ExtCom$, we have 
\begin{align}\label{eq:break_hiding_two}
    \left|\Pr[H_3^{R^*}(\secpar,m_0)=1]-\Pr[H_3^{R^*}(\secpar,m_1)=1]\right|=\negl(\secpar)
\end{align}
for all $m_0,m_1$.
Combining \Cref{eq:hiding_hybrid_one,eq:hiding_hybrid_two,eq:break_hiding_two}, we obtain \Cref{eq:break_hiding}.
\end{itemize}
This completes the proof of \Cref{thm:small-tag:async:classical:hiding}.

\end{proof}

We establish non-malleability by \Cref{thm:nm:small-tag:classical}, whose proof will be presented in \Cref{sec:proof_NM_async}.
\begin{theorem}\label{thm:nm:small-tag:classical}
The commitment scheme $\langle C, R\rangle^{\msf{async}}_{\msf{tg}}$ in \Cref{prot:small-tag:classical} is non-malleable against asynchronous PPT adversaries with tag space $[n]$, with $n$ being any polynomial on $\secpar$.
\end{theorem}

\subsection{Adaptive Schedule vs. Predetermined Schedule}
\label{sec:adaptive-predetermined:two-slot:async}
Before proving the non-malleability of \Cref{prot:small-tag:classical}, we prove a useful lemma that enable us to use {\em predetermined} schedules when proving non-malleability.  
The following \Cref{lem:adaptive-predetermined:async}  should be attributed to \cite[Lemma 5.1]{BLS21}: Although they only show this lemma for their specific construction, their proof technique is general enough. Our proof of \Cref{lem:adaptive-predetermined:async} is identical to theirs except for some notational changes customized to our application.
\begin{lemma}[{\cite[Lemma 5.1]{BLS21}}]\label{lem:adaptive-predetermined:async}
Let $\scheduleset$ be the set of all possible schedules of a MIM adversary against a commitment scheme $\langle C,R \rangle$. 
%\takashi{I'm wondering if we need to write the definition of schedule. Especially, we need to make sure that it contains $t$ and $\tilde{t}$.}
Let $S_1,...,S_N$ be efficiently recognizable subsets of $\scheduleset$ such that $\bigcup_{i\in [N]}S_i=\scheduleset$ for $N=\poly(\secpar)$.
If {\em for every $i\in [N]$}, 
$\langle C,R \rangle$ is non-malleable against non-uniform PPT MIM adversaries whose schedule belongs to $S_i$, then it is also non-malleable against arbitrary non-uniform PPT MIM adversaries.
\end{lemma}
\begin{proof}
We can assume that $S_1,...,S_N$ are disjoint w.l.o.g. (If they are not disjoint, we can define $S'_1\defeq S_1$, $S'_{i}\defeq S_{i}\setminus \left(\bigcup_{j\in [i-1]}S_j\right)$ for $i>1$ and consider disjoint sets $S'_1,...,S'_N$.)

Given an arbitrary non-uniform PPT MIM $\mcal{M}$ and non-uniform PPT distinguisher $\mcal{D}$ that break non-malleability for some
values $m_0, m_1$ with advantage $\delta$, we construct a new non-uniform PPT adversary with a schedule that belongs to $S_i$ for a predetermined $i\in [n]$, which
breaks the scheme with probability $\delta/N$.

Consider an adversary $\mcal{M}'$ that first samples uniformly at random $i\pick [N]$. $\mcal{M}'$ then runs  $\mcal{M}$ to complete the MIM attack. If the transcript generated in the execution belongs to $S_i$, 
$\mcal{M}'$ outputs whatever $\mcal{M}$ outputs, and otherwise outputs $\bot$. 

Since every execution must belong to exactly one of $S_1,...,S_N$, $\mcal{M}'$ breaks non-malleability
with probability exactly $\delta/N$  (with respect to the same distinguisher $\mcal{D}$ and $m_0, m_1$). Finally, by an
averaging argument, we fix the choice of $\mcal{M}'$ for a schedule to be of some type $S_i$ that maximizes
$\mcal{D}$'s distinguishing advantage. We obtain a corresponding MIM with a schedule that belongs to $S_i$ for a predetermined $i\in [n]$ with at least
the same advantage $\delta/N$.

\end{proof}

\subsection{Proving Non-Malleability (Proof of \Cref{thm:nm:small-tag:classical})}\label{sec:proof_NM_async}
%In this subsection, we prove non-malleability. 

For proving \Cref{thm:nm:small-tag:classical}, we consider the following ``bad'' schedules:
\begin{itemize}
    \item{$\Bad~1$:} \label{item:bad_case:1}
    The right message of \Cref{item:small-tag:async:step:OWFs} is sent before the left message of \Cref{item:small-tag:async:step:committing} is sent. \item{$\Bad~2$:} \label{item:bad_case:2}
    The left message of \Cref{item:small-tag:async:step:OWFs} is sent before the right message of \Cref{item:small-tag:async:step:committing} is sent.
    \item{$\Bad~3$:} \label{item:bad_case:3}
    The right {\bf WIPoK-1-A} starts before the left    message of \Cref{item:small-tag:async:step:OWFs} is sent.
 \item{$\Bad~4$:} \label{item:bad_case:4}
    The right {\bf WIPoK-1-B} starts before the left {\bf WIPoK-1-A} finishes.  
 \item{$\Bad~5$:} \label{item:bad_case:5}
    The right {\bf WIPoK-2} starts before the left {\bf WIPoK-1-B} finishes.   
\end{itemize}
In the above, when we say ``{\bf WIPoK-XX} starts", this means that the first message of the first execution of $\WIPoK$ in {\bf WIPoK-XX} is sent. Similarly, when we say ``{\bf WIPoK-XX} finishes", this means that the final message of the final execution of $\WIPoK$ in {\bf WIPoK-XX} is sent. 
We remark that the above bad cases are not disjoint.  

We first prove non-malleability in the case where the schedule does not suffer from any of them (which we call a \emph{good} schedule) in \Cref{sec:NM:good}. Then, we prove it for the remaining schedules (which we call \emph{bad} schedules) in \Cref{sec:NM:bad}. 
We remark that we can analyze them separately due to \Cref{lem:adaptive-predetermined:async}. 

\subsection{Non-Malleability for Good Schedules}\label{sec:NM:good}
Based on the observations made in \Cref{sec:small-tag:classical:async:high-level}, we can show the non-malleability analogously to the synchronous case if none of $\Bad~1$-$5$ occurs. 
\begin{lemma}\label{lem:good_case}
The commitment scheme $\langle C, R\rangle^{\msf{async}}_{\msf{tg}}$ in \Cref{prot:small-tag:classical} is non-malleable against asynchronous PPT adversaries whose schedule does not suffer from any of $\Bad~1$-$5$.
\end{lemma}
\begin{proof}
Similarly to the proof of non-malleability of \Cref{prot:small-tag:sync:classical} in \Cref{sec:small-tag:classical:sync}, we consider the three cases of $t=\tilde{t}$, $t<\tilde{t}$, and $\tilde{t}<t$ separately where $t$ and $\tilde{t}$ are the tags of the left and right sessions, respectively. 
We remark that we can assume that the reduction algorithm knows which of them happens from the beginning because of \Cref{lem:adaptive-predetermined:async}, even if the adversary adaptively determines the schedule.  

Again, the case of $t=\tilde{t}$ is already ruled out by the definition of non-malleability. Next, we analyze the remaining two cases in \Cref{lem:good_case:proof:le,lem:good_case:proof:ge}, which will eventually complete the proof of \Cref{lem:good_case}. 

%\takashi{Similar to the synchronous case. The differences are that we have to chose good slots of {\bf WIPoK-1-A or B} and {\bf WIPoK-2}. Another issue is how to change the committed messages inside \{\bf ExtCom-{1,2,3}\}. Basically, we can do that whenever we rely on WI of the left WI in the original proof. }
\end{proof}

%\para{The case of $t=\tilde{t}$.}
%This case is already ruled out by the definition of non-malleability.

\subsubsection{The Case of $t<\tilde{t}$.}\label{lem:good_case:proof:le}
For a PPT man-in-the-middle adversary $\mcal{M}$ with non-uniform advice $z$ and a message $m$, we define the prefix generation algorithm $H^{\mcal{M}}_{\msf{pre}}(\secpar, m, z)$ as follows similarly to that in \Cref{hybrid:H:reinterpretation}: 
$H^{\mcal{M}}_{\msf{pre}}(\secpar, m, z)$ runs the man-in-the-middle experiment  $\msf{mim}^\mcal{M}_{\langle C, R \rangle^{\msf{async}}_{\msf{tg}}}(\secpar, m, z)$  
until $\mcal{M}$ receives the left message of \Cref{item:small-tag:async:step:committing} \emph{and} sends the right message of \Cref{item:small-tag:async:step:committing}. 
Then, it outputs the prefix $\msf{pref}=(\msf{st}_{\mcal{M}},  \msf{st}_C, \msf{st}_R, \tau, \tilde{\tau})$ where $\msf{st}_{\mcal{M}}$,  $\msf{st}_C$, $\msf{st}_R$ are the states of $\mcal{M}$, $C$, and $R$ at this point, respectively; And $\tau$ (reps.\ $\tilde{\tau}$) is Naor's commitment in the right (resp.\ left) session generated in \Cref{item:small-tag:async:step:Naor-rho,item:small-tag:async:step:committing}. 
We denote by $\A(\st_{\mcal{M}})$ to mean the algorithm that works similarly to $\mcal{M}$ from the point where the prefix is generated. 
We remark that $H^{\mcal{M}}_{\msf{pre}}(\secpar, m, z)$ does not reach to \Cref{item:small-tag:async:step:OWFs} in either of the left or right session since we assume that neither of $\Bad~1$ or $\Bad~2$ occurs. 

We define a machine $\mcal{G}_i$ similarly to that used in the proof of \Cref{lem:small-tag:proof:se:proof:K} (recall it from  \Cref{figure:one-sided:Gi}) except that $C$ commits to $0$ instead of $m$ in the left session. Formally, it works as follows: 

\para{Machine $\mcal{G}_i$:}  For a prefix $\msf{pref}=(\msf{st}_{\mcal{M}},  \msf{st}_C, \msf{st}_R, \tau, \tilde{\tau})$, $\mcal{G}_i(\msf{st}_{\mcal{M}}, \msf{st}_R, \tau, \tilde{\tau})$ runs the rest of $\msf{mim}^\mcal{M}_{\langle C, R \rangle^{\msf{async}}_{\msf{tg}}}(\secpar, m, z)$ except for the following three differences:
\begin{enumerate}
    \item $C$ commits to $0$ instead of $m$ in the left {\bf ExtCom-\{1,2,3\}}. 
    \item $R$ uses $(i,\tilde{x}^A_i)$ instead of $(1,\tilde{x}^A_1)$ in the right {\bf WIPoK-1-A}.
\item It chooses $i^*$ such that the left {\bf WIPoK-1-A-$i^*$} does not interleave with the right Step \ref{item:small-tag:async:step:Naor-rho} to \ref{item:small-tag:async:step:ExtCom:1}, i.e., no message of the right Step \ref{item:small-tag:async:step:Naor-rho} to \ref{item:small-tag:async:step:ExtCom:1} is sent during the execution of the left {\bf WIPoK-1-A-$i^*$}. 
Note that such $i^*$ exists by the pigeonhole principle, and we call the smallest such $i^*$ the \emph{good} index. 
Then, instead of executing the left {\bf WIPoK-1-A-$i^*$} honestly, it uses the witness-extended emulator $\mcal{WE}$ to extract a witness:
\begin{itemize}
\item
 If the left committer accepts the left {\bf WIPoK-1-A-$i^*$} and the extracted witness $(j,x^A_j,r^A_j)$ is valid, $\mcal{G}_i$ uses $(j,x^A_j)$ to finish the left {\bf WIPoK-2} and outputs $\mcal{M}$'s final state and the right receiver's decision bit $b$; 
 \item
 If the left committer accepts the left {\bf WIPoK-1-A-$i^*$}  but the extracted witness is invalid, it aborts immediately and outputs $(\bot, \bot)$;
 \item 
 If the left committer rejects the left {\bf WIPoK-1-A-$i^*$}, it runs the rest of man-in-the-middle experiment to output $\mcal{M}$'s final state and the right receiver's decision bit $b$.  Note that it does not need to run the left {\bf WIPoK-2} in this case since the left committer aborts after the left {\bf WIPoK-1-A-$i^*$}.

%\item
% If the extracted witness $(j,x^A_j,r^A_j)$ is valid, $\mcal{G}_1$ uses $(j,x^A_j)$ to finish the left {\bf WIPoK-2} and outputs $\mcal{M}$'s final state and the right receiver's decision bit $b$; 
% \item
% Otherwise, it aborts immediately and outputs $(\bot, \bot)$.
\end{itemize}
\end{enumerate}
%We denote the above procedure by $(\OUT, b) \gets \mcal{G}_1(\msf{st}_{\mcal{M}}, \msf{st}_R, \tau, \tilde{\tau})$. 
%It is worth noting that $\mcal{G}_1$ does {\em not} need to know $\msf{st}_C$.
We let $p^{\msf{Sim}}_{\msf{pref}}$ be the probability that $\mcal{G}_1(\msf{st}_{\mcal{M}}, \msf{st}_R, \tau, \tilde{\tau})$ returns $b=\top$. 

\begin{remark}\label{rem:adaptive_good_slot}
Strictly speaking, the above description of $\mcal{G}_i$ is not well-defined since the good index depends on the schedule, which is adaptively determined. However, since there are constant number of possibilities for the good index, we only have to prove non-malleability assuming that the good index is fixed at first by \Cref{lem:adaptive-predetermined:async}.
A similar remark applies to later parts of the current proof. We will not repeat it as it should be clear from context. 
\end{remark}

By repeating similar arguments as in \Cref{one-sided:non-malleability:proof:classical,sec:one-sided:core-lemma:proof,lem:small-tag:proof:se:proof},  we can see that it suffices to prove the following lemma, which is an analog of \Cref{lem:small-tag:proof:se:proof:K}. 
\begin{lemma}\label{lem:small-tag:proof:se:proof:K:asyn}
There exists an expected PPT machine $\mcal{K}$ such that for any $\msf{pref}=(\msf{st}_{\mcal{M}},  \msf{st}_C, \msf{st}_R, \tau, \tilde{\tau})$ in the support of $H^{\mcal{M}}_{\msf{pre}}(\secpar, m, z)$ and any noticeable $\epsilon(\secpar)$, the following holds:
\begin{enumerate}
\item \label[Property]{property:small-tag:proof:se:proof:K:syntax:asyn}
{\bf (Syntax.)} $\mcal{K}$ takes as input $(1^\secpar, \msf{st}_R, \tau, \tilde{\tau})$ and makes oracle access to $\Adv(\msf{st}_{\mcal{M}})$. It output a value $\msf{Val}_{\mcal{K}} \in \bits^{\ell(\secpar)} \cup \Set{\bot}$ such that $\msf{Val}_{\mcal{K}} = \msf{val}(\tilde{\tau})$ whenever $\msf{Val}_{\mcal{K}} \ne \bot$.

\item \label[Property]{property:small-tag:proof:se:proof:K:asyn}
If $p^{\msf{Sim}}_{\msf{pref}} \ge \epsilon(\secpar)$, then it holds that
$$\Pr[\Val_{\mcal{K}} = \msf{val}(\tilde{\tau}) : \Val_{\mcal{K}} \gets \mcal{K}^{\A(\msf{st}_{\mcal{M}})}(1^\secpar, \msf{st}_R, \tau, \tilde{\tau})] \ge \frac{\epsilon'(\secpar)}{\tilde{t}},$$
where  $\epsilon'(\secpar) \coloneqq \frac{\epsilon(\secpar)}{10t^2}$.
\end{enumerate}
\end{lemma}
\begin{proof}[Proof of \Cref{lem:small-tag:proof:se:proof:K:asyn}]
We construct $\mcal{K}_i$ and $\mcal{K}$ similarly as in \Cref{sec:proof:nm:small-tag:sync:classical:case2}, which is in turn based on the technique in \Cref{sec:lem:small-tag:proof:se:proof:K:proof}.  

\para{Machine $\mcal{K}_i$ ($i \in [\tilde{t}]$):} 
On input $(1^\secpar, \msf{st}_R, \tau, \tilde{\tau})$, it behaves identically to $\mcal{G}_i(\msf{st}_{\mcal{M}}, \msf{st}_R, \tau, \tilde{\tau})$  except for the following difference. Machine $\mcal{K}_i^{\A(\st_{\mcal{M}})}(1^\secpar, \msf{st}_R, \tau, \tilde{\tau}))$ uses the witness-extended emulator $\mcal{WE}$ to finish the right {\bf WIPoK-2}, instead of playing the role of the honest receiver.  If it extracts a witness for $(\tilde{\msf{com}},\{\tilde{\tau}^c_i\}_{(c,i)\in \extcomindex})\in \Lang_{\tilde{\beta}}$, then it sets $\Val$ to be the first component $\tilde{m}$ of the witness (which is suppose to be the committed message in $\tilde{\tau}$) and outputs $\Val=\tilde{m}$; Otherwise, it outputs $\bot$.

\para{Extractor $\mcal{K}$:}  On input $(1^\secpar, \msf{st}_R, \tau, \tilde{\tau})$, it samples uniformly at random an index $i \pick [\tilde{t}]$, executes $\mcal{K}_i^{\A(\st_{\mcal{M}})}(1^\secpar, \msf{st}_R, \tau, \tilde{\tau})$, and outputs whatever $\mcal{K}_i^{\A(\st_{\mcal{M}})}(1^\secpar, \msf{st}_R, \tau, \tilde{\tau})$ outputs.

We can show that $\mcal{K}$ as described above satisfies \Cref{lem:small-tag:proof:se:proof:K:asyn} by similar analyses to those in \Cref{sec:lem:small-tag:proof:se:proof:K:proof,sec:lem:bound:Ki:proof,sec:proof:claim:K'':non-abort} additionally relying on the computational hiding property of $\ExtCom$ along with the following observations:
\begin{enumerate}
    \item By our choice of the good index $i^*$, the witness-extended emulator for the left {\bf WIPoK-1-A-$i^*$} does not rewind the right {\bf WIPoK-Trap}. Thus, $\mcal{K}_i$ does not extract a ``trapdoor witness" (i.e., a witness for $(\tilde{V}_0,\tilde{V}_1)\in \Lang_f^{\mathrm{OR}}$) except for a negligible probability by the one-wayness of $f$ and the WI property of the right {\bf WIPoK-Trap}.
    \item \label{item:WIPoK-1-A_does_not_rewind_WIPoK-1-B}
    Since we assume that $\Bad~4$ does not occur, the witness-extended emulator for the left {\bf WIPoK-1-A-$i^*$} does not rewind the right {\bf WIPoK-1-B}. Thus, $\mcal{K}_i$ does not extract a witness for {\bf Slot-B}, (i.e., a witness for $\tilde{Y}^B\in \Lang_f^{n-t}$) except for a negligible probability by the one-wayness of $f$ and the WI property of the right {\bf WIPoK-1-B}. This can be proven by a similar analysis to that in \Cref{sec:proof:nm:small-tag:sync:classical:case2}. 
    \item 
    Since we assume that $\Bad~3$ does not occur, we can rely on the WI property of the right {\bf WIPoK-1-A} in a hybrid where we find the committed messages in $\{\com_i^A\}_{i\in [t]}$ by brute-force. This is needed in a step corresponding to the proof of \Cref{eq:relation:K:WI} in \Cref{sec:lem:bound:Ki:proof}.
    \item Since  we assume that $\Bad~5$ does not occur,
    the left {\bf WIPoK-1-A} and the right {\bf WIPoK-2} do not run simultaneously. Thus, there is no nesting of extractors in hybrids where we invoke extractors for both of them (e.g., $\mcal{K}$ and $\mcal{K}_i$).
\end{enumerate}
\begin{remark}
One might wonder how we rely on the computational hiding property of the left {\bf ExtCom-\{1,2,3\}} even though the witness-extended emulator of the right {\bf WIPoK-2} may rewind them. However, one can see by a close inspection of the proof that we only need to use computational hiding of the left {\bf ExtCom-\{1,2,3\}} in hybrids that do not run the witness-extended emulator of the right {\bf WIPoK-2}. This is similar to how we replace the witness used in the left  {\bf WIPoK-2} using the WI property in the proof for the one-sided synchronous case in \Cref{sec:small-tag-one-sided-sync-classical}.
\end{remark}

This completes the proof of \Cref{lem:small-tag:proof:se:proof:K:asyn}, which finishes the proof for the case of $t<\tilde{t}$. 

\end{proof}

\subsubsection{The Case of $t>\tilde{t}$}\label{lem:good_case:proof:ge}
Similarly to the synchronous case (i.e., the proof of \Cref{thm:nm:small-tag:sync:classical}), the proof of this case can be done similarly to that of the case of $t<\tilde{t}$ except that we extract a witness from (the good index of) the left {\bf WIPoK-1-B} instead of from {\bf WIPoK-1-A}. In doing so, one has to be careful that $\Bad~4$ is not symmetric for {\bf Slot-A} and {\bf Slot-B}. In particular, even if we assume that $\Bad~4$ does not occur, it may be still possible that 
the \emph{left} {\bf WIPoK-1-B} and the \emph{right} {\bf WIPoK-1-A} run concurrently. However, this is not an issue since we rewind the left {\bf WIPoK-1-B} of the good index that does not interleave with \emph{any} message of the right session before {\bf WIPoK-1-B}. In particular, its rewinding does not rewind \emph{any} slot of the right {\bf WIPoK-1-A}. 
Except for the above remark, the proof of this case is almost identical to the proof for the case of $t<\tilde{t}$ with the roles of {\bf Slot-A} and {\bf Slot-B} being swapped.

\subsection{Non-Malleability for Bad Schedules}\label{sec:NM:bad}
Before analyzing bad cases, we show a lemma that is used many times in the analysis of bad cases.
\begin{lemma}\label{lem:ExtCom_soundness}
Let $C^*$ be a PPT cheating prover of the protocol in \Cref{prot:small-tag:classical} with advice $z$.  
For $(c,i)\in \extcomindex$ (where $\extcomindex=(\{1\}\times [n_5])\cup (\{2\}\times [n_7])\cup (\{3\}\times [n_9])$ as defined in \Cref{item:small-tag:async:step:WIPoK:2} of \Cref{prot:small-tag:classical}) we define experiments $\Exp$ and $\Exp_{c,i}$ as follows.

\begin{itemize}
    \item{\emph{Experiment} $\Exp(\secpar,z)$:}
It simulates the interaction between $C^*$ and $R$ in the commit stage and outputs $(\OUT_{C^*} ,\val_b(\tau))$ where $\OUT_{C^*}$ is the final output of $C^*$, $b$ is $R$'s decision bit, $\tau$ is Naor's commitment generated in  \Cref{item:small-tag:async:step:Naor-rho,item:small-tag:async:step:committing}, and 
$$
\msf{val}_b(\tau) \coloneqq 
\begin{cases}
\msf{val}(\tau) & b = \top\\
\bot & b = \bot
\end{cases}.
$$
 \item{\emph{Experiment} $\Exp_{c,i}(\secpar,z)$:}
 This experiment works similarly to $\Exp(\secpar,z)$ except that it runs the extractor of {\bf ExtCom-$c$-$i$} to obtain the extracted message $m^*$ and outputs $(\OUT_{C^*},\Gamma_b(m^*))$ where 
 $$
\Gamma_b(m^*) \coloneqq 
\begin{cases}
m^* & b = \top\\
\bot & b = \bot
\end{cases}.
$$
\end{itemize}
Then, for all $(c,i)\in \extcomindex$, it holds that 
$$
\{\Exp(\secpar,z)\}_{\secpar\in \mathbb{N},z\in \bit^*}\sind \{\Exp_{c,i}(\secpar,z)\}_{\secpar\in \mathbb{N},z\in \bit^*}.
$$

\end{lemma}
\begin{proof}
First, we prove the following claim.
\begin{MyClaim}\label{claim:consistency_commitments}
\begin{align}\label{eq:consistency_commitments}
\Pr[b=\top \land (\msf{com},\{\tau^c_i\}_{(c,i)\in \extcomindex})\notin \Lang_{\beta}]=\negl(\secpar)
\end{align}
where the probability is taken over the execution of $\Exp(\secpar,z)$. 
\end{MyClaim}
\begin{proof}[Proof of \Cref{claim:consistency_commitments}]
 We consider a modified experiment $\Exp'(\secpar,z)$ that works similarly to  $\Exp(\secpar,z)$ except that it runs the witness-extended emulator for {\bf WIPoK-2} instead of running it honestly. By the PoK property (as per \Cref{def:WEE}), it suffices to prove \Cref{eq:consistency_commitments} holds in $\Exp'(\secpar,z)$. 

Let $\tilde{w}$ be the extracted witness from {\bf WIPoK-2}. By the PoK property (as per \Cref{def:WEE}), $\tilde{w}$ is a valid witness for $(\msf{com},\{\tau^c_i\}_{(c,i)\in \extcomindex}, Y^A, Y^B, V_0,V_1) \in \Lang_{\beta} \vee \Lang^t_f \vee \Lang^{n-t}_f \vee \Lang_{f}^{\mathrm{OR}}$ 
 except for a negligible probability when $b=\top$. 

We can show that $\tilde{w}$ cannot be a witness of $Y^A\in \Lang^t_f$ except for a negligible probability by using the one-wayness of $f$ and WI of {\bf WIPoK-1-A}. Indeed, this can be seen as follows. 
\begin{itemize}
    \item In $\Exp'(\secpar,z)$, the receiver uses $(1,x^A_1,r^A_1)$ as a witness for {\bf WIPoK-1-A}. Therefore, if the probability that $\tilde{w}=(j,x^A_j)$ such that $y^A_j=f(x^A_j)$ for some $j\neq 1$ is non-negligible, it clearly breaks the one-wayness of $f$ or the computational hiding property of $\Com$. 
    \item If the probability that $\tilde{w}=(1,x^A_1)$ such that $y^A_1=f(x^A_1)$ is non-negligible, by the WI property of {\bf WIPoK-1-A}, a similar statement holds even if the receiver uses $(2,x^A_2,r^A_2)$ instead of $(1,x^A_1,r^A_1)$  as a witness for {\bf WIPoK-1-A}. Then, by a similar argument to the above case, it also breaks the one-wayness of $f$ or the computational hiding property of $\Com$.  
\end{itemize}

By similar arguments, $\tilde{w}$ cannot be a witness of $Y^B\in \Lang^{n-t}_f$ or $(V_0,V_1)\in \Lang_{f}^{\mathrm{OR}}$ except for a negligible probability. 
Thus, $\tilde{w}$ is a witness for $(\msf{com},\{\tau^c_i\}_{(c,i)\in \extcomindex})\in \Lang_{\beta}$ except for a negligible probability when $b=\top$.  
This completes the proof of \Cref{claim:consistency_commitments}.

\end{proof}

\Cref{claim:consistency_commitments} in particular means that we have $\val(\tau)=\val(\tau^i_c)\neq \bot$ whenever $b=\top$ except for a negligible probability. 
Then, the extractability of $\ExtCom$ immediately implies \Cref{lem:ExtCom_soundness}.

\end{proof}

Next, we analyze the bad cases one by one.

\subsubsection{Analysis of $\Bad~1$}
\begin{lemma}\label{lem:bad:1}
The commitment scheme $\langle C, R\rangle^{\msf{async}}_{\msf{tg}}$ in \Cref{prot:small-tag:classical} is non-malleable against asynchronous PPT adversaries whose schedule satisfies $\Bad~1$.
\end{lemma}
\begin{proof}
The proof for this case is easy. When $\Bad~1$ occurs, the right commitment $\tilde{\com}$ (in \Cref{item:small-tag:async:step:committing}) is sent before the left commitment $\com$ (in \Cref{item:small-tag:async:step:committing}) is sent. Then, a straightforward reduction to the computational hiding of the left session works by extracting $\tilde{m}$ from $\tilde{\com}$ by brute-force and treating it as non-uniform advice of an adversary against computational hiding. 
%Specifically, the reduction algorithm first run the man-in-the-middle adversary until it sends $\tilde{\com}$. Then it  and 
%This means that the committed message in the right session cannot depend on the left message $m$, in which case 

\end{proof}

\subsubsection{Analysis of $\Bad~2$}
\begin{lemma}\label{lem:bad:2}
The commitment scheme $\langle C, R\rangle^{\msf{async}}_{\msf{tg}}$ in \Cref{prot:small-tag:classical} is non-malleable against asynchronous PPT adversaries whose schedule satisfies $\Bad~2$.
\end{lemma}
\begin{proof}
Recall that $\Bad~2$ means that 
    the left message of \Cref{item:small-tag:async:step:OWFs} is sent before the right message of \Cref{item:small-tag:async:step:committing} is sent. 
We assume that $\mcal{M}$ receives the right message of \Cref{item:small-tag:async:step:Naor-rho} at first w.l.o.g.
In particular, no massage in the right session is sent during the execution of the left {\bf WIPoK-Trap}.

We consider the following sequence of hybrids.
\begin{itemize}
    \item {Hybrid $H_1^{\mcal{M}}(\secpar, m, z)$:} This hybrid is the real man-in-the-middle experiment $\msf{mim}^\mcal{M}_{\langle C, R \rangle}(\secpar, m, z)$. The output of the experiment can be written as $(\OUT_{\mcal{M}},\val_b(\tilde{\tau}))$ where $\OUT_{\mcal{M}}$ is $\mcal{M}$'s final output, $b$ is the $R$'s decision bit, and $\tilde{\tau}$ is Naor's commitment generated in \Cref{item:small-tag:async:step:Naor-rho,item:small-tag:async:step:committing} in the right session, and
$$
\msf{val}_b(\tilde{\tau}) \coloneqq 
\begin{cases}
\msf{val}(\tilde{\tau}) & b = \top\\
\bot & b = \bot
\end{cases}.
$$
   \item {Hybrid $H_2^{\mcal{M}}(\secpar, m, z)$:}  It is identical to $H_1^{\mcal{M}}(\secpar, m, z)$ except that it runs the witness-extended emulator for the left {\bf WIPoK-Trap} instead of running it honestly. Let $v$ be the extracted witness. If 
   $C$ accepts the left {\bf WIPoK-Trap} but  
   $v$ is not a valid witness for $(V_0,V_1)\in \Lang_f^{\mathrm{OR}}$, it aborts with outputting $(\bot,\bot)$. Otherwise, it works similarly to $H_1^{\mcal{M}}(\secpar, m, z)$. We remark that it does not use $v$ though it extracts it. 
     \item {Hybrid $H_3^{\mcal{M}}(\secpar, m, z)$:}  It is identical to $H_2^{\mcal{M}}(\secpar, m, z)$ except that it uses $v$ extracted from the left {\bf WIPoK-Trap} as a witness in the left {\bf WIPoK-2}. 
      \item {Hybrid $H_4^{\mcal{M}}(\secpar, m, z)$:}  It is identical to $H_3^{\mcal{M}}(\secpar, m, z)$ except that it commits to $0$ instead of $m$ in all slots of {\bf ExtCom-\{1,2,3\}}.  
   %uses $z$ in the left ${\bf WIPoK-2}$.  
\end{itemize}
We prove that each pair of neighboring hybrids is computationally indistinguishable.  
\begin{MyClaim}\label{cla:bad:2:hybrid:1}
It holds that
\begin{align*}
\{H_1^{\mcal{M}}(\secpar, m, z)\}_{\secpar \in \Naturals,m \in \bits^{\ell(\secpar)},z \in \bits^*} \sind \{H_2^{\mcal{M}}(\secpar, m, z)\}_{\secpar \in \Naturals, m \in \bits^{\ell(\secpar)}, z \in \bits^*}.
\end{align*}
\end{MyClaim}
\begin{proof}[Proof of \Cref{cla:bad:2:hybrid:1}]
This follows from the PoK property (as per \Cref{def:WEE}) of the left {\bf WIPoK-Trap} straightforwardly. 

\end{proof}

\begin{MyClaim}\label{cla:bad:2:hybrid:2}
It holds that
\begin{align*}
\{H_2^{\mcal{M}}(\secpar, m, z)\}_{\secpar \in \Naturals, m \in \bits^{\ell(\secpar)}, z \in \bits^*} \cind \{H_3^{\mcal{M}}(\secpar, m, z)\}_{\secpar \in \Naturals, m \in \bits^{\ell(\secpar)}, z \in \bits^*}.
\end{align*}
\end{MyClaim}
\begin{proof}[Proof of \Cref{cla:bad:2:hybrid:2}]
We introduce the following additional hybrids:
\begin{itemize}
    \item {Hybrid $\bar{H}_2^{\mcal{M}}(\secpar, m, z)$:}
    Choose the smallest $i^*\in [n_5]$ such that the right {\bf ExtCom-1-$i^*$} does not interleave with the left {\bf WIPoK-2}, i.e., no message of the left {\bf WIPoK-2} is sent during the execution of the right {\bf ExtCom-1-$i^*$}.\footnote{\Cref{rem:adaptive_good_slot} applies here.} 
    Such $i^*$ exists by the pigeonhole principle since we assume that $n_5$ is larger than the round complexity of {\bf WIPoK-2}.  
    This experiment is identical to $H_2^{\mcal{M}}(\secpar, m, z)$ except that it runs the extractor of the right {\bf ExtCom-1-$i^*$} to extract $\tilde{m}^*$, and outputs $(\OUT_{\mcal{M}},\Gamma_b(\tilde{m}^*))$ instead of $(\OUT_{\mcal{M}},\val_b(\tilde{\tau}))$ where  
$$
\Gamma_b(m^*) \coloneqq 
\begin{cases}
m^* & b = \top\\
\bot & b = \bot
\end{cases}.
$$
\item {Hybrid $\bar{H}_3^{\mcal{M}}(\secpar, m, z)$:}  It is identical to $\bar{H}_2^{\mcal{M}}(\secpar, m, z)$ except that it uses $v$ extracted from the left {\bf WIPoK-Trap} as a witness in the left {\bf WIPoK-2}. 
\end{itemize}
Note that both $\bar{H}_2^{\mcal{M}}(\secpar, m, z)$ and $\bar{H}_3^{\mcal{M}}(\secpar, m, z)$ are efficient since they no longer need to output $\val_b(\tilde{\tau})$. 
Then, by the assumption that the right {\bf ExtCom-1-$i^*$} does not interleave with the left {\bf WIPoK-2}, the WI property of the left {\bf WIPoK-2} gives 
\begin{align}\label{eq:hybrid_barH2_to_barH3}
    \{\bar{H}_2^{\mcal{M}}(\secpar, m, z)\}_{\secpar \in \Naturals, m \in \bits^{\ell(\secpar)}, z \in \bits^*} \cind \{\bar{H}_3^{\mcal{M}}(\secpar, m, z)\}_{\secpar \in \Naturals, m \in \bits^{\ell(\secpar)}, z \in \bits^*}.
\end{align}

By the assumption that $\Bad~2$ occurs and that $\mcal{M}$ receives the right message of \Cref{item:small-tag:async:step:Naor-rho} at first,  hybrids $H_2^{\mcal{M}}(\secpar, m, z)$ and $H_3^{\mcal{M}}(\secpar, m, z)$ do not rewind the right session at all.
Indeed, they only rewind the left {\bf WIPoK-Trap}, but the right message of \Cref{item:small-tag:async:step:Naor-rho} is sent before the left {\bf WIPoK-Trap} starts by our simplifying assumption and the rest of messages in the right session is sent after the left {\bf WIPoK-Trap} finishes by the assumption that $\Bad~2$ occurs. 
Thus, by considering the combination of $\mcal{M}$ and $C$ as a single cheating committer,  \Cref{lem:ExtCom_soundness} gives
\begin{align}\label{eq:hybrid_H2_to_barH2}
\{H_2^{\mcal{M}}(\secpar, m, z)\}_{\secpar \in \Naturals, m \in \bits^{\ell(\secpar)}, z \in \bits^*} \cind \{\bar{H}_2^{\mcal{M}}(\secpar, m, z)\}_{\secpar \in \Naturals, m \in \bits^{\ell(\secpar)}, z \in \bits^*}
\end{align}
and 
\begin{align}\label{eq:hybrid_H3_to_barH3}
\{H_3^{\mcal{M}}(\secpar, m, z)\}_{\secpar \in \Naturals, m \in \bits^{\ell(\secpar)}, z \in \bits^*} \cind \{\bar{H}_3^{\mcal{M}}(\secpar, m, z)\}_{\secpar \in \Naturals, m \in \bits^{\ell(\secpar)}, z \in \bits^*}.
\end{align}
By combining \Cref{eq:hybrid_barH2_to_barH3,eq:hybrid_H2_to_barH2,eq:hybrid_H3_to_barH3}, we complete the proof of \Cref{cla:bad:2:hybrid:2}. 

\end{proof}

\begin{MyClaim}\label{cla:bad:2:hybrid:3}
It holds that
\begin{align*}
\{H_3^{\mcal{M}}(\secpar, m, z)\}_{\secpar \in \Naturals, m \in \bits^{\ell(\secpar)}, z \in \bits^*} \cind \{H_4^{\mcal{M}}(\secpar, m, z)\}_{\secpar \in \Naturals, m \in \bits^{\ell(\secpar)}, z \in \bits^*}.
\end{align*}
\end{MyClaim}
\begin{proof}[Proof of \Cref{cla:bad:2:hybrid:3}]
This follows from a similar argument to that in the proof of \Cref{cla:bad:2:hybrid:2}. 
However, we remark that we cannot prove it at once since the \emph{total} round complexity of {\bf ExtCom-\{1,2,3\}} is larger than $n_5$, in which case we cannot choose $i^*$ such that the right {\bf ExtCom-1-$i^*$} does not interleave with any message of the left {\bf ExtCom-\{1,2,3\}}. 
To resolve this issue, we replace the committed message $m$ with $0$ for each execution of $\ExtCom$ in {\bf ExtCom-\{1,2,3\}} one by one. Since we assume that $n_5$ is larger than the round complexity of $\ExtCom$, a similar argument to the proof of \Cref{cla:bad:2:hybrid:2} works by using the computational indistinguishability of $\ExtCom$.

\end{proof}

\begin{MyClaim}\label{cla:bad:2:hybrid:4}
It holds that
\begin{align*}
\{H_4^{\mcal{M}}(\secpar, m_0, z)\}_{\secpar,m_0,m_1,z} \cind \{H_4^{\mcal{M}}(\secpar, m_1, z)\}_{\secpar,m_0,m_1,z},
\end{align*}
 where both ensembles are indexed by $\secpar \in \Naturals$, $(m_0,m_1) \in \bits^{\ell(\secpar)}\times \bits^{\ell(\secpar)}$, and $z \in \bits^*$.
\end{MyClaim}
\begin{proof}[Proof of \Cref{cla:bad:2:hybrid:4}]
This immediately follows from the computational hiding property of Naor's commitment noting that the left message $m$ is only used for generating Naor's commitment in \Cref{item:small-tag:async:step:committing} in $H_4^{\mcal{M}}(\secpar, m, z)$.
 
\end{proof}

By combining \Cref{cla:bad:2:hybrid:1,cla:bad:2:hybrid:2,cla:bad:2:hybrid:3,cla:bad:2:hybrid:4}, we obtain 
\begin{align*}
\{H_1^{\mcal{M}}(\secpar, m_0, z)\}_{\secpar,m_0,m_1,z} \cind \{H_1^{\mcal{M}}(\secpar, m_1, z)\}_{\secpar,m_0,m_1,z}
\end{align*}
where both ensembles are indexed by $\secpar \in \Naturals$, $(m_0,m_1) \in \bits^{\ell(\secpar)}\times \bits^{\ell(\secpar)}$, and $z \in \bits^*$.
This completes the proof of \Cref{lem:bad:2}.

%\takashi{Extract the trapdoor. Extract $\tilde{m}$ from a slot of right {\bf ExtCom-1} that does not interleave with the left {\bf WIPoK-2}. Change the witness in the left {\bf WIPoK-2} to the trapdoor. We quit extraction of $\tilde{m}$. For each slot of {\bf ExtCom-\{1,2,3\}}, we replace the committed messages. There, we again go through a hybrid where we choose a good slot of {\bf ExtCom-1} from which we can safely extract $\tilde{m}$. For the proof, we will make use of \Cref{lem:ExtCom_soundness} many times.}
\end{proof}

\subsubsection{Analysis of $\Bad~3$}
\begin{lemma}\label{lem:bad:3}
The commitment scheme $\langle C, R\rangle^{\msf{async}}_{\msf{tg}}$ in \Cref{prot:small-tag:classical} is non-malleable against asynchronous PPT adversaries whose schedule satisfies $\Bad~3$.
\end{lemma}
\begin{proof}
Recall that $\Bad~3$ means that 
    the right {\bf WIPoK-1-A} starts before the left message of \Cref{item:small-tag:async:step:OWFs} is sent.

We consider the following sequence of hybrids.
\begin{itemize}
    \item {Hybrid $H_1^{\mcal{M}}(\secpar, m, z)$:} This hybrid is the real man-in-the-middle experiment $\msf{mim}^\mcal{M}_{\langle C, R \rangle}(\secpar, m, z)$. The output of the experiment can be written as $(\OUT_{\mcal{M}},\val_b(\tilde{\tilde{\tau}}))$ where $\OUT_{\mcal{M}}$ is $\mcal{M}$'s final output, $b$ is the $R$'s decision bit, and $\tilde{\tau}$ is Naor's commitment generated in \Cref{item:small-tag:async:step:Naor-rho,item:small-tag:async:step:committing} in the right session, and
$$
\msf{val}_b(\tilde{\tau}) \coloneqq 
\begin{cases}
\msf{val}(\tilde{\tau}) & b = \top\\
\bot & b = \bot
\end{cases}.
$$
\item {Hybrid $H_2^{\mcal{M}}(\secpar, m, z)$:} 
Choose the smallest $i^*\in [n_5]$ such that the right {\bf ExtCom-1-$i^*$} does not interleave with Steps \ref{item:small-tag:async:step:Naor-rho} to \ref{item:small-tag:async:step:OWFs} in the left, i.e., no message of the left {\bf WIPoK-2} is sent during the execution of the right {\bf ExtCom-1-$i^*$}.\footnote{\Cref{rem:adaptive_good_slot} applies here.} 
Such $i^*$ exists by the pigeonhole principle since we assume that $n_5$ is larger than the total round complexity of Steps \ref{item:small-tag:async:step:Naor-rho} to \ref{item:small-tag:async:step:OWFs}.   
This experiment is identical to $H_1^{\mcal{M}}(\secpar, m, z)$ except that it runs the extractor of the right {\bf ExtCom-1-$i^*$} to extract $\tilde{m}^*$, and outputs $(\OUT_{\mcal{M}},\Gamma_b(\tilde{m}^*))$ instead of $(\OUT_{\mcal{M}},\val_b(\tilde{\tau}))$ where  
$$
\Gamma_b(m^*) \coloneqq 
\begin{cases}
m^* & b = \top\\
\bot & b = \bot
\end{cases}.
$$
\end{itemize}

\begin{MyClaim}\label{cla:bad:3:hybrid:1}
It holds that
\begin{align*}
\{H_1^{\mcal{M}}(\secpar, m, z)\}_{\secpar \in \Naturals, m \in \bits^{\ell(\secpar)}, z \in \bits^*} \sind \{H_2^{\mcal{M}}(\secpar, m, z)\}_{\secpar \in \Naturals, m \in \bits^{\ell(\secpar)}, z \in \bits^*}.
\end{align*}
\end{MyClaim}
\begin{proof}[Proof of \Cref{cla:bad:3:hybrid:1}]
This immediately follows from \Cref{lem:ExtCom_soundness}.

\end{proof}

\begin{MyClaim}\label{cla:bad:3:hybrid:2}
It holds that
\begin{align*}
\{H_2^{\mcal{M}}(\secpar, m_0, z)\}_{\secpar,m_0,m_1,z} \sind \{H_2^{\mcal{M}}(\secpar, m_1, z)\}_{\secpar,m_0,m_1,z}
\end{align*}
 where both ensembles are indexed by $\secpar \in \Naturals$, $(m_0, m_1) \in \bits^{\ell(\secpar)} \times \bits^{\ell(\secpar)}$, and $z \in \bits^*$.
\end{MyClaim}
\begin{proof}
By the choice of $i^*$ and the assumption that $\Bad~3$ occurs, $H_2^{\mcal{M}}(\secpar, m, z)$ does not rewind the left session at all. Therefore, by considering the combination of $\mcal{M}$ and $R$ as a single cheating receiver,  \Cref{cla:bad:3:hybrid:2} follows from the computational hiding property of \Cref{prot:small-tag:classical}. 

\end{proof}
By combining \Cref{cla:bad:3:hybrid:1,cla:bad:3:hybrid:2}, we obtain  
\begin{align*}
\{H_1^{\mcal{M}}(\secpar, m_0, z)\}_{\secpar,m_0,m_1,z} \cind \{H_1^{\mcal{M}}(\secpar, m_1, z)\}_{\secpar,m_0,m_1,z}
\end{align*}
 where both ensembles are indexed by $\secpar \in \Naturals$, $m_0,m_1 \in \bits^{\ell(\secpar)}$, and $z \in \bits^*$.
This completes the proof of \Cref{lem:bad:3}.
%When $\Bad~3$ occurs, by the choice of $n_5$ and the pigeonhole principle, there is at least one slot of $\{\ExtCom-1\}$ that does not interleave with Steps \ref{item:small-tag:async:step:Naor-rho} to \ref{item:small-tag:async:step:OWFs} in the left.We consider an experiment where we extract the right message $\tilde{m}$ by running the extractor for that slot of ${ExtCom-1}$. By  \Cref{lem:ExtCom_soundness}, this is the correct committed value in the right if $R$ accepts in the right session except for a negligible probability. Moreover, by our choice of the slot, the experiment does not rewind $C$ in the left session. Thus, it is straightforward to reduce non-malleability to the computational hiding of the left session.\takashi{need to polish.} 

\end{proof}

\subsubsection{Analysis of $\Bad~4$}
\begin{lemma}\label{lem:bad:4}
The commitment scheme $\langle C, R\rangle^{\msf{async}}_{\msf{tg}}$ in \Cref{prot:small-tag:classical} is non-malleable against asynchronous PPT adversaries whose schedule satisfies $\Bad~4$.
\end{lemma}
\begin{proof}
Recall that $\Bad~4$ means that 
    the right {\bf WIPoK-1-B} starts before the left {\bf WIPoK-1-A} finishes.   
\Cref{lem:bad:4} can be proven similarly to \Cref{lem:bad:3} except that we extract $\tilde{m}$ from the right {\bf ExtCom-2-$i^*$} that does not interleave with Steps \ref{item:small-tag:async:step:Naor-rho} to \ref{item:small-tag:async:step:WIPoK:1A} in the left.

\end{proof}

\subsubsection{Analysis of $\Bad~5$}
\begin{lemma}\label{lem:bad:5}
The commitment scheme $\langle C, R\rangle^{\msf{async}}_{\msf{tg}}$ in \Cref{prot:small-tag:classical} is non-malleable against asynchronous PPT adversaries whose schedule satisfies $\Bad~5$.
\end{lemma}
\begin{proof}
 Recall that $\Bad~5$ means that 
    the right {\bf WIPoK-2} starts before the left {\bf WIPoK-1-B} finishes. 
\Cref{lem:bad:5} can be proven similarly to \Cref{lem:bad:3} except that we extract $\tilde{m}$ from the right {\bf ExtCom-3-$i^*$} that does not interleave with Steps \ref{item:small-tag:async:step:Naor-rho} to \ref{item:small-tag:async:step:WIPoK:2} in the left.

\end{proof}

By combining \Cref{lem:good_case,lem:bad:1,lem:bad:2,lem:bad:3,lem:bad:4,lem:bad:5} along with \Cref{lem:adaptive-predetermined:async}, we complete the proof of \Cref{thm:nm:small-tag:classical}.

%!TEX root = ../main.tex
\section{Extract-and-Simulate Lemma}\label{sec:extract_and_simulate}

In this section, we show a generalized version of the extract-and-simulate lemma given in \cite[Lemma 4]{cryptoeprint:2021:1516}.  
%\takashi{I made the logic of the lemma clearer.
%The high-level structure is:\\
%``$\forall \A$, 
%if $\exists \mcal{K}$ s.t. ($\forall \rho$ ...), then $\exists \SimExt$ s.t. ($\forall \rho$ ...)."\\ 
%The previous version was a litle bit ambiguous and could be understood as\\ 
%``$\forall \A(\rho)$, if $\exists \mcal{K}$ s.t. (...), then $\exists \SimExt$ s.t. (...)."\\  
%In the latter version, there's a subtlety about dependence on non-uniform advice. 
%Moreover, I don't think the latter version can be proven. 
%}
\takashi{modified the statement. Especially, $\A$ does not appear in the lemma anymore. When using this lemma, we should think of the description of $(\mcal{M}_\secpar,\rho_\secpar)$ as part of $\rho$.}
\takashi{I added $1^{\gamma^{-1}}$ in the input of $\mcal{G}$. This is needed to deal with the proof of our main step. ($\mcal{G}$ corresponds to the simulated main-thread $\mcal{G}_1$) that internally run the extractor for WIAoK, so it needs to know $\epsilon$ to determine the simulation error.}
\begin{lemma}[Extract-and-Simulate Lemma]
\label{lem:extract_and_simulate}
Let $\mcal{G}$ be a QPT algorithm that takes the security parameter $1^\secpar$, an error parameter $1^{\gamma^{-1}}$, a quantum state $\rho$, {and a classical string $z$} as input,  and outputs  $b\in \{\top,\bot\}$ and a quantum state $\rho_\out$. 

Suppose that there exists a QPT algorithm $\mcal{K}$ (referred to as the simulation-less extractor) that takes as input the security parameter $1^\secpar$, an error parameter $1^{\gamma^{-1}}$, a quantum state $\rho$, {and a classical string $z$}, and outputs $s\in \bit^{\poly(\secpar)}\cup \{\bot\}$   
satisfying the following w.r.t.\ some sequence of classical strings {$\{s^*_{z}\}_{z\in \bit^*}$.}
\takashi{The index of $s^*$ is changed from $\secpar$ to $z$. (Imagine that $s^*$ is the committed message and $z$ is the partial transcript. Then this should be a more natural formalization.)}
%\footnote{Strictly speaking, we consider a sequence $\{s^*_\secpar\}_{\secpar\in \mathbb{N}}$. We simply denote by $s^*$ to refer to $s^*_\secpar$. \label{footnote:sequence_s}}   
\begin{enumerate}
 \item  \label{item:s_star_or_bot}
    For any $\secpar$, $\gamma$,  $\rho_\secpar$, {and $z_\secpar$},  
    the output $s$ of $\mcal{K}(1^\secpar,1^{\gamma^{-1}},\rho_\secpar,{z_\secpar})$ is equal to {$s^*_{z_\secpar}$} whenever $s\neq \bot$.
    \item \label{item:gamma_delta}
 For any noticeable function  $\gamma(\secpar)$, there exists a noticeable function $\delta(\secpar)$ that is efficiently computable from $\gamma(\secpar)$ and satisfies the following. For any sequence $\{\rho_\secpar,{z_\secpar}\}_{\secpar\in\mathbb{N}}$ of polynomial-size quantum states and classical strings, %\footnote{Similarly to \Cref{footnote:sequence_s}, we consider a sequence $\{\rho_\secpar\}_{\secpar\in \mathbb{N}}$ and denote by $\rho$ to mean $\rho_\secpar$.}    
 if 
$$
\Pr[b=\top : (b,\rho_\out) \leftarrow \mcal{G}(1^\secpar,1^{\gamma^{-1}},\rho_\secpar,{z_\secpar})]\geq  \gamma(\secpar), 
$$  
then 
$$
\Pr[\mcal{K}(1^\secpar,1^{\gamma^{-1}},\rho_\secpar,{z_\secpar})=s^*_{{z_\secpar}}]\geq   \delta(\secpar).
$$
\end{enumerate}
Then, there exists a QPT algorithm $\SimExt$ such that for any noticeable function $\epsilon=\epsilon(\secpar)$, there exists a noticeable function $\gamma=\gamma(\secpar)\le \epsilon(\secpar)$ that is efficiently computable from $\epsilon$ and satisfies the following:
For any sequence $\{\rho_\secpar,{z_\secpar}\}_{\secpar\in\mathbb{N}}$ of polynomial-size quantum states and classical strings,  
$$
\{\SimExt(1^\secpar,1^{\epsilon^{-1}},\rho_\secpar,{z_\secpar})\}_{\secpar \in \Naturals}
\statind_{\epsilon} 
\{(\rho_\out,\Gamma_b(s^*_{{z_\secpar}})):(b,\rho_\out)\leftarrow \mcal{G}(1^\secpar,1^{\gamma^{-1}},\rho_\secpar,{z_{\secpar}})\}_{\secpar \in \Naturals},
$$
where $\Gamma_b(s^*_{{z_\secpar}})\defeq 
\begin{cases}
s^*_{{z_\secpar}} & \text{if}~ b=\top \\
\bot & \text{otherwise}
\end{cases}
$.
%Moreover, we have $\gamma=O((\epsilon/\secpar)^c)$ for some constant $c>1$. 
\if0
Then, there exists a QPT algorithm $\SimExt$ and an efficiently computable polynomial $\poly$ such that for any sequence $\{\rho_\secpar,{z_\secpar}\}_{\secpar\in\mathbb{N}}$ of polynomial-size quantum states and classical strings and
noticeable function $\epsilon=\epsilon(\secpar)$,  
$$
\{\SimExt(1^\secpar,1^{\epsilon^{-1}},\rho_\secpar,{z_\secpar})\}_{\secpar \in \Naturals}
\statind_{\epsilon} 
\{(\rho_\out,\Gamma_b(s^*_{{z_\secpar}})):(b,\rho_\out)\leftarrow \mcal{G}(1^\secpar,1^{\gamma^{-1}},\rho_\secpar,{z_{\secpar}})\}_{\secpar \in \Naturals},
$$
where $\gamma=\poly(\epsilon)$ and $
\Gamma_b(s^*_{{z_\secpar}})\defeq 
\begin{cases}
s^*_{{z_\secpar}} & \text{if}~ b=\top \\
\bot & \text{otherwise}
\end{cases}
$.
\fi
\end{lemma}
In \cite{cryptoeprint:2021:1516}, the authors essentially proved a special case of the above lemma where the simulation-less extractor is fixed to Unruh's rewinding extractor~\cite{EC:Unruh12}. 
We observe that essentially the same proof works for general simulation-less extractors.\footnote{Indeed, this observation is also mentioned in the technical overview of \cite{cryptoeprint:2021:1516}.}  
We provide the full proof of \Cref{lem:extract_and_simulate} in \Cref{sec:proof_extract_and_simulate} for completeness.

\section{Post-Quantum Non-Malleable Commitments}\label{sec:full:pq:nmcom}

In this section, we construct a full-fledged post-quantum non-malleable commitment scheme that supports an exponential-size tag space and is secure against QPT asynchronous MIM adversaries.

\subsection{Small-Tag, One-sided, Synchronous, Post-Quantum Setting}\label{summary:pq:sec:small-tag-one-sided-sync-pq}
First, we show that the \Cref{prot:one-sided:classical} given in \Cref{sec:small-tag-one-sided-sync-classical} is post-quantumly secure if we instantiate it with post-quantum building-blocks. Specifically, let $\langle C, R\rangle^{\msf{OneSided}}_{\msf{tg,PQ}}$ be \Cref{prot:one-sided:classical} instantiated with post-quantum Naor's commitment $\Com$ (which is constructed based on post-quantum OWFs), post-quantum injective OWF $f$, and  a post-quantum witness-distinguishable {\emph arguments} of knowledge with $\epsilon$-close emulation as per \Cref{def:PQWIAoK} (see \Cref{pq:prot:one-sided:classical} in \Cref{pq:sec:small-tag-one-sided-sync-pq} for the full description of $\langle C, R\rangle^{\msf{OneSided}}_{\msf{tg,PQ}}$). 
Then, we can prove the following theorem:
\begin{theorem}\label{summary:pq:thm:one-sided:non-malleability}
The commitment scheme $\langle C, R\rangle^{\msf{OneSided}}_{\msf{tg,PQ}}$ is non-malleable against  one-sided synchronous QPT adversaries with tag space $[n]$, with $n$ being any polynomial on $\secpar$. 
\end{theorem}	

We remark that the injectivity assumption for $f$ can be removed in exactly the same way as in the classical setting (as explained in \Cref{sec:removing-injectivity}). 

The proof of \Cref{summary:pq:thm:one-sided:non-malleability} is very similar to that of its classical counterpart (\Cref{thm:one-sided:non-malleability}) except for one step as explained below. Thus, we only give a proof sketch below, and defer the full proof to \Cref{pq:sec:small-tag-one-sided-sync-pq}.

\begin{proof}[Proof of \Cref{summary:pq:thm:one-sided:non-malleability} (sketch)]
We observe that all steps of the proof of \Cref{thm:one-sided:non-malleability}, except for the proof of \Cref{lem:small-tag:proof:se} (in particular, \Cref{lem:small-tag:proof:se:proof:K} in \Cref{lem:small-tag:proof:se:proof}), can be translated into the post-quantum setting easily with the following remarks. The only difference in those steps (except for some superficial differences like PPT vs QPT) is that we only have WIAoK with $\epsilon$-close emulation (as per \Cref{def:PQWIAoK}) instead of WIPoK (with negligibly-close emulation). First, the AoK property instead of the PoK property suffices because we run witness-extended emulators only for (possibly non-uniform) efficient cheating provers. Second, the $\epsilon$-close emulation (in contrast to negligibly-close emulation) also suffices, because we can take a sufficiently small noticeable error parameter such that it is much smaller than the MIM adversary's advantage. With these modifications, the whole proof will go through.  

Thus, the only remaining issue is how to prove the post-quantum version of \Cref{lem:small-tag:proof:se} (using a post-quantum version of \Cref{lem:small-tag:proof:se:proof:K}). Roughly, we have to construct a simulation-extractor $\mcal{SE}$ that extracts the right committed message $\val(\tilde{\tau})$ while simulating $\mcal{M}$'s final state from a simulation-less extractor $\mcal{K}$, which extracts $\val(\tilde{\tau})$ with a noticeable probability {\em without simulating $\mcal{M}$'s final state}. This task is exactly what can be achieved by our new extract-and-simulate lemma~(\Cref{lem:extract_and_simulate}). Indeed, the requirements for $\mcal{K}$ in \Cref{lem:small-tag:proof:se:proof:K} (more precisely, its post-quantum version~\Cref{pq:lem:small-tag:proof:se:proof:K}) exactly correspond to the assumptions of \Cref{lem:extract_and_simulate}. Therefore, this step can be completed by reducing it to \Cref{lem:extract_and_simulate}. (We believe that readers with good understanding of the classical case can directly go to {\em Proof of \Cref{pq:lem:small-tag:proof:se}} in  \Cpageref{pageref:different_part} after checking the statements of \Cref{pq:lem:small-tag:proof:se:proof:K,pq:lem:small-tag:proof:se} and relevant definitions  
to see how this step exactly works.) 

\end{proof}

\subsection{Small-Tag, Asynchronous, Post-Quantum Setting}
\label{sec:full:pq:nmcom:async}
Next, we explain how to make $\langle C, R\rangle^{\msf{OneSided}}_{\msf{tg,PQ}}$ (shown in \Cref{pq:prot:one-sided:classical}) secure against two-sided asynchronous adversaries. We emphasize that this subsection still focuses on the small-tag setting.  
This step is essentially the same as its classical counterpart in \Cref{sec:small-tag:classical:sync,sec:small-tag:async:classical}
except for that we rely on post-quantum building blocks.

We present our construction in \Cref{prot:small-tag:pq}.
%\begin{itemize}
%    \item $n_4$ is...
%    \item For $i=5,...$ $n_i=$...
%\end{itemize}
This protocol makes use of the following building blocks.
 \begin{itemize}
 \item
A post-quantum OWF $f$; 
\item
Naor's commitment $\Com$ that is implemented with a post-quantum OWF;
\item
A post-quantum witness-indistinguishable argument of knowledge with $\epsilon$-close emulation $\WIAoK$ 
(as per \Cref{def:PQWIAoK}); 
\item A post-quantum $\epsilon$-simulatable extractable commitment $\ExtCom$ (as per \Cref{def:epsilon-sim-ext-com:pq}).
We denote by $\Verify_{\ExtCom}$ the verification algorithm of $\ExtCom$ in the decommit stage.  
\end{itemize} 

Similarly to \Cref{sec:small-tag:async:classical}, we recursively define $n_5,...,n_9$ so that 
$n_i$ is greater than the total round complexity of Steps $1$ to $i-1$. We also require that $n_5$ is greater than the round complexities of $\WIAoK$ and $\ExtCom$. It is easy to see that we can set $n_i$ to a constant if both $\WIPoK$ and $\ExtCom$ have constant rounds. In particular, \Cref{prot:small-tag:pq} runs in constant rounds.

\begin{ProtocolBox}[label={prot:small-tag:pq}]{Post-Quantum Small-Tag Asynchronous NMCom \textnormal{$\langle C, R\rangle^{\msf{async}}_{\msf{tg,PQ}}$}}
The tag space is defined to be $[n]$ where $n$ is a polynomial on $\secpar$. Let $t \in [n]$ be the tag for the following interaction. Let $m$ be the message to be committed to. 

\para{Commit Stage:}
\begin{enumerate}
\item\label[Step]{item:small-tag:async:step:Naor-rho:pq}
Receiver $R$ samples and sends the first message $\beta$ for Naor's commitment.
\item\label[Step]{item:small-tag:async:step:committing:pq}
Committer $C$ commits to $m$ using the second message of Naor's commitment. $C$ also sends a $\beta'$ that will be used as the first Naor's message for $R$ to generate a commitment in next step. Formally, $C$ samples $r$ and $\beta'$ and sends the tuple $\big(\msf{com} = \Com_\beta(m;r),~\beta'\big)$. 

\item\label[Step]{item:small-tag:async:step:W-td:pq}  $R$ and $C$ do the following:
\begin{enumerate}
\item {\bf (TrapGen.)}\label[Step]{item:small-tag:async:step:W-td:Z:pq}
$R$ computes $V_0=f(v_0)$ and $V_1=f(v_1)$ with $v_0,v_1\pick \bit^\secpar$ and sends $(V_0,V_1)$ to $C$ 
\item \label[Step]{item:small-tag:async:step:W-td:WIPoK:pq} {\bf (WIAoK-Trap.)}
$R$ and $C$ execute an instance of $\WIAoK$ where $R$ proves to $C$ that he ``knows'' a pre-image of $V_0$ or $V_1$.
Formally, $R$ proves that $(V_0,V_1) \in \Lang_{f}^{\mathrm{OR}}$, where
\begin{equation}\label[Language]{eq:small-tag:async:Lang:TD:pq}
\Lang_{f}^{\mathrm{OR}} \coloneqq \big\{(V_0,V_1)~\big|~\exists v ~s.t.~~f(v)=V_0~\vee~f(v)=V_1
\big\}.
\end{equation}
Note that $R$ uses $v_0$ as the witness when executing this $\WIAoK$.
\end{enumerate}
%\subpara{Comment:}This step is inserted for \Cref{item:trick_trapdoor} in \Cref{sec:small-tag:pq:async:high-level} where a witness for $(V_0,V_1) \in \Lang_{f}^{\mathrm{OR}}$ playes the role of a ``trapdoor". 

\item\label[Step]{item:small-tag:async:step:OWFs:pq}
$R$ performs the following computation:
\begin{enumerate}
\item \label[Step]{item:small-tag:async:step:OWFs:A:pq}
$R$ computes $\Set{y^A_i = f(x^A_i)}_{i \in [t]}$ with $x^A_i \pick \bits^\secpar$ for each $i \in [t]$, and sets $Y^A \coloneqq (y^A_1, \ldots, y^A_t)$. $R$ also computes $\Set{\msf{com}^A_i = \Com_{\beta'}(x^A_i;r^A_i)}_{i\in [t]}$, where $r^A_i$ is a random string for each $i\in [t]$.  
\item \label[Step]{item:small-tag:async:step:OWFs:B:pq}
$R$ computes $\Set{y^B_i = f(x^B_i)}_{i \in [n-t]}$ with $x^B_i \pick \bits^\secpar$ for each $i \in [n-t]$, and sets $Y^B \coloneqq (y^B_1, \ldots, y^B_{n-t})$. $R$ also computes $\Set{\msf{com}^B_i = \Com_{\beta'}(x^B_i;r^B_i)}_{i\in [n-t]}$, where $r^B_i$ is a random string for each $i\in [n-t]$.  
\end{enumerate}
$R$ sends the tuple $\big(Y^A,~\Set{\msf{com}^A_i = \Com_{\beta'}(x^A_i;r^A_i)}_{i\in [t]},~Y^B,~\Set{\msf{com}^B_i = \Com_{\beta'}(x^B_i;r^B_i)}_{i\in [n-t]}\big)$.

\item\label[Step]{item:small-tag:async:step:ExtCom:1:pq}
{\bf (ExtCom-1 $\times~n_5$.)}
$C$ and $R$ sequentially execute $n_5$ instances of $\ExtCom$ where $C$ commits to $m$. Let $\tau^1_i$ and $\decom^1_i$ be the transcript and decommitment (privately obtained by $C$) of the $i$-th execution. 

%\subpara{Comment:}This step is inserted for \Cref{item:trick_extcom}and repeated for \Cref{item:trick_repeat}  in \Cref{sec:small-tag:pq:async:high-level}.

\item\label[Step]{item:small-tag:async:step:WIPoK:1A:pq}
{\bf (WIAoK-1-A $\times~n_6$.)} $R$ and $C$ sequentially execute $n_6$ instances of $\WIAoK$ where $R$ proves to $C$ that he ``knows'' a pre-image of some $y^A_i$ contained in $Y^A$, and $y^A_i$ are consistent with $\msf{com}^A_i$ (as defined in \Cref{item:small-tag:async:step:OWFs:A:pq}). Formally, $R$ proves that $(Y^A, \Set{\msf{com}^A_i}_{i\in[t]}) \in \Lang^t_{f,\beta'}$, where
\begin{equation}\label[Language]{eq:small-tag:async:Lang:OWF:A:pq}
\Lang^t_{f,\beta'} \coloneqq \Bigg\{(y^A_1, \ldots, y^A_t), (\msf{com}^A_1, \ldots, \msf{com}^A_t)~\Bigg|~\exists (i, x^A_i, r^A_i) ~s.t.~~
\begin{array}{l}
i \in [t] ~\wedge \\
y^A_i = f(x^A_i) ~\wedge \\
\msf{com}^A_i = \Com_{\beta'}(x^A_i; r^A_i)
\end{array}\Bigg\}.
\end{equation}
Note that $R$ uses $(1, x^A_1, r^A_1)$ as the witness when executing this $\WIAoK$.

%\subpara{Comment:}This step is repeated for \Cref{item:trick_repeat}  in \Cref{sec:small-tag:pq:async:high-level}.

\item\label[Step]{item:small-tag:async:step:ExtCom:2:pq}
{\bf (ExtCom-2 $\times~n_7$.)}
$C$ and $R$ sequentially execute $n_7$ instances of $\ExtCom$ where $C$ commits to $m$. Let $\tau^2_i$ and $\decom^2_i$ be the transcript and decommitment (privately obtained by $C$) of the $i$-th execution. 

%\subpara{Comment:}This step is inserted for \Cref{item:trick_extcom} and repeated for \Cref{item:trick_repeat}  in \Cref{sec:small-tag:pq:async:high-level}.

\item\label[Step]{item:small-tag:async:step:WIPoK:1B:pq}
{\bf (WIAoK-1-B $\times~n_8$)} $R$ and $C$ sequentially execute $n_8$ instances of $\WIAoK$ where $R$ proves to $C$ that he ``knows'' a pre-image of some $y^B_i$ contained in $Y^B$, and $y^B_i$ are consistent with $\msf{com}^B_i$ (as defined in \Cref{item:small-tag:async:step:OWFs:B:pq}). Formally, $R$ proves that $(Y^B, \Set{\msf{com}^B_i}_{i\in[n-t]}) \in \Lang^{n-t}_{f,\beta'}$, where
\begin{equation}\label[Language]{eq:small-tag:async:Lang:OWF:B:pq}
\Lang^{n-t}_{f,\beta'} \coloneqq \Bigg\{(y^B_1, \ldots, y^B_{n-t}), (\msf{com}^B_1, \ldots, \msf{com}^B_{n-t})~\Bigg|~\exists (i, x^B_i, r^B_i) ~s.t.~~ 
\begin{array}{l}
i \in [n-t] ~\wedge \\
y^B_i = f(x^B_i) ~\wedge \\
\msf{com}^B_i = \Com_{\beta'}(x^B_i; r^B_i)
\end{array}\Bigg\}.
\end{equation}
Note that $R$ uses $(1, x^B_1, r^B_1)$ as the witness when executing this $\WIPoK$.

%\subpara{Comment:}This step is repeated for \Cref{item:trick_repeat}  in \Cref{sec:small-tag:pq:async:high-level}.

\item\label[Step]{item:small-tag:async:step:ExtCom:3:pq}
{\bf (ExtCom-3 $\times~n_9$.)}
$C$ and $R$ sequentially execute $n_9$ instances of $\ExtCom$ where $C$ commits to $m$. Let $\tau^2_i$ and $\decom^2_i$ be the transcript and decommitment (privately obtained by $C$) of the $i$-th execution.

%\subpara{Comment:}This step is inserted for \Cref{item:trick_extcom} and repeated for \Cref{item:trick_repeat}  in \Cref{sec:small-tag:pq:async:high-level}.

\item\label[Step]{item:small-tag:async:step:WIPoK:2:pq}
{\bf (WIAoK-2.)} $C$ and $R$ execute an instance of $\WIAoK$ where $C$ proves to $R$ that he ``knows'' the (same) message committed in $\msf{com}$ (defined in \Cref{item:small-tag:async:step:committing:pq}) and all commitments $\tau^c_i$ of $\ExtCom$ (in \Cref{item:small-tag:async:step:ExtCom:1:pq,item:small-tag:async:step:ExtCom:2:pq,item:small-tag:async:step:ExtCom:3:pq}), {\em or} a pre-image of some $y^A_i$ contained in $Y^A$ (defined in \Cref{item:small-tag:async:step:OWFs:A:pq}), {\em or} a pre-image of some $y^B_i$ contained in $Y^B$ (defined in \Cref{item:small-tag:async:step:OWFs:B:pq})
{\em or} a preimage of either of $V_0$ or $V_1$ (defined in \Cref{item:small-tag:async:step:W-td:Z:pq}). Formally, $C$ proves that $(\msf{com},\{\tau^c_i\}_{(c,i)\in \extcomindex}, Y^A, Y^B, V_0,V_1) \in \Lang_{\beta} \vee \Lang^t_f \vee \Lang^{n-t}_f \vee \Lang_{f}^{\mathrm{OR}}$,
where $\extcomindex=(\{1\}\times [n_5])\cup (\{2\}\times [n_7])\cup (\{3\}\times [n_9])$, 
\begin{align}
& \Lang_{\beta} \coloneqq \Bigg\{(\msf{com},\{\tau^c_i\}_{(c,i)\in \extcomindex}) ~\Bigg|
\begin{array}{l}
~ \exists (m, r,\{\decom^c_i\}_{(c,i)\in \extcomindex})~s.t.~\\ 
\begin{array}{l}
\msf{com} = \Com_\beta(m;r)~\wedge\\
\forall~(c,i)\in \extcomindex,~ \Verify_{\ExtCom}(m,\tau^c_i,\decom^c_i)=\top\\
\end{array}
\end{array}
\Bigg\}
\label[Language]{eq:small-tag:async:Lang:Com:pq}\\
& \Lang^t_f \coloneqq \Set{(y^A_1, \ldots, y^A_t) ~|~ \exists (i, x^A_i)~s.t.~ i \in [t] \wedge y^A_i = f(x^A_i)}, \\
& \Lang^{n-t}_f \coloneqq \Set{(y^B_1, \ldots, y^B_{n-t}) ~|~ \exists (i, x^B_i)~s.t.~ i \in [n-t] \wedge y^B_i = f(x^B_i)},
\end{align}
and $\Lang_{f}^{\mathrm{OR}}$ is defined in \Cref{eq:small-tag:async:Lang:TD:pq}. 
Note that $C$ uses the $(m, r, \{\decom^c_i\}_{(c,i)\in\extcomindex})$ defined in \Cref{item:small-tag:async:step:committing:pq,item:small-tag:async:step:ExtCom:1:pq,item:small-tag:async:step:ExtCom:2:pq,item:small-tag:async:step:ExtCom:3:pq} as the witness when executing this $\WIAoK$.

%\subpara{Comment:} Compared with \Cref{item:small-tag:sync:step:WIPoK:2} of \Cref{prot:small-tag:sync:classical}, we add $\Lang_{f}^{\mathrm{OR}}$ as an additional OR part to the language to be proven for \Cref{item:trick_trapdoor} in \Cref{sec:small-tag:pq:async:high-level}. We also modify $\Lang_{\beta}$ to prove that all commitments $\tau^c_i$ of $\ExtCom$ commit to the same message as $\com$ for \Cref{item:trick_extcom} in \Cref{sec:small-tag:pq:async:high-level}. 
\end{enumerate}
\para{Decommit Stage:} 
$C$ sends $(m, r)$. $R$ accepts if $\msf{com} = \Com_\beta(m;r)$, and rejects otherwise.
\end{ProtocolBox}

We remark that \Cref{prot:small-tag:pq} is identical to \Cref{prot:small-tag:classical} except that we require post-quantum security for building blocks and use WIAoK instead of WIPoK. 

\para{Security.}
Completeness is straightforward. Statistical binding follows from that of Naor's commitment. 

For computational hiding and non-malleability, we observe that the proofs against classical adversaries in \Cref{sec:small-tag:async:classical} is already quantum-friendly, and they can be translated into proofs against quantum adversaries in a straightforward manner. The only difference is that we now have $\epsilon$-close simulation for $\WIAoK$ and $\ExtCom$ though the classical counterparts have negligible simulation errors. Thus, whenever we invoke extractors of $\WIPoK$ or $\ExtCom$ in the original proof in the classical setting, a noticeable error occurs in the post-quantum setting. However, this is not a problem since the error can be an \emph{arbitrarily small} noticeable function. Thus, by repeating the same proofs as those in the classical case, we can show that QPT MIM adversary's advantage is smaller than \emph{any} noticeable function in the security parameter. This means that the advantage is negligible, which is nothing but non-malleability in the standard sense. 
Thus, we obtain the following theorems. 

\begin{theorem}\label{thm:small-tag:async:classical:hiding:pq}
The commitment scheme $\langle C, R\rangle^{\msf{async}}_{\msf{tg,PQ}}$ in \Cref{prot:small-tag:pq} is computationally hiding against QPT adversaries.
\end{theorem}
\begin{theorem}\label{thm:nm:small-tag:pq}
The commitment scheme $\langle C, R\rangle^{\msf{async}}_{\msf{tg,PQ}}$ in \Cref{prot:small-tag:pq} is non-malleable against asynchronous QPT adversaries with tag space $[n]$, with $n$ being any polynomial on $\secpar$.
\end{theorem}

\subsection{Tag Amplification}
\label{sec:full:pq:nmcom:tag-amp}
Finally, we explain how to amplify the tag space of \Cref{prot:small-tag:pq}. For this step, we rely on the following variant of a theorem shown in \cite{BLS21}. 

\begin{theorem}[{\cite[Section 6]{BLS21}}]\label{thm:tag-amp}
Assume the existence of  
\begin{itemize}
\item
a post-quantum, $k_1$-round, $\epsilon$-simulatable extractable commitment (as per \Cref{def:epsilon-sim-ext-com:pq}), and
\item
a post-quantum, $k_2$-round, witness-indistinguishable argument. 
\end{itemize} 
Let $\langle C, R \rangle_{\msf{tg,PQ}}^{\msf{async}}$ be a post-quantumly non-malleable commitment scheme, where the subscript $\msf{tg}$ denotes the {\em length} of the maximum tag it can support. Assume  $\langle C, R \rangle_{\msf{tg,PQ}}^{\msf{async}}$ satisfies the following conditions: 
\begin{enumerate}
\item
$\langle C, R \rangle_{\msf{tg,PQ}}^{\msf{async}}$  supports $\msf{tg} \in \Set{3, 4, \ldots, O(\log \secpar)}$ bit tags, and  
\item \label[Condition]{item:BLS:tag-amp:condition:2}
$\langle C, R \rangle_{\msf{tg,PQ}}^{\msf{async}}$ is also a post-quantum $(k_1+k_2)$-robust $\epsilon$-simulatable extractable commitment scheme (as per \Cref{def:robust-epsilon-sim-ext-com:pq}).
\end{enumerate}
Then, there exists a post-quantum non-malleable commitment scheme $\langle C, R \rangle_{\msf{TG,PQ}}^{\msf{async}}$ which supports tags of $\msf{TG} = 2^{\msf{tg}-1}$ bits. Moreover, $\langle C, R \rangle^{\msf{async}}_{\msf{TG,PQ}}$ has $k_1 + k_2 + O(1)$ rounds.
\end{theorem}
\begin{remark}
In the original version of the above theorem in \cite{BLS21}, they require $\negl$-simulatability instead of $\epsilon$-simulatability for the extractable commitment and the non-malleable commitment schemes (in \Cref{item:BLS:tag-amp:condition:2}). However, we can see that the $\epsilon$-simulatable version suffices due to a similar reason as explained in the end of the previous subsection. That is, if we use extractors with a noticeable error in their proof, we can show that the adversary's advantage is at most a noticeable function in $\secpar$. Since the noticeable error can be chosen to be arbitrarily small, this implies non-malleability in the standard sense. 
\end{remark}
%\xiao{make a remark: in the origianl version of the above theorem, they require that the ExtCom and NMC satisfy $\negl$-simulatable extractability. But as observed in CCLY21, the $\epsilon(\secpar)$-simulatable extractability suffices.}

%\xiao{to do: write our final theorem. Say that we only need to show the $(k_1 + k_2)$-robust, $\epsilon$-simulatable extractability of our $\langle C, R \rangle_{\msf{tg}}$.}

Given the above \Cref{thm:tag-amp}, we only need to prove that $\langle C, R\rangle^{\msf{async}}_{\msf{tg,PQ}}$ in \Cref{prot:small-tag:pq} has $(k_1 + k_2)$-robust  $\epsilon$-simulatable extractability where $k_1$ and $k_2$ are the round complexities of $\ExtCom$ and $\WIAoK$ used in the construction of $\langle C, R\rangle^{\msf{async}}_{\msf{tg,PQ}}$.\footnote{For applying \Cref{thm:tag-amp}, we only need to require computational soundness for the witness indistinguishable argument rather than the witness-extended emulation property. But since the latter is stronger than the former, we can simply instantiate it with any $\WIAoK$ that satisfies \Cref{def:PQWIAoK}.} 
We show this below.
\begin{theorem}\label{thm:robust:pq}
The commitment scheme $\langle C, R\rangle^{\msf{async}}_{\msf{tg,PQ}}$ in \Cref{prot:small-tag:pq} satisfies post-quantum $(k_1+k_2)$-robust $\epsilon$-simulatable extractability (as per \Cref{def:robust-epsilon-sim-ext-com:pq}) where $k_1$ and $k_2$ are round complexities of $\ExtCom$ and $\WIAoK$.  
\end{theorem}
\begin{proof}(sketch.)
We first remark that we have $n_9>k_1+k_2$ since we assume that $n_9$ is larger than the total round complexity of Steps \ref{item:small-tag:async:step:Naor-rho:pq} to \ref{item:small-tag:async:step:WIPoK:1B:pq}, which is clearly larger than $k_1+k_2$ since those steps include repetitions of $\ExtCom$ and $\WIAoK$.  
%We can show this easily if we modify the parameter setting by additionally requiring $n_5>k_1+k_2$. 
%The reason is that if $n_5>k_1+k_2$, 
Then, by the pigeonhole principle, for each session with $O^\infty$, 
there is $i^*$ such that {\bf ExtCom-3-$i^*$} does not interleave with the interaction with the external $(k_1+k_2)$-round machine $B$ in the definition of $r$-robust $\epsilon$-simulatable extractability (in \Cref{def:robust-epsilon-sim-ext-com:pq}). Then, we can extract the committed message by running the extractor for  {\bf ExtCom-3-$i^*$} without rewinding $B$.\footnote{We can assume that such $i^*$ is known in advance w.l.o.g. by \Cref{lem:adaptive-predetermined:async}.} We can show that the extracted message is the correct committed message in $\com$ (in \Cref{item:small-tag:async:step:Naor-rho:pq}) whenever the receiver accepts except for a negligible probability similarly to \Cref{lem:ExtCom_soundness}. This enables us to simulate the oracle $O^{\infty}$ without rewinding $B$. 

%In the above, we implicitly assume that the extractor knows $i^*$ such that {\bf ExtCom-3-$i^*$} does not interleave with the interaction with $\B$. However, since the adversary may adaptively decide the schedule, we cannot assume this in general. To resolve this issue, we rely on by now a standard trick that was introduced in \cite{STOC:BitShm20}. %o that used in many previous works e.g., \cite{STOC:BitShm20,C:ChiChuYam21,cryptoeprint:2021:1516,BLS21}
%Specifically, we first construct an extractor that guesses the smallest $i^*$ such that {\bf ExtCom-3-$i^*$} does not interleave with the interaction with $\B$ for each session with $O^\infty$. It runs the simulation assuming that the guess is  correct. If it turns out to be wrong, it aborts by outputting $\bot$. Since there are polynomially many choices for $i^*$'s (because there are polynomially many sessions with $O^\infty$), this extractor does not abort with an inverse polynomial probability, and conditioned on that the guess is correct, the output is $\epsilon$-close to the real execution by the above argument. Such an extractor can be converted into a full-fledged $\epsilon$-simulatable extractor that does not abort by simply applying Watrous' rewinding lemma (\Cref{lem:quantum_rewinding}) to the above extractor.
%Since there exists constant-round post-quantum $\epsilon$-simulatable extractable commitments and post-quantum witness indistinguishable arguments assuming the existence of post-quantum OWFs, t
\end{proof}

Finally, by combining \Cref{thm:nm:small-tag:pq,thm:tag-amp,thm:robust:pq}, we obtain the following theorem. 

\begin{theorem}\label{thm:nmcom_largetag_pq}
Assuming the existence of post-quantum OWFs, there exist constant-round post-quantum non-malleable commitments supporting tag space $[\Omega(2^\secpar)]$. 
%\takashi{$\Omega$ or $\Theta$ instead of $O$? (Technically, $O$ only gives an upper bound.)} 
\end{theorem}

%!TEX root = ../main.tex

\section{Application: Quantum-Secure MPC in Constant Rounds from QLWE}
\label{sec:app-to-MPC}
\para{MPC for Classical Functionalities.} Agarwal et al.~\cite{EC:ABGKM21} constructed a constant-round post-quantum MPC protocol in the plain model assuming the super-polynomial hardness of QLWE and a QLWE-based circular security assumption. The only reason why they rely on the super-polynomial hardness of QLWE is for their construction of a constant-round post-quantum non-malleable commitment scheme.\footnote{Actually, what they need is the so-called \emph{many-to-one} non-malleable commitments. But it is known that (one-to-one) non-malleability as defined in \Cref{def:NMCom:pq} is equivalent to many-to-one non-malleability (even in the post-quantum setting, as noted in \cite[Lemma 7.3]{EC:ABGKM21}).} Since we have constructed constant-round post-quantum non-malleable commitments from post-quantum OWFs (\Cref{thm:nmcom_largetag_pq}), we can weaken the assumption to the polynomial hardness of QLWE. Thus, we obtain the following theorem.

\begin{theorem}\label{formal:cor-1}
Assuming the polynomial hardness of QLWE and a QLWE-based circular security assumption (as in~\cite{EC:ABGKM21}), there exist constant-round constructions of post-quantum MPC for classical functionalities in the plain model.
\end{theorem}

Recall that ``post-quantum MPC'' means an MPC protocol secure against QPT adversaries where honest parties only need to perform classical computation. This is the first construction of constant-round post-quantum MPC from polynomial hardness assumptions in the plain model. 

\begin{remark}[Synchronous Security Suffices.]
For the construction of post-quantum MPC, we only need post-quantum non-malleable commitments secure against 
\emph{synchronous} adversaries. It is simpler to obtain such non-malleable commitments than those against asynchronous adversaries. For example, this can be done by applying the tag amplification of~\cite[Sec. 7.3]{EC:ABGKM21} to our small-tag, synchronous construction (i.e., \Cref{prot:small-tag:sync:classical} instantiated with post-quantum building blocks).
% the synchronous setting to post-quantum non-malleable commitments for small tags against synchronous adversaries (\Cref{prot:small-tag:sync:classical} instantiated with post-quantum building blocks).  
\end{remark}

\para{MPC for Quantum Functionalities.} Recently, Bartusek et al.\ \cite{C:BCKM21a} built a constant-round quantum-secure MPC for quantum functionalities {\em in the CRS model}\footnote{More accurately, \cite{C:BCKM21a} presented their construction in the OT-hybrid model. But it is known that the OT functionality can be instantiated using the post-quantum straight-line simulatable construction in the CRS model from \cite{C:PeiVaiWat08}, which is also constant-round and based on the hardness of QLWE.}, based on the hardness of QLWE. It is easy to see that the post-quantum MPC from \Cref{formal:cor-1} can be used to instantiate the CRS required by the \cite{C:BCKM21a} protocol. Since the protocol  from \Cref{formal:cor-1} is also constant round, this leads to the first constant-round quantum-secure MPC for quantum functionalities from polynomial hardness assumptions {\em without any trusted setup}.
\begin{theorem}\label{formal:cor-2}
Assuming (polynomial) QLWE and the QLWE-based circular security assumption (as in~\cite{EC:ABGKM21}), there exists a constant-round construction of quantum-secure MPC for quantum functionalities in the plain model.
\end{theorem}

\bibliographystyle{alpha}
\bibliography{main.bbl}
\addcontentsline{toc}{section}{References}

\newpage
\appendix
\section*{Appendix}

%!TEX root = ../main.tex
\section{Proof of Extract-and-Simulate Lemma (\Cref{lem:extract_and_simulate})}\label{sec:proof_extract_and_simulate}

We prove \Cref{lem:extract_and_simulate}. 
The proof follows similar techniques as in the proof of \cite[Lemma 4]{cryptoeprint:2021:1516}.%, and many parts of the following proof are taken verbatim from there with notational adaptation. 

\subsection{Preparation}
We prepare several lemmas that will be used in the proof of \Cref{lem:extract_and_simulate}.

\para{Watrous' Rewinding Lemma.} The following is Watrous' rewinding lemma \cite{SIAM:Watrous09} in the form of  \cite[Lemma 2.1]{STOC:BitShm20}. \takashi{Replaced $\gamma$ with $\alpha$ for avoiding collision.}
\begin{lemma}[Watrous' Rewinding Lemma \cite{SIAM:Watrous09}]\label{lem:quantum_rewinding}
There is a quantum algorithm $\sfR$ that gets as input the following:
\begin{itemize}
\item A quantum circuit $\sfQ$ that takes $n$-input qubits in register $\reginp$ and outputs a classical bit $b$ (in a register outside $\reginp$)  and an $m$-qubit output.  
\item An $n$-qubit state $\rho$ in register $\reginp$.
\item A number $T\in \mathbb{N}$ in unary.
\end{itemize}

$\sfR(1^T,\sfQ,\rho)$ executes in time $T\cdot|\sfQ|$  and outputs a distribution over $m$-qubit states  $D_{\rho}\defeq \sfR(1^T,\sfQ,\rho)$  with the following guarantees.

For an $n$-qubit state $\rho$, denote by $\sfQ_{\rho}$ the conditional distribution of the output distribution $\sfQ(\rho)$,
conditioned on $b = 0$, and denote by $p(\rho)$ the probability that $b = 0$. If there exist $p_0, q \in (0,1)$, $\alpha \in (0,\frac{1}{2})$
such that:
\begin{itemize}
    \item  Amplification executes for enough time: $T\geq \frac{\log (1/\alpha)}{4p_0(1-p_0)}$,
    \item  There is some minimal probability that $b = 0$: For every $n$-qubit state $\rho$, $p_0\leq p(\rho)$,
    \item  $p(\rho)$ is input-independent, up to $\alpha$ distance: For every $n$-qubit state $\rho$, $|p(\rho)-q|<\alpha$, and
    \item  $q$ is closer to $\frac{1}{2}$: $p_0(1-p_0)\leq q(1-q)$,
\end{itemize}
then for every $n$-qubit state $\rho$,
\begin{align*}
    \TD(\sfQ_{\rho},D_{\rho})\leq 4\sqrt{\alpha}\frac{\log(1/\alpha)}{p_0(1-p_0)}.
\end{align*}
%\naihui{The statement about amplification?}
%Moreover, $\sfR(1^T,\sfQ,\rho)$ works in the following manner: 
%It  uses $\sfQ$ for only implementing oracles that perform the unitary part of $\sfQ$ and its inverse, acts on $\reginp$ only through these oracles, and the output of $\sfR$ is the state in the output register of $\sfQ$ after the simulated execution.
%We note that $\sfR$ may directly act on $\sfQ$'s internal registers other than $\reginp$. 
\end{lemma}

\begin{lemma}[{\cite[Lemma 8]{cryptoeprint:2021:1516}}]\label{lem:state-close}
Let $\ket{\phi_b}=\ket{\phi_{b,0}}+\ket{\phi_{b,1}}$ be a normalized quantum state in a Hilbert space $\hil$.  \takashi{I removed the assumption that $\bra{\phi_{b,0}}\ket{\phi_{b,1}}=0$ because this was not used.}
%such that $\bra{\phi_{b,0}}\ket{\phi_{b,1}}=0$ for $b\in \bit$.
%such that $\|\ket{\phi_{0,1}}-\ket{\phi_{1,1}}\|\leq \eta$.
Let $F$ be a quantum algorithm  that takes a state in $\hil$ as input and outputs a quantum state (not necessarily in $\hil$) or a classical failure symbol $\fail$. 
Suppose that we have  
$$\Pr[F\left(\frac{\ket{\phi_{b,0}}\bra{\phi_{b,0}}}{\|\ket{\phi_{b,0}}\|^2}\right)=\fail]\geq 1-\gamma$$
for $b\in \bit$
and  
$\|\ket{\phi_{1,1}}-\ket{\phi_{1,0}}\|\le \delta$. \takashi{Note that this is Euclidean distance instead of trace distance unlike the previous version.}   %where $\gamma< 1/36$.  
%where $u$ is a classical string, $p_b\leq \gamma$ 
%for $b\in \bit$.  
%there exists a classical string $u$ such that 
%$\Pr[M\circ F(\phi_b)=u]=1-\gamma$ where $M\circ F(\phi_b)$ denotes the measurement of $F(\phi_b)$ in the computational basis. 
Then for any distinguisher $D$, it holds that 
\begin{align*}
    |\Pr[D(F(\ket{\phi_0}\bra{\phi_0}))=1]
    -\Pr[D(F(\ket{\phi_1}\bra{\phi_1}))=1]|\leq (12\gamma^{1/2}+2\delta)^{1/2}. 
\end{align*}
%\begin{align*}
%    \|F(\ket{\phi_0}\bra{\phi_0})-F(\ket{\phi_1}\bra{\phi_1}))\|_{tr}\leq \gamma + \|\ket{\phi_{0,1}}\bra{\phi_{0,1}}-\ket{\phi_{1,1}}\bra{\phi_{1,1}}\|_{tr}.
%\end{align*}
\end{lemma}
%Since the proof is similar to the latter half of the proof of \cite[Claim 4.5]{C:ChiChuYam21},  

\begin{lemma}[{\cite[Lemma 3.2]{C:ChiChuYam21}}]\label{lem:gentle_measurement} \takashi{The previous version also used this lemma in the case of $\nu=\negl$ by just saying "the gentle measurement lemma".}
    Let $\ket{\psi}_\regX$ be a (not necessarily normalized) state over register $\regX$ and $U$ be a unitary over registers $(\regX,\regY,\regZ)$.  
    Suppose that a measurement of register $\regZ$ of $U\ket{\psi}_\regX\ket{0}_{\regY,\regZ}$ results in a deterministic value except for probability $\nu$, i.e., 
    there is $z^*$ such that 
    \begin{align*}
        \|(I-\ket{z^*}\bra{z^*})_{\regZ} U\ket{\psi}_\regX\ket{0}_{\regY,\regZ}\|^2\le \nu. 
    \end{align*}
    If we let $R:= (\ket{0}\bra{0})_{\regY,\regZ}U^\dagger (\ket{z^*}\bra{z^*})_{\regZ} U$, 
then we have 
\begin{align*}
    \|
    \ket{\psi}_\regX\ket{0}_{\regY,\regZ}-
   R \ket{\psi}_\regX\ket{0}_{\regY,\regZ}\|\le \sqrt{\nu}.
\end{align*}
\end{lemma}

%where we highlight the difference from \cite[Lemma 3.2]{C:ChiChuYam21} in red.  
%we take the following lemma from \cite[Lemma 3.2]{C:ChiChuYam21}, which is an easy implication of Jordan's lemma.  
\begin{lemma}[A variant of {\cite[Lemma 10]{cryptoeprint:2021:1516}}]\label{lem:amplification}
Let $\Pi$ be a projection over a Hilbert space $\hil_\regX \ot \hil_\regY$. %of $\poly(\secpar)$-qubit.\footnote{More precisely, we consider a sequence of projections $\{\Pi_\secpar\}_{\secpar\in\mathbb{N}}$  where $\Pi_{\secpar}$ is over a Hilbert space $\hil_{\regX_\secpar} \ot \hil_{\regY_\secpar}$. We omit the indexing by the security parameter $\secpar$ for notational simplicity.
%A similar remark applies to other objects that appear in this lemma (e.g., $S_{<\delta}$, $S_{\geq \delta}$, $U_{\amp,T}$, etc.)
%}   
For any noticeable function $\delta=\delta(\lambda)$, there exists an orthogonal decomposition $(S_{<\delta}, S_{\geq \delta})$ of $\hil_\regX \ot \hil_\regY$ that satisfies the following:
\begin{enumerate}
\item(\textbf{$S_{<\delta}$ and $S_{\geq \delta}$ are invariant under $\Pi$ and $(\ket{0}\bra{0})_{\regY}$.}) \label{item:amplification_invariance_projection}
For any $\ket{\psi}_{\regX,\regY}\in S_{<\delta}$, we have 
\begin{align*}
\Pi \ket{\psi}_{\regX,\regY}\in S_{<\delta},~~~~~ (I_{\regX}\otimes(\ket{0}\bra{0})_{\regY})\ket{\psi}_{\regX,\regY}\in S_{<\delta}.
\end{align*}
Similarly, 
 for any $\ket{\psi}_{\regX,\regY}\in S_{\geq \delta}$, we have 
\begin{align*}
\Pi \ket{\psi}_{\regX,\regY}\in S_{\geq \delta},~~~~~ (I_{\regX}\otimes(\ket{0}\bra{0})_{\regY})\ket{\psi}_{\regX,\regY}\in S_{\geq \delta}.
\end{align*}
\item(\textbf{$\Pi$ succeeds with probability $<\delta$ and $\geq \delta$ in $S_{<\delta}$ and $S_{\geq \delta}$.})  \label{item:amplification_success_probability}
%For any quantum state $\ket{\psi_{<\delta}}_{\regX}\in \hil_{\regX}$ s.t.  $\ket{\psi_{<\delta}}_{\regX}\ket{0}_{\regY}\in \hil_{<\delta}$, we have 
%$\|\Pi\ket{\psi_{<\delta}}_{\regX}\ket{0}_{\regY}\|^2< \delta.$
%Similarly, for any quantum state $\ket{\psi_{\geq \delta}}\in  \hil_{\geq \delta}$, we have 
%$\|\Pi\ket{\psi_{\geq \delta}}_{\regX}\ket{0}_{\regY}\|^2\geq \delta.$
For any quantum state $\ket{\phi}_{\regX}\in \hil_{\regX}$ s.t.  $\ket{\phi}_{\regX}\ket{0}_{\regY}\in S_{<\delta}$ we have 
\begin{align*}
\|\Pi\ket{\phi}_{\regX}\ket{0}_{\regY}\|^2< \delta. 
\end{align*}
Similarly, for any quantum state $\ket{\phi}_{\regX}\in \hil_{\regX}$ s.t.  $\ket{\phi}_{\regX}\ket{0}_{\regY}\in S_{\geq \delta}$ we have 
\begin{align*}
\|\Pi\ket{\phi}_{\regX}\ket{0}_{\regY}\|^2\geq \delta. 
\end{align*}
 \item(\textbf{Unitary for amplification.})
For any $T\in \mathbb{N}$, there exists a unitary  $U_{\amp,T}$ over $\hil_\regX \ot \hil_\regY \ot \hil_\regB \ot \hil_\reganc$ where $\regB$ is a register to store a qubit and $\reganc$ is a register to store ancillary qubits with the following properties:
 \label{item:amplification}
 \begin{enumerate}
 \item(\textbf{Mapped onto $\Pi(I_{\regX}\otimes(\ket{0}\bra{0})_{\regY})$ when $\regB$ contains $1$.})
 %For any quantum state $\ket{\psi}_{\regX,\regY}\in \hil_{\regX}\otimes \hil_{\regY}$, suppose that  we apply $U_{\amp,T}$ to $\ket{\psi}_{\regX}$  and measure register $\regB$. If the outcome is $1$, then the resulting state in registers $(\regX,\regY)$ is in the span of $\Pi$.
For any quantum state $\ket{\psi}_{\regX,\regY}\in \hil_{\regX}\otimes \hil_{\regY}$, we can write
 \[
 \ket{1}\bra{1}_{\regB} U_{\amp,T}\ket{\psi}_{\regX,\regY}\ket{0}_{\regB,\reganc}=\sum_{anc}\ket{\psi'_{anc}}_{\regX,\regY}\ket{1}_{\regB}\ket{anc}_{\reganc}
 \]
 by using sub-normalized states $\ket{\psi'_{anc}}_{\regX,\regY}$ that are in the span of $\Pi(I_{\regX}\otimes(\ket{0}\bra{0})_{\regY})$.
 \label{item:amplification_map_to_pi}
\item(\textbf{Amplification of success probability in $S_{\geq \delta}$.})  \label{item:amplification_amplification}
For any noticeable function $\nu=\nu(\secpar)$, there is $T=\poly(\secpar)$ such that  
for any quantum state $\ket{\phi}_{\regX}\in \hil_{\regX}$ s.t.  $\ket{\phi}_{\regX}\ket{0}_{\regY}\in S_{\geq \delta}$, we have 
 \[
% \Pr[M_\regB\circ U_{\amp,T}\ket{\phi}_{\regX}\ket{0}_{\regY}\ket{0}_{\regB,\reganc}=1]
 \|\ket{1}\bra{1}_\regB U_{\amp,T}\ket{\phi}_{\regX}\ket{0}_{\regY}\ket{0}_{\regB,\reganc}\|^2
\geq 1-\nu. %(1-2\delta+2\delta^2)^{T-1}(1-\delta).
 \]
 \item(\textbf{$S_{<\delta}$ and $S_{\geq \delta}$ are invariant under $U_{\amp,T}$}). 
 For any  quantum state $\ket{\psi_{<\delta}}_{\regX,\regY}\in  S_{<\delta}$ and any $b,anc$,
we can write 
  \[
U_{\amp,T}\ket{\psi_{< \delta}}_{\regX,\regY}\ket{b,anc}_{\regB,\reganc}=\sum_{b',anc'}\ket{\psi'_{<\delta,b',anc'}}_{\regX,\regY}\ket{b',anc'}_{\regB,\reganc}
  \]
  by using sub-normalized states $\ket{\psi'_{<\delta, b',anc'}}_{\regX,\regY} \in S_{<\delta}$.
  
  Similarly, 
 for any  quantum state $\ket{\psi_{\geq \delta}}_{\regX,\regY}\in  S_{\geq \delta}$ and any $b,anc$,
we can write 
  \[
U_{\amp,T}\ket{\psi_{\geq \delta}}_{\regX,\regY}\ket{b, anc}_{\regB,\reganc}=\sum_{b', anc'}\ket{\psi'_{\geq \delta, b', anc'}}_{\regX,\regY}\ket{b', anc'}_{\regB,\reganc}
  \]
  by using sub-normalized states $\ket{\psi'_{\geq \delta, b',anc'}}_{\regX,\regY} \in S_{\geq \delta}$. 
  \label{item:amplification_invariance}
 \end{enumerate}
 \item(\textbf{Efficient Implementation of $U_{\amp,T}$}.) 
 There exists a QPT algorithm $\Amp$ (whose description is independent of $\Pi$) that takes as input  $1^T$, a description of quantum circuit that perform a measurement $(\Pi, I_{\regX,\regY}-\Pi)$, and a state $\ket{\psi}_{\regX,\regY,\regB,\reganc}$, and outputs $U_{\amp,T}\ket{\psi}_{\regX,\regY,\regB,\reganc}$.
 Moreover, $\Amp$ uses the measurement circuit for only implementing an oracle that apply unitary to write a measurement result in a designated register in $\reganc$, and it acts on $\regX$ only through the oracle access.
 \label{item:amplification_efficiency}
\end{enumerate}
\end{lemma}

\subsection{Proving \Cref{lem:extract_and_simulate}} 
%Let $\mcal{K}$ and \red{$\{s^*_z\}_{z\in \bit^*}$} be as in \Cref{lem:extract_and_simulate}. 
Since any mixed state can be seen as a distribution over pure states, we assume $\mcal{G}$'s input $\rho_\secpar$ is a pure state  and denote it by $\ket{\psi}$, omitting the dependence on $\secpar$. 
Similarly, we simply write $z$ to mean $z_\secpar$ and $s^*$ to mean $s^*_{z_\secpar}$ for simplicity. 

\takashi{In the following, I added $z$ as the rightmost input of $\Exp$, $\mcal{G}$, $\SimExt$, etc. Please fix it if it is missing somewhere.}
For a noticeable function $\gamma(\secpar)$ and a quantum state $\ket{\psi}$, 
we define an experiment $\Exp(\secpar,1^{\gamma^{-1}},\ket{\psi},z)$ as follows %where $\gamma\defeq \left(\frac{\epsilon}{4}\right)^4$:

\smallskip
\noindent
$\Exp(\secpar,1^{\gamma^{-1}},\ket{\psi},z)$:
Run $(b,\rho_\out)\sample \mcal{G}(1^\secpar,1^{\gamma^{-1}},\ket{\psi},z)$. 
    \begin{itemize}
        \item If $b=\top$, the experiment outputs $(\rho_\out,s^*)$.  We remark that this step may not be done efficiently since we do not assume that $s^*$ can be computed from $z$ efficiently.
        \xiao{we may need to explain how $\Exp$ obtains $s^*$. More accurately, why $\Exp$ cannot get $s^*$ efficiently.}\takashi{Added an explanation. 
        As a very minor issue, I noticed that $\Exp$ even may not be an ``algorithm" if $s^*$ is an uncomputable function of $z$. Should we care about this? Mathematically, I believe this is fine because I don't think we use the fact that $\Exp$ is computable.} \xiao{I think this is fine. As you said, in this lemma itself, we don't require Exp to be an algorithm. Moreover, when we use this lemma, Exp is indeed an algorithm as $s^*$ can be brute-forced from $z$ in our application.}
        \item If $b=\bot$, the experiment outputs $(\rho_\out,\bot)$. 
    \end{itemize}
What we have to do is to construct a QPT $\SimExt$ such that for any polynomial-size $\ket{\psi}$ and noticeable $\epsilon=\epsilon(\secpar)$, 
$$
\{\SimExt(1^\secpar,1^{\epsilon^{-1}},\ket{\psi},z)\}_{\secpar\in \Naturals}
\statind_{\epsilon} 
\{(\Exp(\secpar,1^{\gamma^{-1}},\ket{\psi},z)\}_{\secpar\in \Naturals}
$$
for some noticeable $\gamma$. %$\gamma=\poly(\epsilon)$. 
%In particular, we prove it for $\gamma\defeq \left(\frac{\epsilon}{4}\right)^4$.  

Let $\Exp_\bot(\secpar,1^{\gamma^{-1}},\ket{\psi},z)$ and $\Exp_\top(\secpar,1^{\gamma^{-1}},\ket{\psi},z)$ be the same as $\Exp(\secpar,1^{\gamma^{-1}},\ket{\psi},z)$ except that they output a failure symbol $\fail$ in the cases of $b=\top$ and $b=\bot$, respectively. 
That is, they work as follows.

\smallskip
\noindent
$\Exp_\bot(\secpar,1^{\gamma^{-1}},\ket{\psi},z)$:
Run $(b,\rho_\out)\sample \mcal{G}(1^\secpar,1^{\gamma^{-1}},\ket{\psi},z)$. 
    \begin{itemize}
        \item If $b=\top$, the experiment outputs $\fail$. 
        \item If $b=\bot$, the experiment outputs $(\rho_\out,\bot)$. 
    \end{itemize}
    
    \smallskip
\noindent
$\Exp_\top(\secpar,1^{\gamma^{-1}},\ket{\psi},z)$:
Run $(b,\rho_\out)\sample \mcal{G}(1^\secpar,1^{\gamma^{-1}},\ket{\psi},z)$. 
    \begin{itemize}
        \item If $b=\top$, the experiment outputs $(\rho_\out,s^*)$. 
        \item If $b=\bot$, the experiment outputs $\fail$. 
    \end{itemize}

First, we give simulation extractors for each of these experiments.
\begin{lemma}[Extract-and-Simulate for the Case of $b=\bot$]\label{lem:aborting_case}
For any noticeable $\gamma$ that is efficiently computable from $\epsilon$, 
%efficiently computable $\gamma=\poly(\epsilon)$, 
there is a QPT algorithm $\SimExt_\bot$ such that  for any polynomial-size quantum state $\ket{\psi}$ and
noticeable $\epsilon$, 
$$
\{\SimExt_\bot(1^\secpar,1^{\epsilon^{-1}},\ket{\psi},z)\}_{\secpar\in \Naturals}
\equiv 
\{\Exp_\bot(\secpar,1^{\gamma^{-1}},\ket{\psi},z)\}_{\secpar\in \Naturals}. %\footnote{$\SimExt_\bot$ does not need to take $1^{\epsilon^{-1}}$ as part of its input, but we include it in the input for notational convenience.}
$$ 
\end{lemma}
\begin{proof}[Proof of \Cref{lem:aborting_case}]
Since $\Exp_\bot$ is efficient (because it never outputs $s^*$), $\SimExt_{\bot}$ just needs to run $\Exp_\bot$.

\end{proof}

\begin{lemma}[Extract-and-Simulate for the Case of $b=\top$]\label{lem:non-aborting_case}
There is a QPT algorithm $\SimExt_\top$ such that  for any polynomial-size quantum state $\ket{\psi}$ and
noticeable $\epsilon$, 
$$
\{\SimExt_\top(1^\secpar,1^{\epsilon^{-1}},\ket{\psi},z)\}_{\secpar\in \Naturals}
\sind_{\epsilon} 
\{\Exp_\top(\secpar,1^{\gamma^{-1}},\ket{\psi},z)\}_{\secpar\in \Naturals},
$$ 
where $\gamma \defeq \left(\frac{\epsilon}{5}\right)^4$.
\end{lemma}
\begin{proof}[Proof of \Cref{lem:non-aborting_case}] 
Let  $\delta$ be a noticeable function that satisfies \Cref{item:gamma_delta} of \Cref{lem:extract_and_simulate} for $\gamma = \left(\frac{\epsilon}{5}\right)^4$.

We apply \Cref{lem:amplification} with respect to a projection corresponding to the success of $\mathcal{K}$. 
Let $U_{\mathcal{K}}$ be the unitary that represents $\mathcal{K}(1^\secpar,1^{\gamma^{-1}},\cdot,z)$.  
More precisely, we define $U_{\mathcal{K}}$ over registers the input register $\reginp$, working register $\regW$, and  output register $\regout$ so that $\mathcal{K}(1^\secpar,1^{\gamma^{-1}},\cdot,z)$ can be described as follows:

\smallskip
\noindent\textbf{$\mathcal{K}(1^\secpar,1^{\gamma^{-1}},\cdot,z)$}:
It takes a quantum state $\ket{\psi}$ in the register $\reginp$ and initializes registers $\regW$ and $\regout$ to be $\ket{0}_{\regW,\regout}$. 
Then it applies the unitary $U_{\mathcal{K}}$, measures the register $\regout$ in the standard basis to obtain $s$,  and outputs $s$. 
\smallskip

We define a projection $\Pi$ over $(\reginp,\regW,\regout)$ as
\begin{align}\label{eq:def_pi}
\Pi \defeq U_{\mathcal{K}}^\dagger \left(\sum_{s\neq \fail}\ket{s}\bra{s}\right)_{\regout}  U_{\mathcal{K}}.
 %\Pi \defeq U_{\CCY}^\dagger (\ket{s^*}\bra{s^*})_{\regout}  U_{\CCY}.
\end{align}
Then the following claim immediately follows from the assumption about $\mathcal{K}$ (\Cref{item:s_star_or_bot} of \Cref{lem:extract_and_simulate}). %former half of \Cref{cla:simless}.
\begin{MyClaim}\label{cla:measure_s_star}
Given any state in the span of $\Pi(I_{\reginp}\otimes(\ket{0}\bra{0})_{\regW,\regout})$, if we apply $U_{\mathcal{K}}$ and then measure register $\regout$, then the measurement outcome is always $s^*$
\end{MyClaim}  
We apply \Cref{lem:amplification} for the above $\Pi$ where $\hil_\regX:=\hil_\reginp$, $\hil_\regY:= \hil_\regW \ot \hil_\regout$,  %$\delta$ is chosen in such a way that \Cref{item:gamma_delta} of \Cref{lem:extract_and_simulate}  holds for $\gamma:=\left(\frac{\epsilon}{5}\right)^4$, 
and $T=\poly(\secpar)$ is chosen in such a way that \Cref{item:amplification_amplification} of \Cref{lem:amplification} holds for $\nu:=\left(\frac{\epsilon}{2}\right)^4$.  
Then we have a decomposition $(S_{<\delta}, S_{\geq \delta})$ of $\hil_\regX\ot \hil_\regY$ and a unitary $U_{\amp,T}$ over $\hil_\regX\ot \hil_\regY \ot \hil_\regB \ot \hil_\reganc$ that satisfy the requirements in \Cref{lem:amplification}. 
We denote by $\regother$ to mean the registers $\regW$, $\regout$, $\regB$, and $\reganc$ for brevity.
%We apply \Cref{lem:extraction-variant} where 
%we define $\A_{\CCY}$ to be $\ext_{\mathcal{K}}(1^\secpar,\{\Pi_i\}_{i\in C},\A,\cdot)$ and $\delta\defeq \left(\frac{\epsilon}{4}\right)^{12}-\left(1-\frac{|S|}{|C|^2}\right)$.  
%We note that $\delta$ is noticeable since $\epsilon$ is noticeable and $1-\frac{|S|}{|C|^2}=\negl(\secpar)$ as required in \Cref{item:extract_and_simulate_overwhelming} of \Cref{lem:extract_and_simulate}. 
%Let $\ext_{\CCY}$ be the corresponding extractor and $S_{<\delta}$ and $S_{\geq \delta}$ be the decomposition of $\hil$ as in \Cref{lem:extraction-variant}.  
We construct the extractor $\SimExt_\top$ for \Cref{lem:non-aborting_case} as follows:

\smallskip
\noindent
$\SimExt_\top(1^\secpar,1^{\epsilon^{-1}},\ket{\psi},z)$:
\begin{enumerate}
\item Set $\ket{\psi}$ in register $\reginp$ and initlialize register $\regother$ to be $\ket{0}$. 
\item Apply $U_{\amp,T}$ by using the algorithm $\Amp$ in \Cref{item:amplification_efficiency} of \Cref{lem:amplification}. 
\item \label{step:measure_b}
Measure register $\regB$ and let $b$ be the outcome. 
If $b=0$, output $\fail$ and immediately halt. Otherwise, proceed to the next step. 
\item \label{step:extract_s} Apply $U_{\mathcal{K}}$, measure register $\regout$ to obtain an outcome $s_{\ext}$, and apply $U_{\mathcal{K}}^\dagger$.
\item Apply $U_{\amp,T}^\dagger$ by using the algorithm $\Amp$ in \Cref{item:amplification_efficiency} of \Cref{lem:amplification}. 
\item Measure register $\regother$. If the outcome is not the all $0$'s string, output $\fail$ and immediately halt. Otherwise, let $\ket{\psi'}$ be the state in register $\reginp$ at this point, and proceed to the next step.
\item Run $(b,\rho_{\mathsf{out}})\gets \mathcal{G}(1^\secpar,1^{\gamma^{-1}},\ket{\psi'},z)$
    \begin{itemize}
        \item If $b=\top$, output $(\rho_{\mathsf{out}}, s_{\ext})$. 
        \item If $b=\bot$, output $\fail$.   
    \end{itemize}
\end{enumerate}
\smallskip 

We can easily see the following claim: 
\begin{MyClaim}\label{claim:always_extract_s_star}
Whenever Step \ref{step:extract_s} of $\SimExt_\top$ is invoked, $s_{\ext}$ obtained in the step is always equal to $s^*$. Moreover, the step does not change the state in registers $\reginp$ and $\regother$, that is, the states before and after the step are identical.  
\end{MyClaim}
\begin{proof}[Proof of \Cref{claim:always_extract_s_star}]
Whenever  Step \ref{step:extract_s} is invoked, the bit $b$ obtained in Step \ref{step:measure_b} is equal to $1$. In this case, by \Cref{item:amplification_map_to_pi} of \Cref{lem:amplification}, the state in registers $\reginp$, $\regW$, and $\regout$ is in the span of $\Pi(I_{\reginp}\otimes(\ket{0}\bra{0})_{\regW,\regout})$. Then, \Cref{cla:measure_s_star} implies that $s_\ext$ is always equal to $s^*$. Then the measurement of $\regout$ does not collapse the state and thus  the step does not change the state.
\end{proof}

The rest of the proof is similar to that of \cite[Claim 4.5]{C:ChiChuYam21}. 
Let $R$ be an operator defined as follows:
\begin{align*}
R:=(\ket{0}\bra{0})_{\regother}U_{\amp,T}^{\dagger}(\ket{1}\bra{1})_{\regB}U_{\amp,T}.
\end{align*}
Let $\Pi_{<\delta}$ and $\Pi_{\ge \delta}$ be projections onto $S_{<\delta}$ and $S_{\ge \delta}$, respectively.  
To apply Lemma \ref{lem:state-close}, 
we define states $\ket{\phi_0}=\ket{\phi_{0,0}}+\ket{\phi_{0,1}}$ and $\ket{\phi_1}=\ket{\phi_{1,0}}+\ket{\phi_{1,1}}$ over $(\regD,\reginp,\regother)$ 
where $\regD$ is an additional one-qubit register as follows:   
\begin{align*}
&\ket{\phi_{0}}:= \ket{1}_{\regD}\ket{\psi}_{\reginp}\ket{0}_{\regother},\\
&\ket{\phi_{0,0}}:= \ket{1}_{\regD}\Pi_{< \delta}\ket{\psi}_{\reginp}\ket{0}_{\regother},\\
&\ket{\phi_{0,1}}:= \ket{1}_{\regD}\Pi_{\ge \delta}\ket{\psi}_{\reginp}\ket{0}_{\regother},\\
&\ket{\phi_{1}}:=\ket{1}_{\regD} R \ket{\psi}_{\reginp}\ket{0}_{\regother}+\alpha \ket{0}_{\regD}\ket{0}_{\reginp}\ket{0}_{\regother},\\
&\ket{\phi_{1,0}}:=\ket{1}_{\regD} R \Pi_{< \delta}  \ket{\psi}_{\reginp}\ket{0}_{\regother}+\alpha \ket{0}_{\regD}\ket{0}_{\reginp}\ket{0}_{\regother},\\
&\ket{\phi_{1,1}}:= \ket{1}_{\regD}R \Pi_{\ge \delta} \ket{\psi}_{\reginp}\ket{0}_{\regother}
\end{align*}
for $\alpha:=\sqrt{1-\|R \ket{\psi}_{\reginp}\ket{0}_{\regother}\|^2}$ (so that $\ket{\phi_{1}}$ is a normalized state). 
Let $F$ be a quantum algorithm that works as follows: 
\begin{description}
\item $F\left(\ket{\phi}_{\regD,\reginp,\regother}\right)$:
It measures $\regD$, and outputs $\fail$ if the outcome is $0$. 
Otherwise, for the state $\ket{\psi_{\mathsf{inp}}}$ in register $\reginp$, 
it runs $(b,\rho_{\mathsf{out}})\gets \mathcal{G}(1^\secpar,1^{\gamma^{-1}},\ket{\psi_{\mathsf{inp}}},z)$ and outputs $(\rho_{\mathsf{out}}, s_{\ext})$ if $b=\top$ and otherwise outputs $\fail$.  
\end{description} 

It is easy to see that 
\begin{align*}
   \Exp_\top(\secpar,1^{\gamma^{-1}},\ket{\psi},z)\equiv F(\ket{\phi_0}\bra{\phi_0}). 
\end{align*}

Moreover, by the definition of $\ext_{\siml,\nonabort}$ and \Cref{claim:always_extract_s_star}, we can see that 
\begin{align*}
   \SimExt_\top(1^\secpar,1^{\epsilon^{-1}},\ket{\psi},z)\equiv F(\ket{\phi_1}\bra{\phi_1}). 
\end{align*}
Thus, it suffices to prove that the distinguishing advantage between $F(\ket{\phi_0}\bra{\phi_0})$ and $F(\ket{\phi_1}\bra{\phi_1})$ is at most $\epsilon$.  
To apply \Cref{lem:state-close}, 
we prove the following claim. 
\begin{MyClaim}\label{claim:condition_check}
The following hold:
\begin{enumerate}
 \item $\Pr[F\left(\frac{\ket{\phi_{b,0}}\bra{\phi_{b,0}}}{\|\ket{\phi_{b,0}}\|^2}\right)=\fail]\geq 1-\gamma$
 %1-\left(\frac{\epsilon}{5}\right)^4$ 
for $b\in \bit$.
 \item $\|\ket{\phi_{1,1}}-\ket{\phi_{0,1}}\|\le \nu^{1/2}$.%\left(\frac{\epsilon}{2}\right)^2$.
\end{enumerate}
\end{MyClaim}
\begin{proof}[Proof of Claim \ref{claim:condition_check}]~ \\
\paragraph{First item.}
We can write $\Pi_{< \delta}\ket{\psi}_{\reginp}\ket{0}_{\regother}=\ket{\psi_{< \delta}}_{\reginp}\ket{0}_{\regother}$. 
By $\ket{\psi_{< \delta}}_{\reginp}\ket{0}_{\regother}\in S_{<\delta}$, 
\Cref{item:amplification_success_probability} of \Cref{lem:amplification}, and \Cref{item:s_star_or_bot} of \Cref{lem:extract_and_simulate},  we have 
\begin{align*}
 \Pr[\mathcal{K}\left(1^\secpar,1^{\gamma^{-1}}\frac{\ket{\psi_{<\delta}}}{\|\ket{\psi_{<\delta}}\|},z\right)=s_z^*]<\delta.   
\end{align*}
By the contraposition of \Cref{item:gamma_delta} of \Cref{lem:extract_and_simulate}, 
we have 
\begin{align*}
\Pr[b=\top:(b,\rho_{\mathsf{out}})\gets \mathcal{G}\left(1^\secpar,1^{\gamma^{-1}}\frac{\ket{\psi_{<\delta}}}{\|\ket{\psi_{<\delta}}\|},z\right)]<\gamma.
\end{align*}
Thus, we have 
\begin{align*}
    \Pr[F\left(\frac{\ket{\phi_{0,0}}\bra{\phi_{0,0}}}{\|\ket{\phi_{0,0}}\|^2}\right)\neq \fail]
    =\Pr[b=\top:(b,\rho_{\mathsf{out}})\gets \mathcal{G}\left(1^\secpar,1^{\gamma^{-1}}\frac{\ket{\psi_{<\delta}}}{\|\ket{\psi_{<\delta}}\|},z\right)]\le \gamma.
\end{align*} 
This completes the proof of the first item for the case of $b=0$. 
The case of $b=1$ can be proven similarly noting that $R \Pi_{< \delta}  \ket{\psi}_{\reginp}\ket{0}_{\regother}\in S_{<\delta}$ by  \Cref{item:amplification_invariance_projection,item:amplification_invariance} of Lemma \ref{lem:amplification}. 

\paragraph{Second Item.} 
By 
\Cref{item:amplification_amplification} of \Cref{lem:amplification}, we have 
$$
\|(\ket{1}\bra{1})_{\regB}U_{\amp,T}\Pi_{\ge t}\ket{\psi}_{\reginp}\ket{0}_{\regother}\|^2\le \nu.
$$
Thus, \Cref{lem:gentle_measurement} implies
\begin{align*}
   \|\Pi_{\ge t}\ket{\psi}_{\reginp}\ket{0}_{\regother}-R\Pi_{\ge t}\ket{\psi}_{\reginp}\ket{0}_{\regother}\|\le \nu^{1/2}. 
\end{align*}
%Since $\nu=\frac{\epsilon^4}{16}$, this immediately implies the second item of the claim.
\end{proof}
By \Cref{lem:state-close} and \Cref{claim:condition_check} 
the distinguishing advantage between $F(\ket{\phi_0}\bra{\phi_0})$ and $F(\ket{\phi_1}\bra{\phi_1})$ is at most 
\begin{align}
\left(12\gamma^{1/2}+2\nu^{1/2}\right)^{1/2}.
\end{align}
By using $\gamma=\left(\frac{\epsilon}{5}\right)^4$ and $\nu=\left(\frac{\epsilon}{2}\right)^4$, we can see that this is at most $\epsilon$. 
This completes the proof of \Cref{lem:non-aborting_case}.

We complete the proof of \Cref{lem:non-aborting_case}.

\end{proof}

Given \Cref{lem:aborting_case} and \Cref{lem:non-aborting_case}, the rest of the proof of \Cref{lem:extract_and_simulate} is very similar to the corresponding part of the $\epsilon$-zero-knowledge property of the protocols in \cite{C:ChiChuYam21}.
We give the full proof for completeness.

Let $\SimExt_{\comb}$ be an algorithm that works as follows:

\smallskip
\noindent
$\SimExt_{\comb}(1^\secpar,1^{\epsilon^{-1}},\ket{\psi},z)$:
\begin{enumerate}
    \item Set $\epsilon'\defeq\frac{\epsilon^2}{4\log^4(\secpar)}$.
    \item Choose $\mathsf{mode}\sample \{\top,\bot\}$.
    \item Run and output $\SimExt_{\mathsf{mode}}(1^\secpar,1^{\epsilon'^{-1}},\ket{\psi})$.
\end{enumerate}

\begin{lemma}[$\SimExt_{\comb}$ Simulates $\Exp$ with Probability almost $1/2$]\label{lem:comb}
Let $p_{\comb}^{\mathsf{suc}}(1^\secpar,1^{\epsilon^{-1}},\ket{\psi},z)$ be the probability that $\SimExt_{\comb}(1^\secpar,1^{\epsilon^{-1}},\ket{\psi},z)$ does not return $\fail$, and let 
$$D_{\mathsf{ext},\comb}(1^\secpar, 1^{\epsilon^{-1}}, \ket{\psi},z)$$
 be a conditional  distribution of $\SimExt_{\comb}(1^\secpar,1^{\epsilon^{-1}},\ket{\psi},z)$, conditioned on that it does not return $\fail$.
Then we have 
\begin{align}
    \left|p_{\comb}^{\mathsf{suc}}(1^\secpar,1^{\epsilon^{-1}},\ket{\psi},z)-1/2\right|\leq \epsilon'/2+\negl(\secpar). \label[Inequality]{eq:pcombsuc}
\end{align}
Moreover, we have 
\begin{align}
\{D_{\mathsf{ext},\comb}(1^\secpar, 1^{\epsilon^{-1}}, \ket{\psi},z)\}_{\secpar\in \Naturals}
\statind_{4\epsilon'} 
\{\Exp(\secpar,1^{\gamma^{-1}},\ket{\psi},z)\}_{\secpar\in \Naturals}, \label{eq:simlcomb}
\end{align} 
where $\gamma \defeq \left(\frac{\epsilon'}{5}\right)^4$.
\end{lemma} 
\begin{proof}(sketch.)
The intuition behind this proof is as follows. 
By \Cref{lem:aborting_case,lem:non-aborting_case}, $\SimExt_{\bot}$ and $\SimExt_\top$ almost simulate $\Exp$ conditioned on that $b=\bot$ and $b=\top$, respectively.
Therefore, if we randomly guess  $b$ and runs either of $\SimExt_{\bot}$ or $\SimExt_\top$ that successfully works for the guessed case, the output distribution is close to  the real output distribution of $\Exp$ conditioned on that the guess is correct, which happens with probability almost $1/2$.

A formal proof can be obtained based on the above intuition and is exactly the same as the proof of \cite[Lemma 5.5]{C:ChiChuYam21}
except for  notational adaptations.  

\end{proof}

Then, we convert $\SimExt_{\comb}$ into a full-fledged simulator that does not return $\fail$ by using Watrous' rewinding lemma (\Cref{lem:quantum_rewinding}).
Namely, we let $\sfQ$ be a quantum algorithm that takes $\ket{\psi}$ as input and outputs $\SimExt_{\comb}(1^\secpar,1^{\epsilon^{-1}},\ket{\psi},z)$ where $b\defeq 0$ if and only if  it does not return $\fail$, $p_0\defeq \frac{1}{4}$, $q\defeq \frac{1}{2}$, $\alpha\defeq \epsilon'$, %\footnote{Do not confuse this $\gamma$ with the error parameter in \Cref{lem:extract_and_simulate}; this is the one in \Cref{lem:quantum_rewinding}.} 
and $T\defeq 2\log (1/\epsilon')$.
Then it is easy to check that the conditions for \Cref{lem:quantum_rewinding} is satisfied by \Cref{eq:pcombsuc} in \Cref{lem:comb} (for sufficiently large $\secpar$).
Then, by using \Cref{lem:quantum_rewinding}, we can see that $\sfR(1^T,\sfQ,\ket{\psi})$ runs in time $T=\poly(\secpar)$ and its output (seen as a mixed state) has a trace distance bounded by $4\sqrt{\alpha}\frac{\log(1/\alpha)}{p_0(1-p_0)}$ from  $D_{\mathsf{ext},\comb}(1^\secpar, 1^{\epsilon^{-1}}, \ket{\psi},z)$.
Since we have $\alpha=\epsilon'=\frac{\epsilon^2}{4\log^4(\secpar)}=1/\poly(\secpar)$, we have $4\sqrt{\alpha}\frac{\log(1/\alpha)}{p_0(1-p_0)}< \sqrt{\alpha} \log^2 (\secpar)=\frac{\epsilon}{2}$ for sufficiently large $\secpar$
 where we used $\log(1/\alpha)=\log(\poly(\secpar))=o(\log^2(\secpar))$ and $\frac{4}{p_0(1-p_0)}=O(1)$.
Thus, by combining the above and \Cref{eq:simlcomb} in \Cref{lem:comb}, if we define $\SimExt(1^\secpar,1^{\epsilon^{-1}},\ket{\psi},z)\defeq \sfR(1^T,\sfQ,\ket{\psi})$, %\footnote{$\SimExt$ only needs black-box access to $\mcal{G}(1^\secpar,1^{\gamma^{-1}},\ket{\psi},z)$ since $\sfR$ uses$\sfQ(\ket{\psi})=\SimExt(1^\secpar,1^{\epsilon^{-1}},\ket{\psi},z)$ in a black-box way.} 
then we have 
\begin{align*}
\{\SimExt(1^\secpar,1^{\epsilon^{-1}},\ket{\psi},z)\}_{\secpar\in \Naturals}
\statind_{\frac{\epsilon}{2}+4\epsilon'} 
  \{\Exp(\secpar,1^{\gamma^{-1}},\ket{\psi},z)\}_{\secpar\in \Naturals}
\end{align*}
where $\gamma = \left(\frac{\epsilon'}{5}\right)^4< \epsilon' < \epsilon$. 
We can conclude the proof of \Cref{lem:extract_and_simulate} by noting that  we have
%\begin{align*}
    $\frac{\epsilon}{2}+4\epsilon'< \epsilon$
%\end{align*}
since we have $\epsilon'= \frac{\epsilon^2}{4\log^4(\secpar)} < \frac{\epsilon}{8}$ for sufficiently large $\secpar$.

%!TEX root = ../main.tex
\section{Small-Tag, One-sided, Synchronous, Post-Quantum Setting}\label{pq:sec:small-tag-one-sided-sync-pq}

\subsection{Construction} 
In this section, we prove that \Cref{prot:one-sided:classical} is post-quantumly secure if we rely on post-quantum building-blocks. 
Similarly to the classical case in \Cref{sec:small-tag-one-sided-sync-classical}, we present the construction assuming the existence of post-quantum {\em injective} OWFs.  
The assumption can be relaxed to  {\em any} OWFs by an appropriate modification to the protocol similarly to the classical case.  
Since this is exactly the same as that in the classical setting (i.e., \Cref{sec:removing-injectivity}), we do not repeat it in this section. 

Our construction is based on the following building blocks:
%We need the following building blocks:
\begin{itemize}
\item
A post-quantum {\em injective} OWF $f$; 
\item
Naor's commitment $\Com$ that is implemented with a post-quantum OWF.\footnote{In this way, the Naor's commitment is statistically binding and {\em post-quantumly} computationally hiding.};
\item
A post-quantum witness-indistinguishable argument of knowledge with $\epsilon$-close emulation $\WIAoK$ %with knowledge error $\kappa(\secpar) = \negl(\secpar)$ 
(as per \Cref{def:PQWIAoK}). 
\end{itemize}
\begin{ProtocolBox}[label={pq:prot:one-sided:classical}]{Small-Tag One-Sided Synchronous Post-Quantum NMCom \textnormal{$\langle C, R\rangle^{\msf{OneSided}}_{\msf{tg,PQ}}$}}
The tag space is defined to be $[n]$ where $n$ is a polynomial on $\secpar$. Let $t \in [n]$ be the tag for the following interaction. Let $m$ be the message to be committed to.

\para{Commit Stage:}
\begin{enumerate}
\item\label[Step]{pq:item:one-sided:step:Naor-rho}
Receiver $R$ samples and sends the first message $\beta$ for Naor's commitment; 
\item\label[Step]{pq:item:one-sided:step:committing}
Committer $C$ commits to $m$ using the second message of Naor's commitment. Formally, $C$ samples a random tape $r$ and sends $\msf{com} = \Com_\beta(m;r)$;
\item\label[Step]{pq:item:one-sided:step:OWFs}
$R$ computes $\Set{y_i = f(x_i)}_{i \in [t]}$ with $x_i \pick \bits^\secpar$ for each $i \in [t]$. $R$ sends $Y = (y_1, \ldots, y_t)$ to $C$;
\item\label[Step]{pq:item:one-sided:step:WIAoK:1}
{\bf (WIAoK-1.)} $R$ and $C$ execute an instance of $\WIAoK$ where $R$ proves to $C$ that he ``knows'' a pre-image of some $y_i$ contained in $Y$ (defined in \Cref{pq:item:one-sided:step:OWFs}). Formally, $R$ proves that $Y \in \Lang^t_f$, where \begin{equation}\label[Language]{pq:eq:one-sided:Lang:OWF}
\Lang^t_f \coloneqq \Set{(y_1, \ldots, y_t)~|~\exists (i, x_i) ~s.t.~ i \in [t] \wedge y_i = f(x_i)}.
\end{equation}
Note that $R$ uses $(1,x_1)$ as the witness when executing this $\WIAoK$.
\item\label[Step]{pq:item:one-sided:step:WIAoK:2}
{\bf (WIAoK-2.)} $C$ and $R$ execute an instance of $\WIAoK$ where $C$ proves to $R$ that he ``knows'' {\em either} the message committed in $\msf{com}$ (defined in \Cref{pq:item:one-sided:step:committing}), {\em or} a pre-image of some $y_i$ contained in $Y$ (defined in \Cref{pq:item:one-sided:step:OWFs}). Formally, $C$ proves that $(\msf{com}, Y) \in \Lang_\beta \vee \Lang^t_f$, where $\Lang_\beta \vee \Lang^t_f$ denotes the  OR-composed language (as per \Cref{def:OR-Comp}), $\Lang^t_f$ was defined in \Cref{pq:eq:one-sided:Lang:OWF} and 
\begin{equation}\label[Language]{pq:eq:one-sided:Lang:Com}
\Lang_\beta\coloneqq \Set{\msf{com} ~|~ \exists (m, r)~s.t.~ \msf{com} = \Com_\beta(m;r)}.
\end{equation}
Note that $C$ uses the $(m, r)$ defined in \Cref{pq:item:one-sided:step:committing} as the witness when executing this $\WIAoK$.
\end{enumerate}
\para{Decommit Stage:} 
$C$ sends $(m, r)$. $R$ accepts if $\msf{com} = \Com_\beta(m;r)$, and rejects otherwise.
\end{ProtocolBox}

Remark that the above protocol is exactly the same as \Cref{prot:one-sided:classical} except that we use post-quantum WIAoK with $\epsilon$-close emulation and assume post-quantum security for all other  building-blocks.

\para{Security.} Completeness is straightforward from the description of \Cref{pq:prot:one-sided:classical}. 
The statistical binding property follows from that of Naor's commitment. Computational hiding of any non-malleable commitment follows from its non-malleability. So, we only need to show that \Cref{pq:prot:one-sided:classical} is non-malleable. This is established by the following \Cref{pq:thm:one-sided:non-malleability}, which we prove in subsequent subsections.

\begin{theorem}\label{pq:thm:one-sided:non-malleability}
The commitment scheme $\langle C, R\rangle^{\msf{OneSided}}_{\msf{tg,PQ}}$ in \Cref{pq:prot:one-sided:classical} is non-malleable against  one-sided synchronous QPT adversaries with tag space $[n]$, with $n$ being any polynomial on $\secpar$.
\end{theorem}	

\subsection{Proving Non-Malleability}
\label{pq:one-sided:non-malleability:proof:classical}

% To prove \Cref{pq:thm:one-sided:non-malleability}, we show that
% \begin{equation}\label[Expression]{pq:eq:one-sided:goal}
% \Set{\msf{mim}^\mcal{M}_{\langle C, R \rangle}(\secpar, m_0, z)}_{\secpar, m_0, m_1, z} \cind \Set{\msf{mim}^\mcal{M}_{\langle C, R \rangle}(\secpar, m_1, z)}_{\secpar, m_0, m_1, z},
% \end{equation}
% where $\secpar \in \Naturals$, $m_0, m_1 \in \bits^{\secpar}$, and $z \in \bits^*$.

We prove \Cref{pq:thm:one-sided:non-malleability} in \Cref{pq:one-sided:non-malleability:proof:classical,pq:sec:one-sided:core-lemma:proof,pq:lem:small-tag:proof:se:proof,pq:sec:lem:small-tag:proof:se:proof:K:proof,pq:sec:lem:bound:Ki:proof,pq:sec:proof:claim:K'':non-abort}.
The proof follows the same template used in \Cref{one-sided:non-malleability:proof:classical}.
%Almost all steps in the proof of \Cref{pq:thm:one-sided:non-malleability} are directly translated into the post-quantum setting. 
The only exception is the proof of \Cref{pq:lem:small-tag:proof:se} (which is the post-quantum counterpart of \Cref{lem:small-tag:proof:se}). Recall that in \Cref{lem:small-tag:proof:se}, we amplify the extractor $\mcal{K}$ to the simulation-extractor $\mcal{SE}$ by rewinding. In the current post-quantum setting, we cannot rewind quantum algorithms in general. Therefore, we rely on an alternative argument based on our new extract-and-simulate lemma~(\Cref{lem:extract_and_simulate}).
The rest of the proof is almost the same as its classical counterpart; Thus, many parts of the proof are taken verbatim from there. Essentially, the only difference is that we have to deal with noticeable errors that come from witness-extended $\epsilon$-close emulator (as per \Cref{item:PQWEE} of \Cref{def:PQWIAoK}). We highlight differences from the classical counterpart in {\color{Plum} purple color} throughout this section.

We use the same notation as in the proof of \Cref{thm:one-sided:non-malleability}, with the only difference that now the adversaries are non-uniform QPT (instead of PPT) machines. It is worth noting that the honest committer and receiver are still classical (i.e., non-uniform PPT) machines.

In the sequel, we write $\langle C, R \rangle$ to mean $\langle C, R\rangle^{\msf{OneSided}}_{\msf{tg,PQ}}$ for notational convenience. 
%To prove \Cref{pq:thm:one-sided:non-malleability}, let us first define the game that captures the man-in-the-middle execution corresponding to $\bar{\msf{mim}}^\mcal{M}_{\langle C, R \rangle}(\secpar, m, z)$ w.r.t.\ the $\langle C, R \rangle$ defined in \Cref{pq:prot:one-sided:classical}.

\para{Game $H^{\mcal{M}_\secpar}(\secpar,m,\rho_\secpar)$:\label{pq:gameH:description}} This game is identical to its classical counterpart defined on \Cpageref{gameH:description}, except that $\mcal{M} = \Set{\mcal{M}_\secpar, \rho_\secpar}_{\secpar\in \Naturals}$ now is a (non-uniform) QPT machine.
That is, this is the man-in-the-middle execution of the commit stage of the $\langle C, R \rangle$ defined in \Cref{pq:prot:one-sided:classical}, where the left committer commits to $m$ and $\mcal{M}$'s non-uniform advice is $\rho_\secpar$. 
The output of $H^{\mcal{M}_\secpar}(m,\rho_\secpar)$ is again defined to be $(\OUT_{\mcal{M}}, \tilde{\tau}, b)$, where $\OUT_{\mcal{M}}$ is the (quantum) output of $\mcal{M}_\secpar(\rho_\secpar)$ at the end of this game, $\tilde{\tau}$ consists of the \Cref{pq:item:one-sided:step:Naor-rho,pq:item:one-sided:step:committing} messages exchanged in the right session, and $b \in \Set{\top, \bot}$ is the honest receiver's final decision.

\if0
\para{Game $H^{\mcal{M}}(\secpar,m,z)$:\label{pq:gameH:description}} This is the man-in-the-middle execution of the commit stage of the $\langle C, R \rangle$ defined in \Cref{pq:prot:one-sided:classical}, where the left committer commits to $m$ and $\mcal{M}$'s non-uniform advice is $z$. The output of this game consists of the following three parts:
\begin{enumerate}
\item
$\OUT_{\mcal{M}}$: this is the output of $\mcal{M}$ at the end of this game;
\item
$\tilde{\tau}$: this is defined to be $\tilde{\tau} \coloneqq (\tilde{\beta}, \tilde{\msf{com}})$, where (i.e., $\tilde{\beta}$ and $\tilde{\msf{com}}$ are the \Cref{pq:item:one-sided:step:Naor-rho} and \Cref{pq:item:one-sided:step:committing} messages exchanged between $\mcal{M}$ and the honest receiver $R$;
\item
$b \in \Set{\top, \bot}$: this is the output of the honest receiver $R$, indicating if the man-in-the-middle's commitment (i.e., the right session) is accepted ($b = \top$) or not ($b = \bot$).
\end{enumerate}
\fi

\para{Notation.} 
Recall from \Cref{def:NMCom:pq} that the man-in-the middle game for a QPT adversary $\mcal{M}$ is denoted by $\bar{\msf{mim}}^{\mcal{M}_\secpar}_{\langle C, R \rangle}(m, \rho_\secpar)$. 
Also, note that ``$\cind$'' refers to {\em quantumly} computational indistinguishability throughout this section.

For any $(\OUT_{\mcal{M}}, \tilde{\tau}, b)$ in the support of $H^{\mcal{M}_\secpar}(\secpar,m,z)$, we define $\msf{val}_b(\tilde{\tau})$ similarly to the classical case, i.e., 
$$
\msf{val}_b(\tilde{\tau}) \coloneqq 
\begin{cases}
\msf{val}(\tilde{\tau}) & b = \top\\
\bot & b = \bot
\end{cases},
$$
where $\msf{val}(\tilde{\tau})$ denote the value statistically-bound in \Cref{pq:item:one-sided:step:Naor-rho,pq:item:one-sided:step:committing} of the right session. %(Recall that these two steps constitute a Naor's commitment.)

Similarly to \Cref{eq:classical:mim-H}, we have
\begin{align}
\big\{\bar{\msf{mim}}^{\mcal{M}_\secpar}_{\langle C, R \rangle}(m, \rho_\secpar)\big\} \idind \big\{\big(\OUT, \msf{val}_b(\tilde{\tau})\big): (\OUT, \tilde{\tau}, b) \gets H^{\mcal{M}_\secpar}(m,\rho_\secpar) \big\},
\end{align}
where both ensembles are indexed by $\secpar \in \Naturals$ and $m \in \bits^{\ell(\secpar)}$.

Then, similarly to \Cref{eq:classical:H:m-0:m-1}, the post-quantum non-malleability can be reduced to establishing the following equation:
\begin{align*}
&\big\{\big(\OUT^0, \msf{val}_{b^0}(\tilde{\tau}^0)\big): (\OUT^0, \tilde{\tau}^0, b^0) \gets H^{\mcal{M}_\secpar}(m_0,\rho_\secpar) \big\} \\
\cind ~& 
\big\{\big(\OUT^1, \msf{val}_{b^1}(\tilde{\tau}^1)\big): (\OUT^1, \tilde{\tau}^1, b^1) \gets H^{\mcal{M}_\secpar}(m_1,\rho_\secpar) \big\} \numberthis \label{pq:eq:classical:H:m-0:m-1},
\end{align*}
where both ensembles are indexed by $\secpar \in \Naturals$ and $(m_0, m_1) \in \bits^{\ell(\secpar)} \times \bits^{\ell(\secpar)}$.

\para{Proof by Contradiction.} Similarly to \Cref{eq:one-sided:proof:contra-assump}, we assume for contradiction that there are a (possibly non-uniform) QPT distinguisher $\mcal{D} = \Set{\mcal{D}_\secpar, \sigma_\secpar}_{\secpar \in \Naturals}$ and a function $\delta(\secpar) = 1/\poly(\secpar)$ such that for infinitely many $\secpar \in \Naturals$, it holds that
\begin{equation}\label[Inequality]{pq:eq:one-sided:proof:contra-assump}
\bigg|\Pr[\mcal{D}_\secpar\big(\OUT^0, \msf{val}_{b^0}(\tilde{\tau}^0); \sigma_\secpar\big)=1] - \Pr[\mcal{D}_\secpar\big(\OUT^1, \msf{val}_{b^1}(\tilde{\tau}^1); \sigma_\secpar\big)=1] \bigg|\ge 3 \cdot \delta(\secpar),
\end{equation}
where the first probability is taken over the random procedure $(\OUT^0, \tilde{\tau}^0, b^0) \gets H^{\mcal{M}_\secpar}(m_0,\rho_\secpar)$, %\xiao{Should I say ``and the randomness due to the measurement performed by $\mcal{D}_\secpar$ if any''}, 
and the second probability is taken over the random procedure $(\OUT^1, \tilde{\tau}^1, b^1) \gets H^{\mcal{M}_\secpar}(m_1,\rho_\secpar)$ (and the randomness due to the measurements performed by $\mcal{D}_\secpar$ (if any) for both probabilities).

Then, we show the following \Cref{pq:lem:one-sided:proof:core}, which should be understood as the post-quantum counterpart of \Cref{lem:one-sided:proof:core}. 
\begin{lemma}\label{pq:lem:one-sided:proof:core}
For the above $\delta(\secpar)$,
there exits a hybrid $G$ such that for any QPT $\mcal{M}=\Set{\mcal{M}_\secpar, \rho_\secpar}_{\secpar\in \Naturals}$, the following holds
\begin{enumerate}
\item\label[Property]{pq:item:lem:one-sided:proof:core:1}
$\big\{(\OUT^0, \Val^0):(\OUT^0, \Val^0) \gets G^{\mcal{M}_\secpar}(\secpar,m_0,\rho_\secpar) \big\} \cind \big\{(\OUT^1, \Val^1):(\OUT^1, \Val^1) \gets G^{\mcal{M}_\secpar}(\secpar,m_1,\rho_\secpar) \big\}$,
where both ensembles are indexed by $\secpar \in \Naturals$ and $(m_0, m_1) \in \bits^{\ell(\secpar)} \times \bits^{\ell(\secpar)}$.
\item\label[Property]{pq:item:lem:one-sided:proof:core:2}
\begingroup\fontsize{9.5pt}{0}\selectfont
$\big\{(\OUT^{G}, \Val^{G}):(\OUT^{G}, \Val^{G}) \gets G^{\mcal{M}_\secpar}(\secpar,m,\rho_\secpar) \big\}
\cind_{\delta(\secpar)} 
\big\{\big(\OUT^{H}, \msf{val}_{b^{H}}(\tilde{\tau}^{H})\big): (\OUT^{H}, \tilde{\tau}^{H}, b^{H}) \gets H^{\mcal{M}_\secpar}(\secpar,m,\rho_\secpar) \big\}$,
\endgroup  
where both ensembles are indexed by $\secpar \in \Naturals$ and $m \in \bits^{\ell(\secpar)}$.
\end{enumerate}
\end{lemma}

It is easy to see that if \Cref{pq:lem:one-sided:proof:core} is true, it contradicts our assumption in \Cref{pq:eq:one-sided:proof:contra-assump}. Therefore, it will finish the proof of non-malleability. Indeed, this lemma is the most technically involved part. We prove it in \Cref{pq:sec:one-sided:core-lemma:proof}.

\subsection{Proof of \Cref{pq:lem:one-sided:proof:core}}
\label{pq:sec:one-sided:core-lemma:proof} 
Similarly to \Cref{hybrid:H:reinterpretation}, we provide a new but equivalent interpretation of the game $H^{\mcal{M}_\secpar}(\secpar, m, \rho_\secpar)$ in \Cref{pq:hybrid:H:reinterpretation}. We also provide a picture in \Cref{pq:figure:one-sided:H:re-interpretation} to illustrate it.
\begin{AlgorithmBox}[label={pq:hybrid:H:reinterpretation}]{Re-interpretation of Game \textnormal{$H^{\mcal{M}_\secpar}(\secpar, m, \rho_\secpar)$}}
Game $H^{\mcal{M}_\secpar}(\secpar, m, \rho_\secpar)$ can be split into the following stages:
\begin{enumerate}
\item \label[Stage]{pq:hybrid:H:reinterpretation:1}
{\bf Prefix Generation:}
First, execute \Cref{pq:item:one-sided:step:Naor-rho,pq:item:one-sided:step:committing} of the man-in-the-middle game of \Cref{pq:prot:one-sided:classical}. That is, it plays as the left honest committer committing to $m$ and the right honest receiver, with $\mcal{M}_{\secpar}(\rho_\secpar)$ being the man-in-the-middle adversary.

\subpara{Notation:} Let $\msf{st}_{\mcal{M}}$ denote the state of $\mcal{M}$ at the end of \Cref{pq:item:one-sided:step:committing}; Let $\msf{st}_C$  (resp.\ $\msf{st}_R$) denote the state of the honest committer (resp.\ receiver) at the end of \Cref{pq:item:one-sided:step:committing}; Let $\tilde{\tau}$ denote the tuple $(\tilde{\beta}, \tilde{\msf{com}})$\footnote{Recall that $\tilde{\beta}$ and  $\tilde{\msf{com}}$ are the \Cref{pq:item:one-sided:step:Naor-rho,pq:item:one-sided:step:committing} messages in the right session; they constitutes an execution of Naor's commitment.}. In terms of notation, we denote the execution of this stage by 
\begin{equation}\label[Expression]{pq:expression:hybrid:H:prefix}
(\msf{st}_{\mcal{M}},  \msf{st}_C, \msf{st}_R, \tau,\tilde{\tau})\gets H^{\mcal{M}_\secpar}_{\msf{pre}}(\secpar, m, \rho_\secpar).
\end{equation}
We will call the tuple $(\msf{st}_{\mcal{M}},  \msf{st}_C, \msf{st}_R, \tau,\tilde{\tau})$ the {\em prefix} and denote it by $\msf{pref}$. It is worth noting that this $\msf{pref}$ contains all the information such that a QPT machine can ``complete'' the remaining execution of $H^{\mcal{M}_\secpar}(\secpar, m, \rho_\secpar)$ starting from $\msf{pref}$.

\item \label[Stage]{pq:stage:hybrid:H:remainder}
{\bf The Remainder:}
Next, it simply resumes from where the {\bf Prefix Generation} stage stops, to finish the remaining steps of the man-in-the-middle execution $H^{\mcal{M}_\secpar}(\secpar, m, \rho_\secpar)$.

\subpara{Notation:} We introduce the following notations to describe this stage. Define a QPT machine $\Adv$ that takes as input $(\msf{st}_{\mcal{M}}, \tilde{\tau})$; Machine $\mcal{A}_\secpar$ is supposed to run the residual strategy of $\mcal{M}_\secpar$ starting from $\msf{st}_{\mcal{M}}$. Also, define a QPT machine $\mcal{B}$ that takes as input $(\msf{st}_C, \msf{st}_R, \tilde{\tau})$; Machine $\mcal{B}$ is supposed to run the residual strategies of the honest committer $C$ and receiver $R$, starting from 
$\msf{st}_C$ and $\msf{st}_R$ respectively\footnote{Note that it is not necessary to give $\tilde{\tau}$ as common input to these parties; indeed, it can be included in their respective internal states. We choose to make $\tilde{\tau}$ explicit only to match the syntax of \Cref{lem:extract_and_simulate}.}. With the above notations, we can denote the execution of the remaining steps of $H^{\mcal{M}}(\secpar, m,z)$ by
\begin{equation}\label[Expression]{pq:expression:hybrid:H:remainder}
(\OUT_\mcal{A}, b)\gets \langle \mcal{A}_\secpar(\msf{st}_{\mcal{M}}), \mcal{B}(\msf{st}_C, \msf{st}_R) \rangle(1^\secpar, \tilde{\tau}),
\end{equation}
where $\OUT_\mcal{A}$ is the output of $\mcal{A}_\secpar$, and $b \in \Set{\bot, \top}$ is the output of the honest receiver $R$ (in the right), indicating if the man-in-the-middle's commitment (i.e., the right session) is accepted ($b = \top$) or not ($b = \bot$). (We remark that $\OUT_\Adv$ is nothing but the man-in-the-middle $\mcal{M}$'s final output.)

\item
{\bf Output:} It outputs the tuple $(\OUT_\Adv, \tilde{\tau}, b)$.  
\end{enumerate}
\end{AlgorithmBox}

%\xiao{Say that this is the major lemma for this proof. }
We prove the following lemma.
\begin{lemma}\label{pq:lem:small-tag:proof:se}
Let $H^{\mcal{M}_\secpar}_{\msf{pre}}(\secpar, m, \rho_\secpar)$, $\mcal{A}_\secpar$, and $\mcal{B}$ be as defined in \Cref{pq:hybrid:H:reinterpretation}.
There exists a QPT machine $\mcal{SE}$ (the simulation-extractor) such that for any $(\msf{st}_{\mcal{M}},  \msf{st}_C, \msf{st}_R, \tau,\tilde{\tau})$ in the support of $H^{\mcal{M}_\secpar}_{\msf{pre}}(\secpar, m, \rho_\secpar)$, any noticeable $\epsilon(\secpar)$, it holds that
\begin{align*}
& \big\{ (\OUT_{\mcal{SE}}, \msf{Val}_{\mcal{SE}}) : (\OUT_{\mcal{SE}}, \msf{Val}_{\mcal{SE}}) \gets \SimExt(1^\secpar,1^{\epsilon^{-1}},\A_\secpar,\msf{st}_{\mcal{M}},\msf{st}_{R},\tau, \tilde{\tau})\big\}_{\secpar\in \Naturals} \\
\cind_{\epsilon(\secpar)} ~&
\big\{\big(\OUT_\mcal{A}, \msf{val}_b(\tilde{\tau})\big) : (\OUT_\mcal{A}, b) \gets \langle \mcal{A}_\secpar(\msf{st}_{\mcal{M}}), \mcal{B}(\msf{st}_C, \msf{st}_R) \rangle(1^\secpar, \tilde{\tau})\big\}_{\secpar\in \Naturals} 
\end{align*}
%\takashi{$b_{\mcal{SE}}$ seems redundant. It will make more sense to just remove it or otherwise include $b_{\mcal{SE}}$ and $b$ in the LHS and RHS respectively. } \xiao{Actually, I do need $b_{\mcal{SE}}$ for notational convenience (e.g., in \Cref{pq:eq:one-sided:proof:se:itself}, \Cref{pq:claim:bounding-E1E2,bound:Val-ne-m}). Without this $b_{\mcal{SE}}$, I need to explain several points in words. We can discuss if there is a better solution.}\takashi{I still prefer omitting $b_{\mcal{SE}}$ from the statement of this lemma and introducing it in the proof of this lemma for consistency to the extract-and-simulate lemma. We may explain that  we use the notation $(\OUT_{\mcal{SE}}, \msf{Val}_{\mcal{SE}}, b_{\mcal{SE}}) \gets \SimExt^{\A(\msf{st}_{\mcal{M}})}(1^\secpar,1^{\epsilon^{-1}})$ for convenience even though $b_{\mcal{SE}}$ is not part of the output.} \xiao{After I fixed the bug, I feel it is convenient to have the $b_{\mcal{SE}}$ here, because I used it in several places (due to my way to fix the bug). So, let's talk about what to do with this $b_{\mcal{SE}}$.}
\end{lemma}
{\color{Plum}
\begin{remark}\label{rmk:input_instead_of_oracle}
Compared to the classical counterpart (\Cref{lem:small-tag:proof:se}), $\mcal{SE}$ takes $\mcal{A}_\secpar$ and $\msf{st}_{\mcal{M}}$ as part of its input instead of accessing $\mcal{A}_\secpar(\msf{st}_{\mcal{M}})$. 
%This is due to a subtlety in the definition of oracle access to quantum algorithms. 
Though we can see that $\mcal{SE}$ actually makes only black-box use of $\mcal{A}_\secpar(\msf{st}_{\mcal{M}})$ in a certain sense, we do not try to formally state it because black-box simulation is not our focus. 
Another minor difference is that we removed $b_{\mcal{SE}}$ from the output of $b_{\mcal{SE}}$. This is because $b_{\mcal{SE}}$ was only used in the proof of the classical counterpart of \Cref{pq:lem:small-tag:proof:se}  (i.e., \Cref{lem:small-tag:proof:se}), which will be replaced with a different proof based on \Cref{lem:extract_and_simulate}.   
\end{remark}
}

\Cref{pq:lem:small-tag:proof:se} is the main technical lemma for the current proof of \Cref{pq:lem:one-sided:proof:core}. We present its proof in \Cref{pq:lem:small-tag:proof:se:proof}. In the following, we finish the proof of \Cref{pq:lem:one-sided:proof:core}  assuming that \Cref{pq:lem:small-tag:proof:se} is true.

%\xiao{Say that \Cref{pq:lem:small-tag:proof:se} is the main technical lemma for the current proof of \Cref{pq:item:lem:one-sided:proof:core:2}. We present its proof in \Cref{pq:lem:small-tag:proof:se:proof}. In the following, we finish the proof of \Cref{pq:lem:one-sided:proof:core}  assuming that \Cref{pq:lem:small-tag:proof:se} is true.}

With \Cref{pq:lem:small-tag:proof:se}, we are now ready to present the description of $G$.
\begin{AlgorithmBox}[label={pq:hybrid:G}]{Hybrid \textnormal{${G}^{\mcal{M}_\secpar}(\secpar, m, \rho_\secpar)$}}
This hybrid proceeds as follows:
\begin{enumerate}
\item
{\bf Prefix Generation:}
This stage is identical to \Cref{pq:hybrid:H:reinterpretation:1} of $H^{\mcal{M}_\secpar}(\secpar, m, \rho_\secpar)$. Formally, it executes 
$$(\msf{st}_{\mcal{M}},  \msf{st}_C, \msf{st}_R, \tau,\tilde{\tau}) \gets H^{\mcal{M}_\secpar}_{\msf{pre}}(\secpar, m, \rho_\secpar),$$
where $H^{\mcal{M}_\secpar}_{\msf{pre}}(\secpar, m, \rho_\secpar)$ is defined in \Cref{pq:hybrid:H:reinterpretation:1} of \Cref{pq:hybrid:H:reinterpretation}.
\item
{\bf The Remainder:}
Define $\mcal{A}_\secpar$ in the same way as in \Cref{pq:stage:hybrid:H:remainder} of $H^{\mcal{M}_\secpar}(\secpar, m, \rho_\secpar)$. With this $\mcal{A}_\secpar$  and the $(\msf{st}_{\mcal{M}},  \msf{st}_R, \tau,\tilde{\tau})$ from the previous stage, ${G}^{\mcal{M}_\secpar}(\secpar, m, \rho_\secpar)$ invokes the $\mcal{SE}$ prescribed in \Cref{pq:lem:small-tag:proof:se}. Formally, it executes the following procedure:
$$(\OUT_{\mcal{SE}}, \msf{Val}_{\mcal{SE}}) \gets \SimExt(1^\secpar, 1^{\delta^{-1}},\A_\secpar,\msf{st}_{\mcal{M}},\msf{st}_R,\tau,\tilde{\tau}),$$
where the $\delta$ is the statistical distance that we want to show for \Cref{pq:item:lem:one-sided:proof:core:2} of \Cref{pq:lem:one-sided:proof:core}.

\begin{remark}
We emphasize that in this stage, ${G}^{\mcal{M}_\secpar}(\secpar, m, \rho_\secpar)$ does {\em not} make use of $\msf{st}_C$.
\end{remark}
\item
{\bf Output:}
% $\bar{G}^{\mcal{M}_\secpar}(m,\beta_\secpar)$ sets $\OUT_{\bar{G}}\coloneqq \OUT_{\mcal{SE}}$ and $ \msf{Val}_{\bar{G}}\coloneqq \msf{Val}_{\mcal{SE}}$, and 
It outputs $(\OUT_{\mcal{SE}}, \msf{Val}_{\mcal{SE}})$.
%\footnote{Note that $b_{\mcal{SE}}$ is not included in the output of $G$. Indeed, $b_{\mcal{SE}}$ is not important for the current machine $G^{\mcal{M}_\secpar}(\secpar, m, \rho_\secpar)$. We choose to include it in the output of $\SimExt^{\A(\msf{st}_{\mcal{M}})}(1^\secpar,1^{\delta^{-1}},\msf{st}_R,\tilde{\tau})$ only for notational convenience when we define/prove properties about $\mcal{SE}$ itself.}
\end{enumerate}
\end{AlgorithmBox}

\para{Proving \Cref{pq:item:lem:one-sided:proof:core:1} of \Cref{pq:lem:one-sided:proof:core}.} Observe that hybrid $G^{\mcal{M}_\secpar}(\secpar, m, \rho_\secpar)$ is an efficient machine, since both $H^{\mcal{M}_\secpar}_{\msf{pre}}(\secpar, m, \rho_\secpar)$ and $\SimExt(1^\secpar, 1^{\delta^{-1}},\A_\secpar,\msf{st}_{\mcal{M}},\msf{st}_R,\tau,\tilde{\tau})$ are efficient. Moreover, it does not rewind \Cref{pq:item:one-sided:step:Naor-rho,pq:item:one-sided:step:committing} of the man-in-the-middle execution, and $\SimExt(1^\secpar, 1^{\delta^{-1}},\A_\secpar,\msf{st}_{\mcal{M}},\msf{st}_R,\tau,\tilde{\tau})$ does {\em not} need to know $\msf{st}_C$. Therefore, \Cref{pq:item:lem:one-sided:proof:core:1} of \Cref{pq:lem:one-sided:proof:core} follows immediately from the computational-hiding property of the left Naor's commitment (i.e., \Cref{pq:item:one-sided:step:Naor-rho,pq:item:one-sided:step:committing} in the left session). 

\para{Proving \Cref{pq:item:lem:one-sided:proof:core:2} of \Cref{pq:lem:one-sided:proof:core}.} First, observe that the distribution of the prefix is identical in $G^{\mcal{M}_\secpar}(\secpar, m, \rho_\secpar)$ and $H^{\mcal{M}_\secpar}(\secpar, m, \rho_\secpar)$.
For each fixed prefix, \Cref{pq:lem:small-tag:proof:se} implies that $G^{\mcal{M}_\secpar}(\secpar, m, \rho_\secpar)$ and $H^{\mcal{M}_\secpar}(\secpar, m, \rho_\secpar)$ are $\delta(\secpar)$-computationally indistinguishable (notice that $G^{\mcal{M}_\secpar}(\secpar, m, \rho_\secpar)$ runs $\mcal{SE}$ with the second input $1^{\delta^{-1}}$). 
This immediately implies \Cref{pq:item:lem:one-sided:proof:core:2} of \Cref{pq:lem:one-sided:proof:core}.

\subsection{Proof of \Cref{pq:lem:small-tag:proof:se}}
\label{pq:lem:small-tag:proof:se:proof}
% \takashi{@Xiao Please make new figures where WIAoK is replaced with WIAoK and $H^{\mcal{M}}(\secpar,m,z)$ is replaced with $H^{\mcal{M}_\secpar}(\secpar,m,\rho_\secpar)$. I believe no other change is needed.}
\begin{figure}[!h]
     \begin{subfigure}[t]{0.47\textwidth}
         \centering
         \fbox{
         \includegraphics[width=\textwidth,page=1]{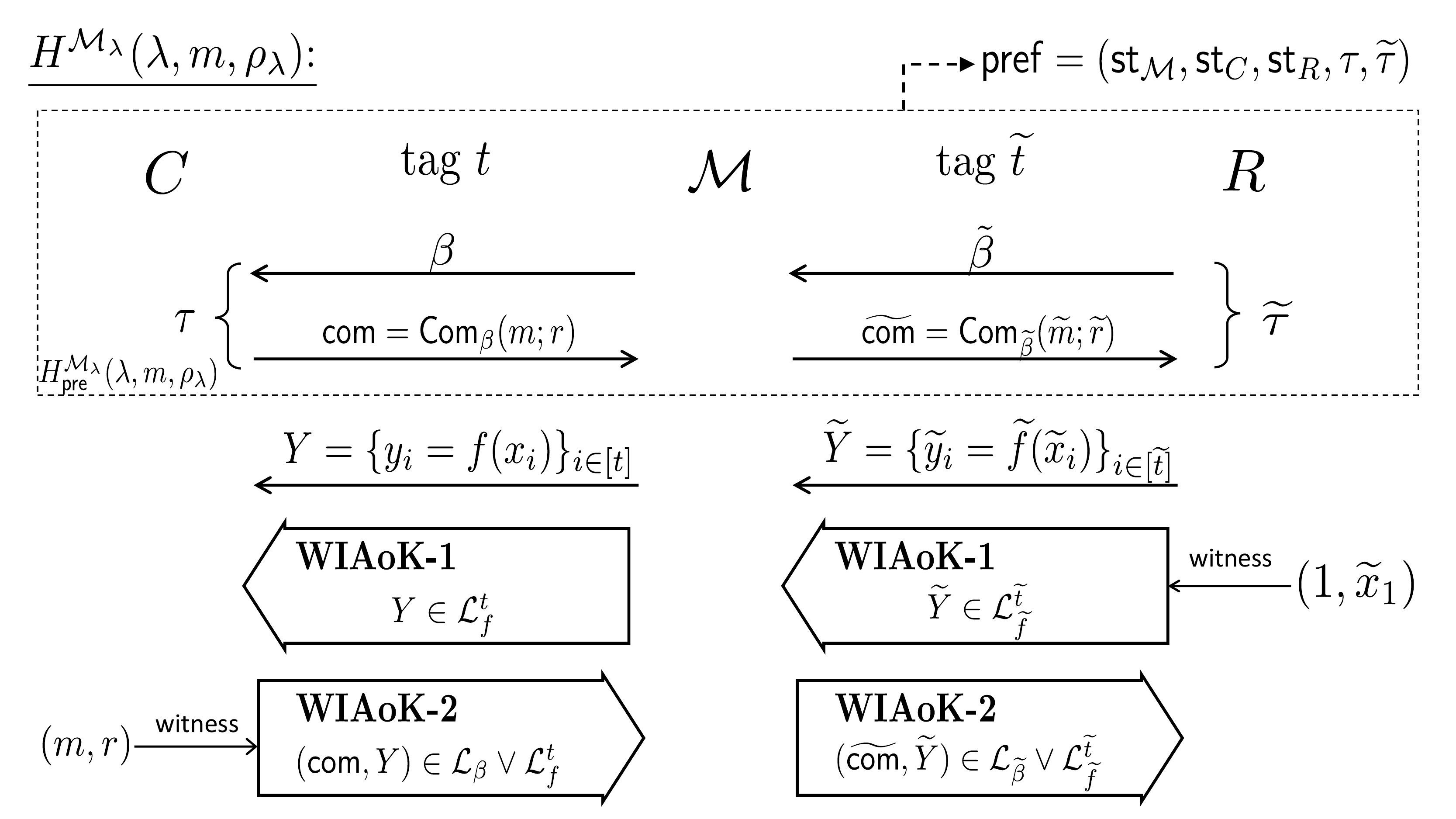}
         }
         \caption{}
         \label{pq:figure:one-sided:H:re-interpretation}
     \end{subfigure}
     \hspace{6.5pt}
     \begin{subfigure}[t]{0.47\textwidth}
         \centering
         \fbox{
         \includegraphics[width=\textwidth,page=2]{figures/figures-PQ.pdf}
         }
         \caption{}
         \label{pq:figure:one-sided:G1}
     \end{subfigure}
     \caption{Machines $H^{\mcal{M}}$ and $\mcal{G}_1$ {\scriptsize (Difference is highlighted in red color)}}
     \label{pq:figure:one-sided:games:M:G1}
\end{figure}

%\xiao{Provide some intuition. Say that our machine $\mcal{SE}$ will make use of two machines $\mcal{G}_1$ that simualtes the main-thread, and an extractor $\mcal{K}$ that extracts the value committed in $\tau$. Thus, in the following, we will first define these two machines, and the present the description of $\mcal{SE}$ (in \Cref{pq:machine:se}).}

%Our machine $\mcal{SE}$ (for \Cref{pq:lem:small-tag:proof:se}) will make use of two machines $\mcal{G}_1$ that simulates the main-thread, and a (simulation-less) extractor $\mcal{K}$ that extracts the value committed in $\tilde{\tau}$. 
%Thus, in the following, we will first define these two machines, and then present the description of $\mcal{SE}$ in \Cref{pq:machine:se}.

First, we describe a machine $\mcal{G}_1$ that simulates the real execution without using $\msf{st}_C$.   
{\color{Plum}
Unlike its classical counterpart, 
$\mcal{G}_1$ is parameterized by a noticeable function $\epsilon_\wiaok(\secpar)$, which is used as the error parameter for witness-extended emulators for $\WIAoK$ (as per \Cref{item:PQWEE} of \Cref{def:PQWIAoK}).  
}

\para{Machine $\mcal{G}_1{\color{Plum}[\epsilon_\wiaok]}$:} (Illustrated in \Cref{pq:figure:one-sided:G1}\label{pq:pageref:one-sided:proof:G1}) For any prefix $\msf{pref}$, $\mcal{G}_1[\epsilon_\wiaok](\msf{st}_{\mcal{M}}, \msf{st}_R, \tau,\tilde{\tau})$ behaves identically to \Cref{pq:stage:hybrid:H:remainder} of $H^{\mcal{M}_\secpar}(\secpar, m, \rho_\secpar)$ shown in \Cref{pq:hybrid:H:reinterpretation} (and depicted in \Cref{pq:figure:one-sided:H:re-interpretation}), except for the following difference: Instead of executing the left {\bf WIAoK-1} honestly, it uses the witness-extended emulator (as per \Cref{item:PQWEE} of \Cref{def:PQWIAoK}) of the left {\bf WIAoK-1} {\color{Plum} with the error parameter $\epsilon_\wiaok(\secpar)$} to extract a witness, and
\begin{itemize}
\item
 If the left committer accepts the left {\bf WIAoK-1} and the extracted witness is valid (i.e., it is a $(j, x_j)$ pair such that $y_j = f(x_j)$ for some $j \in [t]$), $\mcal{G}_1$ uses $(j, x_j)$ to finish the left {\bf WIAoK-2}. Similarly to $H^{\mcal{M}_\secpar}(\secpar, m, \rho_\secpar)$, $\mcal{G}_1$ eventually outputs $\mcal{M}$'s final state and the right receiver's decision bit $b$; 
 \item
 If the left committer accepts the left {\bf WIAoK-1}  but the extracted witness is invalid, it aborts immediately and outputs $(\bot, \bot)$.
 \item If the left committer rejects the left {\bf WIAoK-1}, it runs the rest of execution of $H^{\mcal{M}_\secpar}(\secpar, m, \rho_\secpar)$ to output $\mcal{M}$'s final state and the right receiver's decision bit $b$.  Note that it does not need to run the left {\bf WIAoK-2} in this case since the left committer aborts after the left {\bf WIAoK-1}. 
\end{itemize}
We denote the above procedure by $(\OUT, b) \gets \mcal{G}_1[\epsilon_\wiaok](\msf{st}_{\mcal{M}}, \msf{st}_R, \tau,\tilde{\tau})$. 
It is worth noting that $\mcal{G}_1$ does {\em not} need to know $\msf{st}_C$.

\if0
\begin{remark}[Precise Meaning of Extraction]\label{pq:rem:meaning_extraction}
In the description of $\mcal{G}_1$, we say that ``it uses the witness-extended emulator of the left {\bf WIAoK-1} to extract a witness". 
Precisely, this means the following: 
We consider a cheating prover against the left {\bf WIAoK-1} that takes the states of $\mcal{M}$ and the right receiver right before starting the left {\bf WIAoK-1} as advice, interacts with the left committer while simulating $\mcal{M}$ and the right receiver, and outputs the states of $\mcal{M}$ and the right receiver right after finishing the left {\bf WIAoK-1}. $\mcal{G}_1$ runs the witness-extended emulator of the left {\bf WIAoK-1} (as per \Cref{item:PQWEE} of \Cref{def:PQWIAoK}) w.r.t. the above described cheating prover to simulate the states of $\mcal{M}$ and the right receiver right after finishing the left {\bf WIAoK-1}, the decision bit of the left {\bf WIAoK-1} by the left committer as well as extracting a (candidate of) witness of $\Lang_f^t$.  
We use similar convention many times throughout the paper. 
\end{remark}
\fi

%\xiao{Next, prove a lemma relates the simulated main-thread with the real main-thread.} 
Next, we prove a lemma that shows that  $\mcal{G}_1[\epsilon_\wiaok]$ simulates the real execution {\color{Plum} up to an error $\epsilon_\wiaok$}. 
\begin{lemma}\label{pq:lem:main-thread:sim:real}
Let $H^{\mcal{M}_\secpar}_{\msf{pre}}(\secpar, m, \rho_\secpar)$, $\mcal{A}_\secpar$, and $\mcal{B}$ be as defined in \Cref{pq:hybrid:H:reinterpretation}. For any $\msf{pref}=(\msf{st}_{\mcal{M}},  \msf{st}_C, \msf{st}_R, \tau,\tilde{\tau})$ in the support of $H^{\mcal{M}_\secpar}_{\msf{pre}}(\secpar, m, \rho_\secpar)$, it holds that
\begin{align*}
& \big\{\big(\OUT, ~b \big): (\OUT, b) \gets \mcal{G}_1[\epsilon_\wiaok](\msf{st}_{\mcal{M}}, \msf{st}_R, \tau,\tilde{\tau}) \big\}_{\secpar \in \Naturals} \\
{\color{Plum}\cind_{\epsilon_\wiaok(\secpar)}} ~&
\big\{\big(\OUT_\mcal{A}, ~b\big) : (\OUT_\mcal{A}, b) \gets \langle \mcal{A}_\secpar(\msf{st}_{\mcal{M}}), \mcal{B}(\msf{st}_C, \msf{st}_R) \rangle(1^\secpar, \tilde{\tau})\big\}_{\secpar\in \Naturals}. 
\end{align*}
\end{lemma}
\begin{proof}
We first define an intermediate machine $\mcal{G}'_1  [\epsilon_\wiaok]$ below. 
\begin{figure}[!h]
\centering
\fbox{
     \includegraphics[width=0.55\textwidth,page=3]{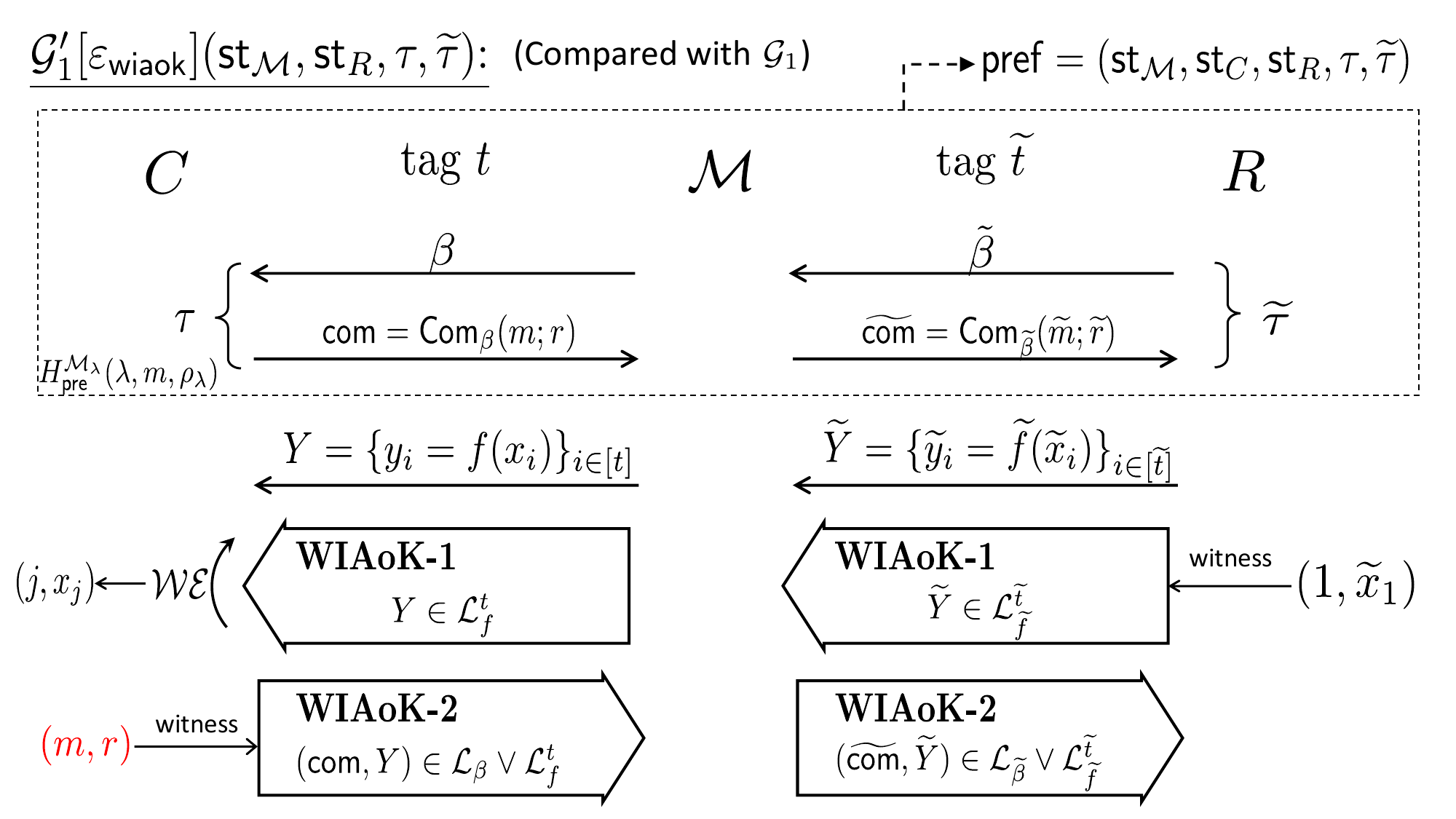}
     }
     \caption{Machine $\mcal{G}'_1$ {\scriptsize (Difference with $\mcal{G}_1$ is highlighted in red color)}}
     \label{pq:figure:one-sided:G'1}
\end{figure}

\para{Machine $\mcal{G}'_1{\color{Plum} [\epsilon_\wiaok]}$:} (Illustrated in \Cref{pq:figure:one-sided:G'1}\label{pq:pageref:one-sided:proof:G'1}) For any prefix $\msf{pref}$, $\mcal{G}'_1[\epsilon_\wiaok](\msf{st}_{\mcal{M}}, \msf{st}_R, \tau,\tilde{\tau})$ behaves identically to $\mcal{G}_1 [\epsilon_\wiaok](\msf{st}_{\mcal{M}}, \msf{st}_R, \tau,\tilde{\tau})$ except that it uses $(m,r)$ as the witness in the left {\bf WIAoK-2} even if it succeeds in extracting a valid witness from the left {\bf WIAoK-1}.

By the WI property of the left {\bf WIAoK-2}, it holds that 
\begin{align} 
\begin{split}\label{pq:eq:G1_to_G'1}
& \big\{\big(\OUT, ~b\big): (\OUT, b) \gets \mcal{G}_1[\epsilon_\wiaok](\msf{st}_{\mcal{M}}, \msf{st}_R, \tau,\tilde{\tau}) \big\}_{\secpar \in \Naturals} \\
\cind ~&
\big\{\big(\OUT, ~b\big): (\OUT, b) \gets \mcal{G}'_1[\epsilon_\wiaok](\msf{st}_{\mcal{M}}, \msf{st}_R, \tau,\tilde{\tau}) \big\}_{\secpar \in \Naturals}
\end{split}.
\end{align}

Note that the only difference between $\mcal{G}'_1[\epsilon_\wiaok](\msf{st}_{\mcal{M}}, \msf{st}_R, \tau,\tilde{\tau})$ and $\langle \mcal{A}_\secpar(\msf{st}_{\mcal{M}}), \mcal{B}(\msf{st}_C, \msf{st}_R) \rangle(1^\secpar, \tilde{\tau})$ is that the former runs the witness-extended emulator of the left {\bf WIAoK-1} {\color{Plum} with the error parameter $\epsilon_\wiaok(\secpar)$} (but does not use the extracted witness at all). Thus, by the AoK property of the left {\bf WIAoK-1}, it holds that 
\begin{align} 
\begin{split}\label{pq:eq:G'1_to_real}
& \big\{\big(\OUT, ~b\big): (\OUT, b) \gets \mcal{G}'_1[\epsilon_\wiaok](\msf{st}_{\mcal{M}}, \msf{st}_R, \tau,\tilde{\tau}) \big\}_{\secpar \in \Naturals} \\
{\color{Plum} \cind_{\epsilon_\wiaok(\secpar)}} ~&
\big\{\big(\OUT_\mcal{A}, ~b\big) : (\OUT_\mcal{A}, b) \gets \langle \mcal{A}_\secpar(\msf{st}_{\mcal{M}}), \mcal{B}(\msf{st}_C, \msf{st}_R) \rangle(1^\secpar, \tilde{\tau})\big\}_{\secpar\in \Naturals}
\end{split}.
\end{align}

By combining \Cref{pq:eq:G1_to_G'1,pq:eq:G'1_to_real}, we obtain \Cref{pq:lem:main-thread:sim:real}. 

\end{proof}

Next, we define the probability of $R$ being convinced in the execution of $\mcal{G}_1[\epsilon_\wiaok]$. This value plays an important role later in our proof.
\begin{definition}\label{pq:def:p-sim-pref}
For any $(\msf{st}_{\mcal{M}},  \msf{st}_C, \msf{st}_R, \tau,\tilde{\tau})$ in the support of $H^{\mcal{M}_\secpar}_{\msf{pre}}(\secpar, m, \rho_\secpar)$ {\color{Plum} and any noticeable $\epsilon_\wiaok$}, we define the following value $p^{\msf{Sim}}_{\msf{pref}}[\epsilon_\wiaok]$:
$$p^{\msf{Sim}}_{\msf{pref}}{\color{Plum}[\epsilon_\wiaok]} \coloneqq \Pr[b = \top : (\OUT, b) \gets \mcal{G}_1{\color{Plum}  [\epsilon_\wiaok]}(\msf{st}_\mcal{M}, \msf{st}_R, \tau,\tilde{\tau})]$$
\end{definition}

Next, we show a technical lemma that gives an extractor $\mcal{K}$ without the simulation property. 
\begin{lemma}\label{pq:lem:small-tag:proof:se:proof:K}
Let $H^{\mcal{M}_\secpar}_{\msf{pre}}(\secpar, m, \rho_\secpar)$, $\mcal{A}_\secpar$, $\mcal{B}$ be as defined in \Cref{pq:hybrid:H:reinterpretation}. There exists a QPT machine $\mcal{K}$ such that for any noticeable $\epsilon(\secpar)$, {\color{Plum} there is a noticeable $\epsilon_\wiaok(\secpar)\le \epsilon(\secpar)$ that is efficiently computable from $\epsilon(\secpar)$} such that the following holds  for any $\msf{pref}=(\msf{st}_{\mcal{M}},  \msf{st}_C, \msf{st}_R, \tau,\tilde{\tau})$:  \takashi{I think only considering honestly generated prefix is insufficient for using \Cref{lem:extract_and_simulate}. We should require the following hold for all $\st_{\mcal{M}}$ that may be outside the support. I believe this is already proven because we do not use any security of primitive in the prefix, but this should be carefully stated.
A simple fix may be to replace this with ``for any $\msf{pref}=(\msf{st}_{\mcal{M}},  \msf{st}_C, \msf{st}_R, \tau,\tilde{\tau})$ such that $\val(\tilde{\tau})\neq \bot$" (in this lemma and also other lemmas in this section.) But then we have to carefully re-examine that they really hold for an arbitrary prefix with $\val(\tilde{\tau})\neq \bot$ instead of honestly generated ones.}
\begin{enumerate}
\item \label[Property]{pq:property:small-tag:proof:se:proof:K:syntax}
{\bf (Syntax.)} $\mcal{K}$ takes as input $(1^\secpar,{\color{Plum} 1^{\epsilon^{-1}}}, \mcal{A}_{\secpar}, \msf{st}_{\mcal{M}}, \msf{st}_R, \tau,\tilde{\tau})$. It outputs a value $\msf{Val}_{\mcal{K}} \in \bits^{\ell(\secpar)} \cup \Set{\bot}$ such that $\msf{Val}_{\mcal{K}} = \msf{val}(\tilde{\tau})$ whenever $\msf{Val}_{\mcal{K}} \ne \bot$.
%(Also see \Cref{pq:rmk:small-tag:proof:se:proof:K:output} for an intuitive explanation.)

\item \label[Property]{pq:property:small-tag:proof:se:proof:K}
If $p^{\msf{Sim}}_{\msf{pref}}{\color{Plum}  [\epsilon_\wiaok]} \ge \epsilon(\secpar)$, then it holds that
$$\Pr[\Val_{\mcal{K}} = \msf{val}(\tilde{\tau}) : \Val_{\mcal{K}} \gets \mcal{K}(1^\secpar,{\color{Plum} 1^{\epsilon^{-1}}}, \mcal{A}_{\secpar}, \msf{st}_{\mcal{M}}, \msf{st}_R, \tau,\tilde{\tau})] \ge \frac{\epsilon'(\secpar)}{\tilde{t}},$$
where  $\epsilon'(\secpar) \coloneqq \frac{\epsilon(\secpar)}{10t^2}$.
\end{enumerate}
\end{lemma}
{\color{Plum}
\begin{remark}
Compared to the classical counterpart (\Cref{lem:small-tag:proof:se:proof:K}), $\mcal{K}$ takes $1^{\epsilon^{-1}}$ as an additional input. 
This is needed because we only assume the AoK property via witness-extended {\em $\epsilon$-close} emulation  (as per \Cref{item:PQWEE} of \Cref{def:PQWIAoK}). 
As a positive effect of the above difference, $\mcal{K}$ in \Cref{pq:lem:small-tag:proof:se:proof:K} runs in \emph{strict} QPT whereas it runs in \emph{expected} PPT in the classical counterpart (\Cref{lem:small-tag:proof:se:proof:K}). 
Also, we give $\mcal{A}_{\secpar}$ and $ \msf{st}_{\mcal{M}}$ as part of input to $\mcal{K}$ instead of giving oracle access to $\mcal{A}_{\secpar}(\msf{st}_{\mcal{M}})$ for a similar reason in \Cref{rmk:input_instead_of_oracle}.
\end{remark}
}

We will prove \Cref{pq:lem:small-tag:proof:se:proof:K} in \Cref{pq:sec:lem:small-tag:proof:se:proof:K:proof}.
In the rest of this subsection, we finish the proof of \Cref{pq:lem:small-tag:proof:se} assuming \Cref{pq:lem:small-tag:proof:se:proof:K} is true.
{\color{Plum} This is the only part of the proof of \Cref{pq:thm:one-sided:non-malleability} that significantly differs from its classical counterpart.
%Since this step is completely different from the classical counterpart, we omit highlighting differences by red color for this proof.

\begin{proof}[Proof of \Cref{pq:lem:small-tag:proof:se}]\label{pageref:different_part}
We apply \Cref{lem:extract_and_simulate} where $\mcal{G}_1[\epsilon_\wiaok]$, $\mcal{K}$, $(\A_\secpar,\msf{st}_{\mcal{M}},\msf{st}_R,\tau)$, $\tilde{\tau}$, and $\val({\tilde{\tau}})$ play the roles of $\mcal{G}$, $\mcal{K}$, $\rho_\secpar$, $z_\secpar$, and $s^*_{z_\secpar}$ in \Cref{lem:extract_and_simulate}, respectively. 
Then, \Cref{pq:lem:small-tag:proof:se:proof:K} ensures that the assumptions of \Cref{lem:extract_and_simulate} are satisfied.\footnote{Remark that $\epsilon(\secpar)$ and $\frac{\epsilon'(\secpar)}{\tilde{t}}$ in \Cref{pq:lem:small-tag:proof:se:proof:K} play the roles of $\gamma(\secpar)$ and  $\delta(\secpar)$ in \Cref{lem:extract_and_simulate}, respectively.
} Thus, there exists a polynomial $\poly$ and a QPT machine $\mcal{SE}'$ such that for any noticeable function $\bar{\epsilon}(\secpar)$,  
\begin{align}
\begin{split}\label{pq:eq:ind_SE_and_G1}
& \big\{ (\OUT_{\mcal{SE}'}, \msf{Val}_{\mcal{SE}'}) : (\OUT_{\mcal{SE}'}, \msf{Val}_{\mcal{SE}'}) \gets \SimExt'(1^\secpar,1^{\bar{\epsilon}^{-1}},\A_\secpar,\msf{st}_{\mcal{M}},\msf{st}_{R},\tau, \tilde{\tau})\big\}_{\secpar\in \Naturals} \\
\sind_{\bar{\epsilon}(\secpar)} ~&
\big\{\big(\OUT, \msf{val}_b(\tilde{\tau})\big) : (\OUT, b) \gets \mcal{G}_1[\epsilon_\wiaok](\msf{st}_{\mcal{M}}, \msf{st}_R, \tau,\tilde{\tau}) \big\}_{\secpar\in \Naturals} 
\end{split}
\end{align}
for a noticeable function $\epsilon_\wiaok(\secpar)\le \bar{\epsilon}(\secpar)$. 

Moreover, since $\val_b(\tilde{\tau})$ is determined by $b$ for each fixed $\tilde{\tau}$, \Cref{pq:lem:main-thread:sim:real} directly implies 
\begin{align} 
\begin{split}\label{pq:eq:ind_G1_and_real}
& \big\{\big(\OUT, ~\val_b(\tilde{\tau})\big): (\OUT, b) \gets \mcal{G}_1[\epsilon_\wiaok](\msf{st}_{\mcal{M}}, \msf{st}_R, \tau,\tilde{\tau}) \big\}_{\secpar \in \Naturals} \\
\cind_{\epsilon_\wiaok(\secpar)} ~&
\big\{\big(\OUT_\mcal{A}, ~\val_b(\tilde{\tau})\big) : (\OUT_\mcal{A}, b) \gets \langle \mcal{A}_\secpar(\msf{st}_{\mcal{M}}), \mcal{B}(\msf{st}_C, \msf{st}_R) \rangle(1^\secpar, \tilde{\tau})\big\}_{\secpar\in \Naturals}
\end{split}.
\end{align}
By setting $\bar{\epsilon}(\secpar):=\epsilon(\secpar)/2$ 
%By taking $\bar{\epsilon}(\secpar)$ to be sufficiently small 
so that $\bar{\epsilon}(\secpar) + \epsilon_\wiaok(\secpar)\leq \bar{\epsilon}(\secpar) +\bar{\epsilon}(\secpar) \le \epsilon(\secpar)$ and defining $\mcal{SE}(1^\secpar,1^{\epsilon^{-1}},...)\defeq \mcal{SE}'(1^\secpar,1^{\bar{\epsilon}^{-1}},...)$, \Cref{pq:eq:ind_SE_and_G1,pq:eq:ind_G1_and_real} give \Cref{pq:lem:small-tag:proof:se}. 
%combining \Cref{pq:eq:ind_SE_and_G1,pq:eq:ind_G1_and_real} and $\epsilon_\wiaok(\secpar)\leq \epsilon(\secpar)$, we obtain \Cref{pq:lem:small-tag:proof:se}.  

\end{proof}

}

\subsection{Extractor $\mcal{K}$ (Proof of \Cref{pq:lem:small-tag:proof:se:proof:K})}
\label{pq:sec:lem:small-tag:proof:se:proof:K:proof}
{\color{Plum} In the following, we fix a noticeable function $\epsilon(\secpar)$ for which we want to prove \Cref{pq:lem:small-tag:proof:se:proof:K}. We show that it suffices to set $\epsilon_\wiaok(\secpar)\defeq \frac{t+1}{t^2+6t+3}\cdot \epsilon'(\secpar)=\frac{t+1}{10t^2(t^2+6t+3)}\cdot \epsilon(\secpar)$. 
Since we fix $\epsilon$ and $\epsilon_\wiaok$, we omit the dependence on $\epsilon_\wiaok$ in our notation, i.e., we simply write $p^{\msf{Sim}}_{\msf{pref}}$ and $\mcal{G}_1$ to mean $p^{\msf{Sim}}_{\msf{pref}}[\epsilon_\wiaok]$ and $\mcal{G}_1[\epsilon_\wiaok]$. 
Similarly, machines introduced in the following also depend on $\epsilon_\wiaok$, but we do not explicitly write it in our notation. 
We also omit writing $(1^\secpar,1^{\epsilon^{-1}},\A_\secpar)$ from inputs of those machines for notational simplicity. 
}

\begin{figure}[!h]
\centering
\fbox{
     \includegraphics[width=0.55\textwidth,page=4]{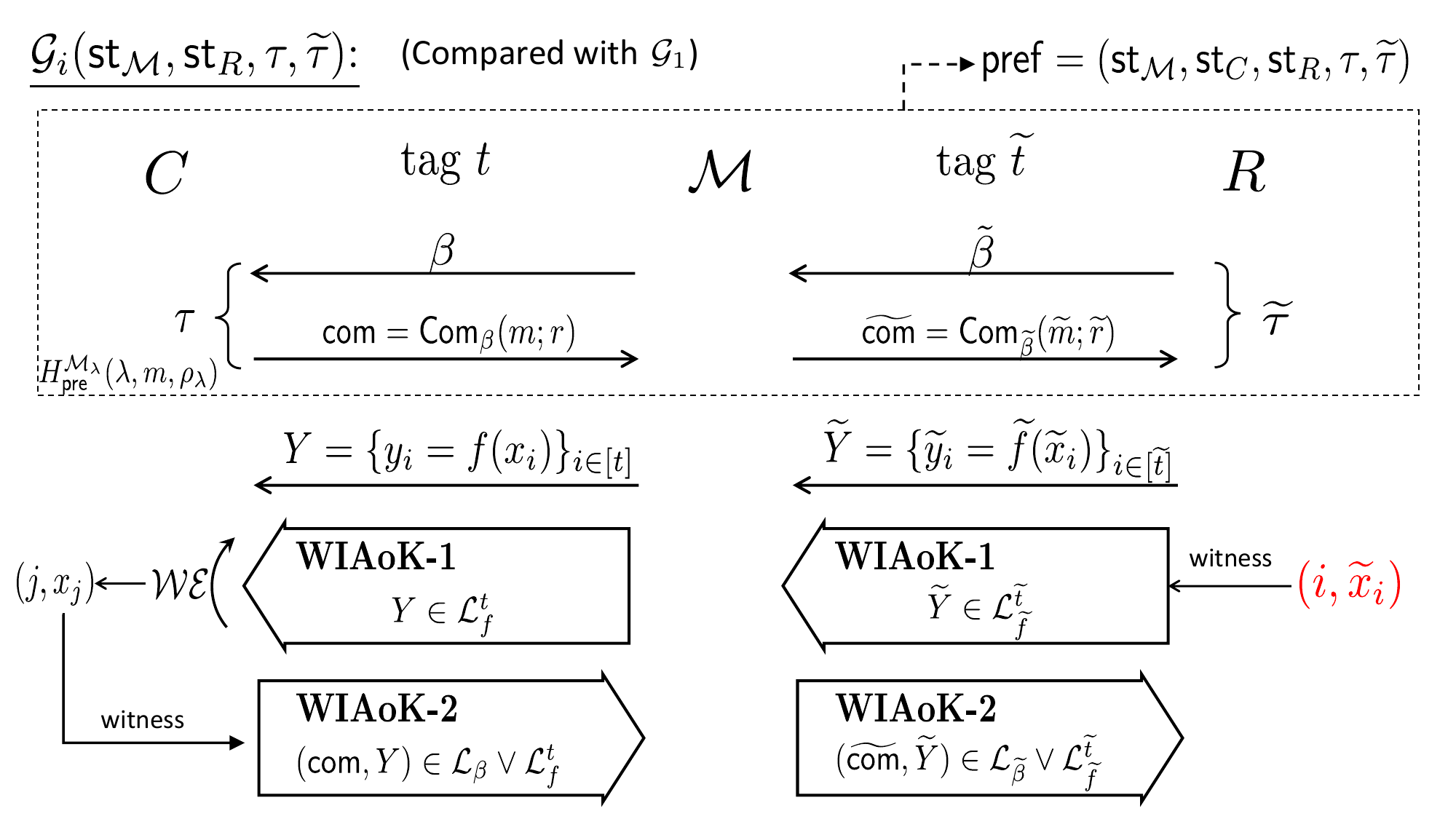}
     }
     \caption{Machine $\mcal{G}_i$ {\scriptsize (Difference with $\mcal{G}_1$ is highlighted in red color)}}\vspace{1em}
     \label{pq:figure:one-sided:Gi}
\end{figure}

\vspace{1em}

\begin{figure}[!h]
     \begin{subfigure}[t]{0.47\textwidth}
         \centering
         \fbox{
         \includegraphics[width=\textwidth,page=5]{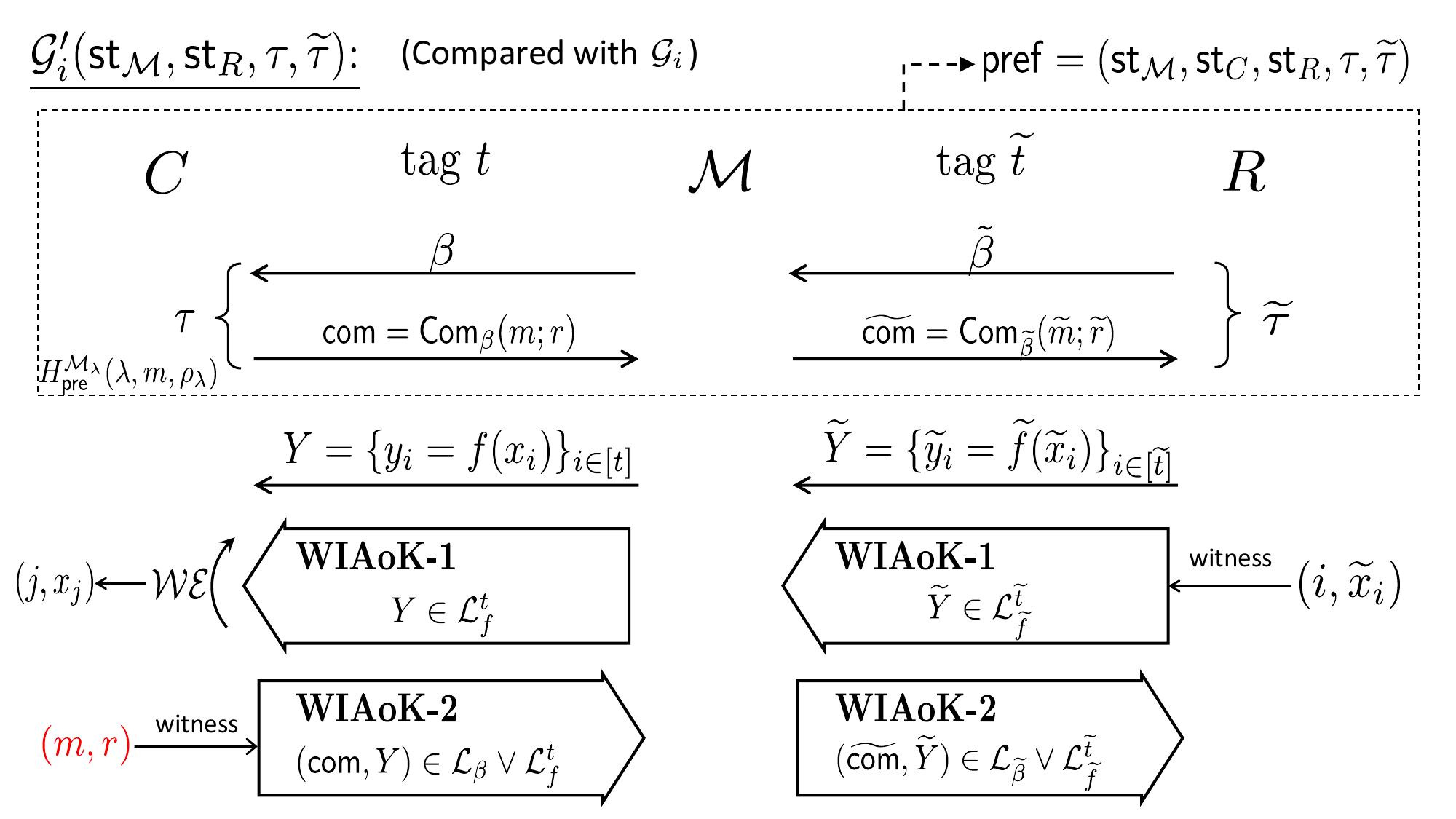}
         }
         \caption{}
         \label{pq:figure:one-sided:G'i}
     \end{subfigure}
     \hspace{6.5pt}
     \begin{subfigure}[t]{0.47\textwidth}
         \centering
         \fbox{
         \includegraphics[width=\textwidth,page=6]{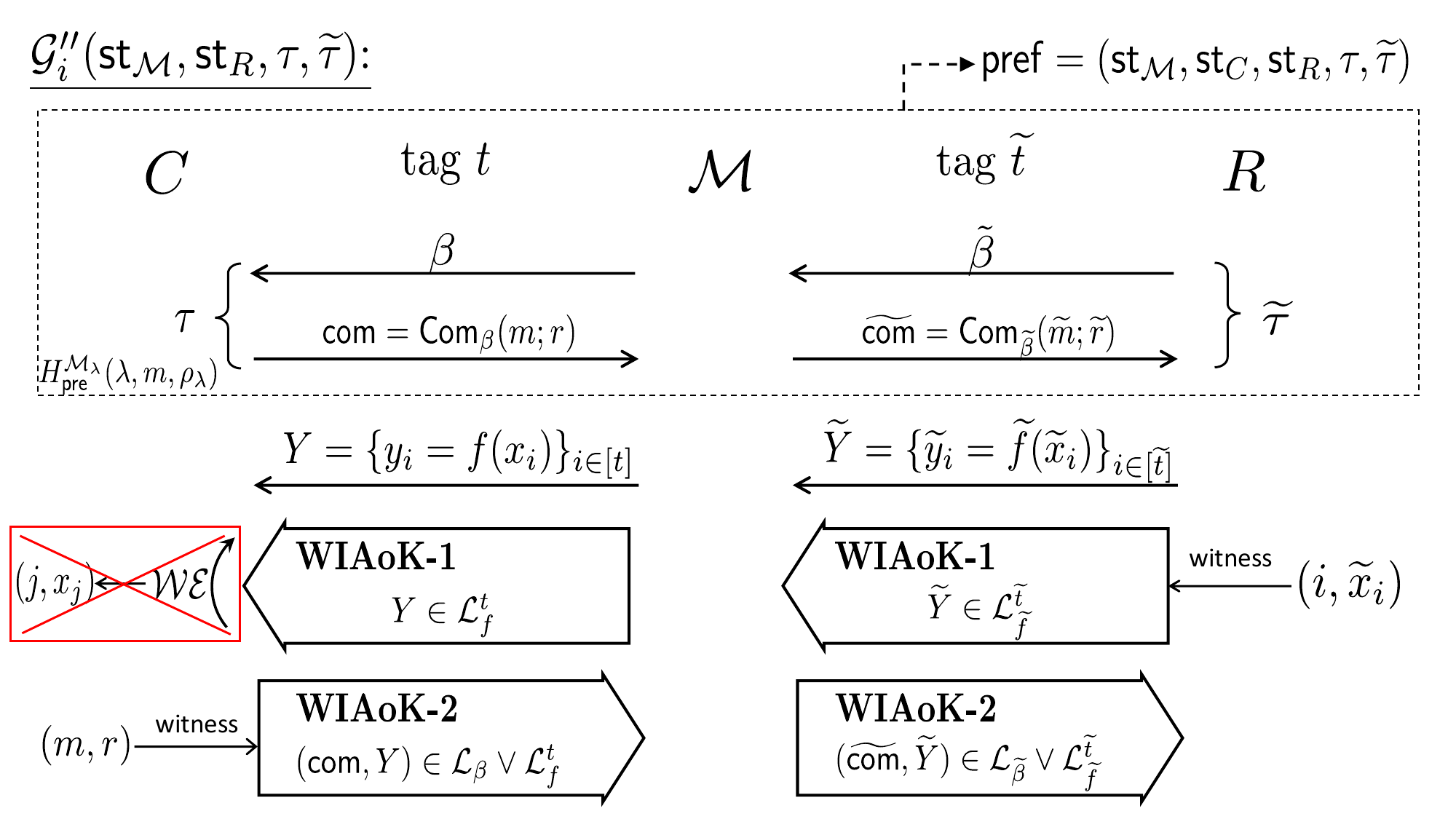}
         }
         \caption{}
         \label{pq:figure:one-sided:G''i}
     \end{subfigure}
     \caption{Machines $\mcal{G}'_i$ and $\mcal{G}''_i$ {\scriptsize (Difference is highlighted in red color)}}
     \label{pq:figure:one-sided:hybrid:G'i:G''i}
\end{figure}

\para{Machine $\mcal{G}_{i}$ ($i \in [\tilde{t}]$):\label{pq:machine:Gi:new}} (Illustrated in \Cref{pq:figure:one-sided:Gi}.)
Recall that we have already defined the machine $\mcal{G}_1(\msf{st}_{\mcal{M}}, \msf{st}_R, \tau,\tilde{\tau})$ on \Cpageref{pq:pageref:one-sided:proof:G1}. Now, for $i \in [\tilde{t}] \setminus \Set{1}$, $\mcal{G}_i(\msf{st}_{\mcal{M}}, \msf{st}_R, \tau,\tilde{\tau})$ behaves identically to $\mcal{G}_1(\msf{st}_{\mcal{M}}, \msf{st}_R, \tau,\tilde{\tau})$ except that it uses $(i, \tilde{x}_i)$ as the witness in the right {\bf WIAoK-1}.

\begin{MyClaim}\label{pq:claim:bound:Gi:new}  
$\forall i \in [\tilde{t}], ~\Pr[b = \top : (\OUT, b) \gets \mcal{G}_i(\msf{st}_{\mcal{M}}, \msf{st}_R, \tau,\tilde{\tau})] \ge p^{\msf{Sim}}_{\msf{pref}}  {\color{Plum} - 2\epsilon_\wiaok(\secpar)} -\negl(\secpar)$.
\end{MyClaim}
\begin{proof}
We define hybrid machines $\mcal{G}'_i$ and $\mcal{G}''_i$ as follows.

\para{Machine $\mcal{G}'_{i}$ ($i \in [\tilde{t}]$):\label{pq:machine:G'i:new}} (Illustrated in \Cref{pq:figure:one-sided:G'i}.)
Recall that we have already defined the machine $\mcal{G}'_1(\msf{st}_{\mcal{M}}, \msf{st}_R, \tau,\tilde{\tau})$ on \Cpageref{pq:pageref:one-sided:proof:G'1}. Now, for $i \in [\tilde{t}] \setminus \Set{1}$, $\mcal{G}_i(\msf{st}_{\mcal{M}}, \msf{st}_R, \tau,\tilde{\tau})$ behaves identically to $\mcal{G}_1(\msf{st}_{\mcal{M}}, \msf{st}_R, \tau,\tilde{\tau})$ except that it uses $(i, \tilde{x}_i)$ as the witness in the right {\bf WIAoK-1}.

\para{Machine $\mcal{G}''_i$ ($i \in [\tilde{t}]$):} (Illustrated in \Cref{pq:figure:one-sided:G''i}\label{pq:pageref:one-sided:proof:G''1}) For any prefix $\msf{pref}$, $\mcal{G}''_i(\msf{st}_{\mcal{M}}, \msf{st}_R, \tau,\tilde{\tau})$ behaves identically to $\mcal{G}'_i(\msf{st}_{\mcal{M}}, \msf{st}_R, \tau,\tilde{\tau})$ except that it honestly runs the left {\bf WIAoK-1} instead of running the witness-extended emulator. In other words, $\mcal{G}''_i(\msf{st}_{\mcal{M}}, \msf{st}_R, \tau,\tilde{\tau})$ behaves identically to $\langle \mcal{A}_\secpar(\msf{st}_{\mcal{M}}), \mcal{B}(\msf{st}_C, \msf{st}_R) \rangle(1^\secpar, \tilde{\tau})$ except that it uses $(i,\tilde{x}_i)$ as the witness in the right {\bf WIAoK-1}. 
In particular, $\mcal{G}''_i(\msf{st}_{\mcal{M}}, \msf{st}_R, \tau,\tilde{\tau})$ is identical to $\langle \mcal{A}_\secpar(\msf{st}_{\mcal{M}}), \mcal{B}(\msf{st}_C, \msf{st}_R) \rangle(1^\secpar, \tilde{\tau})$. 

Then, \Cref{pq:claim:bound:Gi:new} follows from the following sequence of inequalities. 
\begin{itemize}
\item
By the  WI property of the left {\bf WIAoK-2} and the definition of $p^{\msf{Sim}}_{\msf{pref}}$, it holds that:
$$\Pr[b = \top : (\OUT, b) \gets \mcal{G}'_1(\msf{st}_{\mcal{M}}, \msf{st}_R, \tau,\tilde{\tau})] \ge p^{\msf{Sim}}_{\msf{pref}} - \negl(\secpar).$$
\item
By the AoK property of the left {\bf WIAoK-1} and the above inequality, it holds that:
$$\Pr[b = \top : (\OUT, b) \gets \mcal{G}''_1(\msf{st}_{\mcal{M}}, \msf{st}_R, \tau,\tilde{\tau})] \ge p^{\msf{Sim}}_{\msf{pref}} {\color{Plum} - \epsilon_\wiaok(\secpar)} -\negl(\secpar).$$
\item
By the WI property of the right {\bf WIAoK-1} and the above inequality, it holds that:
$$\forall i \in [\tilde{t}],~\Pr[b = \top : (\OUT, b) \gets \mcal{G}''_i(\msf{st}_{\mcal{M}}, \msf{st}_R, \tau,\tilde{\tau})] \ge p^{\msf{Sim}}_{\msf{pref}}{\color{Plum} - \epsilon_\wiaok(\secpar)} - \negl(\secpar).$$
\item
By the AoK property of the left {\bf WIAoK-1} and the above inequality, it holds that:
$$\forall i \in [\tilde{t}],~\Pr[b = \top : (\OUT, b) \gets \mcal{G}'_i(\msf{st}_{\mcal{M}}, \msf{st}_R, \tau,\tilde{\tau})] \ge p^{\msf{Sim}}_{\msf{pref}}{\color{Plum} - 2\epsilon_\wiaok(\secpar)} - \negl(\secpar).$$
\item
By the  WI property of the left {\bf WIAoK-2} and the above inequality, it holds that:
$$\forall i \in [\tilde{t}],~\Pr[b = \top : (\OUT, b) \gets \mcal{G}_i(\msf{st}_{\mcal{M}}, \msf{st}_R, \tau,\tilde{\tau})] \ge p^{\msf{Sim}}_{\msf{pref}} {\color{Plum} - 2\epsilon_\wiaok(\secpar)}- \negl(\secpar).$$
\end{itemize}
\end{proof}

\begin{figure}[!th]
     \begin{subfigure}[t]{0.47\textwidth}
         \centering
         \fbox{
         \includegraphics[width=\textwidth,page=7]{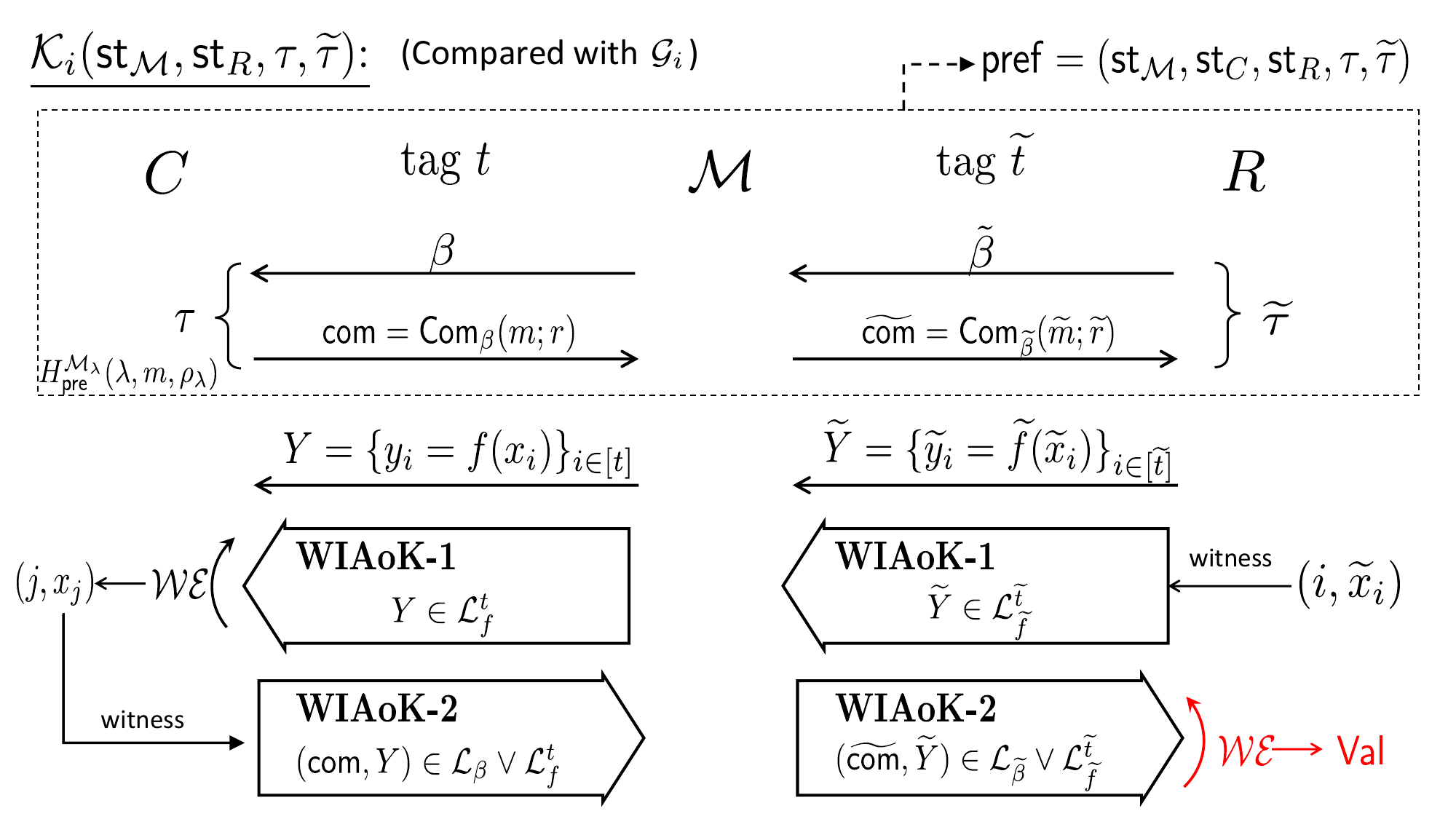}
         }
         \caption{}
         \label{pq:figure:one-sided:Ki}
     \end{subfigure}
     \hspace{6.5pt}
     \begin{subfigure}[t]{0.47\textwidth}
         \centering
         \fbox{
         \includegraphics[width=\textwidth,page=8]{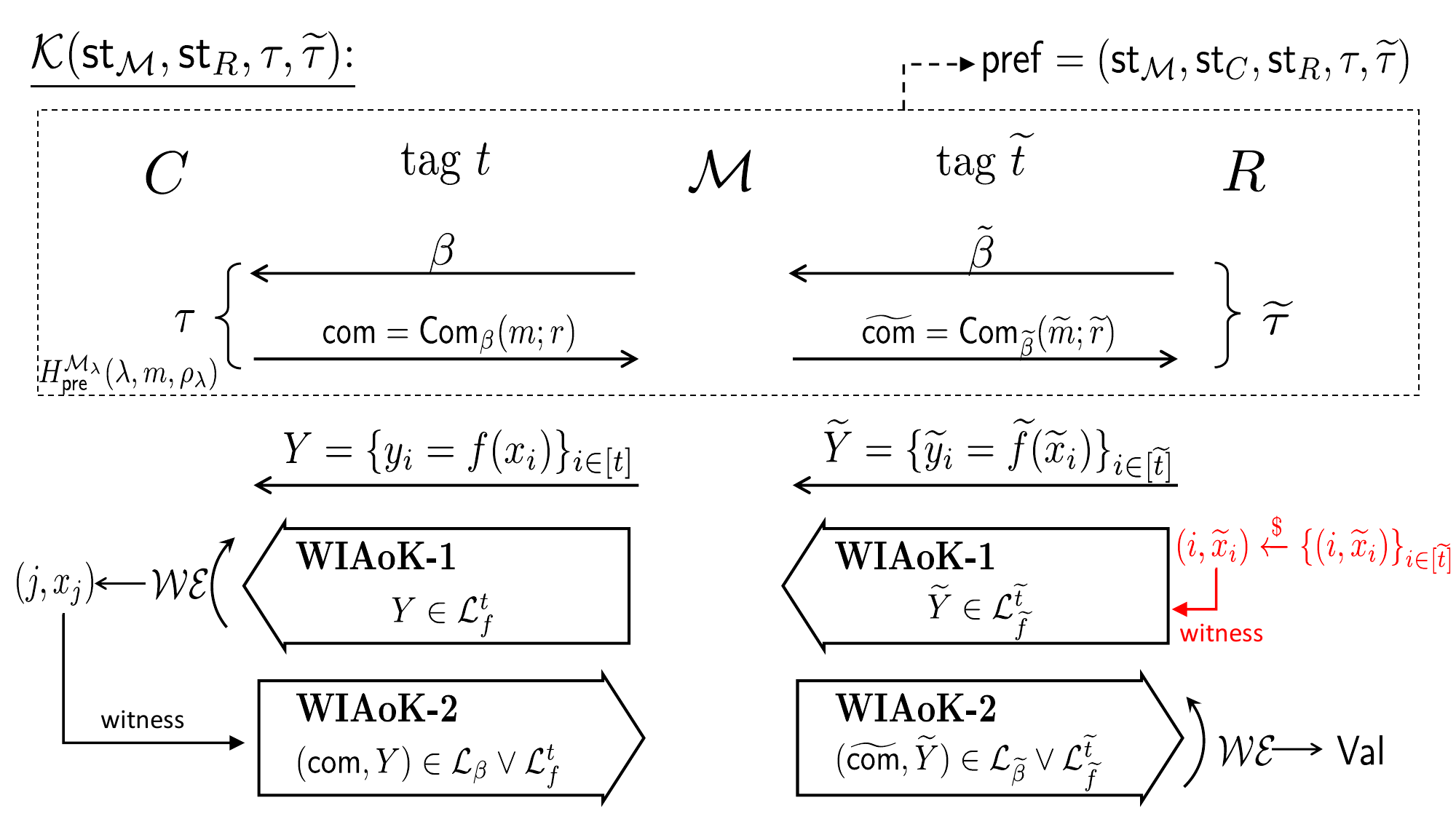}
         }
         \caption{}
         \label{pq:figure:one-sided:K}
     \end{subfigure}
     \caption{Machines $\mcal{K}_i$ and $\mcal{K}$ {\scriptsize (Difference is highlighted in red color)}}
     \label{pq:figure:one-sided:hybrid:Ki:K}
\end{figure}

\para{Machine $\mcal{K}_i$ ($i\in [\tilde{t}]$):} (Illustrated in \Cref{pq:figure:one-sided:Ki}.) For a prefix $\msf{pref}$, $\mcal{K}_i(\msf{st}_{\mcal{M}}, \msf{st}_R, \tau,\tilde{\tau})$ 
behaves identically to the $\mcal{G}_i(\msf{st}_{\mcal{M}}, \msf{st}_R, \tau,\tilde{\tau})$ depicted in \Cref{pq:figure:one-sided:Gi}, except for the following difference. Machine $\mcal{K}_i(\msf{st}_{\mcal{M}}, \msf{st}_R, \tau,\tilde{\tau})$ uses the witness-extended emulator (as per \Cref{item:PQWEE} of \Cref{def:PQWIAoK}) {\color{Plum} with the error parameter $\epsilon_\wiaok(\secpar)$} to finish the right {\bf WIAoK-2}, instead of playing the role of the honest receiver. 

\subpara{$\mcal{K}_i$'s Output:\label{pq:subpara:K-i:output}} Let $w'$ denote the third output of the witness-extended emulator (see \Cref{item:PQWEE} of \Cref{def:PQWIAoK}), which is supposed to be the witness used by $\mcal{M}$ in the right {\bf WIAoK-2} (for the statement $(\tilde{\msf{com}}, \tilde{Y}$) w.r.t.\ the language $\Lang_{\tilde{\beta}}\vee \Lang^{\tilde{t}}_{\tilde{f}}$). Depending on the value of $w'$, we define a value $\msf{Val} \in \bits^{\ell(\secpar)} \cup \Set{\bot_{\tilde{Y}}, \bot_{\msf{invalid}}}$ as follows:
\begin{enumerate}
\item \label[Case]{pq:K-i:output:case:1}
If $w'$ is a valid witness for $(\tilde{\msf{com}}, \tilde{Y}) \in \Lang_{\tilde{\beta}}\vee \Lang^{\tilde{t}}_{\tilde{f}}$. Then, there are tow sub-cases:
\begin{enumerate}
\item \label[Case]{pq:K-i:output:case:1a}
$w'$ is a valid witness for $\tilde{\msf{com}} \in \Lang_{\tilde{\beta}}$. In this case, $w'$ consists of the value $\msf{val}(\tilde{\tau})$, i.e., the value committed in $\tilde{\tau}=(\tilde{\beta}, \tilde{\msf{com}})$, and the randomness $\tilde{r}$. We set $\msf{Val} \coloneqq \msf{val}(\tilde{\tau})$. Importantly, notice that we do {\em not} include the randomness $\tilde{r}$ in $\msf{Val}$. %(as explained in \Cref{pq:rmk:small-tag:proof:se:proof:K:output}.).
\item \label[Case]{pq:K-i:output:case:1b}
$w'$ is a valid witness for $\tilde{Y} \in \Lang^{\tilde{t}}_{\tilde{f}}$: In this case, we set $\msf{Val} \coloneqq \bot_{\tilde{Y}}$.
\end{enumerate}

\item \label[Case]{pq:K-i:output:case:2}
Otherwise, set $\msf{Val} \coloneqq \bot_{\msf{invalid}}$.
\end{enumerate}

The output of $\mcal{K}_i$ is defined to be the above $\msf{Val}$. Notice that this is in contrast to all previous machines, for which the output is defined to be the main-in-the-middle's  output and the honest receiver's decision bit. We emphasize that such a $\msf{Val}$ satisfies the syntax requirement in \Cref{pq:property:small-tag:proof:se:proof:K:syntax} of \Cref{pq:lem:small-tag:proof:se:proof:K}. In particular, {\em $\msf{Val} = \msf{val}(\tilde{\tau})$ whenever $\msf{Val} \ne \bot$}.\footnote{Note that here we defined two types of abortion: $\bot_{\tilde{Y}}$ and $\bot_{\msf{invalid}}$, while \Cref{pq:property:small-tag:proof:se:proof:K:syntax} of \Cref{pq:lem:small-tag:proof:se:proof:K} only allows a single abortion symbol $\bot$. We remark that this is only a cosmetic difference---It can be made consistent using the following rules: $\bot = \bot_{\tilde{Y}}$ {\em and} $\bot = \bot_{\msf{invalid}}$ (i.e., $\msf{Val} \ne \bot  \Leftrightarrow (\msf{Val} \ne \bot_{\tilde{Y}} \wedge \msf{Val} \ne \bot_{\msf{invalid}})$).}

\begin{MyClaim}\label{pq:claim:bound:Ki}
$\forall i \in [\tilde{t}], ~\Pr[\Val \ne \bot_{\msf{invalid}} :\Val \gets \mcal{K}_i(\msf{st}_{\mcal{M}}, \msf{st}_R, \tau,\tilde{\tau})] \ge p^{\msf{Sim}}_{\msf{pref}}{\color{Plum} - 3\epsilon_\wiaok(\secpar)} - \negl(\secpar)$.
\end{MyClaim}
\begin{proof}
This claim follows from \Cref{pq:claim:bound:Gi:new} and the AoK property of the right {\bf WIAoK-2}. 
%witness-extended emulation property (as per \Cref{item:PQWEE} of \Cref{def:PQWIAoK}) of the right {\bf WIAoK-2}.

\end{proof}

Finally, we are ready to define the extractor $\mcal{K}$. Intuitively, $\mcal{K}$ can be conceived as an average-case version of $\Set{\mcal{K}_i}_{i\in[\tilde{t}]}$:

\para{Extractor $\mcal{K}$:\label{pq:extractorK:description}} (Illustrated in \Cref{pq:figure:one-sided:K}.)  For a prefix $\msf{pref}$, $\mcal{K}(\msf{st}_{\mcal{M}}, \msf{st}_R, \tau,\tilde{\tau})$ samples uniformly at random an index $i \pick [\tilde{t}]$, executes $\mcal{K}_i(\msf{st}_{\mcal{M}}, \msf{st}_R, \tau,\tilde{\tau})$, and outputs whatever $\mcal{K}_i(\msf{st}_{\mcal{M}}, \msf{st}_R, \tau,\tilde{\tau})$ outputs.

%\begin{remark}[On the Notation of $\mcal{K}$]\label{pq:rmk:notation-K}It is worth noting that in the above, we write machine $\mcal{K}$ as $\mcal{K}(\msf{st}_{\mcal{M}}, \msf{st}_R, \tau,\tilde{\tau})$, while it was written in \Cref{pq:lem:small-tag:proof:se:proof:K} as $\mcal{K}^{\A(\msf{st}_{\mcal{M}})}(1^\secpar, \msf{st}_R, \tau, \tilde{\tau})$. This is only a cosmetic difference. 
%\end{remark}

Next, we show that the extractor $\mcal{K}$ satisfies the requirements in \Cref{pq:lem:small-tag:proof:se:proof:K}.

\para{Running Time of $\mcal{K}$.} Observe that for each $i \in [\tilde{t}]$, $\mcal{K}_i(\msf{st}_{\mcal{M}}, \msf{st}_R, \tau,\tilde{\tau})$ differs from the real man-in-the-middle execution only in the following places: 
\begin{itemize}
\item
$(i,\tilde{x}_i)$ is used in the right {\bf WIAoK-1};
\item
the witness-extended emulator is used in the right {\bf WIAoK-2} and the left {\bf WIAoK-1}. 
\end{itemize}
Since the witness-extended emulator (as per \Cref{item:PQWEE} of \Cref{def:PQWIAoK}) runs in QPT, so does $\mcal{K}_i$. Thus, $\mcal{K}$ is QPT.

\para{Proving \Cref{pq:property:small-tag:proof:se:proof:K:syntax} of \Cref{pq:lem:small-tag:proof:se:proof:K}.} It is straightforward to see that the $\mcal{K}$ defined above satisfies the syntax requirement in \Cref{pq:lem:small-tag:proof:se:proof:K}. In particular, we have $\msf{Val}_{\mcal{K}} = \msf{val}(\tilde{\tau})$ whenever $\msf{Val}_{\mcal{K}} \ne \bot$, because this is true for each $\mcal{K}_i$ by definition (see the paragraph for ``\underline{$\mcal{K}_i$'s Output}'' on \Cpageref{pq:subpara:K-i:output}).

\para{Proving \Cref{pq:property:small-tag:proof:se:proof:K} of \Cref{pq:lem:small-tag:proof:se:proof:K}.} First, recall that \Cref{pq:property:small-tag:proof:se:proof:K} requires that 
for any $\msf{pref}$ in the support of $H^{\mcal{M}_\secpar}_{\msf{pre}}(\secpar, m, \rho_\secpar)$ , if $p^{\msf{Sim}}_{\msf{pref}} \ge \epsilon(\secpar)$, then it holds that
\begin{equation}\label[Inequality]{pq:eq:bound:averageK}
\Pr[\Val_{\mcal{K}} = \msf{val}(\tilde{\tau}) : \Val_{\mcal{K}} \gets \mcal{K}(\msf{st}_{\mcal{M}}, \msf{st}_R, \tau, \tilde{\tau})] \ge \frac{\epsilon'(\secpar)}{\tilde{t}}.
\end{equation} 
Also recall that $\mcal{K}(\msf{st}_{\mcal{M}}, \msf{st}_R, \tau, \tilde{\tau})$ is defined to execute the machine $\mcal{K}_i(\msf{st}_{\mcal{M}}, \msf{st}_R, \tau, \tilde{\tau})$  with  $i$ uniformly sampled from $[\tilde{t}]$. Therefore, \Cref{pq:eq:bound:averageK} can be reduced to the following \Cref{pq:lem:bound:Ki}. We will prove \Cref{pq:lem:bound:Ki} in \Cref{pq:sec:lem:bound:Ki:proof}, which will eventually finish the current proof of \Cref{pq:lem:small-tag:proof:se:proof:K}.
\begin{lemma}\label{pq:lem:bound:Ki}
Let $\epsilon(\secpar) = \frac{1}{\poly(\secpar)}$ and $\epsilon'(\secpar) = \frac{\epsilon(\secpar)}{t^2}$.
For any $\msf{pref} = (\msf{st}_{\mcal{M}},\msf{st}_C, \msf{st}_R, \tau, \tilde{\tau})$, if $p^{\msf{Sim}}_{\msf{pref}} \ge \epsilon(\secpar)$, then there exists an $i \in [\tilde{t}]$ such that 
$$\Pr[\Val = \msf{val}(\tilde{\tau}) :\Val \gets \mcal{K}_i(\msf{st}_{\mcal{M}}, \msf{st}_R, \tau, \tilde{\tau})] \ge \epsilon'(\secpar).$$
\end{lemma}

\subsection{Proof of \Cref{pq:lem:bound:Ki}}
\label{pq:sec:lem:bound:Ki:proof}

\para{Notation.} We highly recommend reviewing the ``\underline{$\mcal{K}_i$'s Output}'' part on \Cpageref{pq:subpara:K-i:output} (in particular, the meanings of $\msf{Val}$, $\bot_{\tilde{Y}}$, and $\bot_{\msf{invalid}}$) before starting reading this subsection.  Recall that $\mcal{K}_i$'s output $\msf{Val}$ is determined by the $w'$ output by the witness-extended emulator of the right {\bf WIAoK-2}.  %simulation-extractor $\mcal{SE}$.
In this subsection, we will need to refer to this $w'$, although it is not included as a part of $\mcal{K}_i$'s output. Particularly, we will make use of the following notation: whenever we write an expression of the form 
 $$\Pr[\text{Some Event $E_{w'}$ about $w'$}: \msf{Val} \gets \mcal{K}_i(\msf{st}_{\mcal{M}}, \msf{st}_R, \tau, \tilde{\tau})],$$
it should be understood as the probability of $E_{w'}$, where $w'$ refers to the $w'$ generated during the random procedure $\msf{Val} \gets\mcal{K}_i(\msf{st}_{\mcal{M}}, \msf{st}_R, \tau, \tilde{\tau})$, over which the probability is taken. 

Using these notations, we can partition the event $\msf{Val} = \bot_{\tilde{Y}}$ as the following {\em mutually exclusive} events: $w' = (1, \tilde{x}_1)$, or, $\ldots$, or $w' =(\tilde{t}, \tilde{x}_{\tilde{t}})$, where $\tilde{y}_i = \tilde{f}(\tilde{x}_i)$ for each $i \in [\tilde{t}]$. Formally, we express this relation by 
\begin{equation}\label{pq:eq:bot-tilde-Y:partition}
\Pr[\msf{Val} = \bot_{\tilde{Y}}: \msf{Val} \gets \mcal{K}_i(\msf{st}_{\mcal{M}}, \msf{st}_R, \tau, \tilde{\tau})] = \sum_{i = 1}^{\tilde{t}} \Pr[w' = (i,\tilde{x}_i): \msf{Val} \gets \mcal{K}_i(\msf{st}_{\mcal{M}}, \msf{st}_R, \tau, \tilde{\tau})]
\end{equation}

With the above notations, we prove \Cref{pq:lem:bound:Ki} in the following.

\para{Proof of \Cref{pq:lem:bound:Ki}.} We assume for contradiction that for some $\msf{pref}$ satisfying $p^{\msf{Sim}}_{\msf{pref}} \ge \epsilon(\secpar)$, it holds that 
\begin{equation}\label[Inequality]{pq:eq:proof:averageK:contra-assump}
\forall i \in [\tilde{t}], ~\Pr[\Val = \msf{val}(\tilde{\tau}) :\Val \gets \mcal{K}_i(\msf{st}_{\mcal{M}}, \msf{st}_R, \tau, \tilde{\tau})] < \epsilon'(\secpar).
\end{equation}
\begin{MyClaim}\label{pq:cliam:bound:Ki-xi}
Under the assumption in \Cref{pq:eq:proof:averageK:contra-assump}, it holds that 
$$\forall i \in [\tilde{t}], ~\Pr[w' = (i, \tilde{x}_i) :\Val \gets \mcal{K}_i(\msf{st}_{\mcal{M}}, \msf{st}_R, \tau, \tilde{\tau})] \ge p^{\msf{Sim}}_{\msf{pref}}{\color{Plum} - 3\epsilon_\wiaok(\secpar)} - \epsilon'(\secpar) - \negl(\secpar).$$
\end{MyClaim}
\begin{proof}
% First, recall that the event $\Val \ne \bot$ (i.e., $\Val$ is a valid witness) can be partitioned into the following {\em disjoint} events: $\Val = \msf{val}(\tilde{\tau})$, $\Val = \tilde{x}_1$, \ldots, $\Val = \tilde{x}_{\tilde{t}}$. Therefore,
In this proof, all the probabilities are taken over the random procedure $\Val \gets \mcal{K}_i(\msf{st}_{\mcal{M}}, \msf{st}_R, \tau, \tilde{\tau})$.

 First, notice that
\begin{align*}
\forall i \in [\tilde{t}], ~\Pr[\Val \ne \bot_{\msf{invlid}}] 
& = 
\Pr[\Val = \msf{val}(\tilde{\tau})] +  \Pr[\Val = \bot_{\tilde{Y}}] \\
\text{(by \Cref{pq:eq:bot-tilde-Y:partition})} ~& = 
\Pr[\Val = \msf{val}(\tilde{\tau})] + \Pr[w' = (i, \tilde{x}_i)] + \sum_{j \in [\tilde{t}]\setminus \Set{i}} \Pr[w' = (j, \tilde{x}_j)]. \numberthis \label{pq:eq:bound:Ki-xi:1}
\end{align*}
Then, the following holds:
{\begingroup\fontsize{10.5pt}{0pt}\selectfont
\begin{align}
\forall i \in [\tilde{t}], ~\Pr[w' = (i, \tilde{x}_i)] 
& = 
\Pr[\Val \ne \bot_{\msf{invalid}}] - \Pr[\Val = \msf{val}(\tilde{\tau})] -  \bigg(\sum_{j \in [\tilde{t}]\setminus \Set{i}} \Pr[w' = (j, \tilde{x}_j)] \bigg)  \label{pq:eq:bound:Ki-xi:derive:1} \\
& \ge p^{\msf{Sim}}_{\msf{pref}}{\color{Plum} - 3\epsilon_\wiaok(\secpar)} - \negl(\secpar) - \Pr[\Val = \msf{val}(\tilde{\tau})] -  \bigg(\sum_{j \in [\tilde{t}]\setminus \Set{i}} \Pr[w' = (j,\tilde{x}_j)] \bigg)  \label[Inequality]{pq:eq:bound:Ki-xi:derive:2}\\
& \ge p^{\msf{Sim}}_{\msf{pref}}{\color{Plum} - 3\epsilon_\wiaok(\secpar)} - \negl(\secpar) - \epsilon'(\secpar) -  \bigg(\sum_{j \in [\tilde{t}]\setminus \Set{i}} \Pr[w' = (j,\tilde{x}_j)] \bigg), \label[Inequality]{pq:eq:bound:Ki-xi:derive:3}
\end{align}
\endgroup}
where \Cref{pq:eq:bound:Ki-xi:derive:1} follows from \Cref{pq:eq:bound:Ki-xi:1}, \Cref{pq:eq:bound:Ki-xi:derive:2} follows from \Cref{pq:claim:bound:Ki}, and \Cref{pq:eq:bound:Ki-xi:derive:3} follows from \Cref{pq:eq:proof:averageK:contra-assump}.

Now, to prove \Cref{pq:cliam:bound:Ki-xi}, it suffices to show that 
\begin{equation}
\forall i \in [\tilde{t}],~\forall j \in [\tilde{t}]\setminus \Set{i}, ~\Pr[w' = (j,\tilde{x}_j) : \Val \gets \mcal{K}_i(\msf{st}_{\mcal{M}}, \msf{st}_R, \tau, \tilde{\tau})] = \negl(\secpar).
\end{equation}
This can be reduced via standard techniques to the (post-quantum) one-wayness of the OWF $\tilde{f}$ in \Cref{pq:item:one-sided:step:OWFs} of the right execution. In more details, we assume for contradiction that there exist $i^*, j^* \in [\tilde{t}]$ such that $i^* \ne j^*$ and that the probability $\Pr[w' = (j^*, \tilde{x}_{j^*}) : \Val \gets \mcal{K}_{i^*}(\msf{st}_{\mcal{M}}, \msf{st}_R, \tau, \tilde{\tau})]$ is non-negligible, where, by definition, $\tilde{x}_{j^*}$ is the preimage of $\tilde{y}_{j^*}$ under the right OWF $\tilde{f}$. Then, we can build a QPT adversary $\Adv_\textsc{owf}$ breaking one-wayness in the following way: $\Adv_\textsc{owf}$ obtains the challenge $y^*$ from the external one-wayness challenger; it then runs the machine $\mcal{K}_{i^*}(\msf{pref})$ internally, for which $\Adv_\textsc{owf}$ uses $y^*$ in place of $\tilde{y}_{j^*}$ when executing \Cref{pq:item:one-sided:step:OWFs} in the right. Note that the internal execution of $\mcal{K}_{i^*}(\msf{pref})$ is identically to the real execution of $\mcal{K}_{i^*}$, thus the extracted $w' =(j^*, \tilde{x}_{j^*})$ must satisfy $\tilde{f}(\tilde{x}_{j^*}) = \tilde{y}_j^*~(=y^*)$ with non-negligible probability, breaking one-wayness. 

This finishes the proof of \Cref{pq:cliam:bound:Ki-xi}.

\end{proof}

\begin{figure}[!tb]
     \begin{subfigure}[t]{0.47\textwidth}
         \centering
         \fbox{
         \includegraphics[width=\textwidth,page=9]{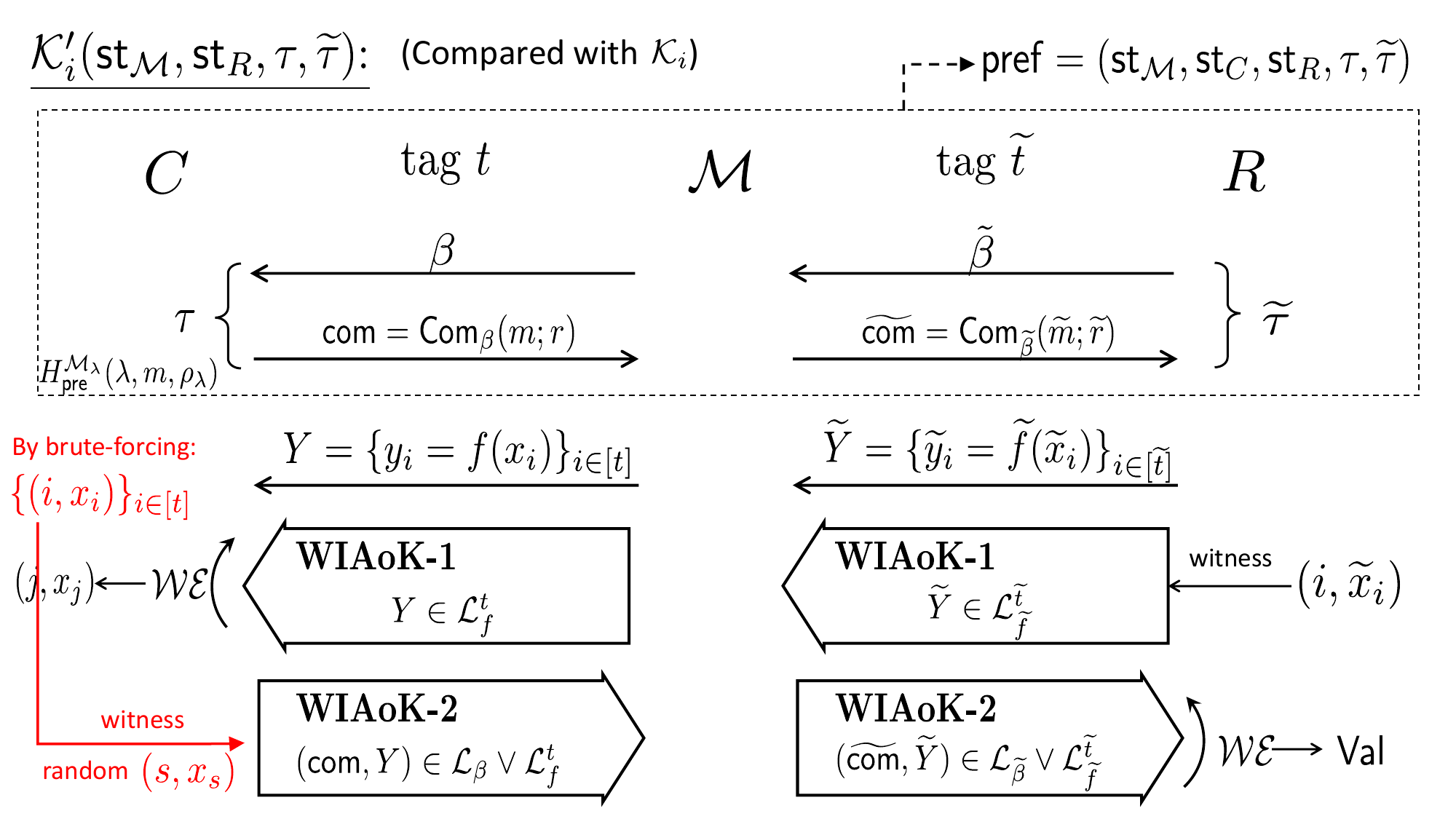}
         }
         \caption{}
         \label{pq:figure:one-sided:K'i}
     \end{subfigure}
     \hspace{6.5pt}
     \begin{subfigure}[t]{0.47\textwidth}
         \centering
         \fbox{
         \includegraphics[width=\textwidth,page=10]{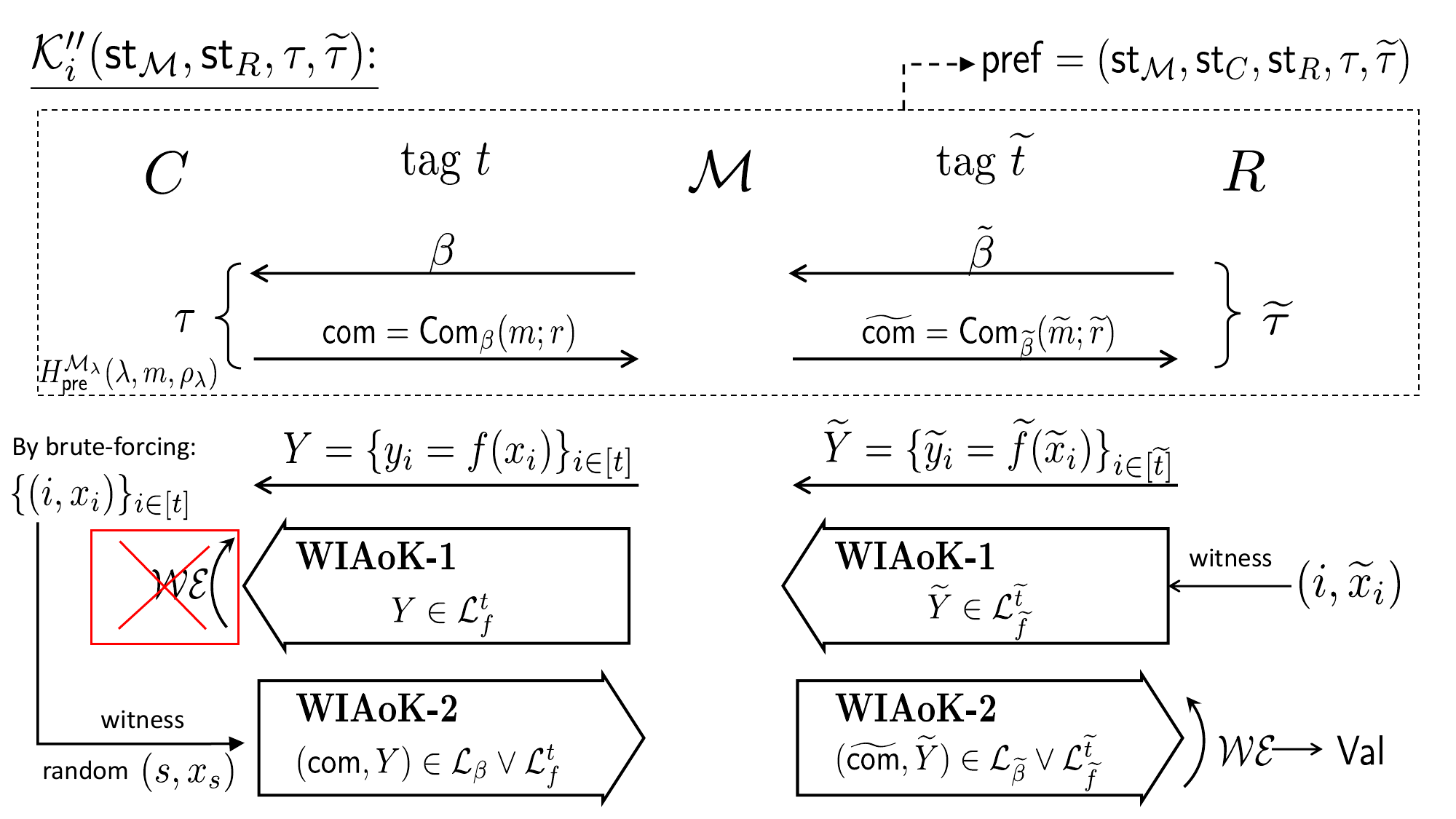}
         }
         \caption{}
         \label{pq:figure:one-sided:K''i}
     \end{subfigure}
     \caption{Machines $\mcal{K}'_i$ and $\mcal{K}''_i$ {\scriptsize (Difference is highlighted in red color)}}
     \label{pq:figure:one-sided:K'i-K''i}
\end{figure}

\para{Machine $\mcal{K}'_i$ ($i \in [\tilde{t}]$):\label{pq:machineK':description}} (Illustrated in \Cref{pq:figure:one-sided:K'i}.) For a prefix $\msf{pref}$, $\mcal{K}'_i(\msf{st}_{\mcal{M}}, \msf{st}_R, \tau, \tilde{\tau})$ proceeds as follows:
\begin{enumerate}
\item
It first finishes \Cref{pq:item:one-sided:step:OWFs} of the man-in-the-middle execution in the same way as $\mcal{K}_i(\msf{st}_{\mcal{M}}, \msf{st}_R, \tau, \tilde{\tau})$. In particular, it will see in the left execution the values $Y = (y_1, \ldots, y_t)$ sent from $\mcal{M}$; 
\item
It then recovers $(x_1, \ldots, x_t)$ by brute-force: Namely, for each $i \in [t]$, it inverts the OWF $f$ to find $x_i$ s.t.\ $f(x_i) = y_i$. It is possible that there exist some ``bad'' $y_i$'s that are not in the range of $f$. For such bad $i$'s, it sets $x_i = \bot$. If all the $x_i$'s are bad, $\mcal{K}'_i(\msf{st}_{\mcal{M}}, \msf{st}_R, \tau, \tilde{\tau})$ halts and outputs $\msf{Fail}$;
\item \label[Step]{pq:item:machineK':3}
If this step is reached, we know that $(x_1, \ldots, x_t)$ cannot be all-$\bot$. $\mcal{K}'_i(\msf{st}_{\mcal{M}}, \msf{st}_R, \tau, \tilde{\tau})$ then picks an $(s, x_s)$ uniformly at random from the good (i.e.\ non-$\bot$) $x_i$'s.
\item
Then, $\mcal{K}'_i(\msf{st}_{\mcal{M}}, \msf{st}_R, \tau, \tilde{\tau})$ continues to finish the execution in the same way as $\mcal{K}_i(\msf{st}_{\mcal{M}}, \msf{st}_R, \tau, \tilde{\tau})$, except that it uses $(s, x_s)$ as the witness when executing the left {\bf WIAoK-2}. 
\end{enumerate}
It is worth noting that the $(j, x_j)$ extracted by the witness-extended emulator for the left {\bf WIAoK-1} (inherited from $\mcal{K}_i(\msf{pref})$) is not used any more by $\mcal{K}'_i(\msf{st}_{\mcal{M}}, \msf{st}_R, \tau, \tilde{\tau})$. 

It is easy to see that if the $(s, x_s)$ picked by $\mcal{K}'_i(\msf{st}_{\mcal{M}}, \msf{st}_R, \tau, \tilde{\tau})$ is equal to the $(j, x_j)$ extracted in $\mcal{K}_i(\msf{st}_{\mcal{M}}, \msf{st}_R, \tau, \tilde{\tau})$ from its left {\bf WIAoK-1}, then $\mcal{K}'_i(\msf{st}_{\mcal{M}}, \msf{st}_R, \tau, \tilde{\tau})$ and $\mcal{K}_i(\msf{st}_{\mcal{M}}, \msf{st}_R, \tau, \tilde{\tau})$ are identical.\footnote{Similar to \Cref{rmk:injectivity-of-OWF}, this is the only place we make use of the injectivity of $f$.} Since $\mcal{K}'_i(\msf{st}_{\mcal{M}}, \msf{st}_R, \tau, \tilde{\tau})$ samples $(s,x_s)$ uniformly from all the good $(i, x_i)$'s, it must hold with probability at least $1/t$ that $(s, x_s) = (j,x_j)$.
Therefore, the following must hold:
{\begingroup\fontsize{10.5pt}{0pt}\selectfont
\begin{equation}\label[Inequality]{pq:eq:bound:K'i}
\forall i \in [\tilde{t}], ~\Pr[w' = (i, \tilde{x}_i) : \Val \gets \mcal{K}'_i(\msf{st}_{\mcal{M}}, \msf{st}_R, \tau, \tilde{\tau})] \ge \frac{1}{t}\cdot \Pr[w' = (i, \tilde{x}_i) : \Val \gets \mcal{K}_i(\msf{st}_{\mcal{M}}, \msf{st}_R, \tau, \tilde{\tau})].
\end{equation}
\endgroup}

%\begin{remark}[On Injectivity of the OWF]\label{pq:rmk:injectivity-of-OWF}
%We emphasize that throughout the proof of non-malleability, this is {\em the only} place where we rely on the injectivity of the OWF. In particular, we rely on the injectivity of the $f$ in the left session to ensure that there is a unique preimage for each $\Set{{y}_i}_{i \in [t]}$. Thus, if both the $x_s$ with $s = j$ (sampled by $\mcal{K}'_i$) and the extracted $x_j$ (in $\mcal{K}_i$) are a valid preimage for the same $y_j$, then the injectvity of $f$ implies that $x_s=x_j$. We will show how to remove injectivity in \Cref{pq:sec:removing-injectivity}.
%\end{remark}
% \xiao{Need to fix. Say that this can be done as in the classical setting}
% \takashi{Simply removed the remark about the injectivity since that is already mentioned at the beginning of this section.}

\para{Machine $\mcal{K}''_i$ ($i\in [\tilde{t}]$):} (Illustrated in \Cref{pq:figure:one-sided:K''i}.) For a prefix $\msf{pref}$,  $\mcal{K}''_i(\msf{st}_{\mcal{M}}, \msf{st}_R, \tau, \tilde{\tau})$ behaves identically to $\mcal{K}'_i(\msf{st}_{\mcal{M}}, \msf{st}_R, \tau, \tilde{\tau})$ except that it plays as the honest committer in the left {\bf WIAoK-1}, instead of running the witness-extended emulator. Recall that starting from $\mcal{K}'_i$, the witness $(j, x_j)$ extracted by the witness-extended emulator from the left {\bf WIAoK-1} is not used any more. Thus, machine $\mcal{K}''_i$ does not need to perform this witness-extended emulation.

By the AoK property %witness-extended emulation property (as per \Cref{item:PQWEE} of \Cref{def:PQWIAoK}) 
of the left {\bf WIAoK-1}, it holds that
\begin{align*}
\forall i \in [\tilde{t}], ~&\big|\Pr[w' = (i,\tilde{x}_i):\Val \gets \mcal{K}''_i(\msf{st}_{\mcal{M}}, \msf{st}_R, \tau, \tilde{\tau})] - \Pr[w' = (i,\tilde{x}_i):\Val \gets \mcal{K}'_i(\msf{st}_{\mcal{M}}, \msf{st}_R, \tau, \tilde{\tau})]\big| \\
&\le {\color{Plum}  \epsilon_\wiaok(\secpar)}+\negl(\secpar). \numberthis \label[Inequality]{pq:eq:relation:K'-K''}
\end{align*}
Next, by the (non-uniform) WI property of the right {\bf WIAoK-1}, it holds that
\begin{align*}
\forall i \in [\tilde{t}], ~&\big|\Pr[w' = (i,\tilde{x}_i):\Val \gets \mcal{K}''_i(\msf{st}_{\mcal{M}}, \msf{st}_R, \tau, \tilde{\tau})] - \Pr[w' = (i,\tilde{x}_i):\Val \gets \mcal{K}''_1(\msf{st}_{\mcal{M}}, \msf{st}_R, \tau, \tilde{\tau})]\big|  \\
 &\le \negl(\secpar). \numberthis \label[Inequality]{pq:eq:relation:K:WI}
\end{align*}
\begin{remark}[Power of Non-Uniform Reductions]\label{pq:rmk:non-uniform_reduction}
Note that we can rely on the AoK and WI properties even though $\mcal{K}'_i$ and $\mcal{K}''_i$ perform brute-force to recover $(x_1,...,x_t)$. This is because the brute-forcing step is done before {\bf WIAoK-1} or {\bf WIAoK-2} starts; Thus, $(x_1,...,x_t)$ can be treated as a non-uniform advice in the reductions. This non-uniform type of argument will be used again in this section later.
\end{remark}

Then, we have the following claim:
\begin{MyClaim}\label{pq:claim:bound:K''1}
It holds that 
% {\begingroup\fontsize{10pt}{0pt}\selectfont
$$\forall i \in [\tilde{t}], ~\Pr[w' = (i,\tilde{x}_i):\Val \gets \mcal{K}''_1(\msf{st}_{\mcal{M}}, \msf{st}_R, \tau, \tilde{\tau})] \ge \frac{1}{t} \cdot \big(p^{\msf{Sim}}_{\msf{pref}} {\color{Plum}-(t+3)\epsilon_\wiaok(\secpar)}-  \epsilon'(\secpar)\big) - \negl(\secpar).$$
% \endgroup}
\end{MyClaim}
\begin{proof} 
This claim follows from \Cref{pq:cliam:bound:Ki-xi} and \Cref{pq:eq:bound:K'i,eq:relation:K'-K'',eq:relation:K:WI}. Formally, (in the following, we omit the input $(\msf{st}_{\mcal{M}}, \msf{st}_R, \tau, \tilde{\tau})$ to $\mcal{K}_i$, $\mcal{K}'_i$, and $\mcal{K}''_i$)
\begin{align}
\forall i \in [\tilde{t}], ~\Pr[w' = (i,\tilde{x}_i):\Val \gets \mcal{K}''_1] 
 & \ge \Pr[w' = (i,\tilde{x}_i):\Val \gets \mcal{K}''_i]  - \negl(\secpar)
\label[Inequality]{pq:claim:bound:K''1:proof:1}\\
 & \ge 
\Pr[w' = (i,\tilde{x}_i):\Val \gets \mcal{K}'_i] {\color{Plum}-\epsilon_\wiaok(\secpar)} - \negl(\secpar) 
\label[Inequality]{pq:claim:bound:K''1:proof:2}\\
 & \ge 
\frac{1}{t}\cdot\Pr[w' = (i,\tilde{x}_i):\Val \gets \mcal{K}_i] {\color{Plum}-\epsilon_\wiaok(\secpar)}  - \negl(\secpar) 
\label[Inequality]{pq:claim:bound:K''1:proof:3}\\
& \ge 
\frac{1}{t} \cdot \big(p^{\msf{Sim}}_{\msf{pref}}{\color{Plum}-3\epsilon_\wiaok(\secpar)}  - \epsilon'(\secpar)\big){\color{Plum}-\epsilon_\wiaok(\secpar)}  - \negl(\secpar), \label[Inequality]{pq:claim:bound:K''1:proof:4}
\end{align}
where \Cref{pq:claim:bound:K''1:proof:1} follows from \Cref{pq:eq:relation:K:WI}, \Cref{pq:claim:bound:K''1:proof:2} follows from \Cref{pq:eq:relation:K'-K''}, \Cref{pq:claim:bound:K''1:proof:3} follows from \Cref{pq:eq:bound:K'i}, and \Cref{pq:claim:bound:K''1:proof:4} follows from \Cref{pq:cliam:bound:Ki-xi}.

\end{proof}
Now, we show the last claim which, together with \Cref{pq:claim:bound:K''1}, will lead to the desired contradiction.
\begin{MyClaim}\label{pq:claim:K'':non-abort}
It holds that
$$\Pr[\Val \ne \bot_{\msf{invalid}}:\Val \gets \mcal{K}''_1(\msf{st}_{\mcal{M}}, \msf{st}_R, \tau, \tilde{\tau})] \le p^{\msf{Sim}}_{\msf{pref}}  {\color{Plum} +2\epsilon_\wiaok(\secpar)} + \negl(\secpar).$$
\end{MyClaim}
\para{Deriving the Contradiction.} Before proving \Cref{pq:claim:K'':non-abort}, we first show why \Cref{pq:claim:bound:K''1,claim:K'':non-abort} are contradictory. In the following, all the probabilities are taken over $\Val \gets \mcal{K}''_1(\msf{st}_{\mcal{M}}, \msf{st}_R, \tau, \tilde{\tau})$:
\begin{align*}
\Pr[\Val \ne \bot_{\msf{invalid}}] 
& = \Pr[\Val = \msf{val}(\tilde{\tau})]  + \Pr[\Val = \bot_{\tilde{Y}}] \\
& = \Pr[\Val = \msf{val}(\tilde{\tau})] + \sum_{i =1 }^{\tilde{t}} \Pr[w' = (i,\tilde{x}_i)] \numberthis \label{pq:eq:bound:Ki:final-contradiction:0}\\
& \ge  \sum_{i =1 }^{\tilde{t}} \Pr[w' = (i,\tilde{x}_i)] \\
&\ge \tilde{t} \cdot \frac{1}{t} \cdot \big(p_{\msf{pref}}{\color{Plum}-(t+3)\epsilon_\wiaok(\secpar)} - \epsilon'(\secpar)\big) - \negl(\secpar) \numberthis \label[Inequality]{pq:eq:bound:Ki:final-contradiction:1}\\
& \ge (1+\frac{1}{t}) \cdot (p_{\msf{pref}}{\color{Plum}-(t+3)\epsilon_\wiaok(\secpar)} - \epsilon'(\secpar)) - \negl(\secpar) \numberthis \label[Inequality]{pq:eq:bound:Ki:final-contradiction:2}\\
& = (1+\frac{1}{t}) \cdot \bigg(p_{\msf{pref}}{\color{Plum}-\frac{t^2+6t+3}{t+1}\cdot\epsilon_\wiaok(\secpar)} - \epsilon'(\secpar)\bigg){\color{Plum}+ 2\epsilon_\wiaok(\secpar)} - \negl(\secpar) \\
& = (1+\frac{1}{t}) \cdot \bigg(p_{\msf{pref}}{\color{Plum}- 2\epsilon'(\secpar)}\bigg){\color{Plum}+ 2\epsilon_\wiaok(\secpar)} - \negl(\secpar) \numberthis \label{pq:eq:bound:Ki:final-contradiction:2.5}\\
& = p_{\msf{pref}}^{\msf{Sim}}{\color{Plum}+ 2\epsilon_\wiaok(\secpar)} + \bigg(\frac{p^{\msf{Sim}}_{\msf{pref}}}{t} {\color{Plum}- 2\epsilon'(\secpar) - \frac{2\epsilon'(\secpar)}{t}}\bigg) - \negl(\secpar)\\
& \ge p^{\msf{Sim}}_{\msf{pref}}{\color{Plum}+ 2\epsilon_\wiaok(\secpar) + \frac{5t^2 - t -1}{5t^3}\cdot \epsilon(\secpar)} - \negl(\secpar)) \numberthis \label[Inequality]{pq:eq:bound:Ki:final-contradiction:3}\\
\end{align*}
where \Cref{pq:eq:bound:Ki:final-contradiction:0} follows from \Cref{pq:eq:bot-tilde-Y:partition}, \Cref{pq:eq:bound:Ki:final-contradiction:1} follows from \Cref{pq:claim:bound:K''1}, \Cref{pq:eq:bound:Ki:final-contradiction:2} follows from the assumption that $\tilde{t} \ge t+1$, 
{\color{Plum} \Cref{pq:eq:bound:Ki:final-contradiction:2.5} follows from our parameter setting $\epsilon_\wiaok(\secpar)=\frac{t+1}{t^2+6t+3}\cdot \epsilon'(\secpar)$,}  
and \Cref{pq:eq:bound:Ki:final-contradiction:3} follows from the assumption that $p^{\msf{Sim}}_{\msf{pref}} \ge \epsilon(\secpar)$ and our parameter setting $\epsilon'(\secpar) = \frac{\epsilon(\secpar)}{10t^2}$.

Recall that $t$ is the tag taking values from $[n]$ with $n$ being a polynomial of $\secpar$. Also recall that $\epsilon(\secpar)$ is an inverse polynomial on $\secpar$. Therefore, \Cref{pq:eq:bound:Ki:final-contradiction:3} can be written as:
$$\Pr[\Val \ne \bot_{\msf{invalid}}:\Val \gets \mcal{K}''_1({\msf{pref}})] \ge p^{\msf{Sim}}_{\msf{pref}} + \frac{1}{\poly(\secpar)} - \negl(\secpar),$$
which contradicts \Cref{pq:claim:K'':non-abort}.

This eventually finishes the proof of \Cref{pq:lem:bound:Ki} (modulo the proof of \Cref{pq:claim:K'':non-abort}, which we show in \Cref{pq:sec:proof:claim:K'':non-abort}).

\subsection{Proof of \Cref{pq:claim:K'':non-abort}}
\label{pq:sec:proof:claim:K'':non-abort}
We first define two extra machines $\mcal{K}^{**}_1$ and $\mcal{K}^*_1$.

\begin{figure}[!tb]
     \begin{subfigure}[t]{0.47\textwidth}
         \centering
         \fbox{
         \includegraphics[width=\textwidth,page=11]{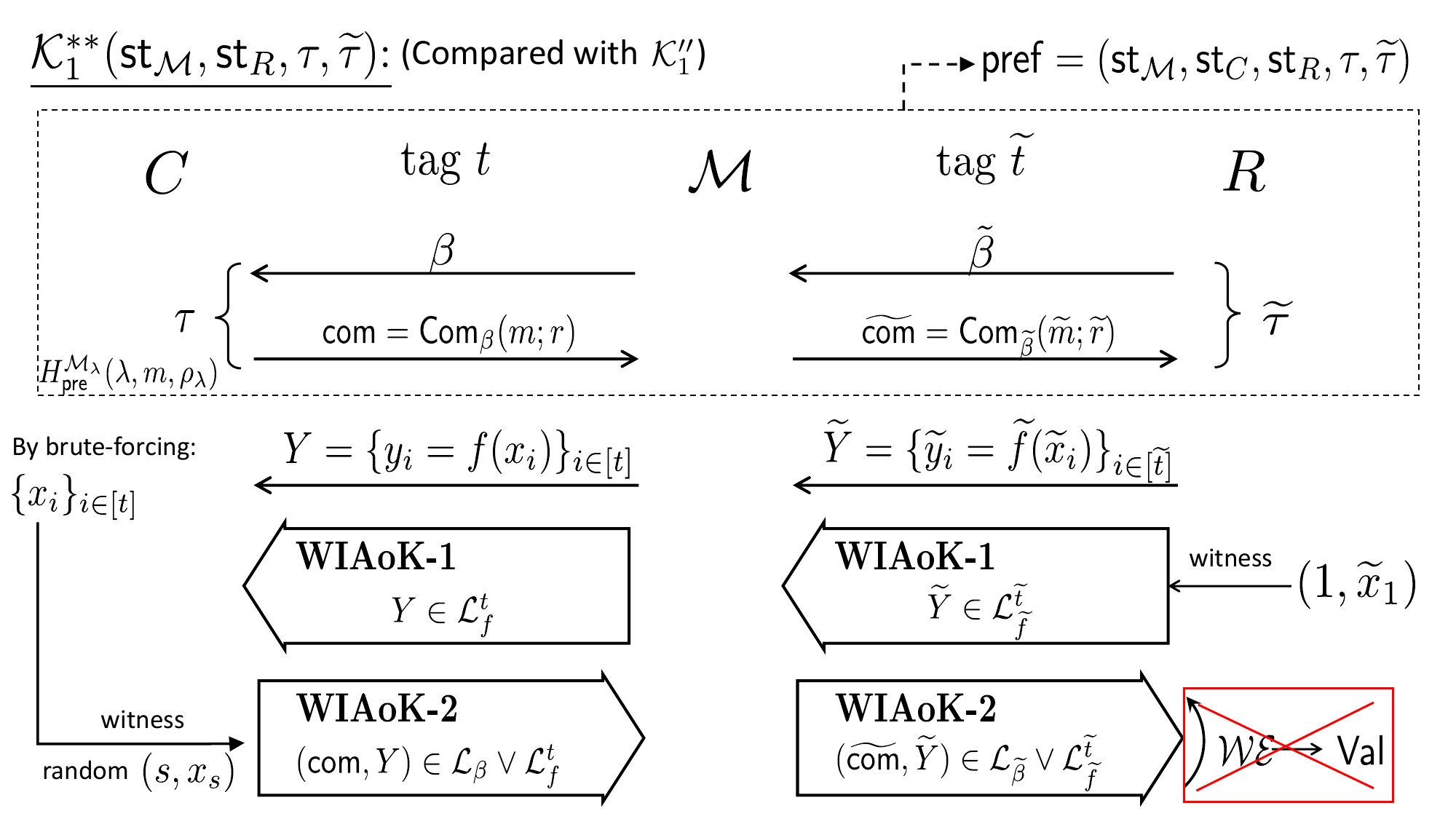}
         }
         \caption{}
         \label{pq:figure:one-sided:K**i}
     \end{subfigure}
     \hspace{6.5pt}
     \begin{subfigure}[t]{0.47\textwidth}
         \centering
         \fbox{
         \includegraphics[width=\textwidth,page=12]{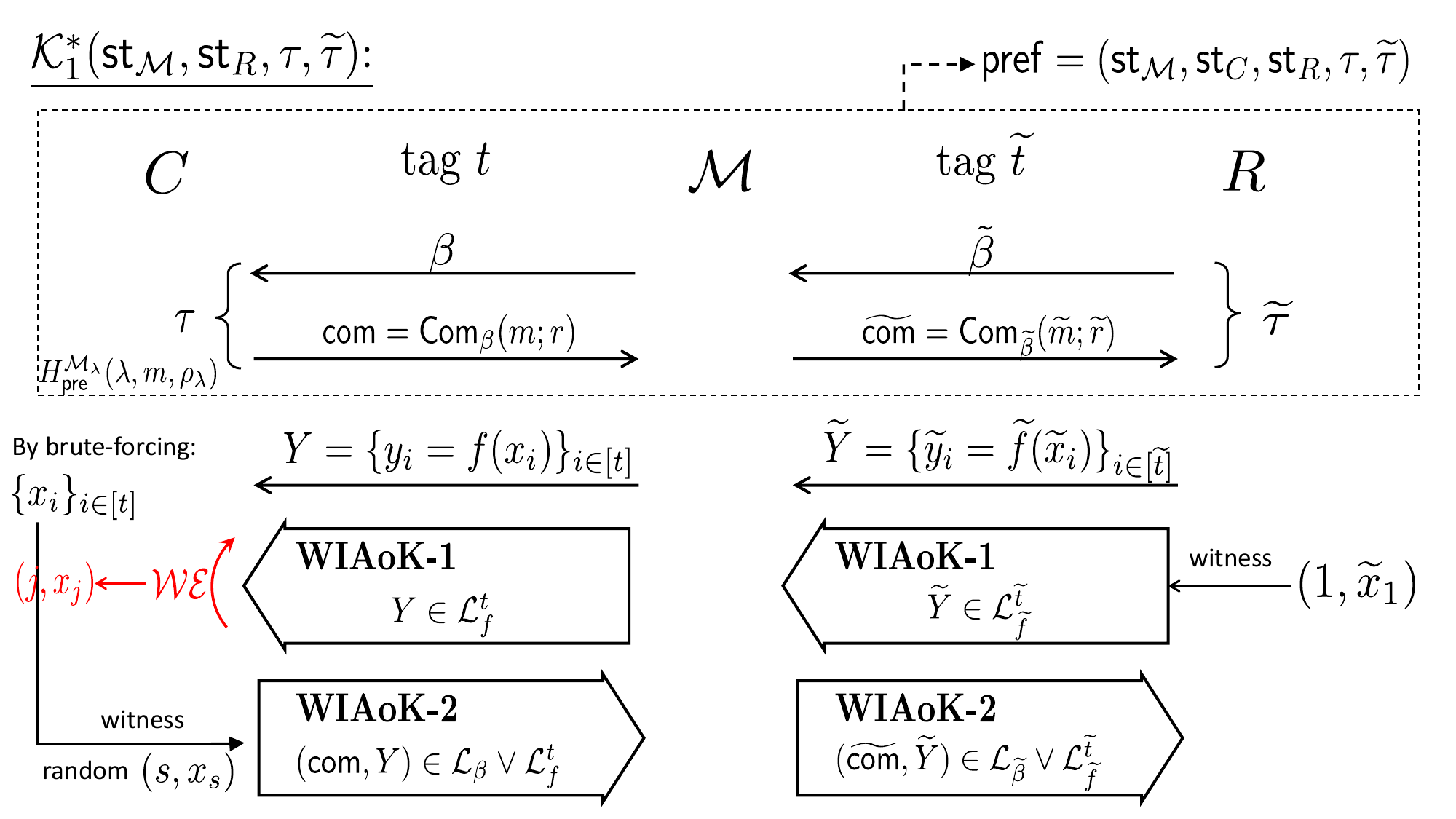}
         }
         \caption{}
         \label{pq:figure:one-sided:K*i}
     \end{subfigure}
     \caption{Machines $\mcal{K}^{**}_1$ and $\mcal{K}^{*}_1$ {\scriptsize (Difference is highlighted in red color)}}
     \label{pq:figure:one-sided:K**i-K*i}
\end{figure}
\para{Machine $\mcal{K}^{**}_1$:} (Illustrated in \Cref{pq:figure:one-sided:K**i}.) For a prefix $\msf{pref}$, $\mcal{K}^{**}_1(\msf{st}_{\mcal{M}}, \msf{st}_R, \tau, \tilde{\tau})$ behaves identically as $\mcal{K}''_1(\msf{st}_{\mcal{M}}, \msf{st}_R, \tau, \tilde{\tau})$ except that $\mcal{K}^{**}_1(\msf{st}_{\mcal{M}}, \msf{st}_R, \tau, \tilde{\tau})$ finishes the right {\bf WIAoK-2} using the honest receiver's algorithm, instead of using the witness-extended emulator. 

\subpara{$\mcal{K}^{**}_1$'s Output.} We define the output of $\mcal{K}^{**}_1(\msf{st}_{\mcal{M}}, \msf{st}_R, \tau, \tilde{\tau})$ to be the honest receiver's decision $b \in \Set{\top, \bot}$. This is in contrast to previous hybrids $\mcal{K}'_i$, $\mcal{K}''_i$, and $\mcal{K}_i$, whose output is defined to be the value $\msf{Val}$ that depends on the value $w'$ extracted by the $\mcal{SE}$ for the right {\bf WIAoK-2}.

 By the AoK property 
 %witness-extended emulation property (as per \Cref{item:PQWEE} of \Cref{def:PQWIAoK}) 
 of the right {\bf WIAoK-2}, it holds that
\begin{equation}\label[Inequality]{pq:eq:K''1-K**1}
\big|\Pr[\Val \ne \bot_{\msf{invalid}} :\Val\gets \mcal{K}''_1(\msf{st}_{\mcal{M}}, \msf{st}_R, \tau, \tilde{\tau})] - \Pr[b = \top : b\gets \mcal{K}^{**}_1(\msf{st}_{\mcal{M}}, \msf{st}_R, \tau, \tilde{\tau})] \big| \le {\color{Plum} \epsilon_\wiaok(\secpar)}+\negl(\secpar).
\end{equation}

\para{Machine $\mcal{K}^{*}_1$:} (Illustrated in \Cref{pq:figure:one-sided:K*i}.) For a prefix $\msf{pref}$, $\mcal{K}^{*}_1(\msf{st}_{\mcal{M}}, \msf{st}_R, \tau, \tilde{\tau})$ behaves identically as $\mcal{K}^{**}_1(\msf{st}_{\mcal{M}}, \msf{st}_R, \tau, \tilde{\tau})$ except the following difference: $\mcal{K}^{*}_1(\msf{st}_{\mcal{M}}, \msf{st}_R, \tau, \tilde{\tau})$ uses the witness-extended emulator {\color{Plum} with the error parameter $\epsilon_\wiaok(\secpar)$} to extract a witness $(j,x_j)$ from the left {\bf WIAoK-1}, and if $x_j$ is not a valid preimage for $y_j$,  $\mcal{K}^{*}_1$ aborts.

 By the AoK property 
 %witness-extended emulation property (as per \Cref{item:PQWEE} of \Cref{def:PQWIAoK}) 
 of the left {\bf WIAoK-1}, it holds that
\begin{equation}\label[Inequality]{pq:eq:K**1-K*1}
\big|\Pr[b = \top : b \gets \mcal{K}^{**}_1(\msf{st}_{\mcal{M}}, \msf{st}_R, \tau, \tilde{\tau})] - \Pr[b = \top : b\gets \mcal{K}^{*}_1(\msf{st}_{\mcal{M}}, \msf{st}_R, \tau, \tilde{\tau})] \big| \le {\color{Plum} \epsilon_\wiaok(\secpar)}+\negl(\secpar).
\end{equation}

\para{Compare $\mcal{K}^{*}_1$ with $\mcal{G}_1$.} Now, let us compare $\mcal{K}^{*}_1$ with the $\mcal{G}_1$ depicted in \Cref{pq:figure:one-sided:Gi}. They only differ in the witness used in the left {\bf WIAoK-2} (and that $\mcal{G}_1$ does not need to perform brute-forcing for $Y$, as it does not use those preimages). Therefore, by the (non-uniform) WI property of the left {\bf WIAoK-2}, it holds that
$$
\big|\Pr[b = \top : b \gets \mcal{K}^{*}_1(\msf{st}_{\mcal{M}}, \msf{st}_R, \tau, \tilde{\tau})] - \Pr[b = \top : (\OUT, b) \gets \mcal{G}_1(\msf{st}_{\mcal{M}}, \msf{st}_R, \tau, \tilde{\tau})] \big| \le \negl(\secpar).
$$
Also, recall (from \Cref{pq:def:p-sim-pref}) that $\Pr[b = \top : (\OUT, b) \gets \mcal{G}_1(\msf{st}_{\mcal{M}}, \msf{st}_R, \tau, \tilde{\tau})]$ is exactly the definition of $p^{\msf{Sim}}_{\msf{pref}}$. Thus, the above implies:
\begin{equation}\label[Inequality]{pq:eq:K*1-G1}
\big|\Pr[b = \top : b \gets \mcal{K}^{*}_1(\msf{st}_{\mcal{M}}, \msf{st}_R, \tau, \tilde{\tau})] - p^{\msf{Sim}}_{\msf{pref}} \big| \le \negl(\secpar).
\end{equation}

Therefore, the following holds:
\begin{align}
\Pr[\Val \ne \bot_{\msf{invalid}} :\Val\gets \mcal{K}''_1(\msf{st}_{\mcal{M}}, \msf{st}_R, \tau, \tilde{\tau})] & \le \Pr[b = \top : b\gets \mcal{K}^{**}_1(\msf{st}_{\mcal{M}}, \msf{st}_R, \tau, \tilde{\tau})]{\color{Plum} +\epsilon_\wiaok(\secpar)} +  \negl(\secpar) \label[Inequality]{pq:eq:K'':non-abort:final:1} \\
& \le \Pr[b = \top : b \gets \mcal{K}^{*}_1({\msf{pref}})]{\color{Plum} +2\epsilon_\wiaok(\secpar)} + \negl(\secpar) \label[Inequality]{pq:eq:K'':non-abort:final:2} \\
& \le p^{\msf{Sim}}_{\msf{pref}} {\color{Plum} +2\epsilon_\wiaok(\secpar)} +\negl(\secpar), \label[Inequality]{pq:eq:K'':non-abort:final:3}
\end{align}
where \Cref{pq:eq:K'':non-abort:final:1} follows from \Cref{pq:eq:K''1-K**1}, \Cref{pq:eq:K'':non-abort:final:2} follows from \Cref{pq:eq:K**1-K*1}, and \Cref{pq:eq:K'':non-abort:final:3} follows from \Cref{pq:eq:K*1-G1}. 

This finishes the proof of \Cref{pq:claim:K'':non-abort}.

\end{document}